\documentclass[a4paper,11pt]{article}
\pdfoutput=1 

\usepackage{jheppub}
\usepackage[T1]{fontenc}
\usepackage{amsmath}

\usepackage[dvipsnames]{xcolor}
\usepackage{tikz}
\usepackage{framed}
\usetikzlibrary{patterns}
\usetikzlibrary{snakes}
\usepackage{subcaption}
\usepackage{amsfonts}
\usepackage{tikz-cd,tikz}
\usepackage{url}
\usepackage{amsmath}

\newcommand{\osp}{\mathfrak{osp}}
\newcommand{\gl}{\mathfrak{gl}}
\newcommand{\so}{\mathfrak{so}}
\newcommand{\spp}{\mathfrak{sp}}

\newcommand{\dalpha}{\mathfrak{d}(2,1;\alpha)}
\newcommand{\ospf}{\mathfrak{osp}(4|2)}
\newcommand{\bP}{\mathbf{P}}
\newcommand{\bQ}{\mathbf{Q}}
\newcommand{\bS}{\mathbf{S}}

\newcommand{\slP}{\dot{\mathbf{P}}}
\newcommand{\slS}{\dot{\mathbf{S}}}
\newcommand{\slQ}{\dot{\mathbf{Q}}}

\newcommand{\eps}{\epsilon}
\newcommand{\del}{\delta}

\newcommand{\ads}{$\text{AdS}_3\times \text{S}^3\times \text{S}^3\times \text{S}^1$}

\newcommand{\beq}{\begin{equation}}
\newcommand{\eeq}{\end{equation}}
\newcommand{\beqa}{\begin{eqnarray}}
\newcommand{\eeqa}{\end{eqnarray}}

\title{On the Quantum Spectral Curve for  $\text{AdS}_3\times \text{S}^3\times \text{S}^3\times \text{S}^1$ strings and the $\mathfrak{d}(2,1;\alpha)$ Q-system}

\author[a]{Andrea Cavagli\`a,}
\author[b]{Rouven Frassek,}
\author[a]{Nicol\`o Primi,}
\author[a]{Roberto Tateo}
\affiliation[a]{Dipartimento di Fisica, Universit\`a di Torino and INFN, Sezione di Torino, Via P. Giuria 1, 10125, Torino, Italy}
\affiliation[b]{Dipartimento di Scienze Fisiche, Informatiche e Matematiche, Università di Modena e Reggio Emilia,
Via G.~Campi 213/B, 41125 Modena and INFN, Sezione di Bologna, Via Irnerio 46, 40126 Bologna, Italy}
\emailAdd{andrea.cavaglia@unito.it}
\emailAdd{rouven.frassek@unimore.it}
\emailAdd{nicolo.primi@unito.it}
\emailAdd{roberto.tateo@unito.it}

\abstract{In this paper, we put forward and discuss a proposal for a Quantum Spectral Curve (QSC) describing the planar spectrum of the holographic CFT dual to strings on AdS$_3\times$ S$^3\times$ S$^3\times$ S$^1$, a theory with global symmetry $\mathfrak{d}(2,1;\alpha)^{\oplus 2}$. We focus mainly on the case when the radii of the two spheres are the same, i.e. $\alpha = 1/2$, where the symmetry reduces to $\mathfrak{osp}(4|2)^{\oplus 2}$. In this case, our proposal is based on two copies of an $\mathfrak{osp}(4|2)$ Q-system, glued through the branch cuts of the Q-functions in a minimal way. We study in detail the ensuing analytic properties of the Q-functions in this proposal. 
 Focusing on purely massive excitations, we consider the large worldsheet limit in which the  QSC leads to a set of Asymptotic Bethe Ansatz (ABA) equations, yielding strong constraints on the (so-far unfixed) dressing factors of the worldsheet S-matrix. 
 In a $\mathbb{Z}_2$-symmetric sector, our proposal is consistent with all previous results on the worldsheet S-matrix. However, in the non-symmetric case, we found a subtle incompatibility between the analytic constraints arising from the proposed QSC, the crossing equations present in the literature, and braiding unitarity.  
 We discuss possible explanations for this mismatch: either our minimal QSC proposal does not hold beyond the symmetric sector, or the crossing unitarity equations receive a nontrivial correction that needs to be understood.\\
 Finally, we also propose a generalisation of the Q-system for the case of $\alpha\neq 1/2$, corresponding to the superalgebra $\mathfrak{d}(2,1;\alpha)$. This novel algebraic structure represents a significant step towards understanding the Quantum Spectral Curve of the entire theory. 
}

\begin{document}

\maketitle
\section{Introduction}
Integrability has been shown to run through some of the most prominent examples of the AdS/CFT duality, in various dimensions: from $\mathcal{N}$=4 SYM to ABJM theory to examples of AdS$_3$/CFT$_2$, see e.g. \cite{Beisert:2010jr,Dorey:2019gkd,Seibold:2024qkh}. 
One way to understand this is to note that the worldsheet theory of the corresponding string model can be formulated as a 1+1-dimensional integrable quantum field theory with factorised scattering, thereby allowing the use of powerful tools from integrability in this framework. 
The presence of integrability has led to a research program aimed at solving these theories, at least in the planar limit. 

The structure arising from integrability that governs the spectrum of these holographic CFTs is known as the Quantum Spectral Curve (QSC). This description was first found for $\mathcal{N}$=4 SYM~\cite{Gromov:2013pga,Gromov:2014caa} and shortly after for ABJM \cite{Cavaglia:2014exa,Bombardelli:2017vhk}, as well as for defects and deformations \cite{Gromov:2015dfa,Kazakov:2018ugh} (and even an application to the Hubbard model \cite{Cavaglia:2015nta}), leading to the possibility of extracting information on the spectrum with in principle arbitrary detail.

The QSC is based on a Q-system, whose origins can be traced back to Baxter's introduction of the Q-operator and the Baxter TQ-equation \cite{Baxter:1982zz}. Q-systems have found applications in many integrable models, such as 2d CFTs \cite{Bazhanov:1994ft, Bazhanov:1996dr, Bazhanov:1998dq}, differential equations \cite{Dorey:2006an, Dorey:2007zx, Masoero:2015lga, Masoero:2015rcz} and spin chains \cite{Gaudin:1992ci, Krichever:1996qd, Antonov:1996ag, Pronko:1999gh, Rossi:2002ed,Korff:2004ev,Bazhanov:2008yc, Bazhanov:2010jq,Kazakov:2010iu,Chicherin:2011sm, Ferrando:2020vzk, Ekhammar:2021myw}.
They are composed of a finite set of Q-operators, whose eigenvalues are the Baxter Q-functions, and functional equations that these must satisfy, known as QQ-relations, determined by the underlying symmetry of the integrable system. To describe the spectrum of AdS/CFT models via the QSC, the Q-functions must be endowed with appropriate analytic properties. 
 
The Quantum Spectral Curves of $\mathcal{N}$=4 SYM and ABJM share important similarities in such analytic properties, while possessing very different algebraic structures -- they are based respectively on the global symmetries $\mathfrak{psu}(2,2|4)$ and $\osp(6|4)$. While for the $\mathfrak{gl}(m|n)$ family of Lie superalgebras the Q-system has been well understood for some time~\cite{Tsuboi:2009ud, Gromov:2010km}, the complete Q-system for $\osp(2m|2n)$ has only been proposed recently \cite{Tsuboi:2023sfs}. An analog of this Q-system for the $m=3,\,n=2$ case was already contained in the functional relations obtained for ABJM theory in \cite{Bombardelli:2017vhk}, which include a subset of the relations found in \cite{Tsuboi:2023sfs} plus further relations that are instead not contained in Tsuboi's work; a detailed mapping between the two frameworks remains to be established. 
 
Recently, a third  Quantum Spectral Curve has been proposed for the string theory with a target space $\text{AdS}_3 \times \text{S}^3 \times \text{T}^4$ and pure Ramond-Ramond flux. 
Currently, this new QSC stands as a conjecture. Still, it passes important consistency checks: a limit of the equations naturally produces the Asymptotic Bethe Ansatz equations~\cite{Cavaglia:2021eqr,Ekhammar:2021pys,Ekhammar:2024kzp}, while also giving a detailed prediction for the dressing phase entering the worldsheet S-matrix. This QSC-compatible dressing phase, in fact, corrects the earlier proposal of \cite{Frolov:2021fmj}, and satisfies a modified crossing equation whose origin was only recently understood~\cite{Frolov:2025ozz}. \\
Furthermore, crucially the $\text{AdS}_3\times \text{S}^3 \times \text{T}^4$ QSC was shown to actually \emph{have nontrivial, isolated  solutions}~\cite{Cavaglia:2022xld}. Thus, it produces analytical and numerical predictions for the spectrum of the still-mysterious CFT$_2$ dual, although these are currently limited to relatively small couplings (far from the stringy regime) and are therefore difficult to test. 
 
Significantly, the QSC for $\text{AdS}_3 \times \text{S}^3 \times \text{T}^4$ has been obtained with a sort of bootstrap method, dubbed ``monodromy bootstrap'' in \cite{Ekhammar:2021pys}. First, the algebraic structure is given by $\mathfrak{psu}(1,1|2)^{\oplus 2} $: this corresponds to two copies of an algebra of the same family as the one governing $\mathcal{N}$=4 SYM. The construction of \cite{Cavaglia:2021eqr,Ekhammar:2021pys} simply postulates the simplest possible analytic properties for the Q-functions, inspired by the SYM and ABJM cases\footnote{This is expected for the pure Ramond-Ramond case, while the cases with mixed fluxes will almost certainly involve more complicated analytic properties. We thank Bogdan Stefanski for many discussions on these points}, by imposing a minimal coupling between the two copies of the system, which are glued via analytic continuation of the Q-functions through their branch cuts. 
This economic construction leads to a surprising consequence: the cuts of the Q-functions in this system cannot be quadratic, thus introducing an important novelty — and nontrivial computational challenges — compared to the previously known cases~\cite{Cavaglia:2022xld}. 
 
The purpose of this paper is to attempt the same strategy for a different string theory expected to be integrable: the one with target space \ads\, with pure Ramond-Ramond flux. In this case, the underlying algebra is 
$\dalpha^{\oplus 2} $: i.e. two copies of an exceptional super-Lie algebra, where $\alpha\in(0,1)$ parametrises the ratio of the radii of the two $\text{S}^3$ factors. 
The problem might seem daunting because  {not even the Q-system} was known for this kind of symmetry. 
However, we rely on an essential piece of information: at $\alpha = \frac{1
}{2}$, which is the symmetric point where the two spheres have equal radii, this exceptional algebra reduces to two copies of $\osp(4|2)$, a case that was analysed in~\cite{Babichenko:2009dk} and which can be seen as a lower rank instance of an orthosymplectic algebra like the one governing the QSC of ABJM theory.

Therefore, at least in this symmetric point, we can proceed in the same spirit as in \cite{Cavaglia:2021eqr}, and we put forward a natural proposal for a Quantum Spectral Curve for \ads\, string theory: it is constructed by gluing two copies of a lower-rank version of the QSC for the ABJM theory. 
In this paper, we analyse the consequences of this construction. 
In particular, we perform some consistency checks in the Asymptotic Bethe Ansatz limit.

We will also propose a conjecture covering the general $\alpha$ case: first proposing a new Q-system, which is entirely new at the purely algebraic level, and then starting to elaborate on the possible structure of the QSC in this case.

\paragraph{Main results, and a puzzle. }
Some of our results directly concern the connection with string theory on \ads and the possible emerging structure of a QSC.  Others are new results on Q-systems in general.

To help the reader, we collect our main results below.
\begin{itemize}
\item We present a reformulation of the Q-system for $\osp(4|2)$, based on a small set of core functional relations which generate the full structure. This construction, which is inspired by the structures found for ABJM theory in \cite{Bombardelli:2018bqz}, is different from the presentation of the Q-system of Tsuboi for $\osp(2m|2n)$ given in \cite{Tsuboi:2023sfs}. However, we prove that it is completely equivalent in Appendix \ref{app:equivalence}. 
\item We then propose to build a QSC by glueing two copies of a Q-system based on the $\osp(4|2)$ symmetry, through branch cuts. We derive the analytic consequences of the construction in detail. This leads to equations that encode the structure of the Riemann surface in a simple way.
\item Focusing on states involving massive\footnote{We expect that the analysis can be extended to massless states along the lines of \cite{Ekhammar:2024kzp}.} worldsheet excitations, we show how the proposed QSC equations can be solved in the large worldsheet limit. This leads to Asymptotic Bethe Ansatz equations, which match the structure of those proposed in the literature for this string theory \cite{Borsato:2012ss}. 
.
\item We conjecture a form of Q-system for the full algebra $\dalpha$, not previously known in the literature. We demonstrate that this algebraic structure naturally gives rise to Bethe equations in all gradings for this algebra, specifically for the zeros of the Q-functions. A natural generalisation of the conjecture for the QSC at general $\alpha$ is briefly discussed.
\end{itemize}

The derivation of ABA equations from the proposed QSC also provides strong constraints on the form of the dressing factors, which were previously unknown.  Potentially, this could resolve the issue completely. Our findings, however, contain some puzzles. 
\begin{itemize}
\item For a symmetric sector of the theory (with symmetry between spinor and antispinor massive particles), our findings seem completely consistent with all the expected properties of the worldsheet S-matrix present in the literature. 
\item
For states that have no such symmetry, the ABA equations are also sensitive to an additional contribution to the dressing factors, which we call $\Sigma_{\text{new}}$. 
From the proposed QSC, we obtain strong constraints on $\Sigma_{\text{new}}$: it is determined by the solution to a linear integral equation, up to some parameters -- including a function of one of the rapidities -- that we are presently not able to fix.  However, for any choice of such parameters, we find a tension with either braiding unitarity or the crossing symmetry equations as written in the literature. 
In the Conclusions, we comment on possible interpretations of this finding.
\end{itemize}

The rest of this paper is organised as follows: in section \ref{sec:QQsys}, we discuss the Q-system for $\osp(4|2)$, reformulating it in a new way. 
Based on this structure, in section \ref{sec:QSCProposal} we present a natural proposal for a QSC for the string theory at $\alpha = 1/2$, built by gluing two copies of the Q-system. The resulting structure is analysed in detail.
In section \ref{sec:ABA}, we study the large worldsheet limit and deduce ABA equations for massive states, leading to a match with the expected form of these equations. 
In section \ref{sec:dressingnew}, we discuss what the proposed QSC implies for the most nontrivial piece of the dressing factors, $\Sigma_{\text{new}}$. Here, we discuss its braiding unitarity and crossing properties, presenting the puzzle.   
Section \ref{sec:alphaQsys} generalises some of the previous considerations to the general case $\alpha\neq \frac{1}{2}$, in particular conjecturing the full Q-system for this case.  
  The paper then closes with a discussion and outlook. 

\paragraph{Note: } At an advanced stage of this project, we learnt that F. Chernikov, S. Ekhammar, N. Gromov and B. Smith had been working on a project with a substantial overlap, also exploring a potential QSC for the \ads \, model in the case $\alpha = \frac{1}{2}$. We decided to coordinate our submission to arXiv. Our proposed QSC equations are  equivalent.
We thank Filipp, Kolya, and Simon for discussions during IGST 2025, particularly for first alerting us to the puzzles emerging in the non-symmetric sector. 

\section{Q-system for $\ospf$}
\label{sec:QQsys} 
In this section, we focus on the algebraic structure of the Q-system for the model at hand, which depends on its underlying symmetry. In particular, string theory on \ads\, background has symmetry $\dalpha_L \oplus \dalpha_R$, hence we will need two copies of a $\dalpha$ Q-system. 

To the best of our knowledge, the Q-system for this exceptional superalgebra is not known in the literature. We will thus advance a proposal for it in section \ref{sec:alphaQsys}. 
In this section, we start from the simpler case where $\alpha=1/2$, which is the main focus of this paper. At this particular value of $\alpha$, the symmetry group of the string theory is the simpler $\osp(4|2)_L \times \osp(4|2)_R$, and thus we will need two copies of an $\osp(4|2)$ Q-system. While the latter has recently been studied in~\cite{Tsuboi:2023sfs}, we will reformulate in a ``covariant'' way, which is akin to the Q-system of the ABJM Quantum Spectral Curve~\cite{Bombardelli:2017vhk}, which possesses $\osp(6|4)$ symmetry. Our formulation differs slightly from that of~\cite{Tsuboi:2023sfs}, since the latter, on the one hand, introduces explicitly the twists in the Q-functions, and on the other hand, uses a different way to label the Q-functions in a non-explicitly covariant way. 
However, in section~\ref{app:equivalence}, we will prove that the two Q-systems are equivalent.
\begin{figure}
\begin{subfigure}{.5\textwidth}
\centering
\includegraphics[width=.5\linewidth]{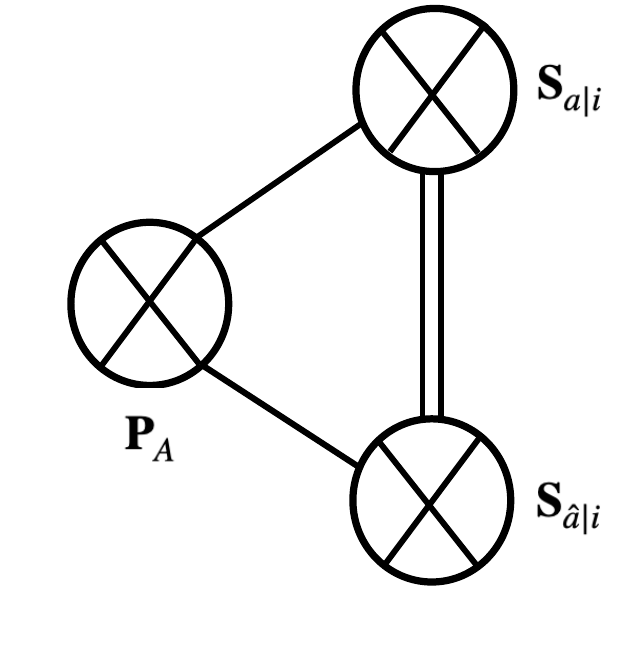}
\caption{Grading $\bar 0\bar 1\bar 0$}
\label{fig:sfig1}
\end{subfigure}%
\begin{subfigure}{.5\textwidth}
\centering
\includegraphics[width=.5\linewidth]{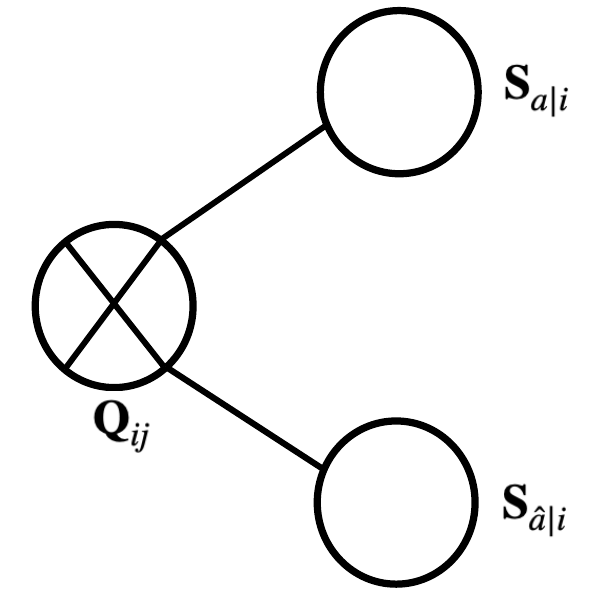}
\caption{Grading $\bar 1\bar 0\bar0$}
\label{fig:sfig2}
\end{subfigure}
\begin{center}
\begin{subfigure}{.6\textwidth}
\centering
\includegraphics[width=.6\linewidth]{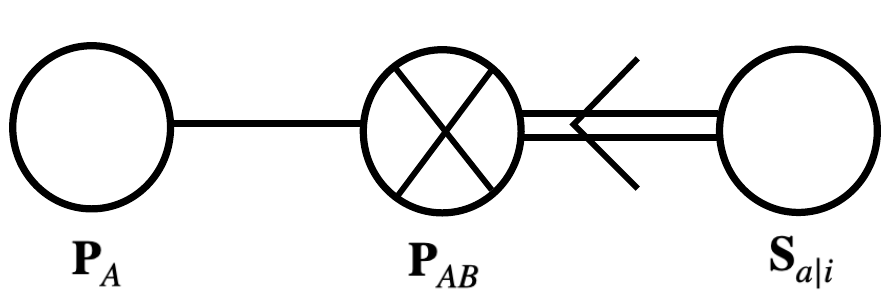}
\caption{Grading $\bar0\bar0\bar1$}
\label{fig:sfig3}
\end{subfigure}
\end{center}
\caption{Dynkin diagrams for the 3 inequivalent gradings of $\osp(4|2)$ with the relevant Q-functions drawn on them. The corresponding simple root systems are built from an $\eps-\del$ sequence, i.e. an ordering of 3 orthonormal unit vectors $\del,\eps_1,\eps_2$ with $\delta$ being odd and $\eps_i$ even. In the gradings, the numbers $\bar 0$ and $\bar 1$ indicate whether the corresponding vector in the $\eps-\del$ sequence is even or odd.}
\label{fig:fig}
\end{figure}

\subsection{Main objects and functional relations}
\label{sec:indices}
In this section, we start by introducing the conventions we use in the covariant formulation of the $\osp(4|2)$ Q-system.
We label its Q-functions by a set of indices taking values in the even subalgebra of $\osp(4|2)$, i.e. $\so(4)\oplus\mathfrak{su}(2)\sim \mathfrak{su}(2)^{\oplus 3}$. These indices will be denoted as follows:\\

\begin{center}\begin{tabular}{|c|c|c|c|}
\hline
     Notation& Representation& Allowed values\\
     \hline
     $A,B\dots$& Vector $\so(4)$& $1,2,3,4$\\
     $a,b\dots$& Spinor $\so(4)$& $1,2$\\
     $\hat a,\hat b\dots$& Antispinor $\so(4)$& $\hat1,\hat2$\\
     $i,j\dots$& Fundamental $\mathfrak{su}(2)$& $1,2$\\
     \hline
\end{tabular}
\end{center}
Notice in particular that we distinguish spinor and anti-spinor representations of $\so(4)$ by a hat on the indices for the latter, a convention that will be kept throughout the paper. 

\paragraph{Invariant tensors.}
To write down our covariant QQ-relations, we will need invariant tensors of $\so(4)\oplus\mathfrak{su}(2)$, i.e. the bosonic subalgebra. First, let us introduce gamma matrices for $
\so(4)$, which are $4\times 4$ matrices
\begin{equation}
\Gamma_A = \left( \begin{array}{cc} 0 & \sigma_A   \\
\bar{\sigma}_A  & 0 \end{array} \right) , 
\end{equation}
where the blocks $\sigma_A$ and $\bar{\sigma}_A$ are $2\times 2$ matrices, with components denoted as 
$\sigma_A = \left( (\sigma_A)_a^{\,\hat{a}} \right)_{1\leq a,\hat{a}\leq 2}$ and  $\bar{\sigma}_A = \left( (\bar{\sigma}_A)_{\hat{a}}^{\,{a}} \right)_{1\leq \hat{a},{a}\leq 2}$. They satisfy:
\begin{eqnarray}
&& \sigma_A \, \bar{\sigma}_B + \sigma_B \, \bar{\sigma}_A  = ( 1 )_{2\times 2} \; \rho_{AB}, \\ &&   \bar\sigma_A \, {\sigma}_B + \bar\sigma_B \, {\sigma}_A  = ( 1 )_{2\times 2}\; \rho_{AB} ,
\end{eqnarray}
where $\rho$ is a $SO(2,2)$ metric. We choose this signature because, based on our experience with the ABJM case, we expect it to yield the simplest reality properties for the Q-functions associated with the string theory we aim to describe.

Our formalism is fully covariant, but  we will take the following choice for the metric $\rho$ when writing equations in components
\begin{equation}
\label{def:sometric}
\rho \equiv \left(\rho_{AB} \right)_{1\leq A,B\leq 4} = \left( \begin{array}{cccc} 0 & 0 & 0 & 1   \\
 0 & 0  & 1 & 0  \\
  0 & 1 & 0 & 0 \\
   1 & 0 & 0 & 0 \end{array} \right) . 
\end{equation}
For the sigma matrices, we will adopt the conventional choice:
{\small
\beqa\label{eq:oursigmas}
\left\{\sigma_A \right\}_{1\leq A \leq 4} &=& \left\{\left( \begin{array}{cc} \
 0 & 0  \\
  1 & 0 \end{array}\right), \, 
  \left( \begin{array}{cc} \
 0 & 0  \\
  0 & 1 \end{array}\right),\,
  \left( \begin{array}{cc} \
 -1 & 0  \\
  0 & 0 \end{array}\right),\,
  \left( \begin{array}{cc} \
 0 & 1  \\
  0 & 0 \end{array}\right)\right\},\\
  \left\{\bar{\sigma}_A \right\}_{1\leq A \leq 4} &=& \left\{\left( \begin{array}{cc} \
 0 & 0  \\
  1 & 0 \end{array}\right), \, 
  \left( \begin{array}{cc} \
 -1 & 0  \\
  0 & 0 \end{array}\right),\,
  \left( \begin{array}{cc} \
 0 & 0  \\
  0 & 1 \end{array}\right),\,
  \left( \begin{array}{cc} \
 0 & 1  \\
  0 & 0 \end{array}\right)\right\}.
\eeqa
}
The gamma matrices are invariant with respect to a charge conjugation operator,
\begin{equation}
(\Gamma_A )^t = - \mathbf{C}_{4\times 4} \cdot \Gamma_A \cdot (\mathbf{C}_{4\times 4})^{-1}\, ,
\end{equation}
which has the decomposition:
\begin{equation}
\mathbf{C}_{4\times 4} \equiv \left( \begin{array}{cc} (\mathbf{C} )_{2\times 2} & 0  \\
 0 & (\hat{\mathbf{C}} )_{2\times 2} \end{array} \right) ,
\end{equation}
where the components of the $2\times 2$ blocks will be indexed as $\mathbf{C} \equiv \left( \mathbf{C}^{ab} \right)_{1\leq a,b\leq 2}$, $\hat{\mathbf{C}} \equiv \left( \hat{\mathbf{C}}^{\hat{a}\hat{b}} \right)_{1\leq \hat{a},\hat{b}\leq 2}$. 
Decomposing the above equation in blocks, we get the following constraints on the $\sigma$ matrices
\begin{equation}
(\sigma_A )_a^{\, \hat b}= - \hat{\mathbf{C}}^{\hat b\hat c} ( \bar{\sigma}_A )_{\hat c}^{\, b}  ({\mathbf{C}}^{-1} )_{ba},\;\;\;\;\;(\bar{\sigma}_A )_{\hat a}^{\,  b}= - {\mathbf{C}}^{ b c} ( {\sigma}_A )_{ c}^{\, \hat b}  (\hat{\mathbf{C}}^{-1} )_{\hat b \hat a}\,.
\end{equation}
With our conventional choice for the sigma matrices, the charge conjugation matrices can be taken to be 
\begin{equation}
\left(\mathbf{C}^{a b} \right)_{1\leq a,b \leq 2}= \left(\hat{\mathbf{C} }^{\hat a \hat b} \right)_{1\leq \hat{a},\hat{b} \leq 2} = \left( \begin{array}{cc} 0 & 1  \\
 - 1 & 0 \end{array} \right) . 
\end{equation}
Finally, for the R-symmetry $SU(2)$ factor, the only invariant tensor is the fully antisymmetric tensor 
\begin{equation}
\epsilon \equiv \left(\epsilon^{ij} \right)_{1\leq i,j \leq 2 } = \left( \begin{array}{cc} 0 & 1  \\
 - 1 & 0 \end{array} \right)\,.
\end{equation}
We also point out that (as in the previous equations) we use the Einstein convention of summing over contracted upper/lower indices, and we will denote the components of any inverse matrix by lowering/raising the indices, for instance
$\epsilon^{-1} \equiv \left( \epsilon_{ij} \right)_{1\leq i,j\leq 2} = \left( \begin{array}{cc} 0 & -1  \\
  1 & 0 \end{array} \right)$. 
\paragraph{Covariant Q-functions. }
The Q-functions are a finite set of complex functions of the spectral parameter $u$. Their number depends on the symmetry of the Q-system, and thus we will label the ones found in the $\ospf$ Q-system via the indices introduced in section~\ref{sec:indices}. 
We will sometimes refer to Q-functions having solely $\so(4)$ indices as ``bosonic'', while those with only $\mathfrak{su}(2)$ indices will be called ``fermionic''. These terms do not imply a graded structure on the Q-functions, but they are a useful shorthand found in the literature \cite{Kazakov:2015efa}.

In the covariant formulation we are proposing, there are actually some extra Q-functions 
compared to the $\ospf$ Q-system appearing in \cite{Tsuboi:2023sfs}, as we will explain. Furthermore, the Q-system contains another complex function, $\bQ_{\emptyset}$, which is state independent and appears in the Wronskian formulas for the T-functions of \cite{Tsuboi:2023sfs} as an overall prefactor. While we keep this function explicit in the QQ-relations presented in this section, for the Quantum Spectral Curve in section \ref{sec:QSCProposal} it will be set to 1.

In our $\osp(4|2)$ Q-system, we have the following Q-functions\footnote{The particular notation used for these functions is inspired by the QSC literature. }:

\begin{center}\begin{tabular}{|c|c|c|c|c|}
\hline
     Notation& Type/Name & Index range& Number\\
     \hline
     $\bP_A$& Vector bosonic & $A=1\dots 4$& 4\\
     $\bS_{a|i}$& Spinorial &$a, i = 1,2$  &4\\
     $\bS_{\hat a|i}$& Antispinorial & $\hat a, i = 1,2$& 4\\
     $\bQ_{ij}$& Fermionic & $i,j = 1,2$&4\\
     $\bP_{AB} $ & Bivector bosonic & $A,B=1\dots 4,\,A\neq B$& 6\\
     \hline
\end{tabular}
\end{center}
It is natural to place the Q-functions on the nodes of the Dynkin diagrams of the Lie algebra underlying the Q-system. Since $\ospf$ possesses 3 inequivalent Dynkin-Kac diagrams, we will place the sets of Q-functions introduced above on each of them, as depicted in figure~\ref{fig:fig}.

From the first three sets of Q-functions we have introduced, i.e. $\bP_A$, $\bS_{a|i}$ and $\bS_{\hat a|{i}}$, we will be able to build explicitly all the other Q-functions in the Q-system. In particular, we can define, with the usual convention $f^{[\pm N]}(u)\equiv f(u\pm N i/2)$, the remaining objects in the list above as follows:
\begin{framed}
\begin{align}\label{eq:defQ}
&\bQ_{\emptyset}^-\bQ_{ij}(u) \equiv\bS_{a|i}^- \, \bP_{\hat b}^{\, a} \,\, \hat{\mathbf{C}}^{\hat b \hat c}\,\, \bS_{\hat c|j}^-=(\bQ_{\emptyset}^-)^2\bS_{\hat a|i}^+ \, {\mathbf{C}}^{b a}\,\bP_{a}^{\, \hat a} \, \bS_{b|j}^+\,,\\
&\label{eq:PABdef}
\bQ_{\emptyset}\bP_{AB}(u) \equiv \bP_A^+ \bP_B^--\bP_A^- \bP_B^+\, .
\end{align}
\end{framed}
We will see that these particular equations are part of the system of QQ-relations described below.
 
In order to write the QQ-relations in a covariant form, it will be convenient to express the 4-component vector $\bP_A$ as $2\times2$ matrices with mixed spinor-antispinor indices, by using the Weyl matrices introduced above as follows:
\begin{eqnarray}
&& \bP_a^{\, \hat b}(u)\equiv (\sigma_A )_a^{\,\hat b} \, \rho^{AB}  \bP_{B}(u) , \label{eq:matP1}\\
&& \bP_{\hat a}^{\,  b}(u)\equiv (\bar\sigma_A )_{\hat a}^{\, b} \, \rho^{AB}  \bP_{B}(u) .\label{eq:matP2}
\end{eqnarray}
With the choice of the metric~\eqref{def:sometric}, 
(\ref{eq:matP1}) and (\ref{eq:matP2}) become:
\begin{equation}\label{eq:Pspinor}
\bP \equiv \left(\bP_a^{\;\hat b} \right)_{1\leq a, \hat{b} \leq 2} =  \left( \begin{array}{cc} -\bP_2 &  \bP_1   \\
  \bP_4 & \bP_3 \end{array} \right) ,\;\;\;\;
 \hat{\bP}\equiv \left(\bP_{\hat a}^{\; b} 
 \right)_{1\leq \hat{a}, b \leq 2} = \left( \begin{array}{cc} -\bP_3 &  \bP_1   \\
  \bP_4 & \bP_2 \end{array} \right)  .
\end{equation}
These two matrices are intertwined by the charge-conjugation tensors $\mathbf{C}$ and $\hat{\mathbf{C}}$:
\begin{equation}\label{eq:Pintertwine} \bP_{a}^{\;\hat a}=-\mathbf{C}_{ab}\bP_{\hat b}^{\;b}\hat{\mathbf{C}}^{\hat b\hat a}\,,\qquad \bP_{\hat a}^{\; a}=-\hat{\mathbf{C}}_{\hat a\hat b}\bP_{ b}^{\;\hat b}\mathbf{C}^{b a}.
\end{equation}
We can also enclose most of the other Q-functions in $2\times 2$ matrices, which we denote as follows:
\beq
\label{def:Q_matrices}
\bS\equiv \left(\bS_{a|i} \right)_{1\leq a,i\leq 2}, \quad \hat\bS\equiv \left(\bS_{\hat a|i} \right)_{1\leq \hat{a}, i\leq 2},\quad \bQ\equiv \left(\bQ_{ij}\right)_{1\leq i,j\leq 2}\,.
\eeq

\paragraph{Core Relations. }
We now introduce some constraints and functional relations between the Q-functions $\bP_A,\,\bS_{a|i},\,\bS_{\hat{a}|i}$ that, along with some basic analyticity requirements, will be enough to reconstruct the full QQ-relations of the $\ospf$ Q-system. 

Firstly, we need the following constraint:
\begin{framed}
\begin{equation}\label{eq:constraintP}
\text{ Constraint 1: }\bP_A(u) \rho^{AB} \bP_B(u) = 2\bQ_{\emptyset}^+\bQ_{\emptyset}^-   . 
\end{equation}
\end{framed}
Using the metric~\eqref{def:sometric}, it reads\footnote{In this form, this constraint is equivalent to the Wronskian expression for the trivial T-function $\mathbf{T}_{1,0}$, see equation (4.140) of \cite{Tsuboi:2023sfs}.}:
\begin{equation}
\label{constraint_detP}
\bP_1(u) \bP_4(u) + \bP_2(u) \bP_3(u) = \bQ_{\emptyset}^+\bQ_{\emptyset}^- .
\end{equation}
Equation (\ref{constraint_detP}) can also be rephrased as a unit determinant condition  for the $2\times 2$ matrices defined in \eqref{def:Q_matrices}:
\begin{equation}
    \text{det}(\bP)= \text{det}(\hat\bP)=-\bQ_{\emptyset}^+\bQ_{\emptyset}^-\,.
\end{equation}
There is also a useful equivalent way to recast these identities: the product of the two matrices in (\ref{eq:Pspinor}) is proportional to the identity matrix:
\begin{equation}
 \label{eq:Punitdet}
    \bP_a^{\;\hat a}\bP_{\hat a}^{\;b}=\delta_a^{\;b}\,\bQ_{\emptyset}^+\bQ_{\emptyset}^-,\qquad \bP_{\hat a}^{\; a}\bP_{ a}^{\;\hat b}=\delta_{\hat a}^{\;\hat b}\, \bQ_{\emptyset}^+\bQ_{\emptyset}^-.
 \end{equation}
Then we demand the validity of the following system of equations, which we dub, for their importance, the {\it Core Relations}:
\begin{framed}
\begin{equation}\label{eq:core_relations}
\text{ Core Relations: \;\;\;\;\;}\bQ_{\emptyset}^-\bS_{a|k}^+(u)  = \bP_a^{\;\hat b}(u)  \bS_{\hat b|k}^-(u)   ,\;\;\; \bQ_{\emptyset}^-\bS_{\hat a|k}^+(u)  = \bP_{\hat a}^{\; b}(u)  \bS_{ b|k}^-(u)  .
\end{equation}
\end{framed}
The latter relations were not written before for the Q-system corresponding to this algebra. They are similar to the ones found in the ABJM QSC~\cite{Bombardelli:2017vhk}. Together with the constraint (\ref{eq:constraintP}), the core relations allow us to deduce the whole Q-system of~\cite{Tsuboi:2023sfs}. 

A second constraint, which is a consequence of (\ref{eq:constraintP}), the core relations, and some simple analyticity assumptions for the Q-functions, takes the following form:
\begin{framed}
\begin{equation}\label{eq:constraintQai}
\text{ Constraint 2: }\;\;\;\;\;\; {\bS_{a|i}(u) \mathbf{C}^{ab} \bS_{b|j}(u)} = - {\bQ_{\emptyset}(u)}\,\epsilon_{ij}= -{\bS_{\hat a|i}(u) \hat{\mathbf{C}}^{\hat a \hat b} \bS_{\hat b|j}(u)} .
\end{equation}
\end{framed}
While we introduced this second constraint here as an independent axiom, it is actually closely related to the previous two conditions once we specify simple analyticity properties for the Q-functions. To see this, consider the two matrices $g_{ij},\hat{g}_{ij}$, which \emph{a priori} could be functions of $u$:
\begin{align}
    &g_{ij}\equiv\frac{\bS_{a|i}\mathbf{C}^{ab}\bS_{b|j}}{\bQ_{\emptyset}}\,,\\
    &\hat g_{ij}\equiv\frac{\bS_{\hat a|i}\hat{\mathbf{C}}^{\hat a\hat b}\bS_{\hat b|j}}{\bQ_{\emptyset}}\, .
\end{align}
Using the core relations \eqref{eq:core_relations} and the constraint \eqref{eq:constraintP}, we can prove the following periodicity properties:
\begin{align}
    &g^{+}_{ij}=\frac{\bS^+_{a|i}\mathbf{C}^{ab}\bS^+_{b|j}}{\bQ_{\emptyset}^+}=-\frac{\hat{\mathbf{C}}^{\hat a \hat b}\bS_{\hat a|i}^-\bS_{\hat b|j}^-}{\bQ_{\emptyset}^-}=- {\hat g^-_{ij}}\,,\\&
    \hat g^{+}_{ij}=\frac{\bS^+_{\hat a|i}\hat{\mathbf{C}}^{\hat a\hat b}\bS^+_{\hat b|j}}{\bQ_{\emptyset}^+}=-\frac{\mathbf{C}^{ a  b}\bS_{ a|i}^-\bS_{ b|j}^-}{\bQ_{\emptyset}^-}=- {g^-_{ij}}\,.
\end{align}
If we assume that Q-functions are free of singularities in either the upper or the lower half plane, and have power-like asymptotics at large $u$ \footnote{These assumptions apply to both rational spin chains and AdS/CFT integrable systems.} all elements of $g,\hat g$ will inherit the same properties. Together with the periodicity property we just established, this implies that these are actually constant functions of $u$. By tuning the normalisation of Q-functions, we are free to choose\footnote{This is consistent with the choice of asymptotics for the Q-functions we specify below, which will imply that some components tend to zero at large $u$.}
\beq
g_{ij}\equiv-\eps_{ij}\,,\qquad \qquad \hat{g}_{ij}\equiv\eps_{ij}\, ,
\eeq
which will then imply (\ref{eq:constraintQai}). 

Notice that (\ref{eq:constraintQai}) are also equivalent to the following unit determinant conditions on the spinorial and antispinorial Q-functions:
\begin{equation} \text{det}\,  \bS(u)   = + {\bQ_{\emptyset}(u)}, \;\;\;\;\; \text{det} \, \hat\bS(u)  = - {\bQ_{\emptyset}(u)}.
\end{equation}
Written in components, these constraints are reminiscent of Wronskian expressions for the trivial spinorial and antispinorial T-functions $\mathbf{T}_{s,0}$ and $\mathbf{T}_{\hat s,0}$. 
Unfortunately, to our knowledge, an explicit formula for $\mathbf{T}_{s,0}$ and $\mathbf{T}_{\hat s,0}$ in terms of Q-functions does not seem to exist in the literature.
\subsection{Deducing the QQ-relations}
\label{sec:proofsQQ}
The bosonic/fermionic QQ-relations can be thought of as the action of even/odd Weyl reflections about a node of the Dynkin diagram on the Q-functions, as explained in \cite{Tsuboi:2023sfs}.

In this section, we show how the core relations and the previously introduced constraints allow us to deduce all the QQ-relations of the $\osp(4|2)$ Q-system. To keep this section concise, we put the full derivations in appendix \ref{appendix:QQproofs}. Furthermore, in the next section, we will see how these QQ-relations imply that the zeros of the Q-functions satisfy the Bethe equations of the $\osp(4|2)$ Lie superalgebra.

We have already encountered the first QQ-relations, i.e. the definition \eqref{eq:PABdef}. These are the bosonic QQ-relations (or equivalently, even Weyl reflection) about the first node of the $\spp$-shaped Dynkin diagram \ref{fig:sfig3}, therefore we shall name it:
\beq
\label{eq:QQrelssp}
\text{\bf Bosonic first node }\spp\qquad {\bQ_{\emptyset}(u)}\bP_{AB}(u) = \bP_A^+ \bP_B^--\bP_A^- \bP_B^+\, .
\eeq
We have also introduced the definition of the object $\bQ_{ij}$ in \eqref{eq:defQ}. These, when manipulated using the core relations, can also be interpreted as bosonic QQ-relations; in particular, they correspond to the even Weyl reflections about the spinor (second) and antispinor (third) nodes of the Dynkin diagram~\ref{fig:sfig1}:
\beqa
&\text{\bf Bosonic spinor node}\qquad &
\label{eq:relQ1}
\bQ_{ij} 
= \bS_{a | i}^- \mathbf{C}^{ba} \,   \bS_{ b | j}^+\,.
\\&
\text{\bf Bosonic antispinor node}\qquad &
\bQ_{ij}
= \bS_{\hat a | i}^+ \; \hat{\mathbf{C}}^{\hat a \hat b}  \bS_{\hat b | j}^- \,.
\label{eq:relQ2}
\eeqa
Note that, from this definition and the constraints \eqref{eq:constraintQai}, it follows immediately that the following determinant condition holds:
\beq
\label{eq:constraintQij}
\text{det}(\bQ)=-\bQ_{\emptyset}^+\bQ_{\emptyset}^-\,.
\eeq
Furthermore, multiplying the core relations by $(\mathbf{S}^-)^{-1}$ and $(\mathbf{\hat S}^-)^{-1}$ we can obtain relations that express the bosonic Q-functions $\bP_A$ in terms of spinors and antispinors\footnote{These are not strictly speaking QQ-relations, in the sense that they are not associated to a standard set of Bethe equations; they are instead a Wronskian-type expression for the $\bP_A$ functions in terms of $\bS,\,\hat\bS$, reminiscent of the ones known for the $\so(2r)$ Q-system \cite{Ferrando:2020vzk, Ekhammar:2021myw}.}:
\begin{eqnarray}
\label{eq:F1}
&\bP_a^{\; \hat b}(u) = - \bS_{a|i}^+(u) \; \epsilon^{ij} \; \bS_{\hat a|j}^-(u) \hat{\mathbf{C}}^{\hat a \hat b} ,\\
& \bP_{\hat a}^{\;  b}(u) = \bS_{\hat a|i}^+(u) \; \epsilon^{ij} \; \bS_{ a|j}^-(u) {\mathbf{C}}^{a  b} \,.
\end{eqnarray}
By defining the Q-function with three indices $Q_{A|ij}$ as
\beq
\label{eq:defQAij}
Q_{A|ij}(u) \equiv \bS_{a|i}(u) \; \bS_{\hat a|j}(u)\; (\sigma_A )_b^{\; \hat a} \;\mathbf{C}^{ab},
\eeq
we can also get QQ-relations involving it, which are:
\begin{eqnarray}
\label{eq:F2QQrel}
&&\bP_A(u) \bQ_{\left\{ij\right\}}(u) = {\bQ_{\emptyset}^-(u)}Q_{A|\left\{ij\right\}}^+(u) - {\bQ_{\emptyset}^+(u)}Q_{A|\left\{ij\right\}}^-(u) ,
\\&&\label{eq:F2QQrel2}
\bP_A(u) \bQ_{ [ij ]}(u) = -{\bQ_{\emptyset}^-(u)}Q_{A|[ij]}^+(u) - {\bQ_{\emptyset}^+(u)}Q_{A| [ij] }^-(u) .
\end{eqnarray}
Here $[i,j]$ and $\{i,j\}$ anti-symmetrise and symmetrise the indices. Taking the first one and setting $i=j$, we get QQ-relations representing the odd Weyl reflection about the first nodes of the Dynkin diagrams \ref{fig:sfig1} and \ref{fig:sfig2}:
\beq
\label{eq:QQrels_node1_fermionic}
\text{\bf Fermionic first node}\qquad \bP_A(u) \bQ_{ii}(u) = {\bQ_{\emptyset}^-(u)}Q_{A|ii}^+(u) - {\bQ_{\emptyset}^+(u)}Q_{A|ii}^-(u) .
\eeq
As we show in the second-to-last paragraph of appendix \ref{appendix:QQproofs}, it follows that the Q-functions $\bP_{AB}$ also satisfy the QQ-relations representing the even reflection about the third node of the Dynkin diagram \ref{fig:sfig3}, which we call \textbf{Bosonic third node $\spp$}:
\begin{eqnarray}
\label{eq:Z2QQrel}
&&2\,\bS_{a|i}^{++} \; \bS_{b|j}^{--} \; \mathbf{C}^{ac}(\sigma_{[A} \cdot \bar{\sigma}_{B]} )_c^{\; b} \epsilon^{ij} = \left( \delta_A^C \delta_B^D-\delta_B^C \delta_A^D  -\epsilon_{C' D' AB} \rho^{C' C} \rho^{D' D} \right) \bP_{CD} ,
\\
\label{eq:Z2starQQrel}
&&2\,\bS_{\hat a|i}^{++} \; \bS_{\hat b|j}^{--} \; \hat{\mathbf{C}}^{\hat a\hat c}(\bar{\sigma}_{[A} \cdot {\sigma}_{B]} )_{\hat c}^{\; \hat b} \epsilon^{ij} = -\left( \delta_A^C \delta_B^D-\delta_B^C \delta_A^D  +\epsilon_{C' D' AB} \rho^{C' C} \rho^{D' D} \right) \bP_{CD} .
\end{eqnarray}
In the next paragraph, we discuss the form these equations take in our representation concretely. We will see that they reduce to a familiar Wronskian-type form only for some particular choices of indices.

Finally, the last QQ-relations that we prove in the last paragraph of \ref{appendix:QQproofs} represent the odd reflections about the spinor and antispinor nodes of the Dynkin diagram \ref{fig:sfig2}:

\begin{footnotesize}
 \begin{eqnarray}
 \label{eq:QQrelZ}
&\text{\bf Fermionic spinor node}\qquad &\bS_{\hat a|i}^{++} \; \bP_{A}^- - \bS_{\hat a|i}^{--} \; \bP_{A}^+ = \left(\bP_A^- (\bP^+)_{\hat a}^{\; b} -  \bP_A^+ (\bP^-)_{\hat a}^{\; b} \right) \bS_{ b|i},
\\&\text{\bf Fermionic antispinor node}\qquad &
\bS_{a|i}^{++} \; \bP_{A}^- - \bS_{a|i}^{--}\; \bP_{A}^+ = \left(\bP_A^- (\bP^+)_a^{\; \hat b}-  \bP_A^+ (\bP^-)_a^{\; \hat b}\right) \bS_{\hat b|i}.
\label{eq:QQrelZ2}
\end{eqnarray}
\end{footnotesize}

\paragraph{Special index pairings. } The equations (\ref{eq:Z2QQrel}), (\ref{eq:Z2starQQrel}) or (\ref{eq:QQrelZ}), (\ref{eq:QQrelZ2}), which we wrote in covariant form valid for any choice of $\rho$ and sigma matrices, might not immediately resemble QQ-relations. In this paragraph, we examine their explicit form using our choice of representations (\ref{def:sometric}) and (\ref{eq:oursigmas}). We show that \emph{only for some choice of indices} the left and right-hand sides take the form of a Wronskian. At these special choices of indices, the relations reproduce precisely the QQ-relations described by Tsuboi in \cite{Tsuboi:2023sfs},   whereas the other, more intricate functional relations extend the system and, to the best of our knowledge, have not been reported before.

Let us start by analysing (\ref{eq:Z2QQrel}), (\ref{eq:Z2starQQrel}). With our choice of metric (\ref{def:sometric}), the tensor structure appearing on the rhs of (\ref{eq:Z2QQrel}) becomes:
\beq
\delta_{[A}^C \delta_{B]}^D -\epsilon_{C' D' AB} \rho^{C' C} \rho^{D' D}  = 
\left\{ 
\begin{array}{ccc} 2 \underbrace{\delta_{[A}^C \delta_{B]}^D }_{\texttt{Wronskian}}  &\text{ if }  &(A,B) \in  \underbrace{\left\{ (1,2),  (3,4)\right\} }_{\texttt{``good'' pairings}},\\
0   &\text{ if }  &(A,B) \in  \left\{ (1,3),  (2,4)\right\}  ,\\
\texttt{not Wronskian} &\text{ if }  &(A,B) \in  \left\{ (1,4),  (2,3)\right\} 
\end{array}
\right. ,
\eeq
where we have listed only pairs of indices with $A<B$, since the expression is antisymmetric in $A$ and $B$.

We can notice that the ``good pairings'' correspond to the two $\bP$ functions appearing in a single row of the matrix $\bP_a^{\, \hat{b}}$: the first row contains $\bP_1$ and $\bP_2$, and the second row contains $\bP_3$ and $\bP_4$. 

For the choice of pairing $(A,B)$ corresponding to the other cases we have, rather than a Wronskian of two $\bP$ functions, a combination of the two Wronskians $\bP_{14}$ and $\bP_{23}$ on the rhs of the equation. 

Instead, for the combination appearing on the rhs of (\ref{eq:Z2starQQrel}) (notice that the combinations of indices are different):
\beq
\delta_{[A}^C \delta_{B]}^D +\epsilon_{C' D' AB} \rho^{C' C} \rho^{D' D}  = 
\left\{ \begin{array}{ccc} 2 \underbrace{\delta_{[A}^C \delta_{B]}^D }_{\texttt{Wronskian}}  &\text{ if }  &(A,B) \in \underbrace{ \left\{ (1,3), (2,4)\right\} }_{\texttt{``good'' pairings}} ,\\
0   &\text{ if }  &(A,B) \in  \left\{ (1,2) , (3,4)\right\}  ,\\
\texttt{not Wronskian} &\text{ if }  &(A,B) \in  \left\{ (1,4), (2,3)\right\} .
\end{array}
\right. 
\eeq
Also in this case, the ``good pairing'', leading to a Wronskian, corresponds to the indices of a couple of $\bP$ functions appearing in a single row of the matrix $\bP_{\hat{a}}^{\;b}$: in our representation the first row contains $\bP_1$ and $\bP_3$, and the second row contains $\bP_2$ and $\bP_4$.

The left-hand side of equations  (\ref{eq:Z2QQrel}),(\ref{eq:Z2starQQrel}) also behaves differently depending on the choice of indices, as dictated by the underlying properties of the matrices. 
$$
\mathbf{C} \cdot \sigma_{[A]} \cdot \bar{\sigma}_{B]} , \;\;\; \hat{\mathbf{C}} \cdot \bar{\sigma}_{[A]} \cdot {\sigma}_{B]} ,
$$
which, corresponding to the index pairings highlighted above, behave as follows:
\begin{itemize}
\item they vanish in the trivial cases where we also get $0$ on the rhs;
\item  in the ``non-Wronskian case'', they are symmetric off-diagonal matrices, so that the lhs will become proportional to the sum of two Wronskians:
$$
\bS_{1|1}^{++} \bS_{2|2}^{--}-\bS_{1|1}^{--} \bS_{2|2}^{++}-\bS_{1|2}^{++} \bS_{2|1}^{--}+\bS_{1|2}^{--} \bS_{2|1}^{++} ,
$$
for (\ref{eq:Z2QQrel}), and the corresponding quantity with anti-spinors for (\ref{eq:Z2starQQrel});
\item
for the special set of “good” index pairings, the matrices become diagonal with a single non-zero eigenvalue. This implies that, for every “good” index pair, there exists a specific spinor index such that the left-hand side is proportional to the corresponding Wronskian combination
$$
\bS_{a|1}^{++} \bS_{a|2}^{--}-\bS_{a|1}^{--} \bS_{a|2}^{++},
$$
for (\ref{eq:Z2QQrel}) (analogously, we would get a Wronskian of anti-spinors from (\ref{eq:Z2starQQrel})). The particular relation between the ``good'' index pair $(A, B)$ and the spinor or anti-spinor index $a$, $\hat{a}$ is very simple: $a$  is the index such that $\bP_A$ and $\bP_B$ appear in row $a$ of $\bP_a^{\;\hat{b}}$ (an analogous definition associates one anti-spinor index $\hat{a}$ to $(A,B)$). 
The index pairings resulting in Wronskian relations are summarised in tables \ref{tab:goodindices} and \ref{tab:goodindicesstar}. 
\end{itemize}
If we focus on such special index pairings, equation (\ref{eq:Z2QQrel}) will give two independent Wronskian equations:
\beq\label{eq:Wronski1}
\bP_{12} = -\bS_{1|1}^{++} \bS_{1|2}^{--} + \bS_{1|2}^{++} \bS_{1|1}^{--} , \;\;\; \bP_{34} = \bS_{2|1}^{++} \bS_{2|2}^{--} - \bS_{2|2}^{++} \bS_{2|1}^{--} ,
\eeq
which obeys the scheme of indices in table \ref{tab:goodindices}. 
\begin{table}
\begin{center}
\begin{tabular}{|c|c|}
\hline
        $(A,B)$ & $a$ \\
        \hline
        (1,2) & 1   \\
        (3,4) & 2  \\
        \hline
    \end{tabular}
    \end{center}
\caption{Choice of ``good'' index pairings resulting in a Wronskian-type relation from (\ref{eq:Z2QQrel}). Notice that $a$ is such that, up to signs and reorderings, $\left\{\bP_A, \bP_B\right\} = \left\{\bP_a^{\;\hat{1}}, \bP_a^{\;\hat{2}}\right\}$. }
\label{tab:goodindices}
\end{table}
From (\ref{eq:Z2starQQrel}), instead, we obtain two independent Wronskian equations:
\beq\label{eq:Wronski2}
\bP_{13} = \hat{\bS}_{\hat{1}|1}^{++} \hat{\bS}_{\hat{1}|2}^{--} - \hat{\bS}_{\hat{1}|2}^{++} \hat{\bS}_{\hat{1}|1}^{--} , \;\;\; \bP_{24} = -\hat{\bS}_{\hat{2}|1}^{++} \bS_{\hat{2}|2}^{--} + \hat{\bS}_{\hat{2}|2}^{++} \bS_{\hat{2}|1}^{--} ,
\eeq
coming from the indices combinations in table \ref{tab:goodindicesstar}. 
\begin{table}
\begin{center}
\begin{tabular}{|c|c|}
\hline
        $(A,B)$ & $\hat{a}$ \\
        \hline
        (1,3) & 1   \\
        (2,4) & 2 \\
        \hline
    \end{tabular}
    \end{center}
\caption{Choice of ``good'' index pairings resulting in a Wronskian-type relation from (\ref{eq:Z2starQQrel}). Here,  $\hat{a}$ is such that, up to signs and reorderings, $\left\{\bP_A, \bP_B\right\} = \left\{\hat{\bP}_{\hat{a}}^{\;{1}}, \hat{\bP}_{\hat{a}}^{\;{2}}\right\}$.}
\label{tab:goodindicesstar}
\end{table}

Such pairings of indices are also seen to be special in equations (\ref{eq:QQrelZ}) and (\ref{eq:QQrelZ2}). In fact, consider the right-hand sides of such equations. 
Consider, for instance, (\ref{eq:QQrelZ2}). For a generic choice of $A$ and $a$, on the right-hand side, we would obtain the sum of two Wronskians, each multiplied by a spinor.
However, precisely when $\bP_A$ appears in row $a$ of $\bP_a^{,\hat{b}}$, an additional cancellation occurs, leaving the right-hand side proportional to a single Wronskian multiplied by a single spinor. In total, the Wronskian cases of (\ref{eq:QQrelZ2}) are as follows (writing the Wronskian of two $\bP$ functions as $\bP_{AB}\mathbf{Q}_{\emptyset}$):
\beqa\label{eq:Wronski3}
&&\bS_{1|i}^{++} \bP_1^- - \bS_{1|i}^{--} \bP_1^+ = \bP_{12}\, \bS_{\hat{1}|i} \, \mathbf{Q}_{\emptyset}, \;\;\; \bS_{1|i}^{++} \bP_2^- - \bS_{1|i}^{--} \bP_2^+ = \bP_{12}\, \bS_{\hat{2}|i} \, \mathbf{Q}_{\emptyset}, \\
&&\bS_{2|i}^{++} \bP_3^- - \bS_{2|i}^{--} \bP_3^+ = -\bP_{34}\, \bS_{\hat{1}|i} \, \mathbf{Q}_{\emptyset}, \;\;\; \bS_{2|i}^{++} \bP_4^- - \bS_{3|i}^{--} \bP_4^+ = \bP_{34}\, \bS_{\hat{2}|i} \, \mathbf{Q}_{\emptyset}.
\eeqa
We see that the indices $(A,B)$ and $a$ of $\bP_{AB}$ appearing on the rhs and $\bS_{a|i}$ appearing on the lhs correspond to the ``good'' cases of table \ref{tab:goodindices}. 
Similarly, the ``Wronskian'' cases of (\ref{eq:QQrelZ}) are:
\beqa
&&\hat{\bS}_{\hat{1}|i}^{++} \bP_1^- - \hat{\bS}_{\hat{1}|i}^{--} \bP_1^+ = \bP_{13}\, \bS_{1|i} \, \mathbf{Q}_{\emptyset}, \;\;\; \hat{\bS}_{\hat{1}|i}^{++} \bP_3^- - \hat{\bS}_{\hat{1}|i}^{--} \bP_3^+ = \bP_{13}\, \bS_{2|i} \, \mathbf{Q}_{\emptyset}, \\
&&\hat{\bS}_{\hat{2}|i}^{++} \bP_2^- - \hat{\bS}_{\hat{2}|i}^{--} \bP_2^+ = -\bP_{24}\, \bS_{1|i} \, \mathbf{Q}_{\emptyset}, \;\;\; \hat{\bS}_{\hat{2}|i}^{++} \bP_4^- - \hat{\bS}_{\hat{2}|i}^{--} \bP_4^+ = \bP_{24}\, \bS_{2|i} \, \mathbf{Q}_{\emptyset} ,\label{eq:Wronski4}
\eeqa
with indices matching the ``good cases'' of table \ref{tab:goodindicesstar}. 

\subsection{Bethe Ansatz equations from the QQ-relations}
\label{sec:Bethe_equations}
In this section, we derive the Bethe Ansatz equations (BAE) for $\ospf$, starting from the QQ-relations obtained above. To do so, we need to write the covariant QQ-relations in components by picking the metric~\eqref{def:sometric}. Each Bethe equation is associated with a node of the Dynkin–Kac diagrams shown in figure~\ref{fig:fig}. The equations are termed bosonic when the corresponding node is even, and fermionic when it is odd.

The BAE for this model are sets of three coupled equations that constrain the zeros of three particular Q functions, chosen appropriately. These equations come in various forms, corresponding to the different Dynkin diagrams.  In the next subsection, we list the BAE associated with each Dynkin diagram separately.

We observe that not all the Q-functions appear in the BAE. In particular, with our particular choice of conventions for the invariant tensors,  it turns out that the Q-functions $\bQ_{12}, \bQ_{21}$ do \emph{not} appear in any simple system of Bethe equations. In fact, consider for example choosing $i=1, \,j=2$ or $i=2, \,j=1$ in the QQ-relations \eqref{eq:relQ1}. This gives the following expression:
\beq
\label{eq:Q12_fromQQ}
\bQ_{12}=\bS_{2|1}^{-}\bS_{1|2}^+-\bS_{1|1}^{-}\bS_{2|2}^+\,,\quad 
\bQ_{21}=\bS_{1|1}^{+}\bS_{2|2}^--\bS_{2|1}^{+}\bS_{1|2}^-\,.
\eeq
The right-hand sides of these expressions are not Wronskian determinants of two Q-functions, but rather involve four different Q-functions. 
Therefore, by manipulating this equation using the standard techniques outlined in the next paragraphs, we do not find an analogous simple system of Bethe equations. A similar phenomenon happens for the other QQ-relations involving $\bQ_{ij}$, namely \eqref{eq:relQ2}, \eqref{eq:F2QQrel} and \eqref{eq:F2QQrel2}.
This already happens in the ABJM case: in that context, there were two additional objects, namely “Q-functions” with incorrect indices, which do not appear in the Bethe equations.
These objects might admit a more rigorous interpretation not as Q-functions, but rather as components of the T-system, as suggested by the form of equation~\eqref{eq:Q12_fromQQ}.
 
Finally, certain Bethe equations will be subject to index restrictions, since, as remarked in the last paragraph of the previous section, the QQ-relations (\ref{eq:Z2QQrel})–(\ref{eq:QQrelZ2}) take a Wronskian form on both sides only for specific index pairings.

\paragraph{Conventions for the indices appearing in the Bethe equations.} 
The sets of Bethe equations derived in the next paragraphs determine the Q-functions placed on a single Dynkin diagram, provided we choose the indices appearing in them consistently. For the diagrams \ref{fig:sfig1} and \ref{fig:sfig2}, this is done by fixing a value for each of the three indices $a,\,\hat a,\,i\in \{1,2\}$. For the diagram \ref{fig:sfig3}, we will need to choose a pattern of indices $A, B$, and $a$ or $\hat{a}$ 
according to tables \ref{tab:goodindices} and \ref{tab:goodindicesstar}. 

\paragraph{BAE for type \ref{fig:sfig1} Dynkin diagram}
To get the BAE associated with the first node of this diagram, we evaluate the fermionic QQ-relations \eqref{eq:QQrels_node1_fermionic} at zeroes of $\bQ_{ii}$, noticing also that the objects $Q_{A|ii}$ appearing on the rhs simply rewrite as a product of a spinor and antispinor according to the definition (\ref{eq:defQAij}). The components of the resulting equations are:
\begin{equation}\label{eq:BAE_1_node1}
    \left(\frac{\bQ_{\emptyset}^-}{\bQ_{\emptyset}^+}\right)\bigg|_{\bQ_{ii}=0}=\left(\frac{\bS_{a|i}^-\bS_{\hat a|i}^-}{\bS_{a|i}^+\bS_{\hat a|i}^+}\right)\bigg|_{\bQ_{ii}=0}\,.
\end{equation}
As anticipated above, the QQ-relations~\eqref{eq:F2QQrel} with $i\neq j$ do not give Bethe equations of the same simple form. The reason is that with this choice of indices, the rhs of (\ref{eq:defQAij}) would not factorise into a product of two spinors. 

Concerning the spinor node in the BAE, we consider the bosonic QQ-relations \eqref{eq:relQ1} with $i=j$, form the ratio of the equations shifted by $\pm i/2$, and evaluate the result at the zeros of $\bS_{a|i}$. We obtain
\begin{align}
\label{eq:BAE_1_node2}
    \left(\frac{\bQ^+_{ii}}{\bQ^-_{ii}}\right)\bigg|_{\bS_{a|i}=0}=\left(-\frac{\bS_{a|i}^{++}}{\bS_{a|i}^{--}}\right)\bigg|_{\bS_{a|i}=0}\,.
\end{align}
For the antispinor node, we do the same with the QQ-relations \eqref{eq:relQ2} with $i=j$, evaluating them at zeros of $\bS_{\hat a|i}$:
\begin{align}
\label{eq:BAE_1_node3}
    \left(\frac{\bQ^+_{ii}}{\bQ^-_{ii}}\right)\bigg|_{\bS_{\hat a|i}=0}=\left(-\frac{\bS_{\hat a|i}^{++}}{\bS_{\hat a|i}^{--}}\right)\bigg|_{\bS_{\hat a|i}=0}\,.
\end{align}
Equations of the form (\ref{eq:BAE_1_node1})-(\ref{eq:BAE_1_node3}) are valid for any $a \in \left\{1,2\right\}$, $\hat{a} = \left\{1,2\right\}$, $i = \left\{1,2\right\}$. 
Choosing one of the 8 possibilities, we would get a set of three coupled Bethe equations for the zeros of three functions of type $\bS_{a|i}$, $\bS_{\hat{a}|i}$ and $\bQ_{ii}$.

\paragraph{BAE for type \ref{fig:sfig2} Dynkin diagram}
To obtain the BAE associated with the first node of this diagram, we evaluate the fermionic QQ-relations \eqref{eq:QQrels_node1_fermionic} at zeroes of $\bP_{A}$. The components of the resulting matrix equation are
\begin{equation}
\label{eq:BAE_2_node1}
     \left(\frac{\bQ_{\emptyset}^-}{\bQ_{\emptyset}^+}\right)\bigg|_{\bP_{\hat a a}=0}=\left(\frac{\bS_{a|i}^-\bS_{\hat a|i}^-}{\bS_{a|i}^+\bS_{\hat a|i}^+}\right)\bigg|_{\bP_{\hat a a}=0}\,,
\end{equation}
where we have defined $\bP_{\hat a a}\equiv \bP_{\hat a}^{\;b}\hat{\mathbf{C}}_{ba}$.

For the spinor node we evaluate the fermionic QQ-relations \eqref{eq:QQrelZ} at the zeros of $\bS_{a|i}$, obtaining:
\begin{equation}
\label{eq:BAE_2_node2}
   \left( \frac{\bS_{\hat a|i}^{++}}{\bS_{\hat a|i}^{--}}\right)\bigg|_{\bS_{a|i}=0}=\left(\frac{\bP_{\hat a a}^+}{\bP_{\hat a a}^-}\right)\bigg|_{\bS_{a|i}=0}\,.
\end{equation}
Similarly, for the antispinor node, we evaluate the fermionic QQ-relations \eqref{eq:QQrelZ2} at zeros of $\bS_{\hat a|i}$, getting:
\begin{equation}
\label{eq:BAE_2_node3}
  1=\left(\frac{\bP_{\hat a a}^+\bS_{a|i}^{--}}{\bP_{\hat a a}^-\bS_{a|i}^{++}}\right)\bigg|_{\bS_{\hat a|i}=0}\,.
\end{equation}
Also for this grading, the equations are valid for any choice of the indices $a$, $\hat{a}$, $i$. Choosing one of the 8 possibilities, we get a set of three coupled Bethe equations for the zeros of three functions of type $\bS_{a|i}$, $\bS_{\hat{a}|i}$ and $\bP_{a\hat{a}}$. 

\paragraph{BAE for type \ref{fig:sfig3} Dynkin diagram}
For the BAE associated with the first node of this diagram, we evaluate the bosonic QQ-relations \eqref{eq:QQrelssp}  at zeros of $\bP_{A}$, shift them by $\pm i/2$ and take the ratio, obtaining:
\begin{equation}
    -\left(\frac{\bQ_{\emptyset}^+}{\bQ_{\emptyset}^-}\right)\bigg|_{\bP_{A}=0}=\left(\frac{\bP_{A}^{++}\bP^-_{AB}}{\bP_{A}^{--}\bP^+_{AB}}\right)\bigg|_{\bP_{A}=0}\,.
\end{equation}
For the second node, we start from the fermionic QQ-relations \eqref{eq:QQrelZ} and \eqref{eq:QQrelZ2}, in the particular case where the rhs contains a single Wronskian, i.e. relations of the form (\ref{eq:Wronski3})-(\ref{eq:Wronski4}).   We evaluate such equations at zeros of $\bP_{AB}$, obtaining:
\begin{equation}
  1=\left(\frac{\bP_{A}^-\bS_{a|i}^{++}}{\bP_{A}^+\bS_{a|i}^{--}}\right)\bigg|_{\bP_{AB}=0}\,, \text{ for indices as in table \ref{tab:goodindices}},
  \end{equation}
  and
  \begin{equation}1=\left(\frac{\bP_{A}^-\bS_{\hat a|i}^{++}}{\bP_{A}^+\bS_{\hat a|i}^{--}}\right)\bigg|_{\bP_{AB}=0}\,, \text{ for indices as in table \ref{tab:goodindicesstar}}.
\end{equation}
Finally, for the third node, we use the bosonic QQ-relations \eqref{eq:Z2QQrel} and \eqref{eq:Z2starQQrel}, again for the particular choices index pairings such that they assume a Wronskian form as in (\ref{eq:Wronski1}),(\ref{eq:Wronski2}). Then, we evaluate them at zeros of $\bS_{a|i}$ and $\bS_{\hat a|i}$ and shift by $\pm i$. Taking the ratio, we get,
\begin{equation}    -1=\left(\frac{\bP^{++}_{AB}\bS_{a|i}^{[-4]}}{\bP^{--}_{AB}\bS_{a|i}^{[+4]}}\right)\bigg|_{\bS_{a|i}=0}\,,  \text{ for indices as in table \ref{tab:goodindices}}\,,
\end{equation}
and
\begin{equation}
    -1=\left(\frac{\bP^{++}_{AB}\bS_{\hat a|i}^{[-4]}}{\bP^{--}_{AB}\bS_{\hat a|i}^{[+4]}}\right)\bigg|_{\bS_{\hat a|i}=0}\,,  \text{ for indices as in table \ref{tab:goodindicesstar}}.
\end{equation}
Equations for this grading are valid only for such specific index pairings. 

We see that we can get either three coupled Bethe equations for functions of the form
$$
\bP_A, \;\; \bP_{AB}, \;\; \bS_{a|i}\,
$$
where we may take $(A,B,a) \in \left\{ (1,2,1),(2,1,1),(3,4,2),(4,3,2)\right\}$, i.e. 4 choices, or we obtain a set of coupled Bethe equations for functions of the kind
$$
\bP_A, \;\; \bP_{AB}, \;\; \bS_{\hat{a}|i},
$$
where we may take $(A,B,\hat{a}) \in \left\{ (1,3,1),(3,1,1),(2,4,2),(4,2,2)\right\}$ (other 4 choices). 

In total, we see that we can obtain 8+8+8 alternative sets of Bethe equations, eight for each grading. This matches the result of \cite{Tsuboi:2023sfs}, see equations (4.133)-(4.135). 

\section{Proposal for the Quantum Spectral Curve at $\alpha = \frac{1}{2}$ }\label{sec:QSCProposal}
In the previous section, we have described the structure of the $\ospf$ Q-system. We will use this as a basis to make a proposal for a QSC for the \ads\, string theory at $\alpha = \frac{1}{2}$. 
The first obvious step is that we should have two copies of the $\osp(4|2)$ Q-system we have just described, since the algebra governing the string theory is $\osp(4|2)_L \oplus \osp(4|2)_R$. 
We will distinguish the two copies by putting dotted indices on one of them, e.g.
\beq
\bP_{A} \equiv ( \bP_A )_L , \;\;\;\slP_{\dot A} \equiv ( \bP_{\dot A} )_R ,
\eeq
and similarly for all the other Q-functions.
Adopting the terminology commonly used in the literature, we refer to the two halves of the system as the left (L) and right (R) wings, also known as the \emph{undotted} and \emph{dotted}  wings, respectively.

Secondly, the $\osp(4|2)$ Q-system in general also contains the Q-function $\bQ_{\emptyset}$, which is expected to determine completely the trivial T-functions on the T-hook $\mathbf{T}_{0,s}$ and $\mathbf{T}_{a,0}$, thus, it is an important object that should be determined from representation-theoretic considerations.

Inspired by all known cases of AdS/CFT QSC, we make the following assumption to describe the \ads \,string theory:
\begin{framed}
\texttt{ Assumption:  $\bQ_{\emptyset}=1$ in both wings.}
 \end{framed}
 This simplifies some of the relations, in particular, we report here the main equations \eqref{eq:defQ}, \eqref{eq:constraintP}, \eqref{eq:constraintQai}, \eqref{eq:constraintQij} in matrix form
 \begin{align}
    & \bQ=(\bS^-)^t\cdot \hat\bP^t\cdot  \hat{\mathbf{C}}\cdot  \hat \bS^-=(\hat\bS^+)^t\cdot  \bP^t \cdot \, {\mathbf{C}}^t \cdot  \bS^+\,,\\&
    \text{det}(\bP) = 1,\quad
    \text{det}(\bS) = -1= -\text{det}(\hat \bS),\quad
    \text{det}(\bQ)=-1\,.
 \end{align}
Finally, to fully specify the QSC, we need to describe the analytic properties and asymptotics of the Q-functions.
Following the logic of \cite{Ekhammar:2021pys,Cavaglia:2021eqr}, we will conjecture these properties, trying to adhere as much as possible to the example of $\mathcal{N}$=4 SYM and ABJM theory, and then deduce all the consequences of the construction. 
\subsection{Analytic properties }
\label{sec:cuts_asymptotics}
\subsubsection{Minimal description of analyticity}

\begin{figure}[htbp]
    \centering
    \begin{minipage}{0.45\textwidth}
        \centering
        \includegraphics[width=\linewidth]{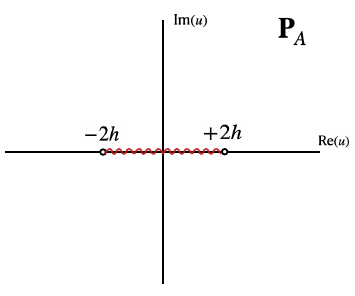}
        \caption{$\bP_A$ in the Riemann sheet with a single short cut.}
        \label{fig:Pcuts}
    \end{minipage}\hfill
    \begin{minipage}{0.45\textwidth}
        \centering
        \includegraphics[width=\linewidth]{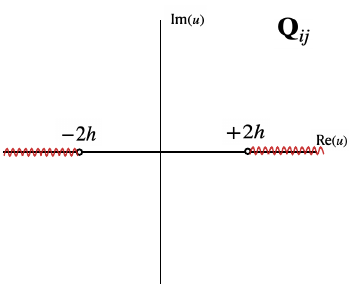}
        \caption{$\bQ_{ij}$ in the Riemann sheet with a single long cut.}
        \label{fig:Qlongcuts}
    \end{minipage}
\end{figure}

\paragraph{The minimal cuts structure. }
Some of the Q-functions, i.e. the ones which we denoted as   $\bP$ and $\bQ$ in the Q-system, are special. They are the ones expected to have the simplest analytic properties, i.e. they are the only ones admitting a Riemann sheet with only two branch points. The position of the branch points will be parametrised by a constant $h$, which should encode the string tension of the model\footnote{We are in the planar limit, so $g_s = 0$, and the only relevant parameter is the string tension/'t Hooft coupling. It is expected that this coupling might be dressed into a nontrivial interpolating function $h$. }.

In particular, we propose the following assumptions, which are valid in both the $L$ and $R$ wings. Here we state them for the $L$ wing:
\begin{itemize}
\item there is a special Riemann sheet where the functions $\bP_A(u)$, $A = 1,\dots,4$ 
 have a single \emph{short} branch cut $u \in [-2 h, + 2 h ]$ on the main Riemann sheet. 
\item on the other hand, the functions $\bQ_{mn}$ admit a Riemann sheet where they have a single \emph{long} branch cut $u \in (-\infty, -2 h]\cup[2h, +\infty]$ 
\item By convention, we identify the two special Riemann sheets described above in the upper half plane.
\item $\bP$ and $\bQ$ should have no other singularities on these sheets.  
\end{itemize}
An illustration of the cuts for $\bP$ and $\bQ$ functions on these special sheets is given in figures \ref{fig:Pcuts} and \ref{fig:Qlongcuts}.   

The parameter $h$ given above should parametrise the coupling (i.e., the string tension) of the theory. 

We also anticipate that Q-functions will have power-like asymptotics for $u\rightarrow + \infty$, of the type
\beq
\bP_A(u) \sim u^{-M_A}, \;\;\;\; \bQ_{ij}(u) \sim u^{\bar{M}_{ij}-1}.
\eeq
 The precise map between the powers in the asymptotics $M_A,\, \dot M_A,$ $\bar{M}_{ij},\, \dot{\bar{M}}_{ij}$ and the quantum numbers of the operators in the dual CFT$_2$ will be described in section \ref{sec:asymptotics}. 

\paragraph{Sections with infinitely many cuts. }
These axioms are still not complete - the next crucial step will be a rule to glue the $L$ and $R$ wings together - but already the previous rules are in some tension with the QQ-relations.

\begin{figure}[htbp]
    \centering
    \begin{minipage}{0.45\textwidth}
        \centering
        \includegraphics[width=\linewidth]{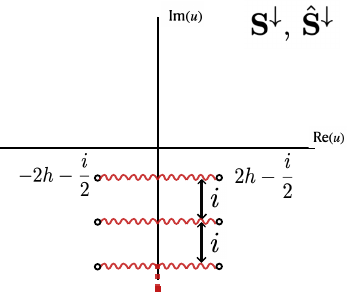}
        \caption{Analytic structure of $\bS^{\downarrow}$, $\hat\bS^{\downarrow}$.}
        \label{fig:SUHP}
    \end{minipage}\hfill
    \begin{minipage}{0.45\textwidth}
        \centering
        \includegraphics[width=\linewidth]{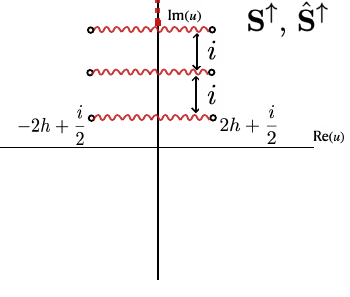}
        \caption{Analytic structure of $\bS^{\uparrow}$, $\hat\bS^{\uparrow}$.}
        \label{fig:SLHP}
    \end{minipage}
\end{figure}

In fact, consider how one could use the core relations and the definition \eqref{eq:defQ} as a system of equations that allows us to \emph{build $\bQ_{ij}(u) $ once we are given $\bP_A(u)$}. We do this in two steps: first, solve \eqref{eq:core_relations} to determine $\bS$ and $\hat{\bS}$, and then use (\ref{eq:defQ}) to compute $\bQ$. The first step is the only one that is nontrivial, since the equation is nonlocal; however, it can be done efficiently, e.g., numerically using the algorithm in \cite{Gromov:2015wca}. This allows us to reconstruct the $\bS$, $\hat{\bS}$ functions\footnote{We are being somewhat schematic here. To fix the functions, one should also specify their asymptotics, which are defined in section \ref{sec:asymptotics}. Once this is done, the functions $\bS$ and $\hat\bS$ would be fixed uniquely, apart for some residual symmetry of the Q-system which can be fixed arbitrarily.} on a Riemann sheet where \emph{necessarily} (due to the Q-system finite difference equations in steps of $i$) the single cut of the $\bP$ functions will be replicated many times into an infinite or semi-infinite ladder of cuts. In fact, we can choose to either have a basis of solutions where all these cuts are in the lower half plane, at $u \in [-2h, 2h] -\frac{i}{2} - i \mathbb{N} $, or in the upper half plane at  $u \in [-2h, 2h] +\frac{i}{2} + i \mathbb{N} $. These two bases of solutions are respectively denoted as $\bS^{\downarrow}$ or $\bS^{\uparrow}$, and they are illustrated in figures \ref{fig:SUHP} and \ref{fig:SLHP}. The connection between these two bases of solutions will be explained below.

\begin{figure}[htbp]
    \centering
    \begin{minipage}{0.45\textwidth}
        \centering
        \includegraphics[width=\linewidth]{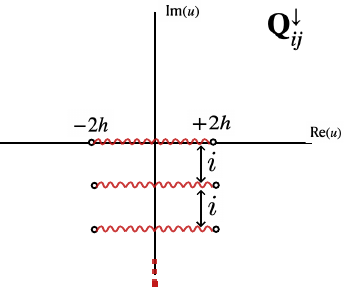}
        \caption{Analytic structure of $\bQ^{\downarrow}_{ij}$.}
        \label{fig:QUHP}
    \end{minipage}\hfill
    \begin{minipage}{0.45\textwidth}
        \centering
        \includegraphics[width=\linewidth]{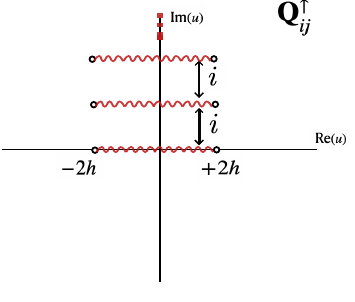}
        \caption{Analytic structure of $\bQ^{\uparrow}_{ij}$.}
        \label{fig:QLHP}
    \end{minipage}
\end{figure}

From these two bases, respectively, we can construct two versions of $\bQ_{ij}$ through (\ref{eq:defQ}). These -- denoted as $\bQ^{\downarrow}$ and $\bQ^{\uparrow}$ -- have respectively an infinite ladder of cuts at $[-2 h, 2h] - i \mathbb{N}$ and $[-2 h, 2h] + i \mathbb{N}$, see figures \ref{fig:QUHP} and \ref{fig:QLHP} respectively. 

At this point, we encounter a tension with the axiom stated above — namely, that the $\bQ$-functions should possess a Riemann sheet with a single \emph{long}  cut. How can this be reconciled with our findings?
\subsubsection{Gluing the two copies}
As in all other examples of the QSC, to resolve this tension, we impose that the $\bQ$-function with long cuts corresponds to a different Riemann sheet of those shown in Figures~\ref{fig:QUHP} and~\ref{fig:QLHP}, possibly up to a redefinition induced by a Q-system symmetry. This identification effectively amounts to gluing together the corresponding Riemann sections.

However, while we have described properties that are valid separately in both wings up to this point, we should remember that the full Q-system for the system we want to describe contains both halves.
Just like in \cite{Ekhammar:2021pys,Cavaglia:2021eqr},  the gluing of Riemann sheets we want to describe also gives us just a minimalistic opportunity to make these two wings communicate with each other, by imposing that they are related by analytic continuation.
\begin{figure}[htbp]
    \centering
    \begin{minipage}{0.45\textwidth}
        \centering
        \includegraphics[width=\linewidth]{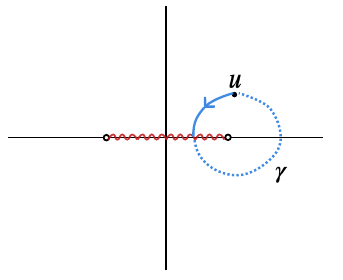}
        \caption{Path of analytical continuation $\gamma$. The dashed line indicates that the path is in the  Riemann sheet connected to the one depicted by the branch cut.}
        \label{fig:gamma}
    \end{minipage}\hfill
    \begin{minipage}{0.45\textwidth}
        \centering
        \includegraphics[width=\linewidth]{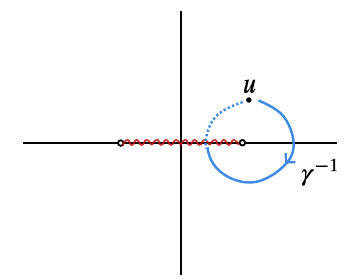}
        \caption{Path of analytical continuation $\gamma^{-1}$. The dashed line indicates that the path is in the  Riemann sheet connected to the one depicted by the branch cut.}
        \label{fig:gammam1}
    \end{minipage}
\end{figure}
\begin{figure}[htbp]
\centering
        \includegraphics[width=\linewidth]{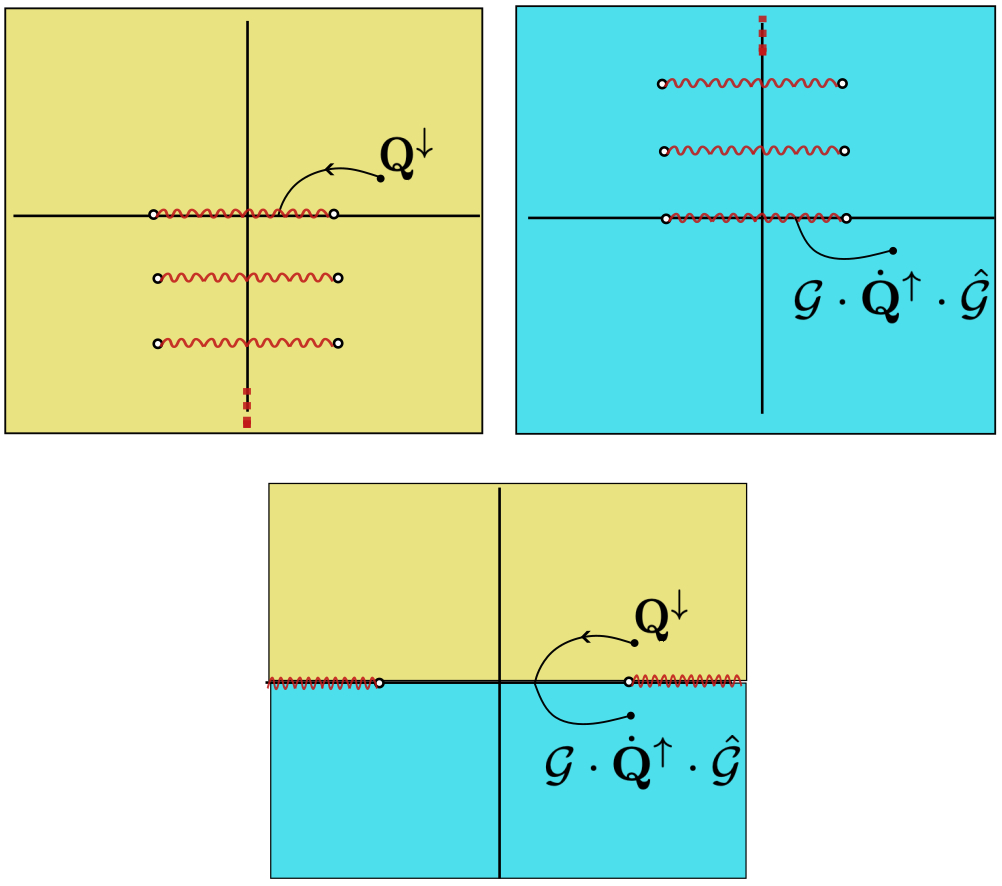}
        \caption{A  pictorial representation of the equivalence of the gluing equation \eqref{eq:gluing1} and having $\bQ$ functions with long branch cuts.}
        \label{fig:Qgluing1}
\end{figure}
This can be described as follows: if we start from $\bQ^{\downarrow}$ and continue around one of its branch points from above, we should emerge \emph{in the other wing} on the sheet of the Q-functions $\slQ^{\uparrow}$. We depicted this in figure \ref{fig:Qgluing1}.
Allowing for a recombination of the components, this is written in matrix form as
\begin{framed}
\beqa 
\label{eq:gluing1}
&(\bQ^{\downarrow}(u))^{\gamma}=\mathcal{G}\cdot \slQ^{\uparrow}(u) \cdot \mathcal{G}^t\,,\\
\label{eq:gluing2}
&(\slQ^{\downarrow}(u))^{\gamma}=\dot{\mathcal{G}}\cdot \bQ^{\uparrow}(u) \cdot \dot{\mathcal{G}}^t\,.
\eeqa
\end{framed}
Here, $\gamma$ is the analytic continuation along a path that encircles the branch points $u=+2h$ anticlockwise or, equivalently, $u=-2h$ clockwise as in figure \ref{fig:gamma} (the opposite path, $\gamma^{-1}$, which will be needed in some of the following, is illustrated in figure \ref{fig:gammam1}).  
The $2 \times 2$ matrices $\mathcal{G}$ and $\dot{\mathcal{G}}$  are called gluing matrices. 
The structure of the above equation deserves some comments. 

First, we allowed for the most general linear transformation of the $\bQ$ functions on the RHS by multiplying them by two independent gluing matrices, which allow for a reshuffling of their asymptotics. In principle, they may even be different for the two indices of $\bQ_{ij}$, e.g., for the first of the equations above, we may write 
 \beq
\bQ^{\downarrow}(u^{\gamma})=\mathcal{G}(u)\cdot \slQ^{\uparrow}(u) \cdot \hat{\mathcal{G}}^t(u)\, ,
\eeq
with independent $\mathcal{G}$ and $\hat{\mathcal{G}}$. 
By compatibility with the analytic structure of the $\bQ$ functions on both sides, we immediately find that these gluing matrices should have no singularities: they are entire functions. 

Moreover, the transformations should leave the Q-system invariant under their action.
In particular, looking at equation (\ref{eq:defQ}), we see that the transformation 
$\bQ \rightarrow  \mathcal{G}\cdot \slQ \cdot \hat{\mathcal{G}}^t$ should be accompanied by $\bS(u) \rightarrow \bS(u) \cdot \mathcal{G}^+(u)$, $\hat{\bS}(u) \rightarrow \hat{\bS}(u) \cdot  \hat{\mathcal{G}}^+(u)$, and the core relations \eqref{eq:core_relations} then give the constraint 
\beq
\label{eq:GGhatrel}
\mathcal{G}(u + i) = \hat{\mathcal{G}}(u) =\mathcal{G}(u - i) , 
\eeq
 furthermore, from (\ref{eq:constraintQai}) 
 we also obtain the constraints \beq\label{eq:constraintsG}
 \text{det}(\mathcal{G}) = \text{det}(\hat{\mathcal{G}})  = 1, \;\;\; \mathcal{G} \cdot \epsilon^{-1}\cdot  \mathcal{G}^t = \hat{\mathcal{G}} \cdot \epsilon^{-1}\cdot   \hat{\mathcal{G}}^t = \epsilon^{-1}  .
 \eeq
In this paper, we are interested in the case of the spectrum of local operators of the dual CFT$_2$. In this case, we expect the gluing matrix to be constant, while it would acquire a $u$ dependence, while remaining an entire periodic function, if we consider analytic continuation in the spin~\cite{Alfimov:2014bwa, Klabbers:2023zdz, Brizio:2024nso}. Under this assumption, equation~\eqref{eq:GGhatrel} simplifies to $\mathcal{G} =     \hat{\mathcal{G}} $, and so the gluing reduces to the one written in~\eqref{eq:gluing1} with a single gluing matrix in each wing. 
The only constraints remaining are (\ref{eq:constraintsG}), valid for both wings. 
As a final comment, we notice that the gluing matrices put in contact one wing with the other. 
If we write them with explicit indices, they would have the forms $\mathcal{G}_a^{\; \dot b}$ and $\mathcal{\dot G}_{\dot a}^{\; b}$. 
\paragraph{Additional requirement on the structure of the gluing matrix.  } We expect the gluing matrices of the theory to be off-diagonal. 
 Given their constraints, they should then have the form
 \beq\label{eq:gluingform}
\mathcal{G} = \left( 
\begin{array}{cc}
0 & \beta \\
-\frac{1}{\beta} & 0  
\end{array} \right), \;\;\; \dot{\mathcal{G}} = \left( \begin{array}{cc}
0 & \dot\beta \\
-\frac{1}{\dot\beta} & 0  
\end{array}\right) .
 \eeq
We do not have a full argument for this, but it is an assumption that seems to simplify the ABA limit. In \cite{Cavaglia:2022xld}, a similar requirement was proposed to be associated with the condition that the Q-functions remain finite at the branch points. 
\subsection{Asymptotics and charges}\label{sec:asymptotics}
So far, we have only prescribed that, in both copies, the Q-functions should have powerlike asymptotics as $u\rightarrow \infty$. We make this precise now, proposing a link between the powers in the asymptotics and the quantum numbers of the $\osp(4|2)_L \oplus \osp(4|2)_R$ multiplet described by the solution to the QSC. 

Each of these multiplets can be labelled by two sets of $\osp$ labels, which we denote as $$
\underbrace{[J_1, J_2, \Delta_L]}_{\texttt{L wing}}, \;\;\; \underbrace{[\dot{J}_1, \dot{J}_2, \Delta_R]}_{\texttt{R wing}},
$$
where $J_1$, $J_2$ (resp., $\dot{J}_1$, $\dot{J}_2$ in the other wing) are (twice) the spins of the $SU(2)\times SU(2)$ R-symmetry group, and $\Delta_L$ (resp., $\Delta_R$) is a weight of the $SL(2, \mathbb{R})$  left/right conformal subgroup.\footnote{The two $SL(2, \mathbb{R})$ subgroups of the two wings are interpreted as holomorphic and anti-holomorphic generators of the global conformal group of the CFT$_2$. Notice that we choose a normalisation such that $\Delta_L$ ($\Delta_R$) is twice the eigenvalue of the Virasoro generator $L_0$ ($\bar{L}_0$) in the CFT$_2$. } 

We defined these weights on the operators with the lowest value of $\Delta_{L/R}$ in the multiplet (the ``top'' component). 
The scaling dimension and spin of the CFT$_2$ operator corresponding to a superconformal primary in both wings are
\beq\label{eq:DeltaSpin}
\Delta_{\text{top}} = \frac{1}{2}\left(\Delta_L + \Delta_R \right), \;\;\; S_{\text{top}} = \frac{1}{2}\left(\Delta_L - \Delta_R \right).
\eeq
We now describe in detail the asymptotic proposed for the QSC and the dictionary for these quantum numbers.
\subsubsection{Structure of the asymptotics}
We begin by describing the asymptotics in a single wing, as the second wing will have the same structure, but with different quantum numbers. 

First, we assume that the $\bP$ functions have powerlike asymptotics
\beq\label{eq:Pasy}
\bP_A(u) \simeq \mathcal{A}_A \; u^{-M_A} , 
\eeq
at $u \rightarrow + \infty$. This is the asymptotic limit considered in this section, which is often implied when left implicit. The quantum numbers  $M_A$ above are
necessarily integers (because the $\bP$ functions have a single cut, and thus $\infty$ on the first sheet is not a branch point), and will be identified with the R-charges of the supermultiplet. To satisfy \eqref{eq:constraintP}, we postulate that these numbers satisfy the constraints
\beq
M_4 = -M_1, \;\;\; M_3 = -M_2 . \;\;\; 
\eeq
The coefficients $\mathcal{A}_A$ in front of the leading power in (\ref{eq:Pasy}) satisfy some constraints, which are described below. Choosing a conventional order for the magnitude of the components of the $\bP$ functions, we will take, without loss of generality
\beq
M_1 > M_2 \geq 0 , \;\;\;\; M_1, \,M_2 \in \mathbb{N} ,
\eeq
where we notice that different Q-functions should have distinguishable asymptotics for physical states at finite coupling. 

It is then simple to fix the asymptotics of the spinorial Q-functions $\bS$ and $\hat{\bS}$, if we demand that the core relations, once expanding the matrix product, give a combination of terms having all the same power behaviour at large $u$. 
 Following this argument, we fix
\beq\label{eq:bSasy}
\bS_{a |i}(u) \simeq B_{(a|i)} u^{{N}_{a} + \bar{{N}}_i } ,\;\;\;\;\;\; 
\hat \bS_{\hat a |i}(u) \simeq B_{(\hat a|i)} u^{{N}_{\hat a} + \bar{{N}}_i }  ,
\eeq
with
\beq
N_a = \frac{1}{2}\left( -M_1 - M_2 , M_1 + M_2 \right) , \;\; \hat N_{\hat a} = \frac{1}{2}\left( -M_1 + M_2 , M_1 - M_2  \right). 
\eeq
In addition, there are constraints on the prefactor coefficients $\mathcal{A}_A$, $B_{(a|i)}$ and $B_{(\hat a|i)}$, in each wing, coming from the core relations (\ref{eq:core_relations}) and the constraints (\ref{eq:constraintP}), (\ref{eq:constraintQai}).
In particular, the unit determinant condition for the spinors imposes
\beq
\bar{N}_2 = -\bar{N}_1 , \;\;\; \dot{\bar{N}}_2 = -\dot{\bar{N}}_1 ,
\eeq
and, furthermore, we have 
\beq\label{eq:Arels}
\mathcal{A}_2 \mathcal{A}_3 = \frac{(M_2^2 - \bar{M}_1^2 )}{(M_2^2-M_1^2)}, \;\; \; \mathcal{A}_1 \mathcal{A}_4 = \frac{(M_1^2 - \bar{M}_1^2 )}{(M_1^2-M_2^2)}.
\eeq
Finally, through (\ref{eq:defQ}) we can determine the resulting asymptotics of the $\bQ$ functions,
\beq
\label{def:Mhats}
\bQ_{11}(u)\simeq \mathcal{B}_{11} u^{ \bar{M}_1-1 }, \;\;\; \bQ_{22}(u)\simeq \mathcal{B}_{22}  u^{-\bar{M}_1 -1 } , \;\;\; \bQ_{12}(u) \simeq -1 
, \;\; \bQ_{21}(u) \simeq 1 ,
\eeq
where the powers in the asymptotics are fixed in terms of
\beq
\bar{M}_1 = 2 \bar{N}_1,
\eeq
and with coefficients constrained as
\beq\label{eq:Brels}
\mathcal{B}_{11} \mathcal{B}_{22} = -  \frac{(\bar{M}_1^2 - M_1^2 ) (\bar{M}_1^2 - M_2^2 )}{4\bar{M}_1^2} .
\eeq
In short, the asymptotics of the first wing are parameterised in terms of three independent numbers, i.e. $M_1, \; M_2 \in \mathbb{N}$ and $\bar{M}_1 \in \mathbb{R}$. 
Moreover, physical solutions are expected to satisfy the constraint
\beq
\bar{M}_1 \geq M_1 > M_2 \geq 0 .
\eeq
We expect the saturation of a BPS bound when $\bar{M}_1 = M_1$. In fact, one can see from (\ref{eq:Arels}), (\ref{eq:Brels}) that such points are special, as they imply the vanishing of some of the $\bP$ and $\bQ$ functions, which usually happens at BPS points, around which the QSC becomes solvable explicitly. 

For the second wing, the structure is formally identical, with the asymptotics being parametrised by three a priori independent quantum numbers, which we shall denote as ${\dot M}_{1} $, ${\dot M}_{2} $, and $\dot{\bar{M}}_1$. 

\paragraph{Conventions for ``pure asymptotics''. }
In this paragraph, we provide some technical comments to define our basis for Q-system solutions precisely. 

By convention, the asymptotic behaviours specified above are valid when we take the limit $u \rightarrow +\infty + i 0^+$, i.e. we approach positive real infinity keeping slightly above the real axis (to avoid potential ambiguities even when we treat functions which have a ``long'' cut on the real axis). When we construct the bases of functions analytic in the upper/lower half of the complex plane, we impose in both cases the asymptotic behaviour along the same line. 

For example, when we construct  $\bS^{\downarrow}$ vs $\bS^{\uparrow}$ as alternative solutions of (\ref{eq:core_relations}), in both cases we demand the asymptotic behaviour given by the first equation in (\ref{eq:bSasy}) in the limit $u \rightarrow + \infty+ i 0^+$. 

Notice also that the coefficients $B_{(a|i)}$ 
are defined up to some freedom. In principle, we could choose them to be different for the two bases of solutions, i.e. $B_{(a|i)}^{\downarrow}$ vs $B_{(a|i)}^{\uparrow}$. However, we can always normalise the solutions such that $B_{(a|i)}^{\downarrow} = B_{(a|i)}^{\uparrow}$, which is what we will always assume.

Finally, and most importantly, to specify the basis of solutions unambiguously, we require that subleading corrections to the asymptotics of the Q-functions appear only in integer powers of $1/u$, a condition referred to as ``pure asymptotics''~\cite{Gromov:2015vua}. 

This eliminates the freedom to redefine $\bS_{(a|i)}$ by recombining different columns of the matrix, and it proves useful for simplifying the asymptotic behaviour of certain quantities that will be introduced later. 

\subsubsection{Identification of the quantum numbers}
\label{sec:quantum_numbers}
In this section, we give the asymptotics of the Q-functions first in terms of the labels of the supermultiplet, and then in terms of the number of Bethe roots appearing in the Asymptotic Bethe Ansatz of \cite{Borsato:2012ss}, which we will analyse in more detail in section \ref{sec:ABA}.
\paragraph{Identification in terms of charges of the supermultiplet. } Taking inspiration from the ABJM case, we identify them as 
\beq
 M_1 \equiv 1 + \frac{1}{2}(J_1 + J_2 ), \;\; M_2 \equiv \frac{1}{2} (J_1 - J_2), \;\;\; \bar{M}_1 \equiv \Delta_L -1 ,
\eeq 
where $J_1$, $J_2$, $\Delta_L$ are defined above. 

A hint that this is the correct identification is that 
the transformation $\bar{M}_1 \rightarrow  -\bar{M}_1$, which reshuffles some of the Q-functions, corresponds to $\Delta_L \rightarrow  2 - \Delta_L$, which is an invariance of the Casimir eigenvalue $\mathcal{C} = - \Delta_L (\Delta_L-2)+ \frac{J_1}{2}(J_1+2)+\frac{J_2}{2}(J_2+2) $, see \cite{Cagnazzo:2014yha}.\footnote{The labels used in this paper are mapped to our labels as $j_1 \equiv \Delta_L/2$, $j_2\equiv J_1/2$, $j_3=J_2/2$.} Similarly, by relabelling the $\bP$ functions, one can get solutions with $J_1\rightarrow -2 - J_1$, $J_2\rightarrow -2 - J_2$. These transformations also leave the Casimir eigenvalue unchanged. Finally, also the saturation of the condition $M_1 = \bar{M}_1$ corresponds precisely to the shortening condition $\Delta_L = 2 + \frac{1}{2}(J_1 + J_2)$~\cite{Cagnazzo:2014yha}.

The asymptotics in the other wing would have the same structure, with the asymptotics labelled by independent quantum numbers, which we mark with a dot, e.g. $\dot M_A$, $\dot{\bar{M}}$. These are identified as follows:
\beq
 \dot M_1 \equiv 1 + \frac{1}{2}(\dot J_1 + \dot J_2 ), \;\; \dot M_2 \equiv \frac{1}{2} (\dot J_1 - \dot J_2), \;\;\; \dot{\bar{M}}_1 \equiv \Delta_R -1 .
\eeq 

\paragraph{Correspondence with Asymptotic Bethe Ansatz quantities. }
In \cite{Borsato:2012ss}, Asymptotic Bethe Ansatz equations for this theory were proposed. In this formalism, which arises in a limit of large quantum numbers, the natural quantities labelling multiplets are the numbers of Bethe roots of various kinds and the integer \emph{length} parameter $L$. 
 Here, we provide a dictionary mapping the ABA labels to the quantum numbers of the supermultiplets introduced above.
 We will focus on the case where only massive-type Bethe roots appear. The ABA equations are reported in section \ref{sec:ABA} (where we will rederive them from our QSC), see for example (\ref{eq:ABA1}). 

They involve the following types of roots. 
The roots in the first wing are denoted as\footnote{The relation with the notation of \cite{Borsato:2012ss,Borsato:2015mma} is:
\beqa &&
s\leftrightarrow 1 , \;\;\hat{s}\leftrightarrow 3 , \;\;\; f \leftrightarrow 2, \nonumber\\
&& \dot s \leftrightarrow \bar{1} , \;\;\hat{\dot{s}}\leftrightarrow \bar{3} , \;\;\; \dot b \leftrightarrow \bar{2} .\nonumber
\eeqa
}
\beq
\left\{ u_{s,i} \right\}_{i=1}^{K_s}, \; \;\; \left\{ u_{\hat{s},i} \right\}_{i=1}^{K_{\hat s}} ,\;\;\; \left\{ w_{f,i} \right\}_{i=1}^{K_f},
\eeq
where $s$ and $\hat{s}$ label {\it momentum-carrying roots} and $f$ is a label for {\it auxiliary roots}. The numbers of roots appearing in the second wing are denoted, analogously, as
\beq
K_{\dot s}, \; K_{\hat{\dot s}} , \; K_{\dot b} ,
\eeq
where $\dot s$ and $\hat{\dot{s}}$ denote {\it momentum-carrying roots} and $\dot b$ denotes {\it auxiliary roots}. They refer to the Bethe Ansatz written in the grading related to the Dynkin diagram \ref{fig:sfig2} for the first wing and \ref{fig:sfig1} for the second wing. 

In terms of these ABA quantities,  our quantum numbers are given by 
  \beqa
M_1 &=& L + K_f + 1 - K_s - K_{\hat{s}} , \;\; M_2 = K_{\hat{s}} - K_s ,\\
\bar{M}_1 &=& L + K_f + {\delta\Delta}  + 1\,,
\eeqa
and
\beqa
\dot M_1 &=& L - K_{\dot b}, \;\; \dot M_2 = K_{\hat{\dot{s}}} - K_{\dot s} ,\\
\dot{\bar{M}}_1 &=& L + K_{\dot s} + K_{\hat{\dot s}} - K_{\dot b} + {\delta\Delta}  ,
\eeqa
where $\delta\Delta$ denotes the \emph{anomalous dimension}, i.e. the part of the scaling dimension which runs with the coupling and vanishes at $  h\rightarrow 0$. In the ABA limit, it is related to the energy of elementary excitations through
$$
\sum_{\bullet = s,\hat{s}, \dot s, \hat{\dot s}} \sum_{i=1}^{K_{\bullet}} \epsilon( u_{\bullet, i} ) \equiv \frac{1}{2} \sum_{\bullet} K_{\bullet} + \delta \Delta,
$$
where $\epsilon(u)$ is the dispersion relation for massive particles, see also section \ref{sec:ABA} for more details.
 
In section \ref{sec:ABA}, we will deduce the ABA equations starting from the limit of our QSC equations in the limit where the charges $J_1, \dot J_1, \Delta_L, \Delta_R $ scale to $\infty$ with the same rate, which corresponds to $L\rightarrow \infty$. 
\subsection{The $\bQ\tau$ and $\bP\nu$ systems}\label{sec:QtauPnu}
In this section, we will derive from our gluing equations \eqref{eq:gluing1} some exact equations describing the analytic continuation of Q-functions through their branch cuts on the real axis, which we will call $\bQ\tau$ and $\bP\nu$ systems in analogy with the cases of \cite{Gromov:2013pga,Cavaglia:2014exa}. 
The former represents the gluing between $\bQ$ functions on different wings but taking the \emph{same} branch cut structure in both wings, while the latter is the gluing equation for the $\bP$ functions. In both cases, the gluing involves new objects, dubbed $\tau$ and $\nu$ functions, respectively\footnote{These are the analogues of $\omega$ and $\mu$ in the case of $\mathcal{N}$=4 SYM~\cite{Gromov:2013pga}.}. They are not Q-functions, but can be seen as nontrivial Q-system transformations that describe the monodromy around the branch points. As such, they are characterised by peculiar periodicity properties which we will deduce (see \cite{Gromov:2014caa} for a discussion).

These formalisms will be very useful for studying the large-volume limit of the QSC equations in section \ref{sec:ABA}. 
\subsubsection{The $\bQ\tau$-system}
The $\bQ\tau$-system is a set of functional equations that allow to relate $\bQ$ functions $\bQ^{\downarrow}$ to $\slQ^{\downarrow}$ (with the same convention to be free of cuts in the upper half plane) via some coefficients built out of a nontrivial $2 \times 2$ matrix denoted as $\tau(u)$. 
We already know that Q-functions on different wings and different branch cut structures, $\bQ^{\uparrow}$ and $\slQ^{\downarrow}$, are connected via the gluing equations (\ref{eq:gluing1}). To construct  $\tau$, we need to consider the composition of the gluing matrices with another type of rotation tensor, that will connect the two bases of solutions with different branch cuts on the \textit{same} wing, i.e. $\bQ^{\uparrow}$ to $\bQ^{\downarrow}$ (or $\slQ^{\uparrow}$ to $\slQ^{\downarrow}$).

\paragraph{Relating the LHPA and UHPA solutions. }
We will build the rotation tensor for $\bQ$ starting from a similar object for spinors and antispinors, which we call $\Omega, \hat{\Omega}$, which are matrices connecting respectively UHPA and LHPA spinors and antispinors:
\beq
\label{def:Omegadef}
\bS^{\uparrow}=\bS^{\downarrow}\cdot(\Omega^+)^t\,,\qquad \hat\bS^{\uparrow}=\hat\bS^{\downarrow}\cdot(\hat{\Omega}^+)^t\,.
\eeq
These objects always exist, given the fact that both $\bS^{\uparrow}$ and $\bS^{\downarrow}$ satisfy the same finite-difference equations, i.e. the core relations \eqref{eq:core_relations}. 
It is also evident from their definition that all the components of the $\Omega$ matrices have an infinite ladder of short branch cuts for $u\in [-2h,2h  ]\pm i \mathbb{N}$.

Furthermore, due to the compatibility with the core relations and constraints \eqref{eq:constraintQai}, we must have that these matrices satisfy the same constraints as the gluing matrices, namely the ones ensuring they leave the Q-system invariant, i.e.
\beqa
&\label{eq:Omegaperiodicity} \Omega=\hat{\Omega}^{++}=\Omega^{[+4]}\,,\\
& \Omega \cdot \epsilon^{-1} \cdot\Omega^t = \epsilon^{-1} , \;\;\; \det\,\Omega= 1.
\eeqa
All the preceding construction has a direct analogue in the other wing, which leads to the construction of two matrices $\dot \Omega,\, \hat{\dot \Omega}$ satisfying identical constraints. Notice that these matrices have indices pertaining to a single wing, e.g. $\Omega$ has two undotted indices, while $\dot \Omega$ has two dotted indices.

Using the QQ-relations \eqref{eq:relQ1} (or equivalently \eqref{eq:relQ2}) and the periodicity properties \eqref{eq:Omegaperiodicity}, it is immediate to see that $\Omega$ and $\hat\Omega$ relate $\bQ^{\uparrow}$ and $\bQ^{\downarrow}$, with $\dot\Omega$ and $\hat{\dot\Omega}$ doing the same on the $R$ wing:
\beqa
&\bQ^{\uparrow} = \Omega \cdot \bQ^{\downarrow} \cdot \hat{\Omega}^t\,,\\
&\slQ^{\uparrow} = \dot{\Omega} \cdot \slQ^{\downarrow} \cdot \hat{\dot{\Omega}}^t\,. 
\eeqa
\paragraph{Rewriting the gluing equations. }
From the $\Omega$ and $\mathcal{G}$ matrices, we can immediately get the gluing equation for $\bQ$ with the same analytical structure, which are
\beqa
&\label{def:gluingQ}
({ \bQ}^{\downarrow} )^{\gamma} =  \tau \cdot \dot{\bQ}^{\downarrow} \cdot \hat{\tau}^t \,,
\\&
({ \slQ}^{\downarrow} )^{\gamma} =  \dot{\tau} \cdot {\bQ}^{\downarrow} \cdot \hat{\dot{\tau}}^t \,,
\eeqa
where we have defined the $\tau$ matrices:
\beq
\label{def:fmatrices}
\tau \equiv \mathcal{G}\cdot  \dot \Omega , \;\;\;\; \hat{\tau} \equiv \mathcal{G}\cdot  \hat{\dot{\Omega} }, \;\;\; 
\dot \tau \equiv \dot{\mathcal{G}} \cdot  \Omega , \;\;\;\; \hat{\dot{\tau}} \equiv \dot{ \mathcal{G}}\cdot \hat{\Omega} .
\eeq
Notice that, just like the $\mathcal{G}$ matrices, the $\tau$ matrices put one wing in contact with the other. For example, $\tau$ would have the index structure $\mathcal{\tau}_i^{\; \dot k}$.\footnote{We note that these objects are named $\tau$ in analogy with some objects appearing in the ABJM case, but have some different features coming from the different algebraic structure. In particular, whereas in ABJM the $\tau$ functions had a single spinor $Sp(4)$ index, here they are matrices carrying two $Sp(2)$ indices, and connect the two wings of the system. In a sense they play a similar role to a quadratic combination of $\tau$ functions called $f_i^j$ in ABJM theory in \cite{Bombardelli:2017vhk} -- but unlike the case in ABJM, they cannot be broken into smaller components. For this reason we use the name $\tau$ for these objects -- in analogy to the ABJM case, their composition gives  $\omega$ which is the monodromy rotating $\bQ$ functions.
}

\paragraph{Periodicity of $\tau$ on the sheet with short cuts. }
By construction the matrices $\tau$ are periodic, with periodicity
\beq\label{eq:perf}
\tau = \tau^{[+4]} = \hat{\tau}^{[+2]}\,,\qquad \dot \tau= \dot{\tau}^{[+4]} = \hat{\dot{\tau}}^{[+2]}\,,
\eeq
and they obey the same constraints as $\mathcal{G}$ and $\Omega(u)$, i.e.
 they have a unit determinant and satisfy
\beq\label{eq:tauconstraint}
\tau(u) \cdot {\dot \epsilon^{-1}} \cdot \tau^t(u) = \epsilon^{-1} ,
\eeq
with the analogous constraint in the other wing.

\paragraph{Repacking the equations. }
Expanding~\eqref{def:gluingQ}  we find relations between the components of $\bQ_{ij}$ and $\slQ_{\dot{m}\dot n}$ -- collecting the coefficients on the RHS in the tensor $\omega$ we find:
\beq\label{eq:Qtildetau}
({ \bQ}^{\downarrow}_{ij} )^{\gamma} = \omega_{(ij)}^{(\dot k\dot l)}\; \dot{\bQ}^{\downarrow}_{\dot k\dot l}  ,\quad \text{where }\;
\omega_{(ij)}^{(\dot k\dot l)} \equiv  \tau_i^{\; \dot k} \tau_j^{\;\dot l[+2]} .
\eeq
The periodicities of the $\tau$ matrices imply that some components of the $\omega$ tensor have an enhanced periodicity with respect to $\tau$ functions, i.e. they are $i$-periodic rather than $2i$-periodic. For example, when indices are pairwise equal, we have that:
\beq
\omega^{(\dot k\dot k)}_{(kk)}=\omega^{(\dot k\dot k)}_{(kk)}\,^{[+2]} .
\eeq
These relations will be useful when analysing the ABA limit. Instead, for other components, the periodicity is more subtle, e.g. we have
\beq
\quad\omega^{(\dot 1\dot 2)}_{(11)}=\omega^{(\dot 2\dot 1)}_{(11)}\,^{[+2]} \,.
\eeq
\paragraph{Analytic continuation of $\tau$. } 
To close the system, we also need to describe what happens to the $\tau$ functions when we go around their branch cuts on the real axis in any direction. This information will let us immediately find the analytic continuation of $\bQ,\,\slQ$ along the path $\gamma^{-1}$, and also what happens when we go around the branch cut $n$ times.

In appendix \ref{appendix:numirror}, we prove that:
\begin{align}
&\tau^{\gamma}=(\bQ^{\downarrow })^{\gamma}\cdot (\tau^{[+2]})^{-t} \cdot (\slQ^{\downarrow-1})^{\gamma}\,,\\&
\dot \tau^{\gamma}=(\slQ^{\downarrow })^{\gamma}\cdot (\dot\tau^{[+2]})^{-t} \cdot (\bQ^{\downarrow-1})^{\gamma}\,.
\end{align}
Continuing these equations along $\gamma^{-1}$ we find that:
\begin{align}\label{eq:taugammam1}
      &( \hat{\tau} )^{\gamma^{-1}} = \bQ^{\downarrow \,t}\cdot    \tau^{-t} \cdot \slQ^{\downarrow -t} ,\\& ( {\tau} )^{\gamma^{-1}}= \bQ^{\downarrow }\cdot \hat{\tau}^{-t} \cdot ( \slQ^{\downarrow} )^{-1} .
\end{align}

\paragraph{Summary of the $\bQ\tau$-system. }
From now on, unless explicitly stated, we omit the label $\bQ^{\downarrow}$ and we will be referring to $\bQ \equiv \bQ^{\downarrow}$. 

In summary, the $\bQ\tau$- system relations are
\begin{align}
&\bQ^{\gamma} = \tau \cdot {\slQ}\cdot  \hat{\tau}^{t} ,\\&   \slQ^{\gamma} = \dot\tau \cdot {\bQ}\cdot  \hat{\dot\tau}^{t} ,
\end{align}
and
\begin{align}&
( \hat{\tau} )^{\gamma^{-1}} = \bQ^t \cdot    \tau^{-t} \cdot \slQ^{ -t} ,\;\;\; ( {\tau} )^{\gamma^{-1}}= \bQ\cdot \hat{\tau}^{-t} \cdot \slQ^{-1} ,\\&
( \hat{\dot\tau} )^{\gamma^{-1}} = \slQ^t \cdot    \dot\tau^{-t} \cdot \bQ^{ -t} ,\;\;\; ( {\dot\tau} )^{\gamma^{-1}}= \slQ\cdot \hat{\dot\tau}^{-t} \cdot \bQ^{-1} ,
\end{align}
with $\tau$ matrices satisfying:
\begin{align}
&\tau \cdot \epsilon^{-1} \cdot \tau^t = \epsilon^{-1}, \;\;\; \tau^{++} = \hat{\tau} = \tau^{--},\\&
\dot \tau \cdot \dot\epsilon^{-1} \cdot \dot \tau^t = \dot \epsilon^{-1}, \;\;\; \dot\tau^{++} = \hat{\dot \tau} = \dot \tau^{--}.
\end{align}
These equations, together with (\ref{eq:Qtildetau}), contain enough information to compute any analytic continuation of $\bQ,\,\slQ$ in terms of the values of $\bQ, \, \slQ$ and $\tau$'s on the first sheet.  For example, suppose we want to obtain
$( \dot\bQ )^{\gamma^{-1}} $, this can be obtained starting from  (\ref{eq:Qtildetau}) and taking its continuation along $\gamma^{-1}$. The resulting expression allows to obtain the desired quantity $\bQ^{\gamma^{-1}}$ in terms of $\tau^{\gamma^{-1}}$ and $\hat{\tau}^{\gamma^{-1}}$, which are in turn given by (\ref{eq:taugammam1}) in terms of quantities on the first sheet.

We will now see that there is a very similar set of equations describing the continuation of $\bP$ functions.

\subsubsection{The $\bP\nu$-system} 
\label{sec:Pmusystem}
In this section, we use the gluing of the $\bQ$ functions we have just computed and the QQ-relations to obtain the $\bP\nu$-system, which describes the analytic continuation of the $\bP_A$ functions through their single branch cut on the real axis. In this case, a new auxiliary object is introduced, a monodromy matrix $\mu$, which has a dual cut structure with respect to $\omega$ found above. 

\paragraph{Definition and properties of the $\nu$ functions. }
We construct the analogues of the matrices $\tau$ introduced above as follows\footnote{As for the $\tau$ functions, the name $\nu$ is an imperfect analogy with the objects named $\nu$ in the case of ABJM theory in \cite{Bombardelli:2017vhk}. These are the simplest objects arising in the theory such that the monodromy of $\bP$ functions will be written in terms of their quadratic combination. However, while in ABJM theory these objects had a single spinor or anti-spinor index, here they are tensors connecting the two wings. Another difference emerging is that in ABJM, these objects were only quasi-periodic (i.e., they picked up a constant factor under the shift); here, we found that they automatically turn out to be $i$-periodic on the Riemann section with long cuts.}
\beq
\label{def:nu_matrices}
 {\nu} \equiv   {\bS}^-\cdot \dot{ {\tau}}^t \cdot (\dot{ {\bS}}^{-} )^{-1} , \;\;\;\;\hat{\nu} \equiv  \hat{\bS}^- \cdot \hat{\dot{\tau}}^t\cdot (\hat{\dot{\bS}}^{-} )^{-1} , \;\;\;\; \dot{\nu} \equiv  \dot{\bS}^-\cdot {{\tau}}^t \cdot ({{\bS}}^{-} )^{-1} , \;\;\;\;\hat{\dot{\nu}} \equiv  \hat{\dot{\bS}}^- \cdot {{\hat{\tau}}}^t\cdot ({{\hat{\bS}}}^{-} )^{-1}\,,
\eeq
where each is a $2 \times 2$ matrix connecting the two wings, e.g. $\nu$ has index structure $\nu_a^{\, \dot b}$ while $\hat{\nu}$ has indices $\hat{\nu}_{\hat a}^{\,\hat{\dot b}}$. The first thing we can establish is that they all satisfy \textbf{mirror periodicity}, which is a property meaning that they are periodic on a Riemann section with long cuts.  When expressed in terms of shifts on the section with short cuts, this property takes the form
\beq
\label{def:mirror_periodicity}
\nu^{++} = \nu^{\gamma} , \;\;\; \hat{\nu}^{++} = (\hat{\nu} )^{\gamma} ,
\eeq
which is proved in appendix \ref{appendix:numirror}. 
The same property holds for $\dot{\nu}$, $\hat{\dot{\nu}}$. 

These matrices satisfy some constraints that, in turn, we can see as a simple statement: the multiplication of the $SO(4)$ spinor (anti)-indices by the appropriate $\nu$ matrices constitutes a symmetry of the Q-system. In particular, as a consequence of (\ref{eq:constraintQai}) and of (\ref{eq:tauconstraint}), they satisfy the constraints 
\beq\label{eq:nuconstraints}
\nu \cdot \dot{\mathbf{C}}^{-1} \cdot \nu^t = \mathbf{C}^{-1} , \;\;\; \hat{\nu} \cdot \hat{\dot{\mathbf{C}}}^{-1}\cdot  \hat{\nu}^t = \hat{\mathbf{C}}^{-1} , \;\;\;\dot \nu  \cdot \mathbf{C}^{-1} \cdot \dot \nu^t = \dot{\mathbf{C}}^{-1} , \;\;\; \hat{\dot{\nu}} \cdot {\hat{\mathbf{C}}}^{-1} \cdot \hat{\dot{\nu}}^t = \hat{\dot{\mathbf{C}}}^{-1} .
\eeq
We can also view them directly as a condition on the invariance of the normalisation (\ref{eq:constraintQai}) under the transformation dictated by the $\nu$'s. 
We also notice another curious identity, which is a direct consequence of the definitions (\ref{def:nu_matrices}) and of the core relations (\ref{eq:core_relations}):
\beq
\bP \cdot \hat{\nu} \cdot \dot{\bP}^{-1} = \nu^{++} = \nu^{\gamma} ,  
\eeq
and similarly
\beq\label{eq:Pnusecond}
\hat{\nu}^{\gamma} = \bP^{-1} \cdot \nu\cdot \dot{\bP} , \;\;\; \dot{\nu}^{\gamma} = \dot{\bP} \cdot \hat{\dot{\nu}} \cdot \bP^{-1} , \;\;\; \hat{\dot{\nu}}^{\gamma} = \dot{\bP}^{-1}\cdot  \dot\nu \cdot {\bP} .
\eeq
\paragraph{Monodromies of $\bP$ functions. }

Having described the main properties of the $\nu$ matrices, it is now easy to compute through them the analytic continuation of the $\bP$ functions. We present the derivation in appendix \ref{appendix:numirror}, obtaining that:

\beq
\label{eq:Pgluing1main}
(\bP)^{\gamma}=(\dot{\nu}^{[+2]})^{-1}\cdot \slP\cdot \hat{\dot{\nu}}^{[+2]}\,,\qquad (\slP)^{\gamma}=({\nu}^{[+2]})^{-1}\cdot \bP\cdot \hat{{\nu}}^{[+2]}\,,
\eeq
and from these equations, taking the opposite analytic continuation along $\gamma^{-1}$ and using mirror periodicity of $\nu$ functions, we get the nice constraints
\beq\label{eq:Pnufirst}
(\bP)^{\gamma^{-1}}=\nu \cdot  {\slP}\cdot \hat{\mathbf{C}}^{-1} \cdot \hat{\nu}^t \cdot \mathbf{C}\,,\qquad (\dot\bP)^{\gamma^{-1}}=\dot\nu \cdot  {\bP}\cdot \hat{\dot{\mathbf{C}}}^{-1} \cdot \hat{\dot{\nu}}^t \cdot \dot{\mathbf{C}} .
\eeq

\paragraph{Summary of the $\bP\nu$-system. } In summary, we found that $\bP$ and $\nu$ functions satisfy a closed system of relations that is a mirror of the ones satisfied by $\bQ$ and $\tau$, i.e. 
\beq\label{eq:Pnufirst2}
(\bP \cdot \mathbf{C}^{-1} )^{\gamma^{-1}}=\nu \cdot \left( {\slP}\cdot \hat{\mathbf{C}}^{-1} \right)\cdot \hat{\nu}^t \,,\qquad (\dot\bP \cdot \dot{\mathbf{C}} ^{-1})^{\gamma^{-1}}=\dot\nu \cdot  \left({\bP}\cdot \hat{\dot{\mathbf{C}}}^{-1} \right)\cdot \hat{\dot{\nu}}^t ,
\eeq
together with 
\beqa\label{eq:Pnusecond2}
 {\nu}^{\gamma}  &=&  {\bP} \cdot \hat{ {\nu}}\cdot  \dot{\bP}^{-1}, \\
 \hat{\nu}^{\gamma}  &=&  \bP^{-1} \cdot \nu \cdot \dot{\bP} , \\
 \dot{\nu}^{\gamma}  &=&  \dot{\bP} \cdot \hat{\dot{\nu}}\cdot  \bP^{-1} , \\ \hat{\dot{\nu}}^{\gamma} &=& \dot{\bP}^{-1} \cdot \dot\nu\cdot   {\bP} ,
\eeqa
and where all four $\nu$ functions are individually mirror periodic and satisfy the constraints (\ref{eq:nuconstraints}). 

We also introduce some useful notation, which we will refer to in the next section: introducing
\beq
\mu_{{a}\hat{ a}}^{\dot a \hat{\dot{a}}} = {\nu}_{{a}}^{\; \dot a} \hat{{\nu}}_{\hat{{a}}}^{\; \hat{\dot{a}}},\qquad \dot{\mu}_{\dot{a}\hat{\dot{ a}}}^{ a \hat{ a}} = \dot{\nu}_{\dot{a}}^{\; a} \hat{\dot{{\nu}}}_{\hat{\dot{{a}}}}^{\; \hat{a}} ,
\eeq
we may rephrase \eqref{eq:Pnufirst2} as
\beq
\label{eq:Pmusystem}
(\bP_{a \hat{a}} )^{\gamma^{-1}} = \mu_{a \hat{a}}^{\dot{a}\hat{\dot a}}\, \dot{\bP}_{\dot a \hat{\dot{a} } } ,
\eeq
where we used the $\mathbf{C},\,\hat{\mathbf{C}}$ tensors (and their dotted versions) to raise and lower indices.

\section{The Asymptotic Bethe Ansatz for massive states}\label{sec:ABA}
In this section, we will consider the Asymptotic Bethe Ansatz (ABA) limit of the QSC equations. This limit consists of taking $L\rightarrow\infty$ in the asymptotics fixed in section \ref{sec:cuts_asymptotics}. 
The way to understand how the QSC simplifies in this limit was first found in \cite{Gromov:2014caa}, and later applied to ABJM theory in \cite{Bombardelli:2017vhk} and to the $AdS_3 \times S^3 \times T^4$ theory in \cite{Cavaglia:2021eqr, Ekhammar:2021pys, Ekhammar:2024kzp}. 
The way to take this limit, as understood today, is partly heuristic, and some normalisation constants are hard to fix. In the following section, we present the main idea.  

\subsection{Scaling of important quantities in the large volume limit}
The crucial observation is that some Q-functions are exponentially suppressed, while others are exponentially large for $L\rightarrow \infty$. 
To highlight this behaviour, we will introduce the useful shorthand\footnote{We hope that the readers do not get confused between $\varepsilon$, the parameter which is small in the large volume limit, and the Levi-Civita tensor $\epsilon$.} 
 $\varepsilon \equiv O(u^{-L})$, such that $\varepsilon \rightarrow 0 $ exponentially in the ABA limit. Then, we apply the useful heuristics found in \cite{Gromov:2013pga}: we determine which functions are large or small in the limit based on their large-$u$ asymptotics, as described in section \ref{sec:cuts_asymptotics}. Namely, we promote this large-$u$ asymptotic as an indication of how they behave for fixed $u$ and large $L$.  

In this way, we  get the following scaling for Q-functions in the ABA limit:
\begin{align}
\label{eq:ABAscaling}
    \bP_A \sim \left( \varepsilon, 1, 1, \varepsilon^{-1} \right) , \;\;\;\; \bQ_{11}\sim \varepsilon^{-1}, \;\;\; \bQ_{22}\sim \varepsilon, \;\;\; \bQ_{12},\bQ_{21}\sim \varepsilon^0, \;\;\; (\bS)_{a|k }\sim (\hat\bS)_{\hat a | k} \sim \left(\begin{array}{cc} 1 & \varepsilon \\ \varepsilon^{-1} & 1 \end{array}\right)\,,
\end{align}
and the same is true in the other wing.

Next, let us consider the scaling of the $\tau$ and $\nu$. From their definitions \eqref{def:fmatrices}, it is easy to see that the scaling of all elements of $\tau$, $\hat{\tau}$ and their dotted counterparts is $O(\varepsilon^0)$ (this also follows from the fact that they are periodic functions on the sheet with short cuts, so they must asymptote to constants). 
 Using the relation (\ref{def:nu_matrices}) and (\ref{eq:ABAscaling}), we then determine the scaling of $\nu$ functions. In particular, 
\beq
\nu_a^{\,\dot{a}} \sim \left(\begin{array}{cc} \varepsilon^{-1} & 1 \\ \varepsilon^{-2} & \varepsilon^{-1}\end{array}\right),
\eeq
 and all the other $\nu$ matrices have the same structure. From this, it is easy to see that  the matrix element $
M \equiv \mu_{1\hat 1}^{\,\dot 2 \hat{\dot 2}} =\nu_1^{\dot 2} \hat{\nu}_{\hat 1}^{\hat{\dot  2} }
$ 
 scales as $O(\varepsilon^0)$. This will serve as the starting point for the subsequent analysis.

\paragraph{Simplification of some important equations.}
In the ABA limit, we have a huge simplification for some equations. Consider for instance the definition of the functions $\nu_1^{\dot{2}}$, given in (\ref{def:nu_matrices}). In general, this involves the sum of several terms, but, due to the scalings of the quantities involved, there is a single dominant term as $L\rightarrow \infty$:
\beq\label{eq:MABA}
\nu_1^{\,\dot{2}} \sim \bS_{1|1}^- \, \dot{\tau}^1_{\dot 2} \,\dot{\bS}^{\dot{2}|2 -} = \bS_{1|1}^- \, \dot{\tau}^1_{\dot 2} \,\dot{\bS}_{\dot{1}|1}^-,
\eeq
where all quantities involved are now and $O(1)$ in the ABA limit. The same happens for the definition of the analogous quantities:
\beq\label{eq:MABA2}
\hat{\nu}_1^{\,\dot{2}} \sim \hat{\bS}_{ {1}|1}^- \, \hat{\dot{\tau}}^{  1}_{\dot 2} \,\hat{\dot{\bS}}_{\dot{ {1}}|\dot 1}^-, \;\;\; \dot{ {\nu}}_{\dot 1}^{\, {2}} \sim \dot{ {\bS}}_{\dot{ {1}}|\dot 1}^- \, { {\tau}}^{ \dot 1}_{ 2} \,{ {\bS}}_{{ {1}}| 1}^- , \;\;\; \hat{\dot{\nu}}_{\dot 1}^{\, {2}} \sim \hat{\dot{\bS}}_{\dot{ {1}}|\dot 1}^- \, {\hat{\tau}}^{ \dot 1}_{ 2} \,{\hat{\bS}}_{{ {1}}| 1}^- .
\eeq
We also obtain important relations from the $\bP\nu$-system. First of all, using \eqref{eq:Pgluing1main}, and then converting $\nu$ functions into $\tau$'s, $\mathbf{S}$ and $\bQ$ using  (\ref{def:nu_matrices}) and (\ref{eq:defQ}), we find
\beq  
 \bP^{\gamma}= {-}\bS^+ \cdot (\tau^{[+2]})^{-t}\cdot \epsilon\cdot \slQ^t \cdot \tau^t \cdot (\hat{\bS}^+)^{-1}.
\eeq
For the matrix element $ (\bP_1^{\, \hat{2} } )^{\gamma}$, in the ABA limit, the RHS of this equation drastically simplifies to a single term:
\beq\label{eq:ABAsimplified2}
 (\bP_1^{\, \hat{2} } )^{\gamma}\sim \bS_{1|1}^+ \, \tau_2^{\;\dot 1}\, (\tau_2^{\;\dot 1})^{[+2]} \, \dot{\bQ}_{\dot 1\dot 1} \, \bS_{\hat{1}|1}^{+},
\eeq
which will be useful in the following. There is a symmetric equation swapping the role of the two wings:
\beq\label{eq:ABAsimplified3}
 (\bP_{\dot 1}^{\, \hat{\dot{2}} } )^{\gamma}\sim \dot\bS_{\dot 1|\dot 1}^+ \, \dot\tau_{\dot 2}^{\; 1}\, (\dot\tau_{\dot 2}^{\; 1})^{[+2]} \, {\bQ}_{1|1} \, \dot\bS_{\hat{\dot{1}}| \dot 1}^{+}.
\eeq
We will present further simplified equations arising in the ABA limit as they become relevant, but (\ref{eq:MABA}), (\ref{eq:MABA2}) and (\ref{eq:ABAsimplified2}), (\ref{eq:ABAsimplified3}) contain already most of the information. They involve a subset of Q-functions, $\nu$ and $\tau$ functions. Those will be the ones we can fix within the ABA limit. 
\subsection{Fixing the $\mu$ and $\omega$ functions}
\paragraph{Fixing $\mu$ functions. }
As the previous examples suggest~\cite{Gromov:2014caa, Bombardelli:2017vhk, Cavaglia:2021eqr}, it is convenient to start the study of the ABA limit by studying some of the $\mu$ functions. 
The first quantity we consider is $\mu_{1 \hat{1}}^{\dot{2} \hat{\dot 2}} $. Among the various components of $\mu$, we expect to be able to compute this quantity because it remains finite in the ABA limit, as we just saw in the previous section. To shorten the notation, we denote it as
\beq
M \equiv\mu_{1 \hat{1}}^{\dot{2} \hat{\dot 2}} = \nu_1^{\dot 2} \hat{\nu}_{\hat{1}}^{\hat{\dot 2}} ,
\eeq
and from~\eqref{eq:MABA} and~\eqref{eq:MABA2} in the ABA limit it has the expression
\beq\label{eq:defMspintau}
M \sim \bS_{1|1}^- \hat{\bS}_{1|1}^- 
\dot{\bS}_{1|1}^- 
\hat{\dot \bS}_{1|1}^- \, \dot{\tau}_2^{1} \hat{\dot{\tau}}_2^{1} .
\eeq
We will make the following
\beq
\texttt{Assumption:    in the ABA limit, $M$ has only square-root type branch points.}
\eeq
Then, suppose this quantity has some zeros on the first sheet. We collect the zeros of $M^+$ (notice the shift!) in a polynomial
\beq
\mathbb{Q}(u) = \prod_i (u - u_i ) ,
\eeq
i.e. $M$ has zeros $u_i - \frac{i}{2}$, where we expect $u_i$ to be real. 

In anticipation of the role of this polynomial in the ABA equations, we split its roots into four sets, so we write
\beq
\mathbb{Q}(u) \equiv \mathbb{Q}_s(u){\mathbb{Q}}_{\dot s}(u)\mathbb{Q}_{\hat{s}}(u){\mathbb{Q}}_{\hat{\dot s}}(u) ,
\eeq
with
\beq
\mathbb{Q}_{\bullet} = \prod_{k=1}^{K_{\bullet}} (u - u_{\bullet, k} ) ,\;\;\;\bullet\in \{s,\hat s,\dot s,\hat{\dot s}\}\,.
\eeq
Later we will assign  the zeros of $\mathbb{Q}_\bullet$ to the functions $\bS_{1|1}$, $\dot{\bS}_{\dot 1|\dot 1}$, $\bS_{\hat{1}|1}$, $\dot{\bS}_{\hat{\dot 1}|\dot 1}$, respectively. Not by chance, we have labelled these four sets of roots according to their role in the ABA with the notation we already introduced. In particular, these will all become momentum-carrying roots. Then, we define the quantity
\beq\label{eq:def_F}
F^2\equiv \frac{M\mathbb{Q}^+ }{ M^{++} \mathbb{Q}^- },
\eeq
which by construction has no poles or zeros on the first sheet. Moreover, due to the mirror periodicity of $M$, it satisfies
\beq\label{eq:FRH}
F \tilde{F} = \frac{\mathbb{Q}^+}{\mathbb{Q}^-} ,
\eeq
where we denoted $\tilde{F} = F^{\gamma} = F^{\gamma^{-1}}$  and used $M^{\gamma} = M^{++}$, $(M^{++} )^{\gamma} = M$, thanks to the assumption that the branch cut is quadratic in the ABA limit. 

Moreover, $F^2$ has no cuts other than the one on the real axis, as can be seen by plugging (\ref{eq:MABA}),(\ref{eq:MABA2}) into the $\nu$ functions entering the definition of $M$. This yields
\beq\label{eq:spinordifference}
F^2 = \frac{\bS_{1|1}^- \,\dot{\bS}_{\dot{1}|\dot 1}^- \bS_{\hat{1}|1}^- \,\dot{\bS}_{\hat{\dot{1}}|\dot 1}^- \mathbb{Q}^+}{\bS_{1|1}^+ \,\dot{\bS}_{\dot{1}|\dot 1}^+ \bS_{\hat{1}|1}^+ \,\dot{\bS}_{\hat{\dot{1}}|\dot 1}^+ \mathbb{Q}^-} ,
\eeq
where the $\tau$ functions cancel due to their periodicity properties. Above, we have found an expression which manifestly has no cuts above the real axis. It is simple to repeat the argument using the basis of $\bS^{\downarrow}$ (one can find that $\nu$ satisfies the same definition (\ref{def:nu_matrices} with $\bS^{\downarrow}\rightarrow \bS^{\uparrow}$ and $\tau \rightarrow \dot{\mathcal{G}}^{-1}\cdot \tau\cdot \mathcal{G}$). This then implies that $F$ also has no cuts in the lower half-plane; therefore, it has a single cut on the real axis. 

Thus, $F$ is a function with a single cut that satisfies the simple Riemann-Hilbert problem (\ref{eq:FRH}), whose solution is
\beq\label{eq:Fsol}
F = \pm \left(\prod_i\sqrt{\frac{x_i^+}{x_i^-}}\right)  \frac{\mathbf{B}_{(+)}}{ \mathbf{B}_{(-)}} ,
\eeq
where we defined
\beq
\mathbf{B}_{(\pm)} = B_{s, (\pm)} B_{\hat{s}, (\pm)} B_{\dot s, (\pm)} B_{\hat{\dot s}, (\pm)} , \eeq 
with $B_{a,(\pm)}$ an explicit function, defined in appendix \ref{appendix:dressing}.

From the large volume solution (\ref{eq:Fsol}), we can now determine the value of $M$, which is related to $F$ by (\ref{eq:def_F}), and which additionally should satisfy mirror periodicity, i.e. $M^{\gamma} = M^{++}$. The solution is given by 
\beq\label{eq:solM}
M  \propto \mathbb{Q}^- \mathbf{f} \bar{ \mathbf{f} }^{--} , 
\eeq
where $\mathbf{f}, \bar{\mathbf{f}}$ are defined in appendix \ref{appendix:dressing}, are analytic in the upper and lower half plane respectively, and satisfy the difference equations 
\beq
{\bar{\mathbf{f}}\over\bar{\mathbf{f}}^{[-2]}}=\frac{\mathbf{B}_{(-)}}{\mathbf{B}_{(+)}}, \;\;\; {{\mathbf{f}}\over{\mathbf{f}}^{[+2]}}=\frac{\mathbf{B}_{(+)}}{\mathbf{B}_{(-)}}.
\eeq
We also note that the expression we have found for $M$ is completely symmetric with respect to the two wings. This means that, repeating the same argument, we would find the exact same result for $\dot M \equiv \dot\nu_{\dot 1}^{ 2} \, \hat{\dot{\nu} }_{\dot 1}^{\hat{\dot 2}}$, i.e. 
\beq
\dot M \propto M \propto \mathbb{Q}^- \mathbf{f} \bar{ \mathbf{f} }^{--} .
\eeq
\paragraph{Zero momentum condition. }
At large $u$, we expect $M/M^{++} \rightarrow 1$, as it exhibits power-like asymptotics. From the solution (\ref{eq:solM}), we find
\beq\label{eq:ratioM}
\frac{M}{M^{++}}=\frac{\mathbf{R}_{(-)} \mathbf{B}_{(+)}}{ \mathbf{R}_{(+)} \mathbf{B}_{(-)} } \sim 1,
\eeq
which at order $u^0$ in the large $u$ limit translates into the zero-momentum condition
\beq
\label{eq:zero_momentum_condition}
1=\prod_{\bullet}\prod_{i=1}^{K_{\bullet}} \frac{x_{\bullet,i}^+}{x_{\bullet,i}^-} .
\eeq

\paragraph{Energy. } From $1/u$ corrections to the large $u$ expansion of $M/M^{++} \sim 1 - \frac{i \#}{u}$, where 
$
M \sim u^{\#}$. 
Here, on the one hand, studying the asymptotics of the QSC we find
$$
\# = \Delta_L+\Delta_R- 4-\frac{J_1+J_2}{2} - \frac{\dot J_1+\dot J_2}{2} ,
$$
while studying the large-$u$ expansion of the expression in (\ref{eq:ratioM}), we obtain 
$$
\# = \sum_{\bullet}\sum_{i=1}^{K_{\bullet}}\left(1+2ih \left(\frac{1}{x^+_{\bullet,i}}- \frac{1}{x^-_{\bullet,i}}\right)\right).
$$
Comparing the two expressions, we find the following condition:
\beq\label{eq:Deltaresulting}
\Delta \equiv \frac{1}{2}(\Delta_L+\Delta_R)=2+\frac{J_1+J_2+\dot J_1+\dot J_2}{4}+\frac{1}{2}\sum_{\bullet}\sum_{i=1}^{K_{\bullet}}\left(1+2ih \left(\frac{1}{x^+_{\bullet,i}}- \frac{1}{x^-_{\bullet,i}}\right)\right).
\eeq
Thus, the anomalous part $\delta\Delta$ has the following expression in the ABA limit:
\beq
\delta\Delta =
ih
\sum_{\bullet}
\sum_{i=1}^{K_{\bullet}} \left(\frac{1}{x^+_i}- \frac{1}{x^-_i}\right).
\eeq
Using the form of the dispersion relation of massive worldsheet particles \cite{Borsato:2015mma}:
\beq
\epsilon(u) \equiv \sqrt{\frac{1}{4}- h^2 \frac{(x^+(u)-x^-(u))^2}{x^+(u)x^-(u)}} = \frac{1}{2} + i h \left( \frac{1}{x^+(u)} - \frac{1}{x^-(u)}\right),
\eeq
we see that (\ref{eq:Deltaresulting}) can be written as
\beq
\Delta = \Delta_{\text{BPS}} +  E ,
\eeq
where
\beq
\Delta_{\text{BPS}}  \equiv 2+ \frac{1}{4}\left(
J_1 + J_2 + \dot{J}_1 + \dot{J}_2 \right) 
\eeq
 can be considered the dimension of a ``vacuum'' state saturating the BPS bound (in both wings), and
 \beq
E = \sum_{\bullet = s,\hat{s}, \dot s, \hat{\dot s}} \sum_{i=1}^{K_{\bullet}} \epsilon( u_{\bullet, i} ) \equiv \frac{1}{2} \sum_{\bullet} K_{\bullet} + \delta\Delta, 
\eeq
is the total energy of excitations upon the vacuum.

\paragraph{Fixing the product of spinor Q-functions. }
We can now solve the difference equation defined by (\ref{eq:spinordifference}) for the product of all spinor Q-functions. Setting $S \equiv \bS_{1|1} \,\dot{\bS}_{\dot{1}|\dot 1} \hat{\dot{\bS}}_{\hat{\dot{1}}|\dot 1} \bS_{\hat{1}|1}  $, (\ref{eq:spinordifference}) implies $\frac{S^{-}}{S^+} = \frac{\mathbf{B}_{(+)}^2 \mathbb{Q}^-}{\mathbf{B}_{(-)}^2 \mathbb{Q}^+}$. The solution with the required analytic properties in the upper half plane is
\beq
\label{eq:sol_prodS}
\bS_{1|1} \,\dot{\bS}_{\dot{1}|\dot 1} \hat{\dot{\bS}}_{\hat{\dot{1}}|\dot 1} \bS_{\hat{1}|1}  \propto \mathbb{Q} ( \mathbf{f}^+ )^2 .
\eeq
So far, we have only been able to determine this product of all spinors from both wings, while we would like to describe individual Q-functions. To do this, let us introduce some \textbf{splitting factors} $W(u)$, $\hat{W}(u)$,  $\dot{W}(u)$, $\hat{\dot W}(u)$, setting
\begin{align}\label{eq:Sfixed}
\bS_{1|1} = \mathbb{Q}_s (\mathbf{f}^+)^{\frac{1}{2}} W, \;\;\;\;
\dot{\bS}_{\dot 1|\dot 1} = \mathbb{Q}_{\dot s} (\mathbf{f}^+)^{\frac{1}{2}}  \dot{W} , \;\;\;\;
\bS_{\hat{1}|1} = \mathbb{Q}_{\hat s} (\mathbf{f}^+)^{\frac{1}{2}} \hat{W} , \;\;\;\;
\dot{\bS}_{\hat{\dot 1}|\dot 1} = \mathbb{Q}_{\hat{\dot{ s}} } (\mathbf{f}^+)^{\frac{1}{2}}  \hat{\dot{W}} ,\end{align}
where, to preserve \eqref{eq:sol_prodS}, the $W$ factors must satisfy
\beq 
W(u) \hat{W}(u) \dot{W}(u) \hat{\dot W}(u) = \text{constant}.
\eeq
The functions $W(u)$, $\hat{W}(u)$,  $\dot{W}(u)$, $\hat{\dot W}(u)$ are, by construction, analytic in the upper half plane and free of poles everywhere. Currently, they are unfixed, but later we will identify some constraints for them.

\paragraph{Fixing the product of $\tau$ functions. }
Finally, from (\ref{eq:defMspintau}) and its analogue in the other wing, we immediately find a result for the product of couples of $\tau$ functions, i.e. 
\beq
\tau_2^{\dot 1} \hat{\tau}_2^{\dot 1} \propto \dot\tau_{\dot 2}^{ 1} \hat{\dot{\tau}}_{\dot 2}^{ 1}  \propto \frac{ \bar{ \mathbf{f} }^{--} }{\mathbf{f}} .
\eeq
We can also verify easily that the result above defines an $i$-periodic function, as we expect for these combinations of $\tau$ functions. Notice that, also in this case, we found two quantities which are symmetric with respect to the two wings in the ABA limit. 

\paragraph{Splitting factors for $\tau$ functions. }
In the following, we will need the values of some individual $\tau$ functions, so we again introduce some new splitting factors $V(u)$ and $\dot V(u)$ as follows:
\beqa
\tau_2^{\dot 1} &=&  K \left(  \frac{ \bar{ \mathbf{f} }^{--} }{\mathbf{f} }\right)^{\frac{1}{2}} V , \;\;\;\;
\hat{\tau}_{2}^{\dot{1}}  = K \left(  \frac{ \bar{ \mathbf{f} }^{--} }{\mathbf{f} }\right)^{\frac{1}{2}}/V , \\
\dot\tau_{\dot 2}^{1} &=& \dot{K}\left(  \frac{ \bar{ \mathbf{f} }^{--} }{\mathbf{f} }\right)^{\frac{1}{2}}\dot{V} , \;\;\;\;
\hat{\dot{\tau} }_{\dot{2}}^{1} =  \dot{K}\left(  \frac{ \bar{ \mathbf{f} }^{--} }{\mathbf{f} }\right)^{\frac{1}{2}}/\dot{V} ,\label{eq:tausplit}
\eeqa
where $K$, $\dot{K}$ are constants unfixed by our argument (but which will play no role in the following). 
Now, since $\tau^{++} = \hat{\tau}$, the splitting functions $V(u)$, $\dot{V}(u)$ should satisfy
\beq
V^{++} = 1/V , \;\;\;\;\dot{V}^{++} = 1/\dot{V} .
\eeq
\paragraph{Splitting $\nu$ functions.}
As a consequence of the previous identities, we can compute individual $\nu$ functions in terms of the splitting factors introduced above:
\beqa\label{eq:nusplit}
\nu_1^{\dot 2} &\simeq& \bS_{1|1}^- \dot{\bS}_{\dot 1|\dot 1}^-  \dot{\tau} \propto 
\dot{V} W^- \dot{W}^- \mathbb{Q}_s^-  \mathbb{Q}_{\dot{s}}^- (\mathbf{f}\bar{\mathbf{f}}^{[-2]} )^{\frac{1}{2}}, \\
\dot{\nu}_{\dot 1}^{2} &\simeq& \bS_{1|1}^- \dot{\bS}_{\dot 1|\dot 1}^-   {\tau} \propto
{V} W^- \dot{W}^- \mathbb{Q}_s^-  \mathbb{Q}_{\dot{s}}^- (\mathbf{f}\bar{\mathbf{f}}^{[-2]} )^{\frac{1}{2}} , \\
\hat{\nu}_{\hat{1}}^{\hat{\dot{2}}} &\simeq& -\bS_{\hat 1|1}^- \dot{\bS}_{\hat{\dot 1}|\dot 1}^-  \hat{\dot{\tau}} \propto -
\dot{ {V}}^{-1} \hat{W}^- \hat{\dot{W}}^- \mathbb{Q}_{\hat s}^- \mathbb{Q}_{\hat{\dot{s}}}^- ( \mathbf{f}\bar{\mathbf{f}}^{[-2]} )^{\frac{1}{2}}, \\
\hat{\dot{\nu}}_{\hat{\dot{1}}}^{\hat{2}}&\simeq& -\bS_{\hat 1|1}^- \dot{\bS}_{\hat{\dot 1}|\dot 1}^-  {\hat \tau} \propto   -
{V}^{-1} \hat{W}^- \hat{\dot{W}}^- \mathbb{Q}_{\hat{s}}^-  \mathbb{Q}_{\hat{\dot{s}}}^- ( \mathbf{f}\bar{\mathbf{f}}^{[-2]} )^{\frac{1}{2}} .
\eeqa
We will get some new constraints on $V$, $\dot V$, $W$, $\dot W$, $\hat W$, $\hat{\dot W}$ from the condition of mirror periodicity of  $\nu$ functions. We will utilise these constraints later, as they will ultimately contribute to the dressing factors appearing in the ABA equations. 
\subsection{Fixing $\bP$ and $\bQ$ functions}
\label{sec:dressings}
From the other known instances of the QSC \cite{Gromov:2014caa, Bombardelli:2017vhk, Cavaglia:2021eqr}, we expect that in the ABA limit, some of the $\bP$ and $\bQ$ functions become simple. Such Q-functions should be the $\bP$'s that are exponentially small and the $\bQ$'s that are exponentially large in the ABA limit $\eps\rightarrow 0$; looking at \eqref{eq:ABAscaling}, we can see that they are respectively $\bP_1$ and $\bQ_{11}$, plus their counterparts on the other wing $\slP_{\dot 1}$ and $\slQ_{\dot 1 \dot 1}$.

We start by parametrising these functions  in the following general way, taking into account their cut structure:
\beq
\label{def:ABA_scalingP}
\bP_{1}(u) \rightarrow x^{-L} \, R_b \; B_{\dot{f}} \; \sigma,\;\;\; \dot{\bP}_{\dot 1}(u) \rightarrow x^{-L} \, R_{\dot b} \; B_{{f}} \; \dot{\sigma} ,
\eeq
where the $R$ and $B$ factors are polynomials in $x$, $1/x$ respectively, defined in \eqref{eq:def_B_R} where we introduce a total of four new sets of roots labelled as $b$, $f$, $\dot f$, $\dot b$. As the notation anticipates, two of these will be roots appearing in the ABA equations.

We assume that the $R_{\bullet}$ factors contain all the zeros of these functions on the first sheet, the factors of $x(u)$ serve to introduce the dependence on $L$ in the large-$u$ asymptotics, 
and the  $B_{\bullet}$ are introduced for later convenience. 
The functions $\sigma(u)$ and $\dot\sigma(u)$ encode the most nontrivial part of the $\bP$ functions. So far, they are unfixed, except for the fact that they should have a single short branch cut on the real axis and no zeros or poles on the first sheet. 
Similarly, we can parametrise the $\bQ$ functions as:
\beq
\label{def:ABA_scalingQ}
\bQ_{11}(u) \rightarrow x^{L} \, R_{\tilde{f}} \; B_{\dot{\tilde{b}}}\,\frac{1}{\sigma'} 
\mathbf{f}
,\;\;\; \dot{\bQ}_{\dot 1\dot 1}(u) \rightarrow x^{L} \, R_{\dot{\tilde{f}}} \; B_{ {\tilde{b}}}\,\frac{1}{\dot{\sigma}'} 
\mathbf{f} ,
\eeq
which introduces yet four -- a priori unrelated -- sets of roots labelled by $\tilde f$, $\tilde b$, $\dot{\tilde f}$, $\dot{\tilde b}$.  
Again, we collected all the zeros of the first sheet into the $R_{\bullet}$ factors, and the $B_{\bullet}$ factors are there for future convenience, as is the factor
$\bf{f}$ 
introduced in the previous paragraphs. 
The nontrivial factors $\sigma'(u)$ and $\dot\sigma'(u)$, encoding the most nontrivial part of these functions, are analytic in the upper half plane (they might possess a ladder of short branch cuts in the lower half plane), and are free of zeros and poles in the first sheet. 

The Ansatze (\ref{def:ABA_scalingP}), (\ref{def:ABA_scalingQ}) are partially ambiguous at the moment, because nothing prevents us from reabsorbing the $B$ factors into the $\sigma$ functions. Now we proceed to constrain them further and link the various types of roots. 

We start using the QQ-relations \eqref{eq:QQrels_node1_fermionic}, with the following choice of indices:
\begin{equation}
\bP_1(u) \bQ_{11}(u) = \bS_{\hat 1|1}^+(u)  \bS_{1|1}^+(u)  - \bS_{\hat 1|1}^-(u)  \bS_{ 1|1}^-(u) \label{eq:forBethe32}\,,
\end{equation}
where we used the definition (\ref{eq:defQAij}) to write $Q_{1|11}$ in terms of spinors and antispinors. In the RHS of this equation, we can plug in our Ansatz for the spinors and antispinors \eqref{eq:Sfixed}, 
while for the LHS we use \eqref{def:ABA_scalingP} and \eqref{def:ABA_scalingQ}. Notice that in the combination of spinors appearing in \eqref{eq:forBethe32} the splitting functions of \eqref{eq:Sfixed} cancel, and we find the following relation between the different sets of Zhukovsky polynomials
\beq
\label{eq:ABA_QQrel32}
 \frac{B_{s,-}B_{\hat s,-}}{\dot B_{\dot s,+}\dot B_{\hat{\dot s},+}}\left(R_{s,+}R_{\hat s,+}\dot B_{\dot s,-}\dot B_{\hat {\dot s},-} -R_{s,-}R_{\hat s,-}\dot B_{\dot s,+}\dot B_{\hat {\dot s},+}\right)= \frac{\sigma}{\sigma'} \,  R_b \; B_{\dot{f}} \, R_{\tilde{f}} \; B_{\dot{\tilde{b}}}\,.
\eeq
By absorbing some of the polynomials with no zeros on the first sheet inside the unfixed $\sigma$ and $\sigma'$ functions, we get a nice simplification. In particular, defining:
\beq
{\varsigma}\equiv \frac{\sigma}{B_{s,-}B_{\hat s,-}}\,,\qquad
{\varsigma'}\equiv \frac{\sigma'}{\dot B_{\dot s,+}\dot B_{\hat{\dot s},+}},
\eeq
we see that \eqref{eq:ABA_QQrel32} becomes
\beq\label{eq:oneside}
R_{s,+}R_{\hat s,+}\dot B_{\dot s,-}\dot B_{\hat {\dot s},-} -R_{s,-}R_{\hat s,-}\dot B_{\dot s,+}\dot B_{\hat {\dot s},+}= \frac{\varsigma}{\varsigma'} \,  R_b \; B_{\dot{f}} \, R_{\tilde{f}} \; B_{\dot{\tilde{b}}}\,.
\eeq
Notice that $\varsigma/\varsigma'$ has no poles or zeros on the first sheet and constant asymptotics at large $u$. Furthermore, from \eqref{eq:oneside} we see that it is a rational functions of the Zhukovski variables $x(u)$ and $x(u)^{-1}$, and by the structure of the equation, since every $R_a$ function can be rewritten as some $B$-type function times a power of $x$, we conclude, excluding trivial factorisations occurring, that  
it can only be a polynomial in $x(u)^{-1}$. 
We then see that, by redefining appropriately the $B$ type factors (themselves polynomials in $x(u)^{-1}$), $\varsigma/\varsigma' $ can be set to a constant.

We can immediately repeat the same argument for the other wing, starting from the QQ-relation
\begin{equation}
\dot{\bP}_1(u) \dot{\bQ}_{\dot 1\dot 1}(u) = \dot{\bS}_{\hat {\dot 1}|\dot1}^+(u)  \dot{\bS}_{\dot 1|\dot 1}^+(u)  - \dot{\bS}_{\hat{\dot1}|\dot1}^-(u) \dot{\bS}_{ \dot 1|\dot1}^-(u)  .\label{eq:forBethe33}
\end{equation}
Plugging our Ansatz for the various Q-functions and redefining \beq
{\dot \varsigma}\equiv \frac{\dot \sigma}{B_{\dot s,-}B_{\hat{\dot s},-}}\,,\qquad
{\dot\varsigma'}\equiv \frac{\dot \sigma'}{  B_{  s,+}  B_{\hat{  s},+}} ,
\eeq
equation \eqref{eq:forBethe33} reduces to
\beq
\label{eq:twoside}
\left( \dot R_{\dot s,+} \dot R_{\hat{\dot s},+} B_{s,-} B_{\hat {  s},-} -\dot R_{\dot s,-} \dot R_{\hat{\dot s},-}  B_{  s,+}  B_{\hat {  s},+}\right)= \frac{\dot \varsigma}{\dot\varsigma'} \,  R_{\dot b} \; B_{ {f}} \, R_{\tilde{\dot f}} \; B_{{\tilde{b}}}\, .
\eeq
Thus again we can take $\dot\varsigma/\dot{\varsigma}' $  to be a constant by redefining the $B$ factors on the RHS. 

Now, we notice that due to the properties of the Zhukovski polynomials $R$ and $B$, the LHS in \eqref{eq:twoside} is (minus) the analytical continuation of the LHS of \eqref{eq:oneside}. Thus, the same must hold for the two RHS. 
This means that the different sets of polynomials appearing in the $\bP$ and $\bQ$ functions are not independent, but related as follows:
\beq\label{eq:RRconstraint}
R_b R_{\tilde{f}} = R_f R_{\tilde{b}}, \;\;\; R_{\dot b} R_{\tilde{\dot f}} = R_{\dot f} R_{\tilde{\dot b}}.
\eeq
Shortly, we will make the further assumption that these sets of roots are pairwise identified. 

\paragraph{Imposing the gluing equations.}
So far, we have reduced our Ansatz for the $\bP$ and $\bQ$ functions in both wings to four sets of auxiliary roots constrained by (\ref{eq:RRconstraint}), the four ``momentum-carrying'' roots already encountered, plus two functions $\varsigma$  and $\varsigma'$. Taking into account the redefinitions described above, the parametrisation is
\beqa
\label{eq:ABA_sol_partial1}
&&\bP_{1}(u) \propto x^{-L} \, R_b \; B_{\dot{f}} \; \; B_{s,(-)} B_{\hat{s}, (-)} \,\varsigma,\;\;\; \dot{\bP}_{\dot 1}(u) \propto x^{-L} \, R_{\dot b} \; B_{{f}} \; B_{\dot s,(-)} B_{\hat{\dot{s}}, (-)} \; \dot{\varsigma} ,\\
&& \bQ_{11}(u) \propto \frac{x^{L} }{B_{\dot s,+} B_{\hat{\dot{s}},+} \varsigma } \, R_{\tilde{f}} \; B_{\dot{\tilde{b}}}\, 
\mathbf{f}
,\;\;\; \dot{\bQ}_{\dot 1\dot 1}(u) \propto \frac{x^{L} }{B_{ s,+} B_{ {\hat{s}},+} \dot{\varsigma} }\, R_{\dot{\tilde{f}}} \; B_{ {\tilde{b}}}  
\mathbf{f} . 
\,.
\eeqa
To constrain the $\varsigma$  and $\varsigma'$, we need to go beyond the first Riemann sheet: in particular, we need to use the gluing conditions for the Q-functions, in particular, the simplified equations (\ref{eq:ABAsimplified2}),(\ref{eq:ABAsimplified3}) we have obtained in the ABA limit. For clarity, we repeat the first of these here:
\beq
 (\bP_1 )^{\gamma}\sim \bS_{1|1}^+ \, \tau  \,\tau^{++} \, \left(\dot{\bQ}_{\dot 1\dot 1} \right)\, \bS_{\hat{1}|1}^{+} .
\eeq
We have found ABA expressions for all the quantities in this equation, see~\eqref{eq:ABA_sol_partial1}, (\ref{eq:Sfixed}),\eqref{eq:tausplit}. Performing analytic continuation for the function $\bP_1$ expressed as in \eqref{eq:ABA_sol_partial1},
we find
\beq
\label{eq:crossing_step1}
\varsigma^{\gamma} \dot{\varsigma} \; B_b \; R_{\dot{f}} \; \; R_{s,(-)} R_{\hat{s}, (-)}  \,\propto\,\mathbb{Q}_s^+ \mathbb{Q}_{\hat{s}}^+ \left( \prod_{\bullet} \mathbf{f}^{++}_{\bullet} \bar{\mathbf{f}}^{--}_{\bullet} \right)\frac{1}{B_{ s,(+)} B_{ {\hat{s}},(+)} }\, R_{\dot{\tilde{f}}} \; B_{ {\tilde{b}}} ,
\eeq
where we notice that, also in this case, the splitting factors of the spinors in (\ref{eq:Sfixed}) cancel out. 

Now, if we make the \textbf{simplifying assumption}, compatible with \eqref{eq:RRconstraint}, that the following sets of roots are identified
\beqa
R_f = R_{\tilde{f}}, \;\; R_{\dot f} = R_{\dot{\tilde{f}}}, \;\;\; R_b = R_{\tilde{b}}, \;\;\;\; R_{\dot b} = R_{\dot{\tilde b}} ,
\eeqa
then we have a particularly simple cancellation and \eqref{eq:crossing_step1} is reduced to
\beq
\label{eq:gluing_varsigma1}
\varsigma^{\gamma} \dot{\varsigma}   = C\,\frac{ R_{s,(+)} R_{\hat s, (+)} }{R_{s,(-)} R_{\hat{s}, (-)}  } \left( \prod_{\bullet} \mathbf{f}^{++}_{\bullet} \bar{\mathbf{f}}^{--}_{\bullet} \right) \,  ,
\eeq
which we can view as a Riemann-Hilbert problem that constrains the functions $\varsigma$, $\dot\varsigma$ in terms of the momentum-carrying roots.

Repeating an analogous calculation starting from the gluing constraint (\ref{eq:ABAsimplified3}), we obtain:
\beq
\label{eq:gluing_varsigma2}
\dot{\varsigma}^{\gamma} {\varsigma}   = \dot C\,\frac{ R_{\dot s,(+)} R_{\hat{\dot s}, (+)}   }{R_{ \dot s,(-)} R_{\hat{\dot{s}}, (-)}   } \left( \prod_{\bullet} \mathbf{f}^{++}_{\bullet} \bar{\mathbf{f}}^{--}_{\bullet} \right) \,.
\eeq
Above, $C$ and $\dot C$ are proportionality constants which are not fixed. Their presence affects the value of the ratio $\varsigma/{\dot {\varsigma}}$, and redefines the product by an overall constant which is not important. We parametrise their ratio as $C/\dot{C} \equiv e^{\alpha_C}$: notice that, although independent of the spectral parameter, this factor can depend on the Bethe roots of the state under consideration.

Equations \eqref{eq:gluing_varsigma1} and \eqref{eq:gluing_varsigma2} have the same form as those known for the QSC of the $\text{AdS}_3\times \text{S}^3 \times \text{T}^4$ model,
in particular (4.36) and (4.37) of \cite{Cavaglia:2021eqr}. Therefore, we will borrow many results from that case to write down the solution to \eqref{eq:gluing_varsigma1} and \eqref{eq:gluing_varsigma2}. The quantities $\varsigma$ and $\dot\varsigma$ will eventually be related to the dressing phases of the theory. 

We also see an immediate consequence of (\ref{eq:gluing_varsigma1}) and (\ref{eq:gluing_varsigma2}), taken together: they are not compatible with the branch cuts being quadratic. This is apparent since, if they were quadratic cuts, then the LHS of one of these two equations would be the analytic continuation inside the cut of the LHS of the other. But this is not true for the RHS. This indicates that, even within the ABA limit, the branch cuts exhibit non-quadratic monodromy.

To write the solution of (\ref{eq:gluing_varsigma1}) and (\ref{eq:gluing_varsigma2}), we introduce the following useful redefinitions,
\beq
\varsigma \equiv \sqrt{\frac{B_{s,(+)} B_{\hat s,(+)}}{ B_{s,(-)} B_{\hat s,(-)} } } 
\prod_{\bullet} \sigma^1_{\bullet, \text{BES}} \; \rho , \;\;\;\; \dot\varsigma \equiv 
\sqrt{\frac{B_{\dot s,(+)} B_{\hat{\dot s} ,(+)}}{ B_{\dot s,(-)} B_{\hat{\dot  s},(-)} } }  \prod_{\bullet} \sigma^1_{\bullet, \text{BES}}  ,
\eeq
where $\sigma^1_{\bullet,\text{BES}}$, parametrised by the momentum-carrying roots of type $\bullet=s,\hat s,\dot s,\hat{\dot s}$, are the building blocks of the BES dressing phase and are defined in appendix \ref{appendix:dressing}, and $\rho$, $\dot \rho$ are new unknown functions, with a single cut on the first sheet.  Thanks to the functional equation satisfied by the BES building blocks \eqref{appendix:bes_functional}, with these redefinitions \eqref{eq:gluing_varsigma1} and \eqref{eq:gluing_varsigma2} become:
\begin{align}
\label{eq:gluing_rho}
& {\rho}^{\gamma} \dot{\rho}  = \sqrt{\frac{ R_{  s,(+)} R_{ {\hat s}, (+)} B_{ \dot s,(-)} B_{\hat{\dot{s}}, (-)}  }{R_{   s,(-)} R_{{\hat{s}}, (-)} B_{ \dot s,(+)} B_{\hat{\dot{s}}, (+)}   } }\; C,
\\&
\label{eq:gluing_rho2}
 {\dot \rho}^{\gamma}  {\rho}  = \sqrt{\frac{ R_{  \dot s,(+)} R_{ \hat{\dot s}, (+)} B_{ s,(-)} B_{ {\hat{s}}, (-)}  }{R_{   \dot s,(-)} R_{\hat{\dot{s}}, (-)} B_{   s,(+)} B_{ {\hat{s}}, (+)}  } }\; \dot C.
\end{align}
To write the solution with the correct analytic properties, it is convenient to introduce 
 $$
 e^{\alpha_C} \equiv \frac{C}{\dot{C}} ,
 $$
and to reparametrise this constant as
\beqa
\label{def:alpha_C}
 \alpha_C &\equiv&  + \frac{1}{4}\sum_{\bullet = s, \hat{s}}\sum_{i=1}^{K_\bullet} \log\left(\frac{(v_{\bullet,i}^+ - 2 h)(v_{\bullet,i}^+ + 2 h)}{(v_{\bullet,i}^- - 2 h)(v_{\bullet,i}^- + 2 h)} \right) +\frac{1}{4}\sum_{\bullet = \dot{s}, \dot{\hat{s}}}\sum_{i=1}^{K_\bullet} \log\left(\frac{(v_{\bullet,i}^+ - 2 h)(v_{\bullet,i}^+ + 2 h)}{(v_{\bullet,i}^- - 2 h)(v_{\bullet,i}^- + 2 h)} \right) \nonumber \\
 &&+\tilde{\alpha}_C ,
\eeqa
where the term in the first line (which is a constant in $u$, depending on the value of the Bethe roots)  serves the purpose of making a connection with the equations \ref{eq:extratermBudapest}. Then, the solution can be written as
\begin{align}
\label{eq:rho_sigma}
&
\rho \propto \sigma^{1,\text{extra}}_s \sigma^{1,\text{extra}}_{\hat s}  \tilde{\sigma}^{1,\text{extra}}_{\dot s} \tilde{\sigma}^{1,\text{extra}}_{\hat{\dot s} } \; b_C , \\&
\dot\rho \propto \tilde{\sigma}^{1,\text{extra}}_s \tilde{\sigma}^{1,\text{extra}}_{\hat s}   {\sigma}^{1,\text{extra}}_{\dot s}  {\sigma}^{1,\text{extra}}_{\hat{\dot s} } \; (b_C )^{-1}, 
\end{align}
where $\sigma_{\bullet}^{1,\text{extra}}$ and $\tilde{\sigma}_{\bullet}^{1,\text{extra}}$ are building blocks of the massive dressing phases appearing in the $\text{AdS}_3\times \text{S}^3 \times \text{T}^4$ model, reviewed in appendix \ref{appendix:dressing}. The subscript here denotes the set of roots at which the second argument of these functions is evaluated.
 The residual factor $b_C(u)$, which is nontrivial only if $\tilde{\alpha}_C\neq 0$, is obtained by solving the equation $b_C^{\gamma} b_C^{-1} = e^{\tilde{\alpha}_C/2}$,
 which gives
\beq\label{eq:residualphase}
b_C(u) = \left(\frac{u-2 h}{u+ 2 h}\right)^{\frac{
\tilde{\alpha}_C}{4 \pi i}} .
\eeq
The constant $\tilde{\alpha}_C$ is unfixed for the moment, but its value will affect the form of the Bethe equations through the factors $b_C(u)$. Hence, we will take $\tilde{\alpha}_C = 0$, since this leads to Bethe equations with a worldsheet S-matrix satisfying all constraints, at least in the symmetric case.\footnote{So far, this is all completely analogous to the situation in the $AdS_3\times S^3\times T^4$ model, see \cite{BudapestTalk}, where a constant playing a similar role is fixed to the same value depending on the roots.}  
We will comment on the possible impact of a different choice in the Conclusions.

\paragraph{Summary of the solution so far.}
The parametrisation  found so far for Q-functions in the ABA limit is:
\beqa
&&\bP_{1}(u) \propto x^{-L} \, R_b \; B_{\dot{f}} \; \; \sqrt{B_{s,(-)} B_{\hat{s}, (-)} B_{ s,(+)} B_{\hat{s}, (+)}  }\,\rho \, \prod_{\bullet} \sigma_{\bullet, \text{BES}} ,\\&& \dot{\bP}_{1}(u) \propto x^{-L} \, R_{\dot b} \; B_{{f}} \; \sqrt{B_{\dot s,(-)} B_{\hat{\dot{s}}, (-)} B_{\dot s,(+)} B_{\hat{\dot{s}}, (+)} } \; \dot{\rho} \,  \prod_{\bullet} \sigma_{\bullet, \text{BES}} ,
\eeqa
and
\beqa
&& \bQ_{11}(u) = \frac{x^{L} }{B_{\dot s,(+)} B_{\hat{\dot{s}},(+)}   } \, \sqrt{\frac{B_{s,(-)} B_{\hat s, (-)}}{ B_{s,(+)} B_{\hat s, (+)}}} \, R_{ {f}} \; B_{\dot{ {b}}}\, \frac{1}{\rho} \prod_{\bullet} \frac{\bf{f}_{\bullet}}{\sigma_{\bullet, \text{BES}} } ,\\&&\dot{\bQ}_{11}(u) = \frac{x^{L} }{B_{ s,(+)} B_{ {\hat{s}},(+)} }\, \sqrt{\frac{B_{\dot s,(-)} B_{\hat{\dot s}, (-)}}{ B_{\dot s,(+)} B_{\hat{\dot s} , (+)}}} \, R_{\dot{ {f}}} \; B_{ { {b}}}  \frac{1}{\dot \rho} \prod_{\bullet} \frac{\bf{f}_{\bullet}}{\sigma_{\bullet, \text{BES}} } ,
\eeqa
where, on top of the four types of momentum-carrying roots, we also have four types of auxiliary roots: two of those at a time will appear in the ABA equations, e.g. $f$ and $\dot b$, while the others are ``dual'' roots $b$ and $\dot f$  enter dual equations where the grading used in the wings is swapped.  

Concerning the spinors, we have found the solution (\ref{eq:Sfixed}), which is not fully fixed yet, since we still need to determine the splitting factors $W$, $\dot W$, $\hat{W}$, $\hat{\dot W}$. This is what we turn to next. 

\subsection{Determining the splitting factors}
The splitting factors $W$, $\dot W$, $\hat{W}$, $\hat{\dot W}$ are the parts of the construction that are sensitive to the possibility that spinors and antispinors carry different sets of Bethe roots. In fact, we will see that if we assume a symmetry between these types of roots, these factors do not contribute to the Bethe equations. In the non-symmetric case, they are responsible for the presence of a new piece of the dressing factor.  

\subsubsection{Symmetric case}
The simplest case is the one where, in each wing, spinors and anti-spinors become equal in the ABA limit, in particular, the sets of roots $s \leftrightarrow \hat{s}$ and $\dot s \leftrightarrow \hat{\dot s}$ are the same, i.e.
\beq
\left\{ u_{s, k} \right\} = \left\{ u_{\hat{s}, k} \right\}  , \;\;\; \left\{ u_{\dot s, k} \right\} = \left\{ u_{\hat{\dot{s}}, k} \right\} .
\eeq
In this case, as we will see later by discussing the general case, there is a very simple solution for the splitting factors appearing in (\ref{eq:Sfixed}), i.e. they all trivialise, $W = \hat{W} =\dot{W} = \hat{\dot{W}} = 1$. Therefore, the solution of the QSC in the ABA limit is completely determined and takes the simple form:
\beqa
&&\bP_{1}(u) \propto x^{-L} \, R_b \; B_{\dot{f}} \; \; B_{s,(-)}  B_{ s,(+)}   \,\rho \, \prod_{\bullet} \sigma_{\bullet, \text{BES}} ,\\&& \dot{\bP}_{1}(u) \propto x^{-L} \, R_{\dot b} \; B_{{f}} \; B_{\dot s,(-)}  B_{\dot s,(+)}   \; \dot{\rho} \,  \prod_{\bullet} \sigma_{\bullet, \text{BES}} ,\\
&& \bQ_{11}(u) = \frac{x^{L} }{(B_{\dot s,(+)})^2  } \, \frac{B_{s,(-)} }{ B_{s,(+)} } \, R_{ {f}} \; B_{\dot{ {b}}}\, \frac{1}{\rho} \prod_{\bullet} \frac{\bf{f}_{\bullet}}{\sigma_{\bullet, \text{BES}} } ,\\&&\dot{\bQ}_{11}(u) = \frac{x^{L} }{(B_{ s,(+)})^2  }\, \frac{B_{\dot s,(-)} }{ B_{\dot s,(+)} } \, R_{\dot{ {f}}} \; B_{ { {b}}}  \frac{1}{\dot \rho} \prod_{\bullet} \frac{\bf{f}_{\bullet}}{\sigma_{\bullet, \text{BES}} } ,
\eeqa
and 
\beq
\bS_{1|1}=\bS_{\hat 1|1}=\mathbb{Q}_s (\mathbf{f}^+)^{1 \over 2},\;\;\;\;\slS_{\dot1|\dot1}=\slS_{\hat{\dot 1}|\dot 1}=\mathbb{Q}_{\dot s} (\mathbf{f}^+)^{1 \over 2}.
\eeq
\subsubsection{Non-symmetric case}\label{sec:simplifynonsym}
In the general case, we have four independent sets of momentum-carrying roots, constrained only by the zero-momentum condition \eqref{eq:zero_momentum_condition}. Now we list two types of constraints on the splitting factors.

\paragraph{Constraints from mirror periodicity of $\nu$. } The splitting factors appear in the ABA expressions for some of the $\nu$ functions, see (\ref{eq:nusplit}). We can now impose mirror periodicity, i.e. 
\begin{equation}\label{eq:mirrornu}
\nu^{\gamma} = \nu^{++}.
\eeq
and similar for the other $\nu$ functions. 
For example, starting from the ABA limit of $\nu_{1}^{\dot{2}}$ in (\ref{eq:nusplit}), we can evaluate easily $\nu^{++}$ appearing on the RHS of the equation. 

Then, we need to evaluate $\nu^{\gamma}$ in the ABA limit. We have $\nu$ in the ABA limit. It is not, a priori, guaranteed that this limit commutes with analytic continuation, i.e. that taking the ABA expression (\ref{eq:nusplit}) and analytically continuing it along $\gamma$, we will get the right result for $\nu^{\gamma}$ in the large-volume limit. However, we note that both $\nu$ and $\nu^{\gamma}$ are quantities expected to be of order $O(1)$ at large $L$. 
We will then assume the following:  analytic continuation of the ABA value gives the correct result for $\nu^{\gamma}$ in the limit, \textbf{modulo a possible multiplicative constant}. 

We can think of this constant as arising from a Stokes-like phenomenon.  
As a simple example of such a phenomenon, we can consider the rational function of the Zhukovski variables $f(u)\equiv 1-\frac{1-C}{1+x^L(u)}$, where $C$ is a constant. Assuming that we are in the region where $|x(u)|>1$ and performing the $L\rightarrow\infty$ limit first, we get $f_{ABA}(u)=1$, whose analytic continuation is trivially $f_{ABA}(u^{\gamma})=1$. On the other hand, if we analytically continue first and then perform the $L\rightarrow\infty$ limit, we arrive at $(f(u^{\gamma}))_{ABA}=C$. The two quantities clearly differ by a constant that cannot be deduced from the ABA limit of $f$ on the first sheet.

Going back to the mirror periodicity equation, we are analogously assuming the following
\beq\label{eq:StokesE}
\left[\nu_1^{\dot{2}}(u^\gamma ) \right]_{\text{ABA}} = \left[\nu_1^{\dot{2}}\right]_{\text{ABA}}\left(u\rightarrow u^\gamma \right) \times \mathcal{E} , 
\eeq
where $\mathcal{E} $ is a constant independent of $u$. Although we cannot compute the value of this constant directly from our construction, it plays a role in the dressing phases of our model. In particular, it seems that imposing some properties on the latter selects a specific form for $\mathcal{E}$, as we discuss in section \ref{sec:dressingnew}. For the moment, however, we will leave it unfixed.

After this long premise, let us come back to the problem of imposing equation (\ref{eq:mirrornu}) in the ABA limit. Considering $\nu_1^{\; \dot 2}$ in (\ref{eq:nusplit}), we then get
\beq\label{eq:functionalVW1}
 {V}(u) {(V(u))}^{\gamma} = \mathcal{A}(u) \times \frac{W^+(u) \dot{W}^+(u)}{(W^-(u) \dot{W}^-(u))^{\gamma}} \; \mathcal{E},
\eeq
where $\mathcal{A}(u)$ is a ``source term'' which is an explicit function of the roots:
\beq\label{eq:defA}
\mathcal{A}(u) = \left(\frac{
\mathbb{Q}_s^+ \mathbb{Q}_{\dot s}^+ \mathbb{Q}_{\hat s}^-\mathbb{Q}_{\hat{\dot s}}^- }{ \mathbb{Q}_s^-\mathbb{Q}_{\dot s}^- \mathbb{Q}_{\hat s}^+\mathbb{Q}_{\hat{\dot s}}^+} \right)^{\frac{1}{2}}.
\eeq
We note that in the symmetric case described above, $\mathcal{A} \equiv 1$. Moreover, in this sector, we know that we should have $W(u) = 1$, and this implies that we should have $\mathcal{E} = 1$ in this sector. 

Repeating the calculation for $\dot{\nu}$, we find an equation of the exact same form, with a priori a different constant $\dot{\mathcal{E}}$, which we keep for generality and is also -- conjecturally -- appearing due to Stokes-like phenomenon in the large volume limit. 
However, considering the ratio $V_{\text{ratio}}\equiv V/\dot{V}$, we find from the above constraints that it should satisfy $V_{\text{ratio}} V_{\text{ratio}}^{\gamma} = \mathcal{E}/\dot{\mathcal{E}}$, and moreover that it should be anti-periodic, i.e. $V_{\text{ratio}}^{++} = 1/V_{\text{ratio}}$
\footnote{The solution of such a problem can be found with Hilbert-transform techniques, see (7-9) in \cite{Gromov:2014eha} for a similar calculation. }. If we do not want to spoil the property that at large $u$ the asymptotics of these functions is $u^{\text{power}}$, rather than exponential, the only possibility is found when $\mathcal{E}/\dot{\mathcal{E}} = 1$ and is given by $V_{\text{ratio}} =1$, namely we should have
\beq
V(u) = \dot{V}(u).
\eeq
Considering the equations for $\hat{\nu}$ and $\dot{\hat{\nu}}$ then gives the constraints
\beq\label{eq:proportionalityone}
\frac{W^+ \dot{W}^+}{(W^- \dot{W}^-)^{\gamma}} {\propto} \left( 
\frac{\hat{W}^+ \hat{\dot W}^+}{(\hat{W}^- \hat{\dot W}^-)^{\gamma}}
\right)^{-1},
\eeq
where the proportionality constant can be traced to the ratio of $\mathcal{E}$ and another constant emerging due to a similar Stokes phenomenon from the analytic continuation of $\hat{\nu}$. 

To fully fix these splitting factors, we also consider another constraint, which comes from the assumption that the gluing matrix is off-diagonal.

\paragraph{Constraints coming from the off-diagonal gluing matrix hypothesis. }
We have been working with the basis of Q-functions analytic in the upper half plane, but the two bases are related by the matrices $\Omega$ defined in the previous section. In the large volume limit, this relation simplifies in some cases, e.g.
$$
\bS_{1|1}^{\uparrow -} = \bS_{1|j}^{\downarrow -} \Omega^j_1 \sim \Omega_1^1 \bS_{1|1}^{\downarrow -} ,
$$
and the same happens for the other spinors. 

In turn, the $\Omega$ matrix elements are related to the gluing matrices and the elements of the $\tau$ matrices by (\ref{def:fmatrices}). Now, under the assumption that the gluing matrix has the form (\ref{eq:gluingform}), at least in the large volume limit, the diagonal elements of $\Omega$  appearing in the previous relation can be determined in terms of the solution (\ref{eq:tausplit}), i.e. we have\footnote{Notice that also in these equations appear some unfixed proportionality constants. However, provided they are finite and non-zero, which follows from our assumption on the form of the gluing matrix, these constants will play no role.}
\beqa
\bS_{1|1}^{\uparrow} &\sim& \frac{\dot{\tau}_2^{1[+1]}}{\dot{\mathcal{G}}_2^1 } \bS_{1|1}^{\downarrow}  \propto 
\mathbb{Q}_s (\bar{\mathbf{f}}^{-} )^{\frac{1}{2}}  W V^+
,\label{eq:glucon0}\\
\bS_{\hat{1}|1}^{\uparrow} &\sim &  \frac{\dot{\tau}_2^{1 [-1]}}{\dot{\mathcal{G}}_2^1 }  \bS_{\hat{1}|1}^{\downarrow}  \propto 
\mathbb{Q}_{\hat s} (\bar{\mathbf{f}}^{-} )^{\frac{1}{2}}  \hat{W}/V^+ ,\label{eq:glucon1}\\
\dot{\bS}_{\dot 1|\dot 1}^{\uparrow} &\sim& \frac{{\tau}_2^{1[+1]}}{{\mathcal{G}}_2^1 }  \bS_{\dot 1| \dot 1}^{\downarrow}  \propto \mathbb{Q}_{\dot s} (\bar{\mathbf{f}}^{-} )^{\frac{1}{2}} \dot{W} V^+ ,\\
\dot{\bS}_{\hat{\dot  1} | \dot 1}^{\uparrow} &\sim& \frac{{\tau}_2^{1 [-1]}}{{\mathcal{G}}_2^1 } 
\bS_{\hat{\dot{1}}| \dot 1}^{\downarrow} \propto \mathbb{Q}_{\hat{\dot{s}}} (\bar{\mathbf{f}}^{-} )^{\frac{1}{2}}  \hat{\dot{W}}/V^+.\label{eq:assumptionsG}
\eeqa
We obtain infinitely many constraints by requiring that these combinations have no cuts in the lower half-plane, as they should.

For instance, consider  (\ref{eq:glucon0}). Imposing the cancellation of the first cut at $\text{Im}(u) = -1/2$ (which is manifestly absent in the LHS) leads to the constraint
\beq\label{eq:discWV}
( V W^- )^{\gamma} = V W^- .
\eeq
We get further conditions from the other apparent cuts lower down. These constraints have a very similar structure, since the polynomial and $\bar{\mathbf{f}}$ factors in (\ref{eq:glucon1}) are analytic in the lower half plane and do not contribute. All in all, imposing the cancellation of all the unwanted cuts of the left-hand side, we find infinitely many relations between the cuts of $W$ and those of $V$ (which are anti-periodic):
\beq\label{eq:cutsW}
\frac{(W^{[-1 - 4 k]})^{\gamma}}{W^{[-1 - 4 k]}}  = \left(\frac{V^{\gamma}}{V} \right)^{{-1}}, \;\;\;\; \frac{(W^{[+1 - 4 k]})^{\gamma}}{W^{[+1 - 4 k]} } = \left(\frac{V}{V^{\gamma}}\right)^{{-1}}, 
\eeq
$k=0,1,2,\dots$, where the alternating pattern comes from the anti-periodicity of $V$.
From the structure of the RHS of (\ref{eq:glucon0})-(\ref{eq:assumptionsG}), we see that the other functions satisfy the relations:
\beqa
&&\frac{(\dot{W}^{[-1 - 4 k]})^{\gamma}}{\dot{W}^{[-1 - 4 k]}}  = \left(\frac{V^{\gamma}}{V} \right)^{{-1}}, \;\;\;\; \frac{(\dot{W}^{[+1 - 4 k]})^{\gamma}}{\dot{W}^{[+1 - 4 k]} } = \left(\frac{V}{V^{\gamma}} \right)^{{-1}}, \\
&&\frac{(\hat{W}^{[-1 - 4 k]})^{\gamma}}{\hat{W}^{[-1 - 4 k]}}  = \left(\frac{V}{V^{\gamma}} \right)^{{-1}}, \;\;\;\; \frac{(\hat{W}^{[+1 - 4 k]})^{\gamma}}{\hat{W}^{[+1 - 4 k]} } = \left(\frac{V^{\gamma}}{V} \right)^{{-1}}, \\
&&\frac{(\hat{\dot{W}}^{[-1 - 4 k]})^{\gamma}}{\hat{\dot{W}}^{[-1 - 4 k]}}  = \left(\frac{V}{V^{\gamma}} \right)^{{-1}}, \;\;\;\; \frac{(\hat{\dot{W}}^{[+1 - 4 k]})^{\gamma}}{\hat{\dot{W}}^{[+1 - 4 k]} } = \left(\frac{V^{\gamma}}{V} \right)^{{-1}}.
\eeqa
The previous relations have the immediate consequence that there is only one independent function among $W$, $\dot{W}$, $\hat{W}$ and $\hat{\dot{W}}$, since their discontinuities are all related to the discontinuity of the single function $V$. 
Combined with their power-like asymptotics and the fact that they have no zeros or poles, the discontinuities of their logarithms completely fix them up to a multiplicative constant. 
 
In short, we can set without loss of generality\footnote{We can use the freedom to redefine each of the $W$ functions by a proportionality constant, which would be invisible in the constraints we have derived as well as in the Bethe equations.}
\beq
W = \dot{W} = 1/\hat{W} = 1/\hat{\dot{W}} ,
\eeq
reducing the whole problem to two single functions, $W$ and $V$. The proportionality constant in (\ref{eq:proportionalityone}) is then seen to be $1$. There is only one constant remaining in the game, $\mathcal{E}$, entering equation (\ref{eq:functionalVW1}). This cannot be removed by a redefinition of $W$ by a multiplicative factor, and on the other hand, the normalisation of $V$ is already fixed by $V^{++} = 1/V$. Therefore, we will carry this constant with us.
\subsubsection{The linear integral equation for the splitting factors}
In the non-symmetric case, we are left with two functions to determine, $V(u)$ and $W(u)$, whose discontinuities are related as described by equation (\ref{eq:discWV}), (\ref{eq:cutsW}). We now demonstrate that the problem of computing these two functions can be reduced to solving a linear integral equation.

\paragraph{The linear integral equation. }
 Let us parametrise the discontinuity of $\log{V}(u)$ on the cut on the real axis by a density function
\beq
\mathcal{D}_V(u) \equiv \log{V(u) }- \log{V^{\gamma^{-1}}(u)} ,
\eeq
for $u \in (-2 h, 2h) + i 0^+$. Then, the constant large-$u$ asymptotics and anti-periodicity of $V(u)$ imply that out of this discontinuity we can reconstruct the full function, as:
\beq\label{eq:Vrepresent}
\log{V}(u) = \int_{-2 h}^{2 h} \mathcal{D}_V(z) \frac{d z }{2 i \sinh(\pi (z - u) )} .
\eeq
Now, the discontinuities of the second function $W$ need to be exactly correlated to those in $V$, as described by (\ref{eq:cutsW}). 
Taking into account the analytic properties of this function, it is simple to see that it also admits a simple integral representation in terms of the same density $\mathcal{D}_V$. The representation is (up to a multiplicative constant in $W$, which we fix arbitrarily and does not have any consequence):
\beq\label{eq:logW}
\log{W}^-(u) = \int_{-2 h}^{2 h} \mathcal{D}_V(z)  \frac{dz}{2i\pi} \mathbb{K}(u-z) , 
\eeq
where the kernel is
\beq
\mathbb{K}(u-z)\equiv  \frac{i}{2}\left[ \psi\left({-}\frac{i}{2}(u-z)\right) -\psi\left({-}\frac{i}{2}(u-z {+}i)\right)  \right],
\eeq
where $\psi(z)$ is the Digamma function with simple poles  at $z = 0,-1,-2,\dots$. In fact, it is simple to see that this integral produces a function with exactly the right discontinuities, correlated to the ones of $V$ as in equations (\ref{eq:discWV}), (\ref{eq:cutsW}), on the cuts at $\text{Im}(u) = -1/2, -3/2, \dots$, and analytic in the upper half plane. 

With these two representations, we have implemented the conditions (\ref{eq:cutsW}), but we still need to impose the remaining relation condition (\ref{eq:functionalVW1}). As we now demonstrate, this translates into a linear integral equation that constrains the density $\mathcal{D}_V$. We start by massaging (\ref{eq:functionalVW1}), by continuation along $\gamma^{-1}$, to the form  
\beq
\label{eq:VVgamma}
{V}(u) {V}^{\gamma^{-1}}(u) = \mathcal{A}(u) \times \left(\frac{ {W}^+(u)}{ W^-(u) }\right)^2 \; \mathcal{E},
\eeq
and in appendix \ref{appendix:detailsLIE} we prove that this implies the linear integral equation
\beq\label{eq:LIE3}
\int_{-2h}^{2h} dz\, \mathcal{D}_V(z) \mathbb{G}(u-z) -\mathcal{D}_V(u)=\log{\mathcal{A}(u)} + \log\mathcal{E}\,,
\eeq
with the kernel
\beqa\label{eq:defG}
\mathbb{G}(u,v)&=&\frac{1}{2i\pi}\left(\frac{2 \pi }{\sinh(\pi (u - v) )}-\frac{2}{(u - v) } - 4 \mathbb{K}(v+i, u) \right)\\&=&\frac{1}{i\pi}\sum_{n=1}^{\infty} (-1)^n\left(\frac{1}{u-v+i n}-\frac{1}{u-v-i n}\right).
\eeqa
We note that this kernel has no singularities on the real axis and is symmetric, i.e. $\mathbb{G}(u-v)=\mathbb{G}(v-u)$. Once we solve this equation for $\mathcal{D}_V$, we can compute $V(u)$ and $W(u)$ through the integral representations (\ref{eq:Vrepresent}), (\ref{eq:logW}) above.
\subsection{Summary of the ABA solution }
At this point, we can list our full solution:
\beqa
&&\bP_{1}(u) \propto x^{-L} \, R_b \; B_{\dot{f}} \; \; \sqrt{B_{s,(-)} B_{\hat{s}, (-)} B_{ s,(+)} B_{\hat{s}, (+)}  }\,\rho \, \prod_{\bullet} \sigma^1_{\bullet, \text{BES}} ,\\&& \dot{\bP}_{1}(u) \propto x^{-L} \, R_{\dot b} \; B_{{f}} \; \sqrt{B_{\dot s,(-)} B_{\hat{\dot{s}}, (-)} B_{\dot s,(+)} B_{\hat{\dot{s}}, (+)} } \; \dot{\rho} \,  \prod_{\bullet} \sigma^1_{\bullet, \text{BES}} ,
\eeqa
and
\beqa
&& \bQ_{11}(u) \propto \frac{x^{L} }{B_{\dot s,+} B_{\hat{\dot{s}},+}   } \, \sqrt{\frac{B_{s,(-)} B_{\hat s, (-)}}{ B_{s,(+)} B_{\hat s, (+)}}} \, R_{ {f}} \; B_{\dot{ {b}}}\, \frac{1}{\rho} \prod_{\bullet} \frac{\bf{f}_{\bullet}}{\sigma^1_{\bullet, \text{BES}} } ,\\&&\dot{\bQ}_{11}(u) \propto  \frac{x^{L} }{B_{ s,+} B_{ {\hat{s}},+} }\, \sqrt{\frac{B_{\dot s,(-)} B_{\hat{\dot s}, (-)}}{ B_{\dot s,(+)} B_{\hat{\dot s} , (+)}}} \, R_{\dot{ {f}}} \; B_{ { {b}}}  \frac{1}{\dot \rho} \prod_{\bullet} \frac{\bf{f}_{\bullet}}{\sigma^1_{\bullet, \text{BES}} } ,
\eeqa
with the factors $\rho$, $\dot \rho$  are given by\footnote{Setting the constant $\tilde{\alpha}_C $ to zero, which simplifies (\ref{eq:rho_sigma}).}
\begin{align}
\label{eq:rho_sigmaupd}
&
\rho \propto \sigma^{1,\text{extra}}_s \sigma^{1,\text{extra}}_{\hat s}  \tilde{\sigma}^{1,\text{extra}}_{\dot s} \tilde{\sigma}^{1,\text{extra}}_{\hat{\dot s} } , \\&
\dot\rho \propto \tilde{\sigma}^{1,\text{extra}}_s \tilde{\sigma}^{1,\text{extra}}_{\hat s}   {\sigma}^{1,\text{extra}}_{\dot s}  {\sigma}^{1,\text{extra}}_{\hat{\dot s} } , 
\end{align}
i.e. they are defined in terms of blocks of the dressing phases of $AdS_3\times S^3\times T^4$.

The spinors Q-functions are instead given by:
\beqa\label{eq:Sfixed2}
\bS_{1|1} &\propto& \mathbb{Q}_s(\prod_{\bullet}\mathbf{f}^+_{\bullet})^{\frac{1}{2}} \, W 
, \\
\dot{\bS}_{\dot 1|\dot 1} &\propto& \mathbb{Q}_{\dot s} (\prod_{\bullet}\mathbf{f}^+_{\bullet})^{\frac{1}{2}} \, W 
, \\
\bS_{\hat{1}|1} &\propto& \mathbb{Q}_{\hat s} (\prod_{\bullet}\mathbf{f}^+_{\bullet})^{\frac{1}{2}} \, W^{-1} 
, \\
\dot{\bS}_{\hat{\dot 1}|\dot 1} & \propto& \mathbb{Q}_{\hat{\dot{ s}} } (\prod_{\bullet}\mathbf{f}^+_{\bullet})^{\frac{1}{2}} 
\, W^{-1} , 
\eeqa
where the factor $W$ is implicitly defined by the linear integral equation discussed above. This is a genuinely new type of function, not seen in the other AdS/CFT models. 
Notice that our solution depends on two unfixed constants: the constant $C $, for which we assumed the specific value given by (\ref{def:alpha_C}) with $\tilde{\alpha}_C = 0$, and the constant $\mathcal{E}$ which we conjectured appears in (\ref{eq:StokesE}). This constant will affect the value of the splitting factor $W$, and we have not yet specified its value.

We notice that in the symmetric sector, which is specified by the equality of spinor and antispinor Q-functions, we should have $W(u) = 1$.

\subsection{Asymptotic Bethe equations in the massive sector}
\label{section:ABA}
In this section, we present one of the main tests of our construction: show that the QSC in the large volume limit reproduces the massive asymptotic Bethe equations obtained from the integrable worldsheet S-matrix bootstrap\footnote{The paper \cite{Borsato:2012ss} actually writes the Asymptotic Bethe equations for the general $\dalpha^{\otimes 2}$ case. Those, however, immediately reduce to our equations when setting $\alpha=1/2$.}. In doing so, we find that the unknown dressing phases of the \ads\,S-matrix of \cite{Borsato:2012ss} are completely given in terms of the functions we have encountered in the construction. 

To obtain the Asymptotic Bethe equations, we need to input in the Bethe equations derived in section \ref{sec:Bethe_equations} the Q-functions found in the last section, choosing the same gradings of \cite{Borsato:2012ss}. In particular, for the L and R wings, we will use respectively the gradings associated with the Dynkin diagrams \ref{fig:sfig2} and \ref{fig:sfig1}.
\subsubsection{ABA for the L wing}
The asymptotic Bethe equations for the L wing come from the Bethe equations \eqref{eq:BAE_2_node1}, \eqref{eq:BAE_2_node2} and \eqref{eq:BAE_2_node3}, which we rewrite here for convenience (setting $\bQ_{\emptyset}=1$):
\begin{eqnarray}
 1&=&  \left. \frac{\bS_{1|1}^- \bS_{\hat 1|1}^-}{\bS_{1|1}^+ \bS_{\hat 1|1}^+}\right|_{\bQ_{11} = 0} \,,\label{eq:zerosofQ11}\\
-1&=&  \left. \frac{\bS_{1|1}^{++}}{\bS_{1|1}^{--} } \frac{\bQ_{11}^-}{\bQ_{11}^+}  \;\right|_{\bS_{1|1} = 0}\,,\\
-1&=& \left. \frac{\bS_{\hat 1|1}^{++}}{\bS_{\hat 1|1}^{--} } \frac{\bQ_{11}^-}{\bQ_{11}^+}  \;\right|_{\bS_{\hat 1|1} = 0}\,.
\end{eqnarray}
\paragraph{Useful conventions for the ABA.}To compare with the asymptotic Bethe equations found in \cite{Borsato:2012ss}, we need to
write explicitly the Zhukovski polynomials in terms of Zhukovski variables. To do so, we will introduce the following notation:
\begin{itemize}
\item We use the notation $y_j\equiv x(u_j)$ if $u_j$ is a Bethe root associated to Q-functions on the first nodes, i.e. $\bQ_{11}$ or $\bP_{1}$ and their dotted counterparts. Those will play the role of \textbf{auxiliary roots}. 
\item Otherwise, we employ $x_k\equiv x(u_k)$ if $u_k$ is a Bethe root associated to Q-functions on the spinor or antispinor nodes, i.e. $\bS_{1|1}$ or $\bS_{\hat 1|1}$ and their dotted counterparts. Those will be \textbf{momentum carrying roots}. The association with the sets of roots labelled by $s$ (spinor L wing), $\hat{s}$ (antispinor L wing), $\dot{s}$ (spinor R wing) or $\dot{\hat s}$ (antispinor R wing) will be marked by writing the index as $k$, $\hat{k}$, $\dot{k}$ or $\dot{\hat k}$, respectively. 
\item We normally use ``$k$'' or its dotted or hatted counterparts for the index of a root at which the BAE is evaluated.  The indices $i$, $m$, $r$ are used as iterators in products over various types of roots.  
Quantities such as $x_i\equiv x(u_i)$, $y_i\equiv x(u_i)$ simply denote the  Zhukovski map evaluated at these Bethe roots.
\end{itemize}
\paragraph{ Bethe equations. }
Let us plug the solution for the Q-functions in the large volume limit into the exact Bethe equations. 
From  equation (\ref{eq:zerosofQ11}), we obtain an auxiliary-type Bethe equation:
\begin{equation}
1= \left. \prod_{i=1}^{K_s}\frac{ (y_k - x_i^+) }{(y_k - x_i^-)} 
\sqrt{\frac{x_i^-}{x_i^+}}
\prod_{\hat{i}=1}^{K_{\hat s}}
\frac{ (y_k - x_{\hat{i}}^+) }{(y_k - x_{\hat{i}}^-)} 
\sqrt{\frac{x_{\hat{i}}^-}{x_{\hat{i}}^+}}  \prod_{\dot{m}=1}^{K_{\dot s}}\frac{ (1/y_k - x_{\dot{m}}^-) }{(1/y_k - x_{\dot{m}}^+)} \sqrt{\frac{x_{\dot{m}}^+}{x_{\dot{m}}^-}}
\prod_{\hat{\dot{m}}=1}^{K_{\hat{\dot s} }}\frac{ (1/y_k - x_{\hat{\dot{m}}}^-) }{(1/y_k - x_{\hat{\dot{m}}}^+)} 
\sqrt{\frac{x_{\dot{\hat{m}}}^+}{x_{\dot{\hat{m}}}^-}} 
\right|_{y_k = y_{f, k}} 
,
\end{equation}
which, using the zero-momentum condition \eqref{eq:zero_momentum_condition},  can also be simplified to:
\beq
1= \left. \prod_{i=1}^{K_s}\frac{ (y_k - x_i^+) }{(y_k - x_i^-)} \prod_{\hat i=1}^{K_{\hat s}}\frac{ (y_k - x_{\hat i}^+) }{(y_k - x_{\hat i}^-)} \prod_{\dot m=1}^{K_{\dot s}}\frac{ \left(1-\frac{1}{y_k x_{\dot m}^-}\right) }{\left(1-\frac{1}{y_k x_{\dot m}^+}\right)} \prod_{\hat{\dot m}=1}^{K_{\hat{\dot  s} }}\frac{ \left(1-\frac{1}{y_k x_{\hat{\dot m}}^-}\right) }{\left(1-\frac{1}{y_k x_{\hat{\dot m}}^+}\right)}   \right|_{y_k = y_{f, k}} .
\eeq
The two equations evaluated on zeros of $\bS_{1|1}$ and $\bS_{\hat{1}|1}$ lead to Bethe equations for the massive nodes:
\begin{eqnarray}
1&=&\left(\frac{x_k^-}{x_k^+}\right)^L \prod_{i\neq k}^{K_s}
\frac{(x_k^+-x_i^-)\left(1-\frac{1}{x_k^+x_i^-}\right)}{(x_k^--x_i^+)\left(1-\frac{1}{x_k^-x_i^+}\right)}
\;
\prod_{\bullet=\dot{s}, \dot{\hat{s}}}\prod_{\dot{{r}}=1}^{K_{\bullet} } \sqrt{\frac{(1-\frac{1}{x_k^+ x_{\dot{{r}}}^-} ) (1-\frac{1}{x_k^+  x_{\dot{{r}}}^+} ) }{(1-\frac{1}{x_k^- x_{\dot{{r}}}^-} ) (1-\frac{1}{x_k^- x_{\dot{{r}}}^+} ) }}  \nonumber \\
 &&\left.\times  \prod_{i=1}^{K_f}\frac{x_k^- - y_i}{x_k^+-y_i} \prod_{\dot{m}=1}^{K_{\dot b}}\frac{\frac{1}{x_k^-} - y_{\dot{m}}}{\frac{1}{x_k^+}-y_{\dot{m}}}\times \underbrace{\prod_{\bullet = s, \hat{s}, \dot{s}, \dot{\hat{s}}} \prod_{i=1}^{K_{ {\bullet}}} \Sigma^{\text{sym}}_{s \bullet }(u_{k} , u_{\bullet, i} ) 
 \times \frac{W^{++}(u_k) }{W^{--}(u_k) } }_{\texttt{dressing 1}}
 \right|_{\substack{u_k=u_{s,k}\\ x_k=x_{s,k}}}, \nonumber\\
1&=&\left(\frac{x_{{k}}^-}{x_k^+}\right)^L \prod_{i\neq {{k}}}^{K_{\hat s}}
\frac{(x_k^+-x_i^-)\left(1-\frac{1}{x_k^+x_i^-}\right)}{(x_k^--x_i^+)\left(1-\frac{1}{x_k^-x_i^+}\right)}\; \prod_{\bullet=\dot{s}, \dot{\hat{s}}}\prod_{\dot{{r}}=1}^{K_{\bullet} } \sqrt{\frac{(1-\frac{1}{x_k^+ x_{\dot{{r}}}^-} ) (1-\frac{1}{x_k^+  x_{\dot{{r}}}^+} ) }{(1-\frac{1}{x_k^- x_{\dot{{r}}}^-} ) (1-\frac{1}{x_k^- x_{\dot{{r}}}^+} ) }} \nonumber \\
 &&\times \prod_{i=1}^{K_f}\frac{x_k^- - y_i}{x_k^+-y_i} \prod_{\dot m=1}^{K_{\dot b}}\frac{\frac{1}{x_k^-} - y_{\dot m}}{\frac{1}{x_k^+}-y_{\dot m}} \times \left. \underbrace{\prod_{\bullet = s, \hat{s}, \dot{s}, \dot{\hat{s}}} \prod_{i=1}^{K_{ {\bullet}}} \Sigma^{\text{sym}}_{\hat s \bullet }(u_{k} , u_{\bullet, i} ) \times \frac{W^{--}(u_k) }{W^{++}(u_k) } }_{\texttt{dressing 2}}
 \right|_{\substack{u_k=u_{\hat s,k}\\ x_k=x_{\hat s,k}}}. \\
 \label{eq:ABA1}
\end{eqnarray}
Here we have defined the following functions (all depending on two rapidities $u$, $v$):
\beqa\label{def:ABA_Sigma_01}
\Sigma_{ss}^{\text{sym}} &=& \Sigma^{\text{sym}}_{\hat s \hat s} =\Sigma_{\text{BES}} \times \Sigma^{\text{extra}} 
,\\
\Sigma^{\text{sym}}_{s \dot s}&=& \Sigma^{\text{sym}}_{\hat s  \hat{\dot s}}  =\Sigma_{\text{BES}} \times \tilde{\Sigma}^{\text{extra}} 
\nonumber ,\\
\Sigma^{\text{sym}}_{s\hat{s}} &=& \Sigma^{\text{sym}}_{\hat s  s} =\Sigma_{\text{BES}} \times \Sigma^{\text{extra}} 
\nonumber,\\
\Sigma^{\text{sym}}_{s \hat{\dot s}} &=& \Sigma^{\text{sym}}_{\hat s  \dot{ s}}=\Sigma_{\text{BES}} \times \tilde{\Sigma}^{\text{extra}} 
\nonumber .
\eeqa
The product of these  terms, with the second variable $v$ evaluated at some Bethe roots $u_{\bullet,i}$, arises in the Bethe equations from evaluating $\rho^+/\rho^-$ (where $\rho$ is given in (\ref{eq:rho_sigmaupd})), leading to the
functions $\Sigma_{\text{BES}},\,\Sigma^{\text{extra}},\,\tilde{\Sigma}^{\text{extra}}$ defined in appendix \ref{appendix:dressing}. 

These Asymptotic Bethe equations match perfectly the ones present in the literature in \cite{Borsato:2012ss} for this sector, provided we identify appropriately the dressing phases, i.e. we would like to interpret the non-rational part of the Bethe equations as
\beq\label{eq:identifydressing}
\texttt{dressing 1} = \prod_{\bullet = s, \hat{s}, \dot{s}, \dot{\hat{s}}} \prod_{i=1}^{K_{ {\bullet}}} \Sigma_{ s \bullet }(u_{k} , u_{\bullet, i} ) ,\;\;\; \texttt{dressing 2} = \prod_{\bullet = s, \hat{s}, \dot{s}, \dot{\hat{s}}} \prod_{i=1}^{K_{ {\bullet}}} \Sigma_{ \hat{s} \bullet }(u_{{k}} , u_{\bullet, i} ) ,
\eeq
with massive-massive dressing factors $\Sigma_{\bullet \bullet}$, for all the possible combinations of channels $\bullet \in \left\{s, \hat{s}, \dot{s}, \dot{\hat{s}} \right\}$. 

In the symmetric sector, where $W(u) = 1$, and the roots are paired, the Bethe equations actually only see the \emph{product} of dressing phases.  Inspecting our results for the dressing phases, we can identify this product as:
\beq
\Sigma_{\bullet s} \times \Sigma_{\bullet \hat{s}} = (\Sigma^{\text{sym}}_{\bullet s} )^2 , \;\;\;\; \Sigma_{\bullet \dot{s}} \times \Sigma_{\bullet \dot{\hat{s}}} = (\Sigma^{\text{sym}}_{\bullet \dot{s}} )^2 ,
\eeq
which can be seen as a \emph{prediction} of the QSC approach. 
Such a product of dressing phases is compatible with the crossing, parity and unitarity relations present in the literature~\cite{Borsato:2015mma}.

In the generic, non-symmetric sector, what seems to emerge is a more general structure for the dressing factors: 
\beqa\label{def:ABA_Sigma_1}
\Sigma_{ss} &=& \Sigma_{\hat s \hat s} =\Sigma_{\text{BES}} \times \Sigma^{\text{extra}} \times \Sigma_{\text{new}}  ,\\
\Sigma_{s \dot s} &=& \Sigma_{\hat s  \hat{\dot s}}  =\Sigma_{\text{BES}} \times \tilde{\Sigma}^{\text{extra}} \times\Sigma_{\text{new}} \nonumber ,\\
\Sigma_{s\hat{s}} &=& \Sigma_{\hat s  s} =\Sigma_{\text{BES}} \times \Sigma^{\text{extra}} \times \Sigma_{\text{new}}^{-1} \nonumber,\\
\Sigma_{s \hat{\dot s}} &=& \Sigma_{\hat s  \dot{ s}}=\Sigma_{\text{BES}} \times \tilde{\Sigma}^{\text{extra}} \times \Sigma_{\text{new}}^{-1} \nonumber,
\eeqa
where the new factor $\Sigma_{\text{new}}$ is constructed  in the next chapter  by identifying:
\beq\label{eq:fromWtoSigma}
\frac{W^{++}(u)}{W^{--}(u)} = \prod_{\bullet = s, \dot{s}} \prod_{i=1}^{K_{ {\bullet}}} \Sigma^{\text{new}}(u , u_{\bullet, i} ) \times \prod_{\bullet = \hat{s}, \dot{ \hat{s}} } \prod_{i=1}^{K_{ {\bullet}}} \left( \Sigma^{\text{new}}(u, u_{\bullet, i} )  \right)^{-1}.
\eeq
This would be a genuinely new part of the dressing factors, specific to this model. 
We will study what our derivation indicates for this function, and some puzzling properties, in section \ref{sec:dressingnew}. 

Furthermore, in appendix \ref{app:square_roots}, we will show that the factors containing square roots in the ABA are only apparent, as they are compensated by similar factors hidden in the dressing phases.
\subsubsection{ABA for the R wing}
The asymptotic Bethe equations for the R wing come from the Bethe equations \eqref{eq:BAE_1_node1}, \eqref{eq:BAE_1_node2} and \eqref{eq:BAE_1_node3}, with dotted Q-functions, which we rewrite here for convenience (setting $\dot\bQ_{\emptyset}=1$):
\begin{eqnarray}
\label{eq:BAE_R1}
1 &=& \left. \frac{\slS_{\hat{\dot 1}|\dot 1}^{--}}{\slS_{\hat{\dot 1}|\dot 1}^{++} } \frac{\slP_{\dot 1}^+}{\slP_{\dot 1}^-}  \;\right|_{\slS_{\dot 1|\dot 1} = 0}\,,\\
\label{eq:BAE_R2}
1 &=&\left. \frac{\slS_{ \dot 1|\dot 1}^{--}}{\slS_{ \dot 1|\dot 1}^{++} } \frac{\slP_{\dot 1}^+}{\slP_{\dot 1}^-}  \;\right|_{\slS_{\hat{\dot 1}|\dot 1} = 0}\,,\\ 1&=& \left. \frac{ \slS_{ \dot 1|\dot 1}^- \slS_{\hat{\dot 1}|\dot 1}^-}{\slS_{\dot 1|\dot 1}^+ \slS_{\hat{\dot 1}|\dot 1}^+}\right|_{\slP_{\dot 1} = 0} \,.
\end{eqnarray}
The last equation is identical in form to the first equation for the L wing \eqref{eq:BAE_2_node1}, with the only difference being that it is evaluated at the Bethe roots corresponding to $\slP_{\dot 1}$. Using the same conventions explained in the previous section, it is given by
\beq
1= \left. \prod_{\dot m=1}^{K_{\dot s}}\frac{(y_{\dot k} - x_{\dot m}^-)}{ (y_{\dot k} - x_{\dot m}^+) } \prod_{\hat{\dot m}=1}^{K_{\hat{\dot s}}}\frac{(y_{\dot k} - x_{\hat{\dot m}}^-)}{ (y_{\dot k} - x_{\hat{\dot m}}^+) } \prod_{i=1}^{K_{ s}}\frac{\left(1-\frac{1}{y_{\dot k} x_i^+}\right)}{ \left(1-\frac{1}{y_{\dot k} x_i^-}\right) } \prod_{\hat i=1}^{K_{ {\hat  s} }}\frac{\left(1-\frac{1}{y_{\dot k} x_{\hat i}^+}\right)}{ \left(1-\frac{1}{y_{\dot k} x_{\hat i}^-} \right) }   \right|_{y_{\dot k} = y_{\dot{b}, k}} .
\eeq
Plugging in the large volume solution in the remaining two equations, we find from \eqref{eq:BAE_R1}, after applying the zero momentum condition:  
\begin{eqnarray}
1&=&\left(\frac{x_{\dot k}^-}{x_{\dot k}^+}\right)^L 
\prod_{\hat{\dot{m}}=1}^{K_{\hat{ \dot s}}}  \frac{x_{\dot k}^- - x_{\hat{\dot{m}}}^+}{x_{\dot k}^+-x_{\hat{\dot{m}}}^-} \, \prod_{{\dot{m}}=1 }^{K_{\dot s}} \frac{1-\frac{1}{x_{\dot k}^+x_{\dot{m}}^-}}{1-\frac{1}{x_{\dot k}^-x_{\dot{m}}^+}} \; \prod_{\bullet={s}, {\hat{s}}}\prod_{i=1}^{K_{\bullet} } \sqrt{\frac{(1-\frac{1}{x_{\dot{k}}^+ x_{i}^-} ) (1-\frac{1}{x_{\dot{k}}^-  x_{i}^-} ) }{(1-\frac{1}{x_{\dot{k} }^+ x_{i}^+} ) (1-\frac{1}{x_{\dot{k}}^- x_{i}^+} ) }} \\
&&\times \prod_{i=1}^{K_f}\frac{\frac{1}{x_{\dot k}^+} - y_i}{\frac{1}{x_{\dot k}^-}-y_i} \prod_{i=1}^{K_{\dot b}}\frac{x_{\dot k}^+ - y_{i}}{x_{\dot k}^--y_{i}}\times \left.
\underbrace{  \prod_{\bullet = s, \hat{s}, \dot{s}, \dot{\hat{s}}} \prod_{i=1}^{K_{ {\bullet}}} \Sigma^{\text{sym}}_{\dot s \bullet }(u_{\dot k} , u_{\bullet, i} ) \times \frac{W^{++}(u_{\dot k}) }{W^{--}(u_{\dot k}) } 
}_{\texttt{dressing 3}}
\right|_{\substack{u_{\dot k}=u_{{\dot s}, k}\\ x_{\dot k}=x_{{\dot  s}, k}}} ,
\eeqa
while from \eqref{eq:BAE_R2} we find:
\begin{eqnarray}
1&=&\left(\frac{x_{\dot{\hat{k}} }^-}{x_{\dot{\hat{k}}}^+}\right)^L 
\prod_{\dot{m}=1}^{K_{{ \dot s}}}  \frac{x_{ \dot{\hat{k}} }^- - x_{\dot{m}}^+}{x_{ \dot{\hat{k}} }^+-x_{\dot{m}}^-} \, \prod_{{\dot{\hat m}}=1 }^{K_{\dot{\hat{s}}}} \frac{1-\frac{1}{x_{\dot{\hat{k}}}^+x_{\dot{\hat m}}^-}}{1-\frac{1}{x_{\dot{\hat{k}}}^-x_{\dot{ \hat m }}^+}} \; \prod_{\bullet={s}, {\hat{s}}}\prod_{i=1}^{K_{\bullet} } \sqrt{\frac{(1-\frac{1}{x_{\dot{\hat{k}}}^+ x_{i}^-} ) (1-\frac{1}{x_{\dot{\hat{k}}}^-  x_{i}^-} ) }{(1-\frac{1}{x_{\dot{\hat{k}} }^+ x_{i}^+} ) (1-\frac{1}{x_{\dot{\hat{k}}}^- x_{i}^+} ) }} \\
&&\times \prod_{i=1}^{K_f}\frac{\frac{1}{x_{\dot{\hat{k}}}^+} - y_i}{\frac{1}{x_{\dot{\hat{k}}}^-}-y_i} \prod_{i=1}^{K_{\dot b}}\frac{x_{\dot{\hat{k}}}^+ - y_{i}}{x_{\dot{\hat{k}}}^--y_{i}}\; \left. \underbrace{ \prod_{\bullet = s, \hat{s}, \dot{s}, \dot{\hat{s}}} \prod_{i=1}^{K_{ {\bullet}}} \Sigma^{\text{sym}}_{\dot{\hat s} \bullet }(u_{\dot{\hat{k}}} , u_{\bullet, i} ) \times \frac{W^{--}(u_{\dot k}) }{W^{++}(u_{\dot{\hat  k}}) }
}_{\texttt{dressing 4}}
\right|_{\substack{u_{\dot{\hat{k}}}=u_{{\dot{\hat s}}, k}\\ x_{\dot{\hat{k}}}=x_{{\dot  \hat{s}}, k}}} ,
\eeqa
where we defined (these are all functions of two rapidities $u$, $v$):
\beqa\label{def:ABA_Sigma_02}
\Sigma_{\dot s\dot s}^{\text{sym}} &=& \Sigma^{\text{sym}}_{\hat{\dot s} \hat{\dot s}} =\Sigma^{\text{sym}}_{\dot s\hat{\dot s}} = \Sigma^{\text{sym}}_{\hat{\dot s}  \dot s} =\Sigma_{\text{BES}} \times \Sigma^{\text{extra}} 
,\\
\Sigma^{\text{sym}}_{ \dot s s}&=& \Sigma^{\text{sym}}_{  \hat{\dot s}\hat s} = \Sigma^{\text{sym}}_{s \hat{\dot s}} = \Sigma^{\text{sym}}_{\hat s  \dot{ s} } =\Sigma_{\text{BES}} \times \tilde{\Sigma}^{\text{extra}} 
\nonumber . 
\eeqa
As in the other wing, these ABA equations match the form of those in \cite{Borsato:2012ss}. The would-be dressing factors, emerging as
\beq\label{eq:identifydressing2}
\texttt{dressing 3} = \prod_{\bullet = s, \hat{s}, \dot{s}, \dot{\hat{s}}} \prod_{i=1}^{K_{ {\bullet}}} \Sigma_{ \dot{s} \bullet }(u_{\dot k} , u_{\bullet, i} ) ,\;\;\; \texttt{dressing 4} = \prod_{\bullet = s, \hat{s}, \dot{s}, \dot{\hat{s}}} \prod_{i=1}^{K_{ {\bullet}}} \Sigma_{ \dot{\hat{s}} \bullet }(u_{\dot{\hat{k}}} , u_{\bullet, i} ) ,
\eeq
where the predicted form of the dressing factors is
\beqa\label{def:ABA_Sigma_2}
\Sigma_{\dot s s} &=& \Sigma_{\hat{\dot s}  {\hat s}} =\Sigma_{\text{BES}} \times \tilde{\Sigma}^{\text{extra}} \times \Sigma_{\text{new}} ,\\
\Sigma_{\dot s \dot s} &=& \Sigma_{\hat{\dot s}  \hat{\dot s}}  = \Sigma_{\text{BES}} \times  {\Sigma}^{\text{extra}} \times\Sigma_{\text{new}} ,\nonumber\\
\Sigma_{\dot s\hat{s}} &=& \Sigma_{\hat{\dot s}  s} =\Sigma_{\text{BES}} \times \tilde{\Sigma}^{\text{extra}} \times \Sigma_{\text{new}}^{-1} \nonumber,\\
\Sigma_{\dot s \hat{\dot s}} &=& \Sigma_{\hat{\dot s}  \dot{ s}}=\Sigma_{\text{BES}} \times  {\Sigma}^{\text{extra}} \times \Sigma_{\text{new}}^{-1} \nonumber,
\eeqa
in terms of the same function $\Sigma_{\text{new}}(u,v)$ appearing in the other wing, which will be discussed at length in the next section. 

Similarly to the other wing, the square roots present in some of the ABA equations are also compensated by similar factors in the dressing phases.
\section{The new dressing phase and its properties}\label{sec:dressingnew}
In this section, we analyse in more detail the structure of the dressing phases appearing in the ABA equations of the last section. In particular, we focus on the $\Sigma_{\text{new}}$ factor defined in \eqref{def:Sigma_new}, a novelty of the \ads\, case, and on the integral equation \eqref{eq:LIE3} that determines it.

\subsection{Defining $\Sigma_{\text{new}}$}
\label{sec:def_sigmanew}
Let us show how to define the function $\Sigma_{\text{new}}$ in such a way that we have the identity (\ref{eq:fromWtoSigma}). The possibility of doing this depends on an assumption on the form of the constant $\mathcal{E}$ appearing in the linear integral equation (\ref{eq:LIE3}). This will provide some extra degrees of freedom, which turn out to be needed when discussing braiding unitarity.

\paragraph{Decomposing the linear integral equation. }
Let us first describe how the solution to \eqref{eq:LIE3} can be decomposed into elementary building blocks. The source term of the integral equation $\log \mathcal{A}(u)$ is already naturally given by a sum over the contributions of many would-be Bethe roots (see (\ref{eq:defA})):
\beq
\log \mathcal{A}(u) = \sum_{\bullet = s, \dot s} \sum_{k=1}^{K_{\bullet}} \mathbf{a}(u - u_{\bullet, k} ) - \sum_{\bullet = \hat{s}, \hat{\dot{s}}} \sum_{k=1}^{K_{\bullet}} \mathbf{a}(u - u_{\bullet, k} ) ,
\eeq
where we have defined an elementary source term as:\footnote{Here, we assumed that the total number of roots of all kinds is even in order to change the sign inside the argument of the logarithm. We take this form in order to have $\mathbf{a}(0) = 0$. }
\beq
\mathbf{a}(u-v) \equiv \frac{1}{2}\log{\left(\frac{u-v + \frac{i}{2}}{v-u+\frac{i}{2}}\right)} .
\eeq
Moreover, we will also \textbf{assume} that the constant $\mathcal{E}$, arising in (\ref{eq:StokesE}), also has a form depending on the roots with a similar pattern:
\beq
\log \mathcal{E} = \sum_{\bullet = s, \dot s} \sum_{k=1}^{K_{\bullet}} \mathbf{e}(u_{\bullet, k} ) - \sum_{\bullet = \hat{s}, \hat{\dot{s}}} \sum_{k=1}^{K_{\bullet}} \mathbf{e}( u_{\bullet, k} ) .
\eeq
We have not found a way to \emph{derive} the form of $\mathcal{E}$ or to prove that it should factorise in this way on the roots. We could even have $\mathcal{E} = 1$, which would correspond to $\mathbf{e}(v) \equiv 0$. 
For the moment, we will not commit to the form of the function $\textbf{e}(v)$. As we explain later, the presence of this degree of freedom seems necessary in order to guarantee braiding unitarity for the new dressing phase.\footnote{We are grateful to P. Chernikov, S. Ekhammar and  N. Gromov for communications and calling our attention to this fact.}

If we assume this factorisation, the solution of the linear integral equation (\ref{eq:LIE3}) will also decompose into elementary building blocks. The elementary equation for each of these blocks is:
\begin{framed}
\beq\label{eq:LIE2elementary}
\int_{-2h}^{2h} \frac{dz}{2i\pi}\mathbf{d}(z|v)\left[2\sum_{n=1}^{\infty} (-1)^n\left(\frac{1}{u-z+i n}-\frac{1}{u-z-i n}\right)\right] - \mathbf{d}(u|v) =
 \mathbf{a}(u-v) + \mathbf{e}(v) ,
\eeq
\end{framed}
where the second rapidity plays the role of a spectator parameter. The full solution to (\ref{eq:LIE3}) can be reconstructed as a sum of solutions to \eqref{eq:LIE2elementary}, i.e.,
\beq\label{eq:decDV}
\mathcal{D}_V(u) = \sum_{\bullet = s, \dot s} \sum_{k=1}^{K_{\bullet}} \mathbf{d}(u | u_{\bullet, k} ) - \sum_{\bullet = \hat{s}, \hat{\dot{s}}} \sum_{k=1}^{K_{\bullet}} \mathbf{d}(u | u_{\bullet, k} ) .
\eeq

\paragraph{Building the function $W(u)$. }
From the solution of \eqref{eq:LIE2elementary} we can build the function $W(u)$ using the integral representation in (\ref{eq:logW}). 
The decomposition (\ref{eq:decDV}) implies that also $W(u)$ will take a factorised form\footnote{Notice that for the symmetric sector, where the set of roots labelled $s \& \hat{s}$, as well as $\dot{s} \& \dot{\hat s}$, coincide, then we have $W(u) =1$ as expected. }:
\beq\label{eq:Wdecomp}
W(u) = \frac{\mathbf{w}_s(u) \mathbf{w}_{\dot s}(u)}{ \mathbf{w}_{\hat s}(u) \mathbf{w}_{\dot{\hat{s}}}(u) },
\eeq
where
\beq
\label{eq:w_definition}
\mathbf{w}_{\bullet}(u) \equiv \prod_{i=1}^{K_{\bullet}}\mathbf{w}(u, u_{\bullet,i} ) ,
\eeq
where the elementary function $\mathbf{w}(u,v)$ is defined as:
\beq\label{eq:defwsmall}
\log \mathbf{w}(u, v) \equiv \int_{-2 h}^{2 h} ds\;\mathbb{K}(u+\frac{i}{2}, s) \mathbf{d}(s, v) .
\eeq

\paragraph{The new part of the dressing factor.  }
Finally, the block of the dressing factor  $\Sigma_{\text{new}}$ is defined as
\beq\label{eq:defnewSigma}
\log\Sigma_{\text{new}}(u,z) \equiv \mathbf{w}^{++}(u,z)/\mathbf{w}^{--}(u,z) .
\eeq
Its expression can be simplified significantly, noting the identity between the shifted kernels:
\beq
\mathbb{K}(u+\frac{3}{2} i , z) - \mathbb{K}(u-\frac{i}{2} , z) = \frac{1}{z - u+i/2}-\frac{1}{z - u-i/2} ,
\eeq
which, together with (\ref{eq:defwsmall}), 
 leads to the form:
\begin{framed}
\beq
\label{def:Sigma_new}
\log\Sigma_{\text{new}}(u,v) \equiv \frac{1}{2 \pi i}\int_{-2 h}^{2 h}dz\,\left(  \frac{1}{z - u+i/2}-\frac{1}{z - u-i/2} \right) \mathbf{d}(z|v) .
\eeq
\end{framed}
In summary, the prescription to build the new block of the dressing phase is to
\begin{itemize}
    \item  compute the density $\textbf{d}$ as the solution to the linear integral equation (\ref{eq:LIE2elementary}). The solution also depends on the form of $\mathbf{e}(v)$, which enters the source term in the linear integral equation.
\item build $\Sigma_{\text{new}}$ using (\ref{def:Sigma_new}). 
    \end{itemize}
Unfortunately, we were unable to find an explicit solution to the linear integral equation, and therefore, we cannot yet write a fully explicit solution $\Sigma_{\text{new}}$. 
However, the equations above specify it uniquely (for any given choice of $\mathbf{e}(v)$), as we describe in the next paragraph.

\paragraph{Existence and uniqueness of solutions. } Since the dressing phase $\Sigma_{\text{new}}$ is directly defined in terms of the solutions to the integral equation \eqref{eq:LIE2elementary}, we need to check that the latter admits a unique solution for a given source, i.e. that it does not possess zero modes. One strategy to do so is to show that the Fredholm determinant of the kernel appearing in the integral equation is non-zero:
\beq
\text{det}_{u,v \in [-2h, 2h]}\left[ \delta(u-v) - 
\mathbb{G}(u-v) \right] \neq 0,
\eeq
where $\mathbb{G}$ is the kernel defined in (\ref{eq:defG}).  
We can compute this quantity numerically by discretisation on a lattice of spacing $a$.  By taking a smaller grid step, the discretised Fredholm determinant appears to converge to a nonzero value, which signals the absence of zero modes in the integral equation. 

We have also solved equation \eqref{eq:LIE2elementary} numerically by discretising the integral on a grid of $N$ points, setting the unknown function $\mathbf{e}(v)=0$ for simplicity. We display some of our results in figure \ref{fig:plots_d}. We verified that at any value of the coupling $h$, the solutions to our integral equation appear to converge as the number of discretisation points increases. We depict this property in figure \ref{fig:d_converges}.

\begin{figure}[ht!]
\centering
\begin{minipage}{0.4\textwidth}
    \centering
    \includegraphics[width=\textwidth]{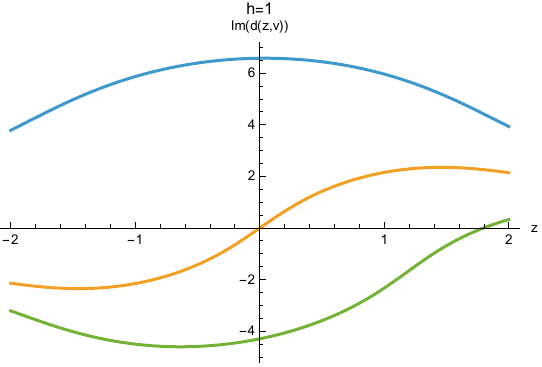}
    \caption*{(b) $h=1$}
\end{minipage}
\hfill
\begin{minipage}{0.55\textwidth}
    \centering
    \includegraphics[width=\textwidth]{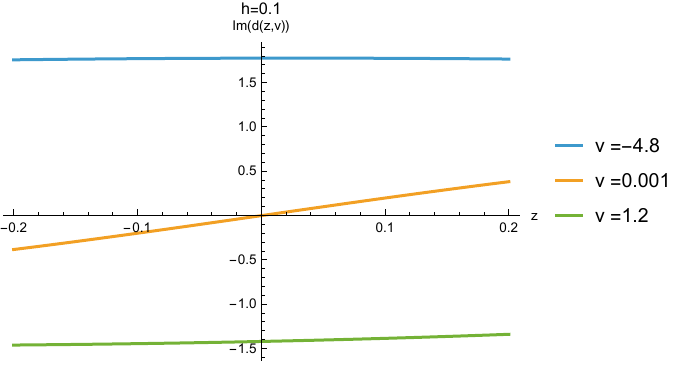}
    \caption*{(c) $h=0.1$}
\end{minipage}

\caption{Plots of $\mathbf{d}(z|v)$ at three real values of $v$ and with $z$ on the branch cut $(-2h,2h)$. For $z,v\in \mathbb{R}$, $\mathbf{d}(z|v)$ is purely imaginary. These solutions correspond to the integral equation with $\mathbf{e}(v) = 0$.}
\label{fig:plots_d}
\end{figure}

\begin{figure}[ht!]
\centering
\begin{minipage}{0.4\textwidth}
    \centering
    \includegraphics[width=\textwidth]{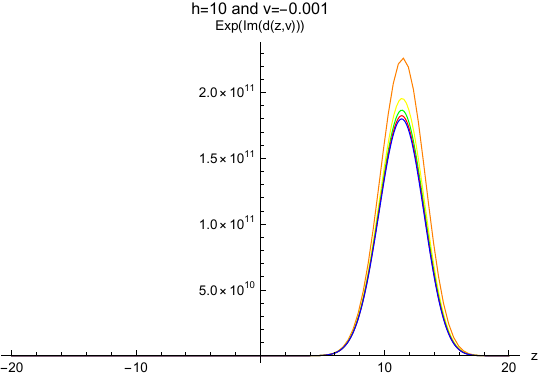}
    \caption*{$h=10$ and $v=-0.001$}
        \end{minipage}
    \hfill
    \begin{minipage}{0.59\textwidth}
    \centering
    \includegraphics[width=\textwidth]{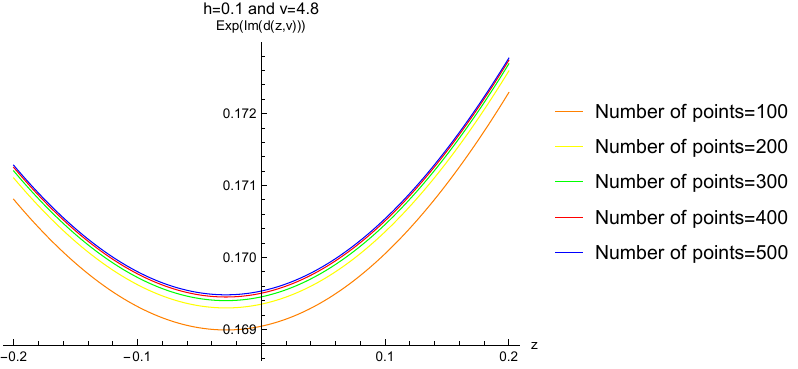}
    \caption*{$h=0.1$ and $v=4.8$}
\end{minipage}
\caption{Exponential plots of the solution to \eqref{eq:LIE2elementary} where the integral has been discretised with a different number of grid points. The solution converges quickly to a single value as the number of points is incremented.}
\label{fig:d_converges}
\end{figure}
\subsection{Crossing equations}
\begin{figure}[htbp]
\centering
\includegraphics[scale=0.3]{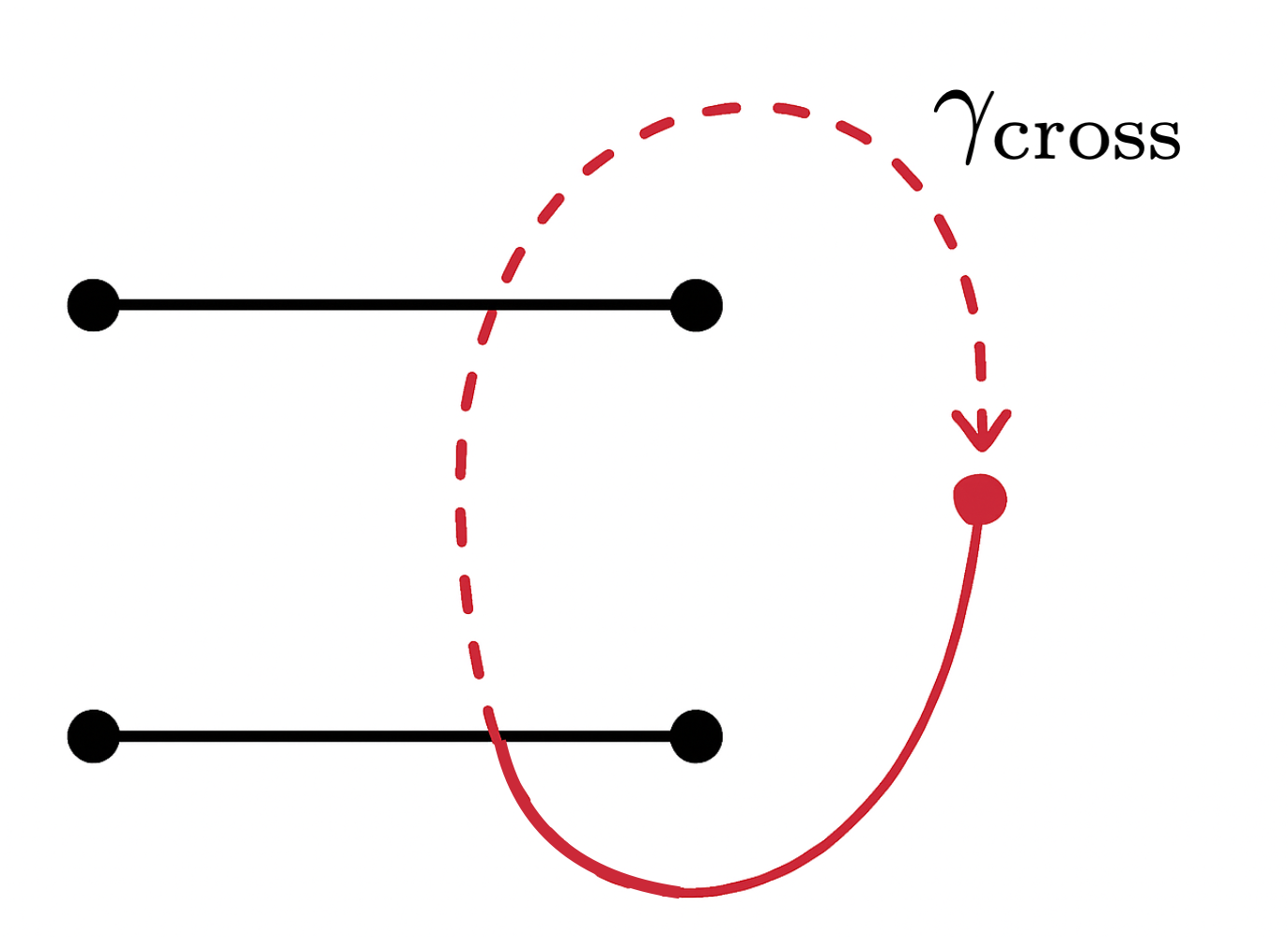}
\caption{The crossing path. In the figure, we also show the two cuts of matrix elements between massive particles, which are located at $\pm i/2 + (-2h, 2h)$. 
    \label{fig:gammacross}}
\end{figure}
\paragraph{Deriving the crossing equation for the new dressing phase. }
The crossing transformation is described by the analytic continuation path in figure \ref{fig:gammacross}.
 Let us compute the effect of this analytic continuation on the first variable of the new dressing phase, i.e. we want to evaluate
  $\log\Sigma_{\text{new}}(u^{\gamma_{\text{cross}}} , v)$. We will see that a crossing equation follows purely from the linear integral equation, even without solving it explicitly.

We start from the expression in (\ref{def:Sigma_new}). This is an integral representation for the dressing factor, where the integrand has a pole at $ z=u+i/2$ with residue $-1$, plus a pole at $z=u-i/2$ with residue $+1$. 
The analytic continuation path $\gamma_{\text{cross}}$ 
leads both poles to pass through the integration contour, in both cases from below. Thus, we need to pick two residues from the integral in \eqref{def:Sigma_new}:
\beq
\log\Sigma_{\text{new}}(u^{\gamma_{\text{cross}}} , v) = \log\Sigma_{\text{new}}(u, v) + \text{residues}.
\eeq
To evaluate the residues, we will need to understand the analytic continuation in $u$ of the density $\mathbf{d}(u|v)$. From the integral equation, it can be established that $\textbf{d}(u|v)$ is analytic in $u$ in a strip around the real axis, i.e. for $|\text{Im}(u)|<1$. At the boundaries of the strip, branch points will appear: we can expect that $\textbf{d}(u|v)$ will have two branch cuts for $u \in (-2 h, 2h) \pm  i$.\footnote{ We also expect $\textbf{d}(u|v)$ to have cuts at $u \in (-2h, 2h) \pm i \mathbb{N}^+$, but only the closest cuts to the real axis will be relevant in the analysis of the crossing equations.}
Taking this structure into account, we find
\beq
\label{eq:LIEstep2}
\log\Sigma_{\text{new}}(u^{\gamma_{\text{cross}}},v)=\log\Sigma_{\text{new}}(u,v) +\underbrace{\mathbf{d}\left( (u+i)^{\gamma^{-1}} -\frac{i}{2}\,|v\right) -\mathbf{d}\left(u-\frac{i}{2}\,|v\right)}_{\text{residues}},
\eeq
where the first of the two residues, i.e. the term 
\beq\label{eq:termcontinued}
\mathbf{d}\left( (u+i)^{\gamma^{-1}} -\frac{i}{2}\,|v\right) ,
\eeq
is analytically continued through a further branch cut of the density.  This is the residue that is picked when $\gamma_{\text{cross}}$ crosses the line $\text{Im}(u) = -1/2$. This frees up a residue proportional to  $\mathbf{d}(u+i/2|v)$, which is then carried along the rest of the path, entering the first cut of the density in the upper half plane from below, before being brought back to the position at $\text{Re}(u) = 1/2$. 

To simplify the expression we found, let us study this term in more detail.  
 To analyse it, we start from the linear integral equation \eqref{eq:LIE2elementary}
and analytically continue the $u$ variable along the same path as in (\ref{eq:termcontinued}): i.e. we shift by $i$, enter the cut at $\text{Im}(u) = +1$ from below, and then come back by $-i/2$ avoiding the cuts. In the process, a pole of the kernel $\mathbb{G}(u-z)$ with  residue $-2$ crosses the integration contour from below, so that we find:
\beqa
\mathbf{d}\left((u+i)^{\gamma^{-1}} -\frac{i}{2}|v\right) &=&\underbrace{ 2 \mathbf{d}(u -\frac{i}{2} |v)}_{\text{residue}} \nonumber\\
&&+  \int_{-2h}^{2h} \frac{dz}{2i\pi}\mathbf{d}(z|v)\left[2\sum_{n=1}^{\infty} (-1)^n\left(\frac{1}{z-u+i n-\frac{i}{2}}-\frac{1}{z-u-i n-\frac{i}{2}}\right)\right] \nonumber \\&&- 
 \mathbf{a}(u+\frac{i}{2}-v) - \mathbf{e}(v) .
\eeqa
Now, using again \eqref{eq:LIE2elementary} we see that all the terms in the last two lines  can be assembled to give a shifted value of $\mathbf{d}$(u+i/2|v), which leads to 
\beq
\mathbf{d}\left((u+i)^{\gamma^{-1}} -\frac{i}{2}|v\right)  = 2 \mathbf{d}\left(u-\frac{i}{2} |v \right) + \mathbf{d}\left(u+\frac{i}{2} |v \right) .
\eeq
Plugging this in \eqref{eq:LIEstep2}, we obtain a simpler relation:
\beq
\label{eq:LIEstep3}
\log\Sigma_{\text{new}}(u^{\gamma_{\text{cross}}},v)=\log\Sigma_{\text{new}}(u,v) +\mathbf{d}\left(u-\frac{i}{2} |v \right) + \mathbf{d}\left(u+\frac{i}{2} |v \right).
\eeq
However, we can still simplify this. To do this, we again start from \eqref{eq:LIEstep2}, and use it to evaluate the two shifted terms on the rhs of the previous equation. Applying the combination of the two shifts, we see that  the sum of poles in the kernel on the right-hand side simplifies drastically: they cancel two by two until only two terms remain:
\begin{small}\beqa
\mathbf{d}\left(u+\frac{i}{2}\bigg|v\right) + \mathbf{d}\left(u-\frac{i}{2}\bigg|v\right) &=& -\int_{-2h}^{2h} \frac{dz}{2i\pi}\mathbf{d}(z|v)\left(\frac{2}{z-u+\frac{i}{2}}-\frac{2}{z-u-\frac{i}{2}}\right) \nonumber\\
&&- 
 \mathbf{a}\left(u-v+\frac{i}{2}\right) -  \mathbf{a}\left(u-v-\frac{i}{2}\right) - 2 \mathbf{e}(v)\nonumber \\
 &=& -2 \mathbf{e}(v) -\mathbf{a}\left(u+\frac{i}{2} - v\right) - \mathbf{a}\left(u-\frac{i}{2} - v\right) -2 
\Sigma(u, v) ,
\eeqa
\end{small}
which finally, together with (\ref{eq:LIEstep3}), leads to the crossing equation:
\begin{framed}
\beq
\label{eq:LIEstep4}
\log\Sigma_{\text{new}}(u^{\gamma_{cross}},v)= -\log\Sigma_{\text{new}}(u,v) -2 \mathbf{e}(v) -\mathbf{a}(u+i/2 - v) - \mathbf{a}(u-i/2 - v) ,
\eeq
which can be written as
\beq
\label{eq:newcrossing}
\Sigma_{\text{new}}(u^{\gamma_{cross}},v)\Sigma_{\text{new}}(u,v) =e^{-2 \mathbf{e}(v)} \sqrt{\frac{u-v-i}{u-v+i}}= e^{-2 \mathbf{e}(v)} \times \sqrt{\frac{(x_u^- - x_v^+)(1-\frac{1}{x_u^- x_v^+})}{(x_u^+ - x_v^-)(1-\frac{1}{x_u^+ x_v^-})}}.
\eeq
\end{framed}
We see that the function $\mathbf{e}(v)$, which until now is unfixed, affects the form of the crossing relation satisfied by $\Sigma_{\text{new}}$. 

\subsubsection{Crossing equations for the full dressing factors }
In this section, we list the crossing equations satisfied by the full dressing factors appearing in the Asymptotic Bethe equations we obtained in section \ref{sec:ABA}, which are defined in \eqref{def:ABA_Sigma_1} and \eqref{def:ABA_Sigma_2}. 

These dressing factors are constructed in terms of the building blocks $\Sigma_{\text{BES}}, \,\Sigma^{\text{extra}},\,\tilde{\Sigma}^{\text{extra}}$ and $\Sigma_{\text{new}}$. 
The blocks $\Sigma_{\text{BES}}, \,\Sigma^{\text{extra}}$ and $\tilde{\Sigma}^{\text{extra}}$ already appeared in other QSC models~\cite{Beisert:2006ib,Beisert:2006ez,Frolov:2021fmj,Borsato:2016xns} (see also \cite{Gromov:2014caa, Cavaglia:2021eqr}) ), and the crossing equations for them read
\beqa
\Sigma_{\text{BES}}(u^{\gamma_{\text{cross}}} , v) \Sigma_{\text{BES}}(u , v)  &=&\frac{x_v^-}{x_v^+} \frac{x_u^- - x_v^+}{x_u^--x_v^-}\frac{1 - \frac{1}{x_u^+ x_v^+}}{ 1 - \frac{1}{x_u^+ x_v^-} },\\
\Sigma^{\text{extra}}(u^{\gamma_{\text{cross}}} , v) \tilde{\Sigma}^{\text{extra}}(u , v)  &=& \sqrt{\frac{x_u^+ - x_v^+}{x_u^--x_v^+}\frac{x_u^- - x_v^-}{x_u^+-x_v^-}},\\
\tilde{\Sigma}^{\text{extra}}(u^{\gamma_{\text{cross}}} , v) {\Sigma}^{\text{extra}}(u , v)  &=& \sqrt{\frac{1 - \frac{1}{x_u^+ x_v^-}}{ 1 - \frac{1}{x_u^+ x_v^+} } \frac{1 - \frac{1}{x_u^- x_v^+}}{ 1 - \frac{1}{x_u^- x_v^-} }} .
\eeqa
For the new piece $\Sigma_{\text{new}}$, we have derived the crossing equation (\ref{eq:newcrossing}). 

Combining everything, we obtain the following crossing equations for the dressing phases:
\begin{small}
\beqa
\label{eq:crossings}
\Sigma_{s s}(u^{\gamma_{\text{cross}}} , v) \Sigma_{\dot{s}  s}(u , v) &=&\frac{x_v^-}{x_v^+}\frac{x_u^--x_v^+}{x_u^+-x_v^-}\frac{\sqrt{\left(1-\frac{1}{x_u^-x_v^+}\right)\left(1-\frac{1}{x_u^+x_v^+}\right)\left(1-\frac{1}{x_u^-x_v^-}\right)}}{\left(1-\frac{1}{x_u^+x_v^-}\right)^{3/2}}e^{-2\mathbf{e}(v)} ,\\
\Sigma_{\dot{s} s}(u^{\gamma_{\text{cross}}} , v) \Sigma_{ {s}  s}(u , v) &=& \frac{x_v^-}{x_v^+}\frac{\left(x_u^- - x_v^+ \right)^{\frac{3}{2}}  }{\sqrt{(x_u^+ - x_v^- )(x_u^+ -x_v^+) (x_u^- -x_v^-) } }
\frac{\left(1 - \frac{1}{x_u^- x_v^+} \right)}{ \left(1 - \frac{1}{x_u^+ x_v^-} \right)} e^{-2 \mathbf{e}(v) }, \\
\Sigma_{s \hat{s}}(u^{\gamma_{\text{cross}}} , v) \Sigma_{\dot{s}  \hat{s} }(u , v) &=& {\frac{x_v^-}{x_v^+}}\sqrt{\frac{\left(1-\frac{1}{x_u^+x_v^+}\right)\left(1-\frac{1}{x_u^-x_v^-}\right)}{\left(1-\frac{1}{x_u^-x_v^+}\right)\left(1-\frac{1}{x_u^+x_v^-}\right)}}e^{2\mathbf{e}(v)}, \\
\Sigma_{\dot{s} \hat{s}}(u^{\gamma_{\text{cross}}} , v) \Sigma_{ {s}  \hat{s}}(u , v) &=&\frac{x_v^-}{x_v^+} \frac{ \sqrt{(x_u^+ -x_v^-) (x_u^- -x_v^+)}  }{\sqrt{(x_u^+ -x_v^+) (x_u^- -x_v^-)} } e^{2 \mathbf{e}(v) } .
\eeqa
\end{small}
In appendix \ref{sec:comparison}, we compare our results with the crossing equations predicted for this model in \cite{Borsato:2015mma}. We find that the two results match perfectly if the function $\mathbf{e}(v)$ is set to zero! 
It would therefore be natural to assume that this is the case, which would mean that, in the derivation of the ABA limit, there is no Stokes phenomenon and the factor $\mathcal{E}$ in (\ref{eq:StokesE}) is simply trivial, i.e.  $\mathcal{E}=1$. However, as we discuss in section \ref{sec:braiding}, $\mathbf{e}(v)$ cannot be zero if we want to have a dressing phase $\Sigma_{\text{new}}$ that satisfies braiding unitarity.

\subsection{Parity of $\Sigma_{\text{new}}$} In this and the next  section we analyze  the various symmetries of the dressing factors. We start by discussing worldsheet parity. This is one of the discrete symmetries relating the dressing phases to one another or to themselves. The dressing phases that appear in the symmetric sector $\Sigma_{\text{BES}}$ and $\Sigma^{\text{extra}}$ are already known to transform nicely under parity~\cite{Frolov:2021fmj}. In this section, we prove that the same property holds for $\Sigma_{\text{new}}$, under a simple assumption for $\mathbf{e}(v)$, namely, assuming that
\beq
\mathbf{e}(v)=-\mathbf{e}(-v).
\eeq
Under this assumption, we show that
\beq
\label{def:parity_Sigma}
\Sigma_{\text{new}}(u,v)\Sigma_{\text{new}}(-u,-v)=1\,.
\eeq
To prove this statement, we start from the integral equation \eqref{eq:LIE2elementary}. Evaluating it at $(-u,-v)$ and relabeling the variable of integration $z\rightarrow-z$, we find
\beq
\int_{-2h}^{2h} \frac{dz}{2i\pi}\mathbf{d}(-z|-v)\left[2\sum_{n=1}^{\infty} (-1)^n\left(\frac{1}{u-z+i n}-\frac{1}{u-z-i n}\right)\right] - \mathbf{d}(-u|-v) =
 \mathbf{a}(v-u) + \mathbf{e}(-v).
\eeq
Adding to this the integral equation \eqref{eq:LIE2elementary} evaluated at $(u,v)$, the source terms $\mathbf{a}$ cancel out. Furthermore, as we are assuming $\mathbf{e}(v)=-\mathbf{e}(-v)$, its contribution also cancels. Therefore, we remain with the following equation:
\beq
\int_{-2h}^{2h} \frac{dz}{2i\pi}\left[2\sum_{n=1}^{\infty} (-1)^n\left(\frac{1}{u-z+i n}-\frac{1}{u-z-i n}\right)\right](\mathbf{d}(z|v)+\mathbf{d}(-z|-v)) = \mathbf{d}(u|v)+\mathbf{d}(-u|-v)\,.
\eeq
This is a linear integral equation for $\mathbf{d}(u|v)+\mathbf{d}(-u|-v)$ with no source term and the same kernel as \eqref{eq:LIE2elementary}; therefore its solution is unique and it is given by $\mathbf{d}(u|v)+\mathbf{d}(-u|-v)=0$, implying that 
\beq
\label{def:parity_d}
\mathbf{d}(u|v)=-\mathbf{d}(-u|-v)\,.
\eeq
Now let us evaluate $\log{\Sigma_{\text{new}}}(-u,-v)$. Using the definition \eqref{def:Sigma_new} and after relabeling $z\rightarrow -z$, we see that:
\beq
\log{\Sigma_{\text{new}}}(-u,-v)=\frac{1}{2 \pi i}\int_{-2 h}^{2 h}dz\,\left(  \frac{1}{z - u+i/2}-\frac{1}{z - u-i/2} \right) \mathbf{d}(-z|-v)\,.
\eeq
Thanks to the property \eqref{def:parity_d} and the definition of $\log{\Sigma_{\text{new}}}$, this immediately implies
\beq
\log{\Sigma_{\text{new}}}(-u,-v)=-\log{\Sigma_{\text{new}}}(u,v)\,,
\eeq
which is the expected parity \eqref{def:parity_Sigma} for the new dressing phase.

\subsection{Unitarity properties of $\Sigma_{\text{new}}$ }\label{sec:braiding}
Other expected properties of S-matrices of integrable models are physical and braiding unitarity.
\paragraph{Physical unitarity. }
Physical unitarity implies that, when we are scattering
particles on the worldsheet theory with real energy, and we are considering a physical kinematics regime, the S-matrix
is a unitary matrix. It translates into the following constraint for the dressing factors:
\beq
\label{def:physical_unitarity}
\log{\Sigma(u^*,v^*)^*}+ \log{\Sigma(u,v)}=0\,,
\eeq
where $\,^*$ denotes complex conjugation.
Both $\Sigma_{\text{BES}}$ and $\Sigma^{\text{extra}}$ are known to satisfy this property \cite{Frolov:2021fmj}. For the novel dressing phase $\Sigma_{\text{new}}$, 
we have verified that, \emph{for purely imaginary $\mathbf{e}(v)$},  \eqref{def:physical_unitarity} holds for any value of $u,v$ thanks to the numerical and perturbative analysis of $\Sigma_{\text{new}}$ that we describe in the next section.

\paragraph{Braiding unitarity: fixing the source term $\mathbf{e}(v)$. }
Braiding unitarity, which descends from the underlying Yangian symmetry of the model, is much more subtle in our construction. The braiding unitarity constraints for S-matrices are the following:
\beq
\mathcal{S}_{ab}(u,v)\mathcal{S}_{ba}(v,u)=1\,,
\eeq
where $\mathcal{S}_{ab}$ and $\mathcal{S}_{ba}$ are two generic S-matrix elements.

It can be easily seen from the Asymptotic Bethe Ansatz equations obtained in section \ref{sec:ABA} that the rational parts of the S-matrices exactly satisfy this requirement. Moreover, this property is true for the building blocks of the dressing factors $\Sigma_{\text{BES}}$,  $\Sigma^{\text{extra}}$ and $\tilde{\Sigma}^{\text{extra}}$~\cite{Frolov:2021fmj}. Therefore, in the symmetric sector, where we only see the product of S-matrices and $\Sigma_{\text{new}}$ cancels, our ABA is fully compatible with braiding unitarity.

For the non-symmetric sector, the form of the dressing factors emerging from our argument is given by \eqref{def:ABA_Sigma_1} and \eqref{def:ABA_Sigma_2}. Braiding unitarity then translates in the following condition for $\Sigma_{\text{new}}$:
\beq\label{eq:braidingSigma}
\log{\Sigma_{\text{new}}}(u,v)+\log{\Sigma_{\text{new}}}(v,u)=0\,.
\eeq
Unfortunately, we could not find an efficient way to manipulate the integral equation to study the limits of validity of this property completely analytically.\footnote{One of the reasons this is complicated is that, while the kernel $\delta - \mathbb{G}$, appearing in the linear integral equation, is of difference form, its inverse is not.} 
However, by numerical or perturbative analysis, we find that if we take $\mathbf{e}(v)=0$ -- which is the value giving us a perfect match between our crossing equation and the one in \cite{Borsato:2015mma} (see section \ref{sec:comparison})-- then the braiding unitarity equation is violated!  A numerical illustration is shown in figure \ref{fig:unitaritymismatch}.

\begin{figure}
    \centering
    \includegraphics[width=1\linewidth]{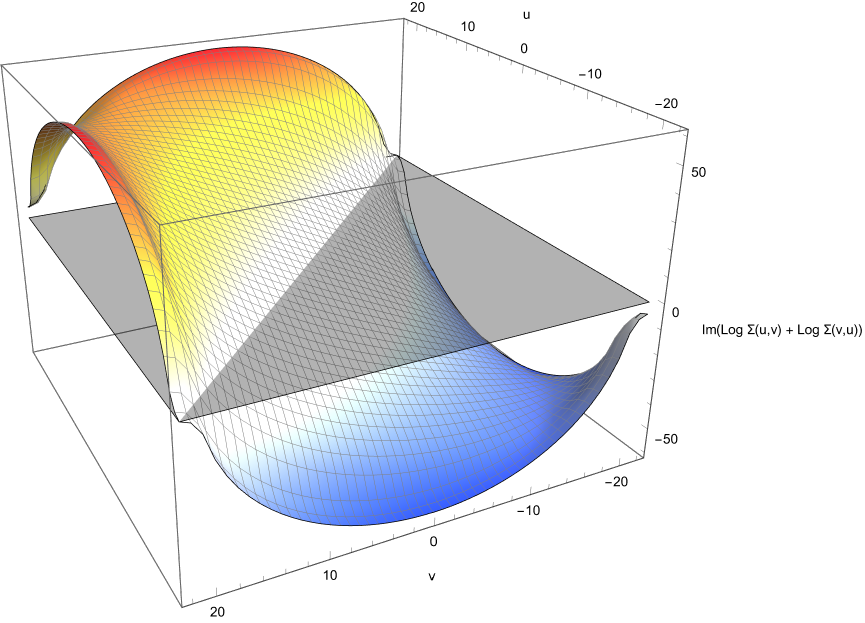}
    \caption{Braiding unitarity mismatch for the numerical solution of $-i\log\Sigma_{\text{new}}$ in the case where we assume $\mathbf{e}(v) = 0$ and at coupling $h=10$. The gray plane represents the plane where  $\log\Sigma(u,v)+\log\Sigma(v,u)=0$. We clearly see that our solution intersects it only for $u=-v$, which is a consequence of parity \eqref{def:parity_Sigma}.}
    \label{fig:unitaritymismatch}
\end{figure}
As we detail in the next section, studying the linear integral equation perturbatively at weak coupling we see that it is possible to \textit{impose} unitarity of $\Sigma_{\text{new}}$ order by order in $h$,  and that this fixes a specific perturbative expansion for the function $\mathbf{e}(v)$, which in this way is specified completely (notice that the weak coupling expansions in $h$ are expected to be convergent, as in $\mathcal{N}$=4 SYM). 
Although we were unable to resum the perturbative series, our results suggest that the exact form of this function, which enforces braiding unitarity, will be quite complicated, as its expansion includes terms with different orders of transcendentality.

\subsection{Perturbative and numerical results}
\subsubsection{Perturbative results at weak coupling}
In this section, we compute the new dressing phase perturbatively at weak coupling, starting from the integral equation \eqref{eq:LIE2elementary} and assuming that both $\mathbf{d}$ and $\mathbf{e}$ admit a series expansion in $h$:
\beq
\label{def:small_h_expansion_d}
\mathbf{d}(u|v)=\sum_{n=0}^{\infty} h^n\, \mathbf{d}_n(u|v),\quad \mathbf{e}(v)=\sum_{n=0}^{\infty} h^n\, \mathbf{e}_n(v)\,.
\eeq
\paragraph{Tree level.} 
While we have computed the dressing phase up to 12 loops at small $h$, some of the important features already emerge at the first order. Therefore, treating this simple case is already quite illustrative: in the integral equation \eqref{eq:LIE2elementary} at order $O(h)$, the integral will not contribute as its first nontrivial contribution is proportional to $h$. Hence the density $\mathbf{d}_0$ is given by the very simple expression
\beq
\label{eq:d_treelevel}
\mathbf{d}_0(u|v)=-\mathbf{a}(u-v)-\mathbf{e}_0(v)\,.
\eeq
Plugging this in \eqref{def:Sigma_new}, we can immediately evaluate the integrals to obtain that at first order:
\beq
\log{\Sigma_{\text{new}}}(u,v)=\frac{h}{\pi}\frac{1}{(u+\frac{i}{2})(u-\frac{i}{2})}\left(\log{\left(\frac{\frac{i}{2}-v}{\frac{i}{2}+v}\right)}+2\mathbf{e}_0(v)\right)+O(h^2)\,.
\eeq
This dressing phase satisfies parity symmetry \eqref{def:parity_Sigma} if we assume that $\mathbf{e}_0$ is an odd function of $v$; however, it is unitary at order $O(h^2)$ if and only if we take the first order of the weak coupling expansion of $\mathbf{e}(v)$ to be:
\beq
\label{eq:e0_unitary}
\mathbf{e}_0(v)=-\frac{1}{2}\log{\left(\frac{\frac{i}{2} -v}{\frac{i}{2}+v}\right)}\, ,
\eeq
which fixed
$$
\mathbf{d}_0(u|v) = -\frac{1}{2}\log{\left[\left(\frac{u-v + \frac{i}{2}}{v-u+\frac{i}{2}} \right)\left(\frac{\frac{i}{2} +v}{\frac{i}{2}-v}\right)\right]} .
$$
Note that $\mathbf{d}_0(z|v) = O(z) $, for small $z$,  this immediately implies that  $\log{\Sigma^{\text{unit}}_{\text{new}}}=O(h^2)$.

\paragraph{NLO.} To obtain $\mathbf{d}$ at order $O(h)$, in general, we cannot discard the integral in the integral equation anymore. 
However, since the endpoints depend on $h$, we can expand the integrand in the integration variable $z$ and keep only the $O(z^0)$ term, as higher-order terms would yield contributions of order $O(h^2)$ upon integration.

In the present case, it is easy to see that, with the leading order $\mathbf{d}_0$ given -- fixing braiding unitarity -- by \eqref{eq:d_treelevel} (and $\mathbf{e}_0$ by \eqref{eq:e0_unitary}), the integrand is already $O(z)$, so the integral will not contribute to $O(h)$. It follows that:
\beq
\mathbf{d}_1(u|v)=-\mathbf{e}_1(v)\,.
\eeq
The resulting leading contribution to the dressing phase is obtained by integrating (\ref{def:Sigma_new}). At order $O(h^2)$, this integral does not receive contributions from $\mathbf{d}_0$\footnote{In fact, $\mathbf{d}_0(z,v) \sim O(z)$, which integrates to zero over the symmetric interval $(-2 h, 2h)$. Higher powers of $z$ contribute from $O(h^3)$.}, and we find simply:

\beq
\log{\Sigma_{\text{new}}}(u,v)=\frac{2h^2}{\pi(u+i/2)(u-i/2)}\mathbf{e}_1(v)+O(h^3)\,.
\eeq
This clearly satisfies braiding unitarity if and only if $\mathbf{e}_1(v)=0$, resulting in $\log{\Sigma_{\text{new}}^{\text{unit}}}=O(h^3)$.
\paragraph{NNLO.} Solving the integral equation for the density at the next order, we find that also at order $O(h^3)$ the integral term in (\ref{eq:LIE2elementary}) trivialises (since at previous orders we have fixed the density to vanish). Therefore, also in this case, the density is given by just the source term:
\beq
\mathbf{d}_2(u|v)=-\mathbf{e}_2(v)\,.
\eeq
Now, from (\ref{def:Sigma_new})  we see that, at order $O(h^3)$ the dressing phase receives contribution from both $\mathbf{d}_2(u,v)$ and $\mathbf{d}_0(u,v)$, resulting in the combination of two terms:
\beqa
\log{\Sigma_{\text{new}}}(u,v)=&&\frac{-ih^3}{3\pi}\frac{2u+v+4uv(u+2v)}{(u+i/2)^2(u-i/2)^2(v+i/2)^2(v-i/2)^2}+\\&&
\frac{-2h^3}{\pi}\frac{\mathbf{e}_2(v)}{(u+i/2)(u-i/2)}+O(h^4)
\,.
\eeqa
It is easy to see that,  to impose braiding unitarity, we need to take:
\beq
\mathbf{e}_2(v)=\left( \frac{1}{(v-i/2)^2}-\frac{1}{(v+i/2)^2}\right)\,.
\eeq
The resulting dressing phase is given by:
\beq
\log\Sigma^{\text{unit}}_{\text{new}}(u,v)=
h^3\, \left[ \frac{2i}{3 \pi}\frac{(u-v)(4uv-1)}{(v-i/2)^2(v+i/2)^2(u-i/2)^2(u+i/2)^2}\right]+O(h^4)\,,
\eeq
which is now free of any ambiguities. 
Interestingly, this is proportional to the first non-zero order of the weak coupling expansion of the BES dressing phase~\cite{Beisert:2006ez}. It would be interesting to try to find a finite-coupling expression by generalising the double integral DHM representation of the BES dressing factor~\cite{Dorey:2007xn}. 
\paragraph{Higher orders.}
At higher orders, the expressions for $\mathbf{d}_i$ become significantly more complicated. Thus we have solved the integral equation \eqref{eq:LIE2elementary} order by order in $h$ on a computer, obtaining an expression for the coefficients $\mathbf{d}_i$ appearing in \eqref{def:small_h_expansion_d} given in terms of the unknown coefficients $\mathbf{e}_j(v),\, j=0\dots n$. To achieve this, we apply the same considerations that are already emerging at lower orders. In particular, the integral term in the linear integral equation takes the form $\int_{-2 h}^{+2 h} dz f(z)$,  where the integrand is (in our problem) always analytic for real $z$. To obtain the weak coupling expansion of the resulting integral, we replace $f(z)$ by its Taylor series around $z=0$, to high enough order: terms up to $z^k$ are sufficient to get the result up to $O(h^{k+1})$ after integration. Since this gives a simple, solvable integral at all orders, the method can be easily made algorithmic.

From this, we can easily obtain a perturbative expansion for the dressing phase via \eqref{def:Sigma_new}, again given in terms of the unknown functions $\mathbf{e}_j(v)$. As a cross check, we have verified that the resulting dressing phase satisfies the parity symmetry \eqref{def:parity_Sigma} order by order if we assume that $\mathbf{e}_j(-v)=-\mathbf{e}_j(v)\;\forall j$. 

Out of this solution, we can then build a perturbative expansion of $\Sigma_{\text{new}}$ by evaluating the last defining integral in (\ref{def:Sigma_new}), which we also do perturbatively. As we have already seen at the first non-vanishing order, the dressing phase violates braiding unitarity for a generic choice of $\mathbf{e}(v)$, and in particular if 
 we take $\mathbf{e}_i(v)=0$. 
 Instead, if we \textit{impose} that $\log{\Sigma}_{\text{new}}$ should satisfy braiding unitarity order by order, we obtain a set of constraints on the functions $\mathbf{e}_j(v)$ in \eqref{def:small_h_expansion_d}. The solution to these constraints seems to be unique, and can be given, at least up to order $O(h^{13})$, in terms of simple rational functions of $v$ and transcendental numbers. For example, the first five nontrivial orders we have found are the following:
\beqa
\label{eq:e_unitary}
\mathbf{e}(v)=&&\frac{1}{2}\log{\frac{-v+i/2}{v+i/2}}+h^2\left( \frac{1}{(v-i/2)^2}-\frac{1}{(v+i/2)^2}\right)+\\&& \nonumber h^3 \left(-\frac{16}{3 \pi}\log{2} \right)\left( \frac{1}{(v-i/2)^2}-\frac{1}{(v+i/2)^2}\right)+\\&&\nonumber
h^4\,\left(\frac{2}{(v-i/2)^4}-\frac{2}{(v+i/2)^4}\right)+\\&&
h^5 \left[-\frac{32}{5\pi} \log{2} \left(\frac{2}{(v-i/2)^4}-\frac{2}{(v+i/2)^4}\right)+\frac{96}{5 \pi}\zeta_3\left( \frac{1}{(v-i/2)^2}-\frac{1}{(v+i/2)^2}\right)\right]+O(h^6)\nonumber
\,;
\eeqa
at higher loops, we also find other odd Riemann zetas $\zeta_{2 n+1}\equiv \zeta(2n+1)$. 
Substituting these values of the source terms, the corresponding solution for the dressing phase, fixed by imposing braiding unitarity, is given by:
\begin{small}
\beqa\label{eq:Sigma_unitary}
\log\Sigma^{\text{unit}}_{\text{new}}(u,v)=&&
h^3\, \left[ \frac{2i}{3 \pi}\frac{(u-v)(4uv-1)}{(v-i/2)^2(v+i/2)^2(u-i/2)^2(u+i/2)^2}\right]+
\\
 && h^5 {\frac{8192 i }{15 \pi \left(4
   u^2+1\right)^4 \left(4 v^2+1\right)^4}} \bigg(2 \left(4
   u^2+1\right)^2 \left(4 v^2+1\right) (2 u (6 u v+1)+3 v)-\nonumber
   \\&&\left(4 u^2+1\right) \left(4 v^2+1\right)^2 (3 u (4 u
   (u-v)+1)+v)-6 u \left(4 u^2-1\right) \left(4 v^2+1\right)^3-12 \left(4 u^2+1\right)^3 v\bigg)\nonumber \\
   &&+h^6\, \frac{16}{\pi}\zeta_3\left[ \frac{2i}{3 \pi}\frac{(u-v)(4uv-1)}{(v-i/2)^2(v+i/2)^2(u-i/2)^2(u+i/2)^2}\right] + O(h^7)\nonumber\,.
\eeqa
 \end{small}
For comparison, had we chosen $\mathbf{e}(v) \equiv 0$ we would have found the following dressing phase, which violates braiding unitarity:
For the dressing phase $\log{\Sigma_{\text{new}}}$, the non-unitary one with $\mathbf{e}(v)=0$ reads, at the first orders:
\beqa
\label{eq:sigma_loop_nonunit}
\left. \log\Sigma_{\text{new}}(u,v) \right|_{\text{ source } \mathbf{e}(v) \equiv 0} =&&h\, \frac{1}{\pi}\frac{\log{\frac{i/2-v}{i/2+v}}}{(u+i/2)(u-i/2)}+h^2\, \frac{8 \log{2}}{\pi^2}\frac{\log{\frac{i/2-v}{i/2+v}}}{(u+i/2)(u-i/2)}+\\&&
h^3\, \left[ -\frac{i}{8 \pi}\frac{2u+v+4u^2v+8uv^2}{(v-i/2)^2(v+i/2)^2(u-i/2)^2(u+i/2)^2}\right]+\nonumber\\&&
h^3 \, \left[ \frac{64}{3 \pi^3}\frac{\pi^2(12u^2-1)+12\log{2}(1+4u^2)^2}{(1+4u^2)^3}\log{\frac{i/2-v}{i/2+v}}\right]+O(h^4)\,\nonumber .
\eeqa
Clearly, this has a more complicated functional form than the one where we impose unitarity.

\paragraph{Tension with the crossing equation. }
The unitarity-corrected dressing phase (\ref{eq:Sigma_unitary}) therefore satisfies almost all the crucial properties -- braiding unitarity, parity and physical unitarity.  However, the downside of introducing a nonzero source $\mathbf{e}(v)$ is that it will enter the crossing equations as in  (\ref{eq:newcrossing}).  
While this modification of the crossing equations would affect only the dependence on one of the two rapidities, it would still be significant, 
since the special choice of $\mathbf{e}(v)$ which enforces unitarity is a quite nontrivial function, containing transcendental numbers in its perturbative expansion.  Presently, we have no understanding of how this modification of the crossing equations could be justified. 
\begin{figure}[ht!]
\centering

\begin{minipage}{0.45\textwidth}
    \centering
    \includegraphics[width=\textwidth]{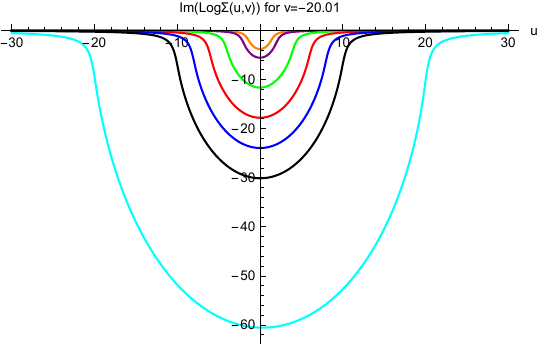}
    \caption*{(a) v=-20.01}
\end{minipage}
\hfill
\begin{minipage}{0.45\textwidth}
    \centering
    \includegraphics[width=\textwidth]{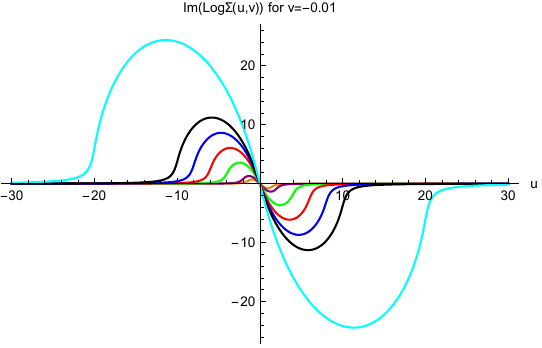}
    \caption*{(b) v=-0.01}
\end{minipage}
\hfill
\begin{minipage}{0.64\textwidth}
    \centering
    \includegraphics[width=\textwidth]{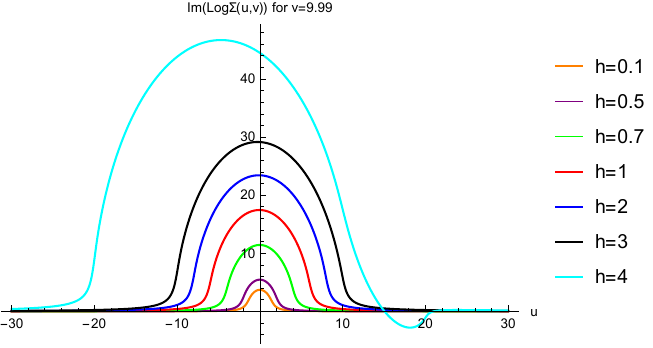}
    \caption*{(c) v=9.99}
\end{minipage}

\caption{Numerical plots of the imaginary part of $\log{\Sigma}_{\text{new}}(u,v)$ for a few values of the coupling $h$ and $v$. The solutions in this plot have $\mathbf{e}(v)=0$. }
\label{fig:sigma_unit_plots}
\end{figure}

\begin{figure}[ht!]
\centering

\begin{minipage}{0.9\textwidth}
    \centering
    \includegraphics[width=\textwidth]{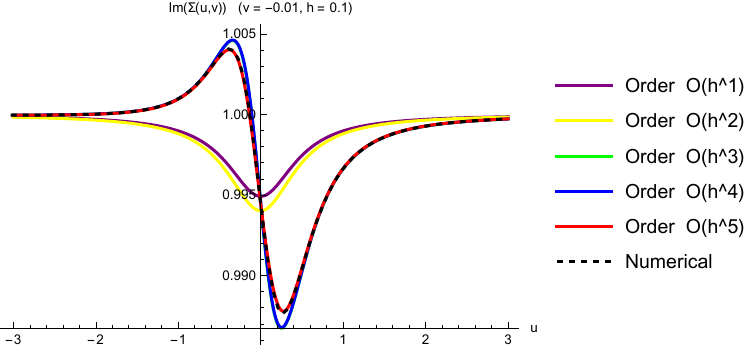}
\end{minipage}
\hfill
\begin{minipage}{0.9\textwidth}
    \centering
    \includegraphics[width=\textwidth]{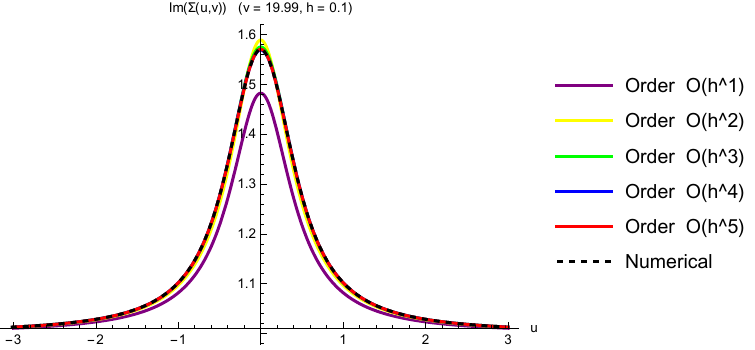}
\end{minipage}
\caption{Comparison between the numerical and perturbative dressing phases at weak coupling. At $h=0.1$, the perturbative solution already matches very precisely the numerical one at 5 loops.}
\label{fig:sigma_pert_vs_num}
\end{figure}

\paragraph{Comments on the source term $\mathbf{e}(v)$ enforcing unitarity. }
 It is interesting to notice that all the functions of $v$ appearing order by order in $\mathbf{e}(v)$ to impose unitarity (cf. (\ref{eq:e_unitary})) are nothing but the coefficients of the series expansion in $h$ of
\beq\label{eq:simpleguess}
-\frac{1}{4}\log{\frac{(v-i/2+2h)(v-i/2-2h)}{(v+i/2+2h)(v+i/2-2h)}}\,.
\eeq
The terms in $\mathbf{e}(v)$  appear with different numerical coefficients as compared to the function above; in particular, they also contain zeta values. 
We are tempted to speculate that the non-perturbative $\mathbf{e}(v)$ would be related to the simple function (\ref{eq:simpleguess}).\footnote{This function is also related to some natural variables appearing in the other AdS$_3$ QSC, e.g. see the ``relativistic variable'' called ``$\gamma$'' in \cite{Frolov:2021fmj}.} Unfortunately, however, we were not able to guess the expression resumming this perturbative series. We leave this for future investigations.

\subsubsection{Numerical results} 
We have also confirmed the results of our perturbative analysis by numerically solving the equations.

In particular, we have computed numerically the solution for the dressing phase, for a few values of the coupling $h$, in the non-unitary case with $\mathbf{e}(v)=0$. 
For simplicity, we concentrated on various values of the rapidities $u,v$  on the real axis, which gives a purely imaginary $\log{\Sigma}_{\text{new}}(u,v)$, as it should be to satisfy physical unitarity. 
In figure \ref{fig:unitaritymismatch}, we plot the ``unitarity mismatch'', i.e. $\log{\Sigma}_{\text{new}}(u,v)+\log{\Sigma}_{\text{new}}(v,u)$. This quantity is the deviation from braiding unitarity of our numerical solution with $\mathbf{e}(v)=0$. We clearly see that in this case, braiding unitarity does not hold.

In figure \ref{fig:sigma_unit_plots}, the result for this ( non-braiding-unitary) phase at various values of the coupling is shown. In figure  \ref{fig:sigma_pert_vs_num} we compare our numerical solution at weak coupling ($h=0.1$) with the perturbative solution \eqref{eq:sigma_loop_nonunit}, for $v=-0.01$ and $v=19.99$, showing that the weak coupling expansion approximates better and better the numerical data with increasing number of terms, which is a test of our weak coupling solution.

In figure \ref{fig:sigma_unit_plots}, we plot the dressing phase both at weak and strong coupling, for some fixed values of $v$ and for $u\in(-30,30)$.
\section{The $\dalpha$ Q-system}\label{sec:alphaQsys}
As we have discussed in the introduction, the full symmetry of the \ads\, string theory is composed of two copies of the $\dalpha$ super Lie algebra, which only in the special case $\alpha={1 \over 2}$ reduces to $\osp(4|2)$. So far, we have only treated this simpler case, as the Yangian for the family of Lie algebras $\osp(2m|2n)$ is rather well known in the literature \cite{Arnaudon,Galleas:2004zz,Molev:2021jgh,Molev:2023vgm,Frassek:2023ghu,Molev:2024xvz}, and the associated Q-system was written down in \cite{Tsuboi:2023sfs}, which allowed us to be confident in the algebraic part of our construction.

The main difficulty in the analysis of the $\alpha\neq {1 \over 2}$ case is the fact that, to the best of our knowledge, the Yangian of the $\dalpha$ algebra has not been studied. In particular, the Q-system, on which the Quantum Spectral Curve is based, has not been written down. In this section, we will advance a proposal for the system of QQ-relations, motivated by the expected general form of a Q-system from Lie-theoretical grounds. In this way, we will see that we can derive Bethe Ansatz equations that perfectly match the ones known in the literature for this superalgebra \cite{Ogievetsky:1986hu, OhlssonSax:2011ms}. We conclude this section by briefly discussing what we expect to be the analytic structure of the Q-functions for the \ads\, model at generic $\alpha$, opening the way to a discussion of the QSC for the whole model.

Note that our $\dalpha$ Q-system naturally reduces to the $\osp(4|2)$ of section \ref{sec:QQsys} when setting $\alpha={1 \over 2}$. However, while in the singular limits $\alpha\rightarrow 0,1$ the algebra $\dalpha$ should reduce to $\mathfrak{psu}(2|2)$, the Q-system we will propose does not seem to reproduce completely the one of \cite{Cavaglia:2021eqr, Ekhammar:2021pys} in the same limit. We further comment on this in section \ref{sec:gllimit}.
\subsection{$\dalpha$ QQ-relations}
To describe the $\dalpha$ Q-system, we employ the same notation that we used in section \ref{sec:QQsys}. 
In particular, we will use the same types of indices, and the Q-functions from which we derive all the other ones will still be denoted as $\bP,\,\bS,\,\hat{\bS}$. The equations will still be written in a covariant way with respect to the bosonic groups $SO(4)\sim SU(2)\otimes SU(2)$ and $SL(2)$, whose direct product is isomorphic to the bosonic subgroup of $D(2,1;\alpha)$.

While we could introduce $\bQ_{\emptyset}$ into these equations, for simplicity, we report them here with $\bQ_{\emptyset} = 1$. 

As a useful shorthand, we define $\hat \alpha\equiv 1-\alpha$; since exchanging $\hat \alpha \leftrightarrow \alpha$ is a symmetry of the $\dalpha$ Lie algebra, our Q-system is invariant under it, after relabeling some of the Q-functions appropriately. Furthermore, thanks to this symmetry, we may assume without loss of generality that $0<\alpha\leq\hat\alpha$ or equivalently that $0<\alpha\leq\frac{1}{2}$.

Our strategy to obtain the QQ-relations for the $\dalpha$ Lie algebra is to make an Ansatz for some ``core relations'' and derive all the other QQ-relations from them, in analogy with our presentation of the $\alpha = \frac{1}{2}$ case. We will not repeat the derivations at length here, as they are essentially the same as for the $\osp(4|2)$ case except for having different shifts. The most natural Ansatz is  to start from a simple modification of \eqref{eq:core_relations} that reduces to them when setting $\alpha={1 \over 2}$:
\begin{framed}
\begin{equation}
\text{Core Relations:}\;\;\; \bS_{a|k}^{[+2\alpha]}(u)  = \bP_a^{\;\hat b}(u)  \bS_{\hat b|k}^{[-2\hat \alpha]}(u)   ,\;\;\; \bS_{\hat a|k}^{[+2\hat \alpha]}(u)  = \bP_{\hat a}^{\; b}(u)  \bS_{ b|k}^{[-2\alpha]}(u)  .
\end{equation}
\end{framed}
It is then immediate to derive the following relations (which will be useful later to study the possible cut structure of the $\bS,\,\hat \bS$ Q-functions):
\beqa
\label{eq:alpha_core_rels}
&\bS_{a|k}^{[+2\alpha]}=\bP_{a}^{\;\hat b}\bP_{\hat b}^{\;c}\,^{[-4\hat\alpha]}\bS^{[+2\alpha-4]}_{c|k}\,,
\\
&\bS_{\hat a|k}^{[+2\hat\alpha]}=\bP_{\hat a}^{\; b}\bP_{ b}^{\;\hat c}\,^{[-4\alpha]}\bS^{[+2\hat \alpha-4]}_{\hat c|k}.
\eeqa
We also assume the constraint \eqref{eq:constraintP} to hold without modifications, i.e.
\begin{equation}\label{eq:constraintPalpha}
\text{ Constraint 1:  }\;\;\; \bP_A(u) \rho^{AB} \bP_B(u) = 2   . 
\end{equation}
In the same way as the $\osp(4|2)$ case, we will also assume a second constraint, analogue to \eqref{eq:constraintQai}, for which we can also develop a similar argument under simple analyticity assumptions:
\begin{equation}
\text{ Constraint 2: }\;\;\;\;\;\; {\bS_{a|i}(u) \mathbf{C}^{ab} \bS_{b|j}(u)} = -\epsilon_{ij}= -{\bS_{\hat a|i}(u) \hat{\mathbf{C}}^{\hat a \hat b} \bS_{\hat b|j}(u)} .
\end{equation}
Other consequences of Constraint 1 together with the core relations above are the Wronskian expressions for the $\bP$ functions, which generalise (\ref{eq:F1}) and read as:
\begin{eqnarray}
&&\bP_a^{\; \hat b} =  \bS_{a|i}^{[+2\alpha]} \; \epsilon^{ij} \; \bS_{\hat a|j}^{[-2\hat\alpha]} \hat{\mathbf{C}}^{\hat a \hat b} ,\\
&&\bP_{\hat a}^{\;  b} =  \bS_{\hat a|i}^{[+2\hat \alpha]} \; \epsilon^{ij} \; \bS_{ a|j}^{[-2\alpha]} {\mathbf{C}}^{a  b} .
\end{eqnarray}
We also give definitions of the $\bQ$ functions in terms of $\bS$ and $\hat \bS$, generalising the QQ-relations \eqref{eq:relQ1} and \eqref{eq:relQ2}:
\begin{eqnarray}
\label{eq:d21a_QQrels_1}
&&\bQ_{ij}=\bS_{a|i}^{[+2\alpha]} \bS_{b|j}^{[-2\alpha]} \; {\mathbf{C}}^{a b},\\
&& \bQ_{ij}=\bS_{\hat a|i}^{[+2\hat \alpha]} \bS_{\hat b|j}^{[-2\hat \alpha]} \; \hat{\mathbf{C}}^{\hat a \hat b} .
\end{eqnarray}
Then we can derive all the following QQ-relations in the same way as we did in section \ref{sec:proofsQQ}.
In particular, we can derive:
\begin{eqnarray}
&&\bP_A \bQ_{[ij]}= -Q_{A|[ij]}^{P} - Q_{A|[ij]}^M ,\\
&&\bP_A \bQ_{\{ij\}} = Q_{A|\{ij\}}^{P} - Q_{A|\{ij\}}^M ,\\
&&\text{where}\;\;\;  Q^M_{A|ij}\equiv  \bS_{a|i}^{[-2\alpha]} \;  \bS_{\hat a|j}^{[-2\hat\alpha]}\;(\sigma_A )_b^{\; \hat a} \;\mathbf{C}^{ab},\\
&&\text{and}\;\;\;\;\;\,\,  Q^P_{A|ij}\equiv  \bS_{a|i}^{[+2\alpha]} \; \bS_{\hat a|j}^{[+2\hat\alpha]}\; (\sigma_A )_b^{\; \hat a} \;\mathbf{C}^{ab},
\end{eqnarray}
which are the analog of \eqref{eq:F2QQrel} and \eqref{eq:F2QQrel2}. Taking $i=j$, the second equation reduces to:
\beq
\label{eq:d21_QQ1}
\bP_A \bQ_{ii} = Q_{A|ii}^{P} - Q_{A|ii}^M=\; (\sigma_A )_b^{\; \hat a} \;\mathbf{C}^{ab}\left(\bS_{a|i}^{[+2\alpha]} \; \bS_{\hat a|i}^{[+2\hat\alpha]}-\bS_{a|i}^{[-2\alpha]} \; \bS_{\hat a|i}^{[-2\hat\alpha]}\right)\,,
\eeq
which can also be rewritten as 
\beq
\label{eq:d21_QQ1alt}
\bP_{a\hat{a}} \bQ_{ii} =  \bS_{a|i}^{[+2\alpha]} \; \bS_{\hat a|i}^{[+2\hat\alpha]}-\bS_{a|i}^{[-2\alpha]} \; \bS_{\hat a|i}^{[-2\hat\alpha]},
\eeq
with $\bP_{a\hat{a}}\equiv \bP_a^{\,\hat{b}}  \hat{\mathbf{C}}_{\hat{b}\hat{a}}$.  Notice that in the general $\alpha$ case, in the RHS we do not have the difference of a single combination of spinors with different shifts, but the difference of two different combinations of a spinor and antispinor with different shifts.

We can also write down QQ-relations that generalise \eqref{eq:QQrelssp} at $\alpha=1/2$, but now it is natural to split this into two relations:
\begin{eqnarray}
\label{eq:d21_QQ3}
&&{\bP}_{AB} = \bP_A^{[+2\alpha]} \bP_B^{[-2\alpha]}-\bP_A^{[-2\alpha]} \bP_B^{[+2\alpha]}\,,\\
\label{eq:d21_QQ32}
&&\hat{\bP}_{AB} = \bP_A^{[+2\hat\alpha]} \bP_B^{[-2\hat\alpha]}-\bP_A^{[-2\hat\alpha]} \bP_B^{[+2\hat\alpha]}\, ,
\end{eqnarray}
where the objects $\bP_{AB}$ and $\hat{\bP}_{AB}$  coincide at $\alpha = \frac{1}{2}$, as we can easily see from the definitions above. This is due to the fact that these Q-functions are associated with the two different Dynkin diagrams, depicted in figures \ref{fig:d21sfig3} and \ref{fig:d21sfig4}, which collapse to the same $\osp(4|2)$ diagram \ref{fig:sfig3} when $\alpha=1/2$.

In particular, it can be proven that these two Q-functions satisfy the following QQ-relations
\begin{eqnarray}
\label{eq:d21_QQ4}
&&2\,\bS_{a|i}^{++} \; \bS_{b|j}^{--} \; \mathbf{C}^{ac}(\sigma_{[A} \cdot \bar{\sigma}_{B]} )_c^{\; b} \epsilon^{ij} = \left( \delta_A^C \delta_B^D-\delta_B^C \delta_A^D  -\epsilon_{C' D' AB} \rho^{C' C} \rho^{D' D} \right) \hat\bP_{CD} ,
\\
&&2\,\bS_{\hat a|i}^{++} \; \bS_{\hat b|j}^{--} \; \hat{\mathbf{C}}^{\hat a\hat c}(\bar{\sigma}_{[A} \cdot {\sigma}_{B]} )_{\hat c}^{\; \hat b} \epsilon^{ij} = -\left( \delta_A^C \delta_B^D-\delta_B^C \delta_A^D  +\epsilon_{C' D' AB} \rho^{C' C} \rho^{D' D} \right) \bP_{CD} .
\end{eqnarray}
which reduce to \eqref{eq:Z2QQrel} and \eqref{eq:Z2starQQrel}. Like in the $\osp(4|2)$ case, the right-hand sides of these equations 
takes a drastically simplified form\footnote{These simplified forms are the analogues of \eqref{eq:Wronski1} and \eqref{eq:Wronski2}.}, with only a single term of the form $\bP_{AB}$ or $\hat{\bP}_{AB}$ on the rhs and a single Wronskian of spinors on the lhs) for some selected choices of the indices. With our convention for the form of the metric and sigma matrices,  the ``special choices'' are the ones in tables \ref{tab:goodindices} for the first equation and \ref{tab:goodindicesstar} for the second equation. 

The last QQ-relations we find are the analogues of \eqref{eq:QQrelZ} and \eqref{eq:QQrelZ2}:
\begin{eqnarray}
\label{eq:d21_QQ2}
   &\bS_{\hat a|i}^{++} \; \bP_{A}^{[-2\alpha]} - \bS_{\hat a|i}^{--} \; \bP_{A}^{[+2\alpha]} = \left(\bP_A^{[-2\alpha]} (\bP_{\hat a}^{[+2\alpha]})^{\; b} -  \bP_A^{[+2\alpha]} (\bP_{\hat a}^{[-2\alpha]})^{\;  b} \right) \bS_{ b|i}.\\
&\bS_{a|i}^{++} \; \bP_{A}^{[-2\hat\alpha]} - \bS_{a|i}^{--} \; \bP_{A}^{[+2\hat\alpha]} = \left(\bP_A^{[-2\hat\alpha]}(\bP_a^{[+2\hat\alpha]})^{\; \hat b} -  \bP_A^{[+2\hat\alpha]} (\bP_a^{[-2\hat\alpha]})^{\; \hat b} \right) \bS_{\hat b|i},\nonumber
\end{eqnarray}
where, in this case, the combinations on the right-hand side can also be seen as containing Wronskians of the form $\bP_{AB}$ for the first equation and $\hat{\bP}_{AB}$ for the second. Just like we discussed in the case at $\alpha = 1/2$, in general, the rhs will contain the sum of two Wronskians, each multiplied by a spinor. However, precisely when the indices are selected in a special way, an additional cancellation occurs, leaving only a single Wronskian on the right-hand side. The selection rule is the same: the indices should be paired as summarised in tables \ref{tab:goodindicesstar} for the first equation and \ref{tab:goodindices} for the second equation. 
\subsection{$\dalpha$ Bethe equations}
\begin{figure}
\begin{subfigure}{.5\textwidth}
  \centering
  \includegraphics[width=.5\linewidth]{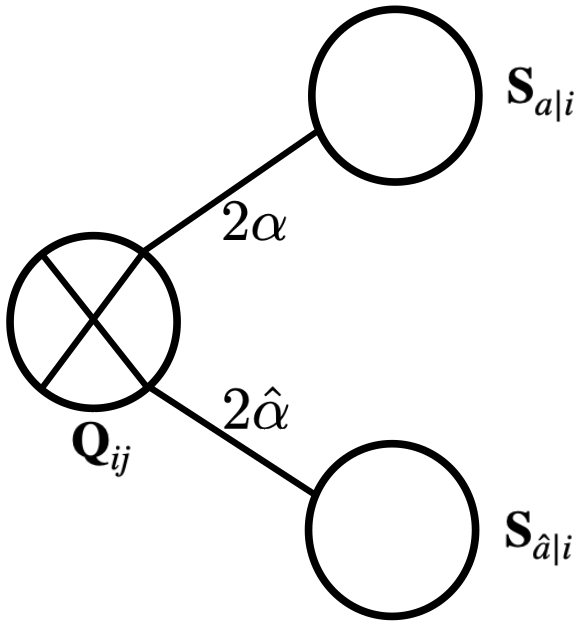}
  \caption{Dynkin-Kac diagram 1}
  \label{fig:d21sfig1}
\end{subfigure}%
\begin{subfigure}{.5\textwidth}
  \centering
  \includegraphics[width=.5\linewidth]{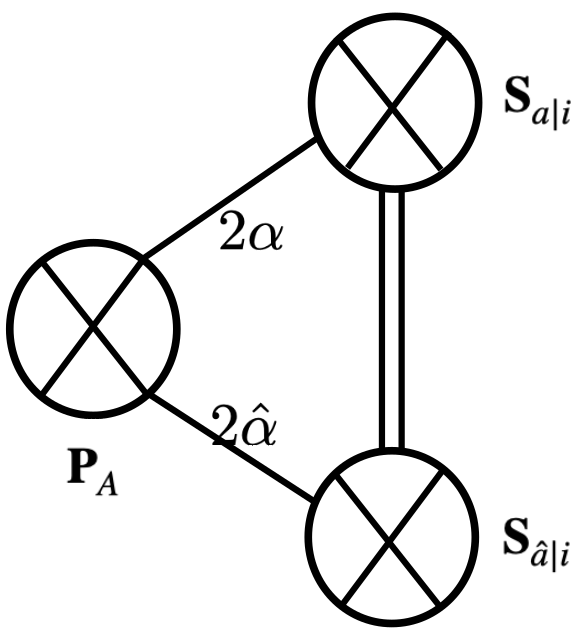}
  \caption{Dynkin-Kac diagram 2}
  \label{fig:d21sfig2}
\end{subfigure}
\begin{center}
    \begin{subfigure}{.6\textwidth}
  \centering
  \includegraphics[width=.6\linewidth]{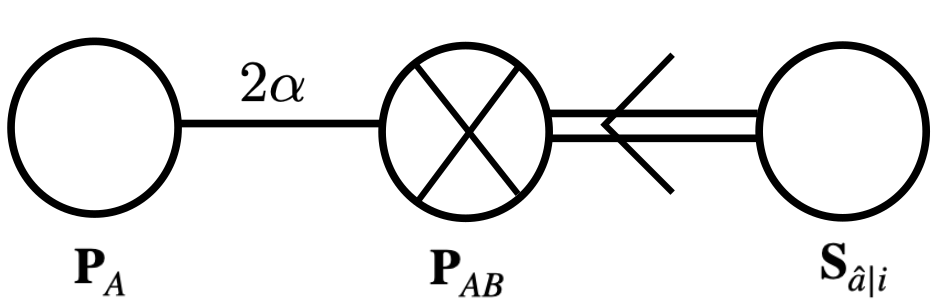}
  \caption{Dynkin-Kac diagram 3}
  \label{fig:d21sfig3}
\end{subfigure}
\end{center}
\begin{center}
    \begin{subfigure}{.6\textwidth}
  \centering
  \includegraphics[width=.6\linewidth]{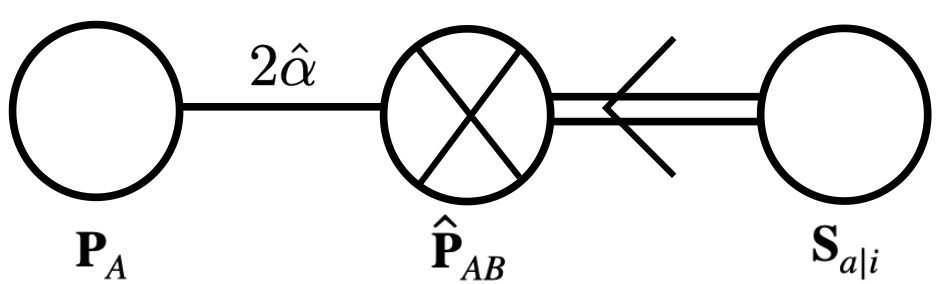}
  \caption{Dynkin-Kac diagram 4}
  \label{fig:d21sfig4}
\end{subfigure}
\end{center}
\caption{The 4 Dynkin-Kac diagrams for $\dalpha$. Note that at $\alpha=1/2$, \ref{fig:d21sfig3} and \ref{fig:d21sfig4} reduce to the same Dynkin-Kac diagram for $\osp(4|2)$ \ref{fig:sfig3}. It is shown which Q-functions are associated with the various nodes.}
\label{fig:d21dynkin}
\end{figure}
From the QQ-relations obtained in the last section, we can easily derive the exact Bethe equations in the standard way explained for $\osp(4|2)$ in section \ref{sec:Bethe_equations}. Once again, each Bethe equation corresponds to a node of the Dynkin-Kac diagrams of $\dalpha$, which we depict in figure \ref{fig:d21dynkin}. To fix the indices consistently in the Bethe equations, we need to employ the same rules explained at the beginning of section \ref{sec:Bethe_equations}.
In this section, we will write down the Bethe equations assuming the choice of conventions for the sigma matrices given in the first section of the paper.

\paragraph{BAE for type \ref{fig:d21sfig1} Dynkin diagram}
For the first node we evaluate \eqref{eq:d21_QQ1} at zeros of $\bQ_{ii}$, obtaining
\begin{equation}\label{eq:BAE_1_node1_alpha}
    1=\left(\frac{\bS_{a|i}^{[-2\alpha]}\bS_{\hat a|i}^{[-2\hat\alpha]}}{\bS_{a|i}^{[+2\alpha]}\bS_{\hat a|i}^{[+2\hat\alpha]}}\right)\bigg|_{\bQ_{ii}=0}\,.
\end{equation}
For the spinor node BAE, we start from the first of \eqref{eq:d21a_QQrels_1}. Since these equations have a Wronskian on the rhs (just with shift modulated by $\alpha$), we can manipulate it in a standard way to get:
\begin{align}
\label{eq:BAE_1_node2_alpha}
    \left(\frac{\bQ^{[+2\alpha]}_{ii}}{\bQ^{[-2\alpha]}_{ii}}\right)\bigg|_{\bS_{a|i}=0}=\left(-\frac{\bS_{a|i}^{[+4\alpha]}}{\bS_{a|i}^{[-4\alpha]}}\right)\bigg|_{\bS_{a|i}=0}\,.
\end{align}
While for the antispinor node we use in the same way the second of \eqref{eq:d21a_QQrels_1}, yielding:
\begin{align}
\label{eq:BAE_1_node3_alpha}
    \left(\frac{\bQ^{[+2\hat\alpha]}_{ii}}{\bQ^{[-2\hat\alpha]}_{ii}}\right)\bigg|_{\bS_{\hat a|i}=0}=\left(-\frac{\bS_{\hat a|i}^{[+4\hat\alpha]}}{\bS_{\hat a|i}^{[-4\hat\alpha]}}\right)\bigg|_{\bS_{\hat a|i}=0}\,.
\end{align}
These Bethe equations at $\alpha=1/2$ reduce to those for the Dynkin diagram \ref{fig:sfig2} of $\osp(4|2)$. For each choice of $a$, $\hat{a}$ and $i$, we get an independent set of three coupled Bethe equations. In total, there are 8 possibilities.  

\paragraph{BAE for type \ref{fig:d21sfig2} Dynkin diagram}
For the first node we evaluate \eqref{eq:d21_QQ1alt} at zeros of $\bP_{a\hat{a}}$, obtaining:
\begin{equation}
\label{eq:BAE_2_node1_alpha}
    1=\left(\frac{\bS_{a|i}^{[-2\alpha]}\bS_{\hat a|i}^{[-2\hat\alpha]}}{\bS_{a|i}^{[+2\alpha]}\bS_{\hat a|i}^{[+2\hat\alpha]}}\right)\bigg|_{\bP_{a\hat{a}}=0}\,.
\end{equation}
For the spinor node we evaluate the first of \eqref{eq:d21_QQ2} at zeros of $\bS_{a|i}$, obtaining:
\begin{equation}
\label{eq:BAE_2_node2_alpha}
   1=\left(\frac{\bP_{\hat a a}^{[+2\alpha]}\bS_{\hat a|i}^{--}}{\bP_{\hat a a}^{[-2\alpha]}\bS_{\hat a|i}^{++}}\right)\bigg|_{\bS_{a|i}=0}\,.
\end{equation}
Analogously for the antispinor node we evaluate the second of \eqref{eq:d21_QQ2} at zeros of $\bS_{\hat a|i}$:
\begin{equation}
\label{eq:BAE_2_node3_alpha}
  1=\left(\frac{\bP_{\hat a a}^{[+2\hat\alpha]}\bS_{a|i}^{--}}{\bP_{\hat a a}^{[-2\hat\alpha]}\bS_{a|i}^{++}}\right)\bigg|_{\bS_{\hat a|i}=0}\,.
\end{equation}
These Bethe equations at $\alpha=1/2$ reduce to those for the Dynkin diagram \ref{fig:sfig1} of $\osp(4|2)$. For this grading, we also have 8 possibilities, one for each choice of $a$, $\hat{a}$ and $i$. Every choice gives an independent set of three coupled Bethe equations.

\paragraph{BAE for type \ref{fig:d21sfig3} Dynkin diagram}
For the first node we start from \eqref{eq:d21_QQ3} and obtain, in the standard way:
\begin{equation}
\label{eq:BAE_3_node1_alpha}
    -1=\left(\frac{\bP_{A}^{[-4\alpha]}\bP_{AB}^{[+2\alpha]}}{\bP_{A}^{[+4\alpha]}\bP_{AB}^{[-2\alpha]}}\right)\bigg|_{\bP_{A}=0}\,.
\end{equation}
For the second node, we evaluate the first QQ-relation of the ones in \eqref{eq:d21_QQ2} -- in the case where the indices are taken as in table \ref{tab:goodindicesstar} --  at zeros of $\bP_{AB}$. This gives:
\begin{equation}
\label{eq:BAE_3_node2_alpha}
   1=\left(\frac{\bP_{A}^{[+2\alpha]}\bS_{\hat a|i}^{--}}{\bP_{A}^{[-2\alpha]}\bS_{\hat a|i}^{++}}\right)\bigg|_{\bP_{AB}=0}\,.
\end{equation}
Finally, for the third node, we start from the second line of \eqref{eq:d21_QQ4} for the same choice of indices to obtain
\begin{equation}
\label{eq:BAE_3_node3_alpha2}
  -1=\left(\frac{\bP_{AB}^{++}\bS_{\hat a|i}^{[-4]}}{\bP_{AB}^{--}\bS_{\hat a|i}^{[+4]}}\right)\bigg|_{\bS_{\hat a|i}=0}\,.
\end{equation}
At $\alpha=1/2$, these Bethe equations reduce to half of those for the Dynkin diagram \ref{fig:sfig3} of $\osp(4|2)$. 

There are in total 4 possible choices of indices for $(A,B, \hat{a})$, which are the same as for the $\alpha = 1/2$ case.

\paragraph{BAE for type \ref{fig:d21sfig4} Dynkin diagram} The derivation in this case is analogous. We will assume the indices chosen as in table \ref{tab:goodindices}, and use the QQ-relations \eqref{eq:d21_QQ32}, \eqref{eq:d21_QQ2} and  \eqref{eq:d21_QQ4}. The latter two equations simplify to take the form of an equality of two Wronskians (with some shifts depending on $\alpha$) at this special choice of indices.

For the first node, we start from \eqref{eq:d21_QQ32} and obtain:
\begin{equation}
\label{eq:BAE_4_node1_alpha}
    -1=\left(\frac{\bP_{A}^{[-4\hat\alpha]}\hat\bP_{AB}^{[+2\hat\alpha]}}{\bP_{A}^{[+4\hat\alpha]}\hat\bP_{AB}^{[-2\hat\alpha]}}\right)\bigg|_{\bP_{A}=0}\,.
\end{equation}
For the second node, we evaluate the second QQ-relation written in \eqref{eq:d21_QQ2} at the zeros of $\hat\bP_{AB}$ and get
\begin{equation}
\label{eq:BAE_3_node2_alpha2}
   1=\left(\frac{\bP_{A}^{[+2\hat\alpha]}\bS_{ a|i}^{--}}{\bP_{A}^{[-2\hat\alpha]}\bS_{ a|i}^{++}}\right)\bigg|_{\hat\bP_{AB}=0}\,.
\end{equation}
While for the third node, we start from the first line of equations \eqref{eq:d21_QQ4} to obtain
\begin{equation}
\label{eq:BAE_3_node3_alpha}
  -1=\left(\frac{\hat\bP_{AB}^{++}\bS_{ a|i}^{[-4]}}{\hat\bP_{AB}^{--}\bS_{ a|i}^{[+4]}}\right)\bigg|_{\bS_{ a|i}=0}\,.
\end{equation}
At $\alpha=1/2$, these equations reduce to half of those for the Dynkin diagram \ref{fig:sfig3} of $\osp(4|2)$.

There are in total 4 possible choices of indices for $(A,B, {a})$, which are the same as for the $\alpha = 1/2$ case. 

It is remarkable that all these different Bethe equations match precisely those present in the literature for this superalgebra, with the shifts coming from the structure of the Cartan matrix of $\dalpha$. We take this as strong evidence that our Q-system for this algebra is correct. 
\subsubsection{The $\alpha\rightarrow 1$ limit}
\label{sec:gllimit}
It is a well-known fact that for $\alpha= 0,1$ the $\dalpha$ algebra is singular \cite{Kac:1977em}. However, one can take the limit $\alpha\rightarrow 1$ (or equivalently $\alpha\rightarrow 0$), in which $\dalpha$ contracts to $\mathfrak{psu}(2|2)$ plus a central extension, originating from the contraction of one of the $\mathfrak{su}(2)$ factors in the bosonic subalgebra of $\dalpha$ to an abelian group. Let us briefly discuss what happens to the Bethe equations and QQ-relations in this limit.

The singularity of the limit is reflected in the Dynkin diagrams of $\dalpha$. In particular, it is evident that at $\alpha\rightarrow 1$\footnote{For $\alpha\rightarrow 0$, we need to exchange the roles of the diagrams \ref{fig:d21sfig3} and \ref{fig:d21sfig4}.} the diagrams \ref{fig:d21sfig1} and \ref{fig:d21sfig3} become disconnected, and the determinant of the associated Cartan matrix is 0. On the other hand, the diagrams \ref{fig:d21sfig2} and \ref{fig:d21sfig4} become the same as two diagrams for $\mathfrak{psu}(2|2)$, once we rescale its simple roots. $\mathfrak{psu}(2|2)$ also has a third type of Dynkin diagram, where two nodes are fermionic: this does not arise from one of the gradings of $\dalpha$.

For the Bethe equations, the situation is similar, reflecting the connection between them and Dynkin diagrams: in fact, in each of the BAEs associated with \ref{fig:d21sfig1} and \ref{fig:d21sfig3}, one of the equations becomes trivial. On the other hand, the BAEs associated with \ref{fig:d21sfig2} and \ref{fig:d21sfig4} do reduce to the ones for $\mathfrak{psu}(2|2)$. It is not possible to reproduce from these BAEs alone the ones for the remaining Dynkin diagram of $\mathfrak{psu}(2|2)$, which also appear in the Asymptotic Bethe Ansatz for the AdS$_3\times$ S$_3\times$ T$^4$ string theory.

A similar thing happens to the QQ-relations in the $\alpha\rightarrow 1$ limit: some of them do reduce to the ones found in the $\mathfrak{psu}(2|2)$ Q-system, while others become trivial. We note in particular that some of the QQ-relations contained in the $\mathfrak{psu}(2|2)$ Q-system cannot apparently be reproduced starting from the Q-system we propose for $\dalpha$.
\subsection{Towards the QSC in the general case}
The QSC for the full string theory on \ads, which has a symmetry based on two copies of the $\dalpha$ super Lie algebra, should be based on two copies of the $\dalpha$ Q-system we have just presented. A natural conjecture is that this is the only difference with the $\alpha = \frac{1}{2}$: the way to glue the two copies would be the same. In this final section, we begin exploring the consequences of this idea, reserving a deeper analysis for future work.

\paragraph{The minimal conjecture. }

According to this minimal conjecture, the basic analytic properties of $\bP_A$ and $\bQ_{ij}$ functions would be the same as we have postulated for the $\alpha = 1/2$ case. Thus, we would assume that the four $\bP_A$ functions possess a short branch cut on the real axis at $u\in [-2h,2h]$ on their simplest Riemann sheet, as depicted in figure \ref{fig:Pcuts}, while the $\bQ_{ij}$ functions admit a sheet with a long cut as in \ref{fig:Qlongcuts} on their simplest sheet. 

The asymptotics of these functions would be identified in terms of weights with respect to the Cartan subalgebra of $\left(\so(4)\oplus \mathfrak{su}(2) \right)_L\oplus \left(\so(4)\oplus \mathfrak{su}(2) \right)_R$ algebra, where each factor is isomorphic to the even subalgebra of $\dalpha$, which is also the same as the even subalgebra for the $\alpha = \frac{1}{2}$ case.

Finally, we would postulate a gluing condition relating the two wings exactly as in (\ref{eq:gluing1}),(\ref{eq:gluing2}).  In principle, if this proves to be correct for the $\alpha = 1/2$ case\footnote{As we explained in the previous section, in the non-symmetric sector we have some puzzling findings, but we hope the construction is valid at least in the symmetric case.}, this description might also contain all the analytic information for the general QSC at any $\alpha$, when supplemented now with a new Q-system.

It also seems likely that we could repeat the study of the large $L$ limit and obtain Bethe Ansatz Equations matching the structure expected for the full string theory~\cite{Borsato:2012ss}.

In the rest of this section, we analyse what the consequences of these simple assumptions would be for the cut structure of the rest of the Q-functions. At $\alpha \neq \frac{1}{2}$, we will see that the structure starts to become quite nontrivial. 

\paragraph{Consequences: cut structure of various Q-functions. }
Let us use the Q-system that we have just introduced to understand the analytical properties of the Q-functions, based on the simple assumption that $\bP_A$ has a single cut. For this analysis, we focus on the L wing only, since the R wing will have an identical structure.

From the single-cut assumption for $\bP_A$ and the core relations \eqref{eq:alpha_core_rels}, we immediately see that for consistency $\bS$ must have a ladder of short branch cuts with either:
\begin{itemize}
\item $\text{Im}(u)=- \alpha \text{ and } (\alpha,\,-\alpha)-2\mathbb{N}$ for $\bS^{\downarrow}$, depicted in figure \ref{fig:SUHPalpha};
    \item $\text{Im}(u)=+ \alpha \text{ and } (-\alpha,\,\alpha)+2\mathbb{N}$ for $\bS^{\uparrow}$, depicted in figure \ref{fig:SLHPalpha}.
     
\end{itemize}
On the other hand, $\hat \bS$ will have a ladder of short branch cuts with either:
\begin{itemize}
\item $\text{Im}(u)=- \hat\alpha \text{ and } (\hat\alpha,\,-\hat\alpha)-2\mathbb{N}$ for $\hat\bS^{\downarrow}$, depicted in figure \ref{fig:ShatUHPalpha};
     \item $\text{Im}(u)=+ \hat\alpha\text{ and } (-\hat\alpha,\,\hat\alpha)+2\mathbb{N}$ for $\hat\bS^{\uparrow}$, depicted in figure \ref{fig:ShatLHPalpha}.
\end{itemize}
For generic $\alpha$, all the ladders of cuts now have alternating spacing.  The cut structure of  $\bS$ and $ \hat \bS$ is different, as their first cut appears at distance $\alpha$ vs $\hat{\alpha}$ from the real axis. This is the first key difference with the simpler $\osp(4|2)$ case, in which spinors and antispinors had the same branch cuts. 

The situation is even more peculiar for the $\bQ$ functions. From the analytical structure of the $\bS$ and $\hat\bS$ functions plus the QQ-relations \eqref{eq:d21a_QQrels_1}, it is easy to see that the $\bQ$ functions have cuts at either:
\begin{align}
& (0,-2\alpha,-2\hat\alpha) -2 \mathbb{N} \text{ for } \bQ^{\downarrow}\,,\\
&(0,+2\alpha,+2\hat\alpha) +2 \mathbb{N} \text{ for } \bQ^{\uparrow}\,.
\end{align}
These two possibilities are depicted in figures \ref{fig:QUHPalpha} and \ref{fig:QLHPalpha}. 
We can think of this analytical structure as deriving from one of the $\osp(4|2)$ $\bQ$-functions, where each cut \emph{except the one on the real axis},  splits into two cuts for the $\dalpha$ case. Namely, each of the cuts situated at $\text{Im}(u) \in \pm (\mathbb{N}+1)$ split into one series of cuts at $\text{Im}(u)=\pm(\mathbb{N}+2\alpha)$ and one series at $\text{Im}(u)=\pm( \mathbb{N}+2\hat\alpha)$ (the two series of cuts coincide at $\alpha = 1/2$, since at this value $\hat\alpha=\alpha$.). 
The cut on the real axis instead remains isolated: this is important as we can then hope that the gluing conditions, which pertain to continuation through this cut, remain valid in the same form (\ref{eq:gluing1}),(\ref{eq:gluing2}). 

\begin{figure}[htbp]
    \centering
    \begin{minipage}{0.45\textwidth}
        \centering
        \includegraphics[width=\linewidth]{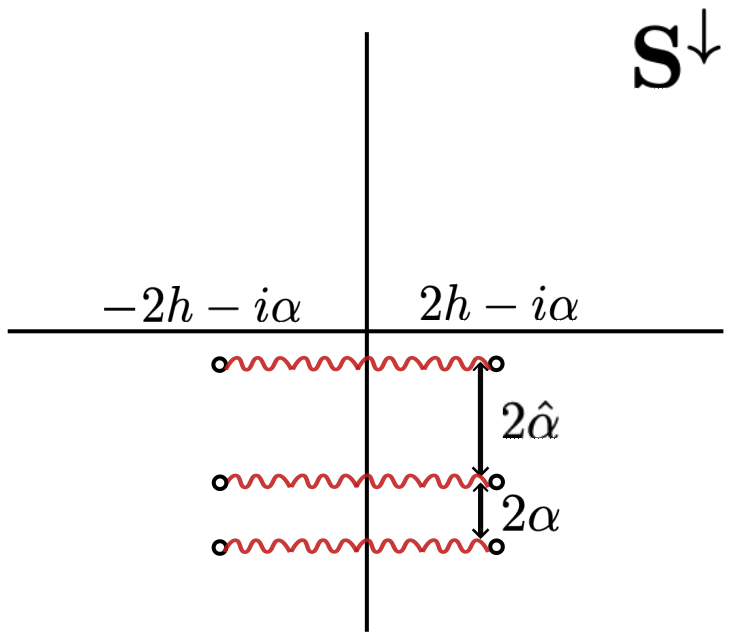}
        \caption{Cuts of $\bS^{\downarrow}$. The ladder of cuts continues in the lower half plane, with alternating spacing.}
        \label{fig:SUHPalpha}
    \end{minipage}\hfill
    \begin{minipage}{0.45\textwidth}
        \centering
        \includegraphics[width=\linewidth]{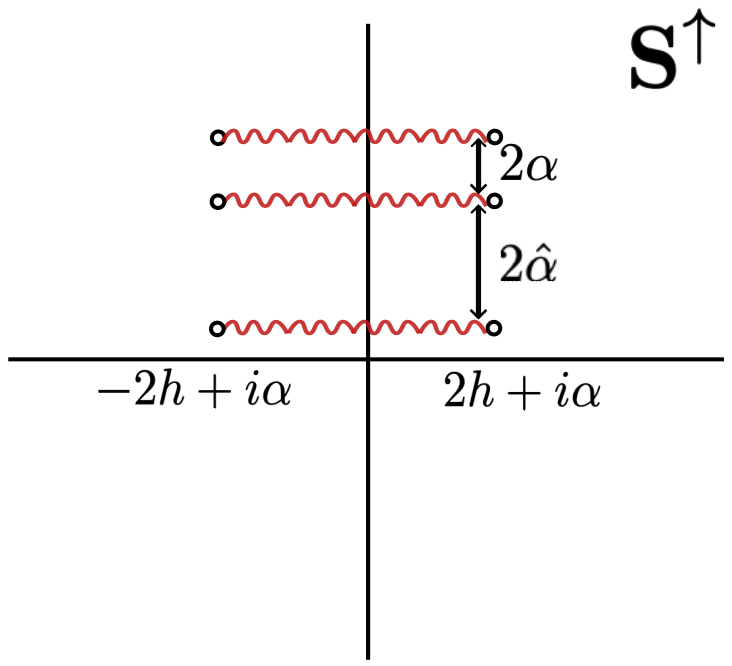}
        \caption{Cuts of $\bS^{\uparrow}$. The ladder of cuts continues in the upper half plane, with alternating spacing.}
        \label{fig:SLHPalpha}
    \end{minipage}
\end{figure}
\begin{figure}[htbp]
    \centering
    \begin{minipage}{0.45\textwidth}
        \centering
        \includegraphics[width=\linewidth]{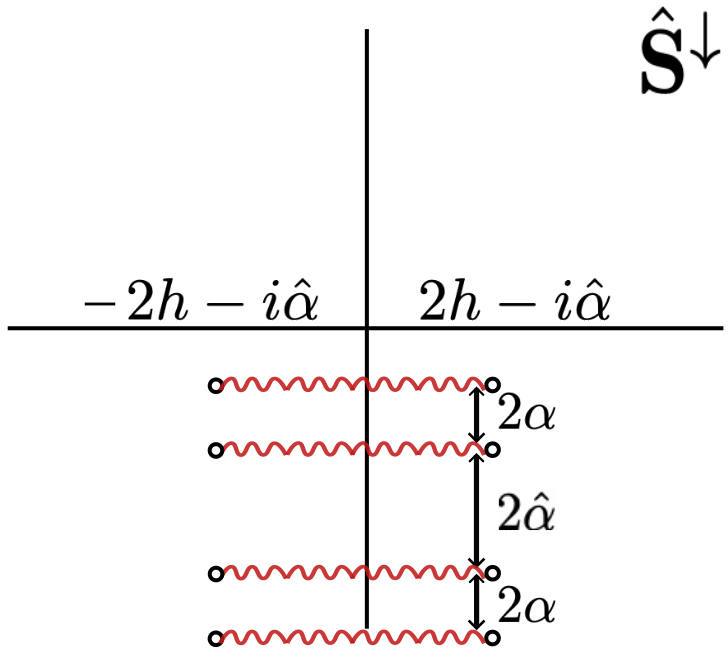}
        \caption{Cuts of  $\hat\bS^{\downarrow}$. The ladder of cuts continues in the lower half plane, with alternating spacing.}
        \label{fig:ShatUHPalpha}
    \end{minipage}\hfill
    \begin{minipage}{0.45\textwidth}
        \centering
        \includegraphics[width=\linewidth]{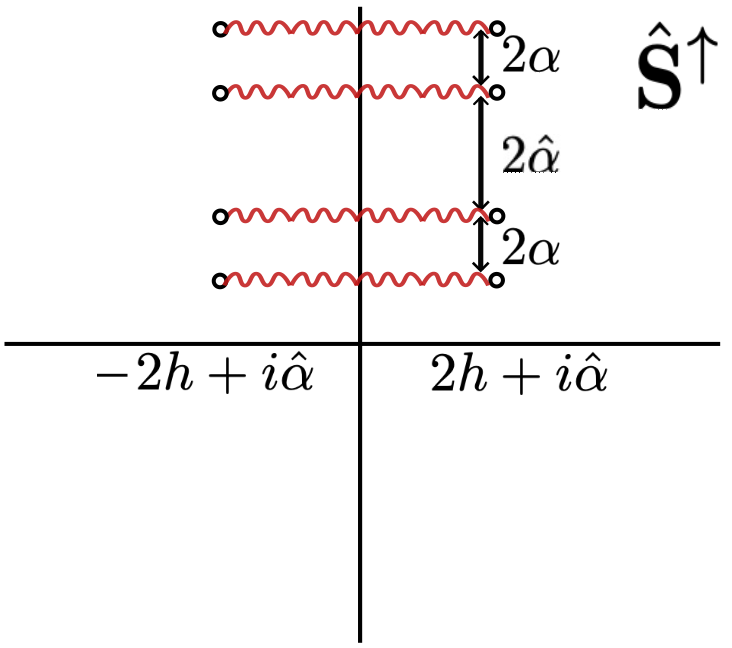}
        \caption{Cuts of $\hat\bS^{\uparrow}$. The ladder of cuts continues in the upper half plane, with alternating spacing.}
        \label{fig:ShatLHPalpha}
    \end{minipage}
\end{figure}
\begin{figure}[htbp]
    \centering
    \begin{minipage}{0.45\textwidth}
        \centering
        \includegraphics[width=\linewidth]{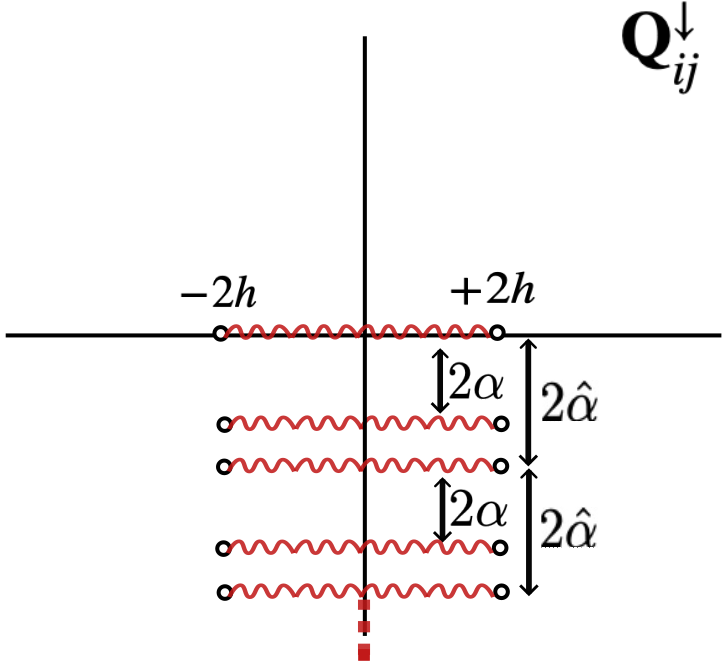}
        \caption{Cuts of $\bQ^{\downarrow}$. The ladder of cuts continues in the lower half plane, with alternating spacing. In the figure is assumed $\alpha <\hat{\alpha}$. }
        \label{fig:QUHPalpha}
    \end{minipage}\hfill
    \begin{minipage}{0.45\textwidth}
        \centering
        \includegraphics[width=\linewidth]{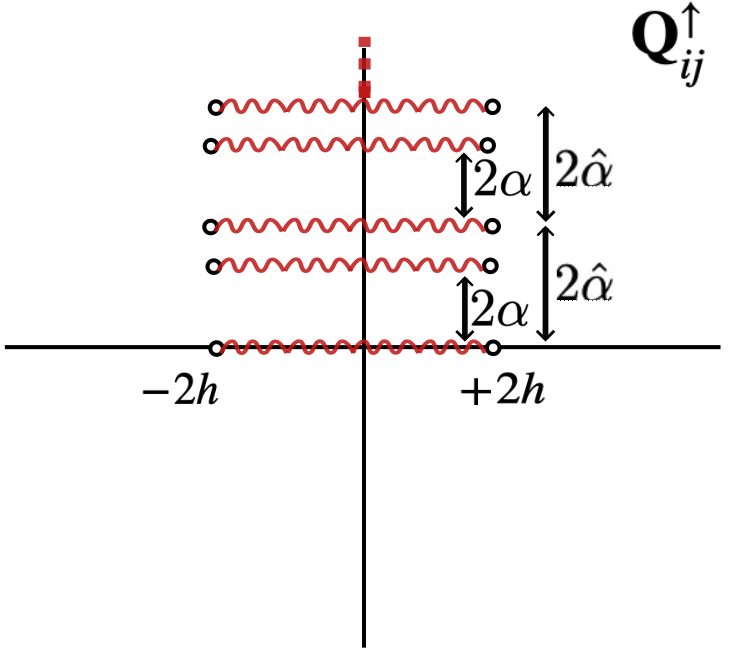}
        \caption{Cuts of $\bQ^{\uparrow}$. The ladder of cuts continues in the upper half plane, with alternating spacing. In the figure is assumed $\alpha <\hat{\alpha}$.}
        \label{fig:QLHPalpha}
    \end{minipage}
\end{figure}

A peculiarity of the analytical structure that we described so far is that the spacings between short branch cuts of the $\bP,\,\bS,\,\hat{\bS}$ functions alternate between two values, determined by the parameters $\alpha$ and $\hat\alpha$. This is radically different from any other known QSC model. 
We can contrast this situation, in particular, with the case of ABJ theory, which extends ABJM theory by having an extra coupling constant. Since the couplings enter the QSC through the positions of the cuts, it would have been a natural conjecture to posit that ABJ  theory extends the QSC of ABJM by introducing such alternating ladders of cuts. However, it was found in \cite{Cavaglia:2016ide} that in that case the QSC seems to be too rigid to allow this consistently!\footnote{Instead, in ABJ it is conjectured that the two couplings must recombine into a single interpolating function, parametrising the position of the cuts \cite{Cavaglia:2016ide}. }
The hope is that in the $\dalpha \oplus \dalpha$ case, thanks to the extra flexibility gained by having two copies of the system, such a structure is instead nontrivial and consistent. 

It would be very interesting to study further the analytic consequences of the gluing conditions in the theory at generic $\alpha$: in particular, one should extend the analysis of section \ref{sec:QtauPnu} to see whether conditions similar to the $\bQ\tau$ and $\bP\nu$ systems arise in the general case, which could give further indication that we are on the right track. We leave these investigations for future work.
\section{Discussion and Conclusions}
In this paper, we have discussed what shape could be taken by a novel Quantum Spectral Curve describing string theory on \ads\, background with pure Ramond-Ramond fluxes. We mainly analysed the case where the radii of the two spheres are the same, corresponding to the global symmetry $\osp(4|2)^{\otimes 2}$. In this case, we presented a proposal for this new QSC. This was based on minimal analyticity assumptions coupled to the algebraic structure of two copies of the known Q-system for $\osp(4|2)$~\cite{Tsuboi:2023sfs}. Our results also include a reformulation of this algebraic structure, re-derived in a covariant way starting from a small set of core relations, while also proposing novel relations between the Q-functions. The analytic properties for the Q-functions in our QSC  essentially consist of gluing the two copies of the system through their branch cuts. This is analogous to a similar construction recently introduced in \cite{Ekhammar:2021pys,Cavaglia:2021eqr} for the AdS$_3\times $S$^3\times $T$^4$ model. 

We have also proposed a new Q-system for $\dalpha$, which would be needed to formulate a QSC for \ads\, strings in the case where the radii of the spheres are different, which enjoys $\dalpha^{\oplus 2}$ global symmetry. We performed a preliminary analysis of the analytical structure of the Q-functions in this case, discovering novel properties while reserving a deeper analysis of the QSC for this general case for future work.
 
Studying the large volume limit in the case $\alpha = \frac{1}{2}$, we find encouraging indications that the QSC might be correct at least in a symmetric sector, where, probing the sector of massive worldsheet excitations, we find full compatibility with all that is known of the worldsheet theory.

However, a puzzle emerged for general non-symmetric solutions, where a tension arises between our QSC equations and either braiding unitarity or crossing symmetry for the worldsheet theory.  Our current understanding regarding this puzzle is discussed in detail in the next paragraph. 

\paragraph{Discussion of the new dressing factors and the crossing/unitarity mismatch. } 
Our analysis of the large volume limit yields a set of Asymptotic Bethe Ansatz equations, which provide strong constraints for the previously unfixed dressing factors entering the worldsheet S-matrix, as discussed in sections \ref{sec:ABA} and \ref{sec:dressingnew}. 

The most nontrivial part of the dressing factors, which we called $\Sigma_{\text{new}}$, is only visible for non-symmetric states where spinor and antispinor Q-functions are different. 
For this factor, our results suggest an incompatibility between the following three conditions: the constraints coming from the QSC (encoded in a linear integral equation with a partially unfixed source term), braiding unitarity, and the crossing equations found in \cite{Borsato:2015mma}. The problem is not present in the sector where spinors and anti-spinors are symmetric, where $\Sigma_{\text{new}}$ cancels from the ABA equations. 

Our results suggest that the QSC we propose may need to be modified in the non-symmetric sector. However, we do not see any natural way to do this at present. To understand what possible modifications or other reasons for the puzzle we are encountering, it is worth summarising the assumptions we made in this paper and the possible loose ends. 

First, the assumptions we made in conjecturing the QSC are the following:
\begin{itemize}
\item two Q-systems (L and R) corresponding to two copies of the superalgebra.
\item the structure of a single ``short'' or ``long'' cut for the Q-functions of type $\bP$ or $\bQ$ on their simplest Riemann sheet 
\item the  minimal gluing prescription (\ref{eq:gluing1}),(\ref{eq:gluing2}) to the relate the two copies,
\item the fact that the ``quantum determinant'' Q-functions are constant in $u$, so we can take $\bQ_{\emptyset}=1$ in both wings\footnote{If $\mathbf{Q}_{\emptyset}$ is a constant, independent of $u$, it can always be set to $1$ by a trivial rescaling of the Q-functions.}
\item the fact that the gluing matrix takes an exactly off-diagonal form (\ref{eq:gluingform}) (at least in the ABA limit).
\end{itemize}
This seems to us the most natural proposal, which follows the paradigm of all other known QSC's. In particular, it is hard to imagine any natural alternative to the first four assumptions, including the condition  $\mathbf{Q}_{\emptyset} = \text{constant}$. The last assumption above is perhaps less fundamental. In this paper, it was employed in section~\ref{sec:simplifynonsym} in the analysis of the large-$L$ limit. It appears to us that this hypothesis provides the only viable way to impose a sufficient number of constraints on the dressing factors in the nonsymmetric case.

There are also some loose ends and hypotheses made during the study of the large volume limit (which, however, do not affect the proposal for the QSC):
\begin{itemize}
\item  First, we are currently not able to fix the value of the proportionality factors in the first line of (\ref{eq:ABA_sol_partial1}). These constants already affect the part of the dressing factors that is visible in the symmetric sector. 
In fact, the way these constants appear in the ABA limit is completely analogous to what happens in the $T^4$ case~\cite{Ekhammar:2024kzp,BudapestTalk}. Also in that model, the value of these proportionality constants affects the dressing phases. While a derivation of their value is not known, there is only one possible value (which depends on the Bethe roots) that yields braiding unitarity for the dressing factors. This is assumed to be the correct value, which then fixes the dressing phases completely. 
 In this work, we fixed these constants in exactly the same way, corresponding to $\tilde{\alpha}_C = 0$ in (\ref{def:alpha_C}). This imposes constraints on the symmetrised combination of dressing factors, which completely fixes them in terms of the dressing factors of the $T^4$ theory.
 
\item Second, in this model we have introduced another consequential constant, i.e. $\mathcal{E}$ in (\ref{eq:StokesE}).  
The mechanism by which this constant appears is, admittedly, more speculative: we had to conjecture an order of limits problem between the ABA limit and the continuation around the branch points when evaluating the quantity $\nu^{\gamma}$. 
The constant $\mathcal{E}$ affects $\Sigma_{\text{new}}$, i.e. the part of the dressing factor that is only visible in the non-symmetric sector. In particular, to impose its braiding unitarity, we had to fix the form of $\mathcal{E}$ in a very specific way. This is related to the function $\mathbf{e}(v)$, which we computed perturbatively to several orders in section \ref{sec:dressingnew}. 
\end{itemize}
While the first type of ambiguity in the ABA derivation also happens in the $T^4$ model, the appearance of the second constant $\mathcal{E}$ is arguably less natural: we did not find any other way to make such a constant appear than by imagining such an order-of-limits phenomenon. 

Notice that the constant appears when we impose the condition $\nu^{\gamma} = \nu^{++}$. Therefore, one might imagine that it could arise if the $\nu$ functions were quasi-periodic, rather than periodic.\footnote{We thank Nikolay Gromov for this suggestion.} In fact, a similar quasi-periodicity \emph{does occur} in the non-symmetric sector in ABJM theory~\cite{Bombardelli:2017vhk}. However, starting from the minimal assumptions at the basis of our QSC, we did not find any consistent way to impose a nontrivial quasi-periodicity:  we \emph{derived}  the standard mirror periodicity $\nu^{\gamma} = \nu^{++}$ from our minimal axioms, with no room to add a multiplicative constant, unlike in ABJM.  
Could this be a sign that our equations are not general enough in the non-symmetric sector?
We do not have a definite answer. However, our current understanding is that, given the algebraic structure of this model, it is challenging to introduce nontrivial quasi-periodicity: the simple property $\nu^{\gamma} = \nu^{++}$ follows from very basic assumptions.

Setting aside the question of how the constant $\mathcal{E}$ appears in the large volume solution, and accepting the results of the analysis of the ABA limit, we have the puzzling situation that to impose braiding unitarity for the worldsheet S-matrix, we have to introduce a specific tuning for the function $\mathbf{e}(v)$.  However, this function would then \textbf{modify the crossing relations} (\ref{eq:newcrossing}) in a nontrivial way!

We note that, recently, the necessity to write ``unconventional'' modified crossing equations has arisen in different contexts in 2D QFTs, e.g. in \cite{Copetti:2024rqj} and in the $AdS_3\times S^3\times T^4$ theory itself~\cite{Frolov:2025ozz}, due to the presence of nontrivial exchange relations between particle lines.\footnote{In the latter case, the correct dressing phase corresponding to the modified crossing was first predicted by deriving the ABA limit from the QSC~\cite{Ekhammar:2024kzp}.} However, there are key differences with the present case:
\begin{itemize}
    \item For the $\text{AdS}_3\times \text{S}^3\times \text{T}^4$ case, this modified crossing arises only when considering the massless sector, while in our case it is already found in the massive one.
    \item In the $\text{AdS}_3\times \text{S}^3\times \text{T}^4$ and in the cases studied in \cite{Copetti:2024rqj}, the modification of the crossing equations is much simpler than what we are finding. In \cite{Frolov:2025ozz}, it basically amounts to an extra minus sign, while the perturbative results of our analysis suggest that $\mathbf{e}(v)$ is a much more complicated function of one variable.  
\end{itemize}
Given these differences, we are wary that this analogy might be a mirage. 
 In the case that it is not, one should be able to \emph{derive} the form of the modified crossing equation, including the exact form of $\mathbf{e}(v)$, but we do not see how this could be done at the moment.  

We also attempted to find a simpler resolution of the puzzle by changing some of the assumptions made in deriving the large $L$ limit. In particular, one can imagine that the constant $\tilde{\alpha}_C$, discussed above, takes a nonzero value, which would introduce an additional contribution (\ref{eq:residualphase}) in the large volume $\bP$ and $\bQ$ functions and thus modify the ABA equations. One can then enforce braiding unitarity by tuning at the same time this constant as well as $\mathbf{e}(v)$ (which is the only one that would affect the crossing equations). This can be done to simplify $\mathbf{e}(v)$ as much as possible, while maintaining braiding unitarity. This can be done to some extent, but we found that, even introducing these new degrees of freedom, the transcendental part of $\mathbf{e}(v)$ cannot be fully eliminated.\footnote{Furthermore, the simplest hypothesis, i.e. setting $\mathbf{e}(v) = 0$ and trying to use only the freedom in $\tilde{\alpha}_C$ to fix braiding unitarity, fails after order $O(h^6)$.The puzzle remains, and we do not see this as a step in the right direction. }

In closing this discussion, we stress again that the explanation could be very prosaic: maybe something is missing in the QSC in the non-symmetric sector, and a modified version will not show any of the puzzles discussed above.  Currently, we do not see a clear way to modify the equations. In the next paragraph, we discuss further studies that could help test the consistency of the proposed QSC.

\paragraph{Future directions. }
To clarify some of the mysteries discussed above, it would be interesting to explore the QSC and its large volume limit in situations where more structure will appear. This includes the situation where we include massless modes (along the lines of \cite{Ekhammar:2024kzp}), as well as the situation for general $\alpha$, which we started to discuss in section \ref{sec:alphaQsys}. In the latter case, the symmetric sector simply disappears since spinors and antispinors will be distinguished by different shifts related to $\alpha$ and $1-\alpha$, respectively. We expect, therefore, that if there is a problem in the QSC equations, this should become manifest more clearly. If the structure holds, the study of this more general case should be illuminating in resolving the paradoxes. 

If we gain confidence in the correctness of the QSC also in the non-symmetric case, it would, of course, be very interesting to understand the mismatch with the crossing equations of \cite{Borsato:2015mma}.  It would be also very nice to deduce a finite-coupling representation for the source term $\mathbf{e}(v)$ giving braiding unitarity, which here we were able to only study perturbatively, and correspondingly for the new piece of the dressing phase $\Sigma_{\text{new}}^{\text{unit}}$. 

Ultimately, the real test of the new QSC equations (including the symmetric case) would be showing that they have a spectrum of isolated solutions corresponding to the expected spectrum of string states at strong coupling. As for the theory with AdS$^3\times$ S$^3 \times$ T$^4$ target space, this QSC is likely to be significantly more difficult to solve than the ones for $\mathcal{N}$=4 SYM and ABJM theory, especially because the branch points are now non-quadratic. Some initial perturbative and numerical studies were performed in \cite{Cavaglia:2022xld} for the T$^4$ model. While currently the strong coupling regime has not been studied, the fact that the equations converge to a discrete set of solutions, with a rich structure at weak coupling, is evidence for the correctness of the QSC.\footnote{From experience with other models such as e.g. ABJM theory, it is very easy to ``break'' the nonlinear QSC equations in such a way that they would not have any solutions. For this it is enough, for instance, to assume the wrong structure for the gluing matrix. } 
It would be very important to attempt to solve the new QSC equations proposed in this paper both numerically and perturbatively. 
 
Concluding the discussion of AdS/CFT applications, it is fair to say that several important open problems remain concerning the integrability description of the spectrum of AdS$_3$/CFT$_2$ models. Among those we have not mentioned yet, the generalisation to mixed-flux cases is certainly very interesting: significant recent advances in the S-matrix bootstrap and TBA for the T$^4$
 model~\cite{Lloyd:2014bsa, Frolov:2023lwd, Baglioni:2023zsf,Frolov:2024pkz, Frolov:2025uwz, Frolov:2025tda} make it increasingly important to determine how the QSC description is altered in the presence of mixed flux on the worldsheet.

\paragraph{Open questions on Q-systems. }

From a more mathematical perspective, several directions merit further investigation.

First, our covariant formulation of the $\ospf$ Q-system yields new constraints among the Q-functions, as seen in \eqref{eq:constraintP}, \eqref{eq:constraintQij}, and \eqref{eq:constraintQai}. The first of these relations already appears in \cite{Tsuboi:2023sfs} as the expression for a T-function. It would therefore be natural to explore whether the remaining constraints can also be interpreted as T-functions in the $\ospf$ T-system. 

This reformulation could be generalisable to other orthosymplectic superalgebras. In particular, it would be interesting to clarify the connection between currently known QQ-relations for ABJM theory~\cite{Bombardelli:2017vhk} and those formulated for the same algebra in \cite{Tsuboi:2023sfs}.

Another promising line of work is to introduce a twist in our Q-system. This would not only complete the correspondence with Tsuboi’s formulation and allow for the construction of  Q-operators from Lax matrices for the underlying spin chain  \cite{Frassek:2023tka}, but should also serve as a stepping stone toward the twisted QSC \cite{Kazakov:2015efa}. 

Further, in this paper, we proposed a novel Q-system for the exceptional super Lie algebra $\dalpha$, starting from a simple modification of the core relations of the orthosymplectic case. We find that this Q-system contains the correct Bethe Ansatz equations, and reduces nicely to the one for $\osp(4|2)$ if we take $\alpha=1/2$, as expected. In the singular limit $\alpha\rightarrow1$ (or equivalently $\alpha\rightarrow0$), we do not, however, find a simple reduction to the $\mathfrak{psu}(2|2)$ Q-system of \cite{Ekhammar:2021pys, Cavaglia:2021eqr}. It is possible that this limit has to be taken with some special prescription from the point of view of the Q-system and the Bethe equations.

The Q-system at general $\alpha$ provides a framework for studying the representation theory of the super Yangian $Y(\dalpha)$ \cite{Lin:21255}. A compelling question is whether the QQ-relations we derived admit solutions that are polynomial in the spectral parameter $u$, as such solutions would be implied by the existence of compact representations of $Y(\dalpha))$. Progress along these lines could also offer insight into the construction of the associated R-matrix, which would represent the first known example for an exceptional Lie superalgebra.

\section*{Acknowledgements}
We thank R.~Borsato,  S.~Frolov, D.~Polvara, M.~Preti, L.~Quintavalle, A.~Sfondrini, B.~Smith,  B.~Stefanski jr, A.~Torrielli, K.~Zarembo and especially F.~Chernikov, S.~Ekhammar and N.~Gromov for discussions on this problem. NP and RT also wish to thank the organisers of the workshop “Higher-$d$ Integrability” held in Favignana, Italy (3–13 June 2025) for their invitation, financial support, and stimulating atmosphere, and in particular S.~Frolov, D.~Polvara and A.~Sfondrini for helpful discussions during the workshop. We also thank them and R. Borsato for useful correspondence.

AC, NP and RT participate in the project  HORIZON-MSCA-2023-SE-01-101182937-HeI.   
RF acknowledges support from Istituto Nazionale di Alta Matematica (GNFM).
The work was also supported by the INFN specific initiatives GAST and SFT, the FAR
UNIMORE project CUP-E93C23002040005, and by the PRIN project “2022ABPBEY” CUP-E53D23002220006.

\newpage
\appendix
\section{Equivalence with an alternative form of the QQ-relations}\label{app:equivalence}
In a recent 
paper~\cite{Tsuboi:2023sfs}, Tsuboi wrote down the Q-system for a large class of Lie superalgebras, employing the reductions by automorphisms proposed in~\cite{Tsuboi:2011iz} of the well-known Q-system of $\gl(M|N)$.
In this section, we propose a map between his results for a spin chain based on the Lie superalgebra $\osp(4|2)$ and our covariant Q-system of section \ref{sec:QQsys}; this amounts to proving the core relations \eqref{eq:core_relations} starting from the QQ-relations found in~\cite{Tsuboi:2011iz}. 
\subsection{Deriving the Core Relations from Tsuboi's QQ-relations}
In this section, we will show that the QQ-relations of Tsuboi plus the constraints on the normalisations of the Q-functions \eqref{eq:constraintP} and \eqref{eq:constraintQai}\footnote{These should follow from the Wronskian solution to the T-system, although only part of it is written in \cite{Tsuboi:2023sfs}.} are equivalent to the two core relations \eqref{eq:core_relations}. 
In particular, we will only need to assume the following three constraints:
\beq
\label{def:constr-tsuboi}
\bP_A\bP^A=2 \bQ_{\emptyset}^+\bQ_{\emptyset}^-,\quad \text{det}(\bS_{a|i})=-\bQ_{\emptyset},\quad \text{det}(\bS_{\hat a|i})=\bQ_{\emptyset}\,,
\eeq
plus the following two QQ-relations, that are the analogue of (4.120) and (4.121) of \cite{Tsuboi:2023sfs}:  
\beq
\label{def:QQ-tsuboi}
\bP_A \bQ_{ii}=\bQ_{\emptyset}^-\bQ_{A|ii}^+-\bQ_{\emptyset}^+\bQ_{A|ii}^-\,, \quad  \bS_{a|i}^+(u) \bS_{b|j}^-(u) \; {\mathbf{C}}^{ab }  = \bQ_{ji}(u)\, ,
\eeq
where the second relation may be seen as a definition of the four elements of $\mathbf{Q}_{ij}$. 
We start from the first QQ-relation in \eqref{def:QQ-tsuboi}; recalling that $\bQ_{A|ii}\equiv \bS_{a|i}\bS_{\hat a|i}(\sigma_A)_b^{\;\hat a}\mathbf{C}^{ab}$, we can use the second QQ-relation in \eqref{def:QQ-tsuboi} to substitute all the $\bS_{a|i}^+$ and $\bS_{\hat a|i}^+$ with combinations of $\bS_{a|i}^-, \,\bS_{\hat a|i}^+$, $\bQ_{ij}$ and $\bQ_{\emptyset}$. Thus, we get a linear system of 8 equations for the 4 $\bP_A$; it is possible to see that it has a unique solution, provided that the determinant condition $\bQ_{11}\bQ_{22}-\bQ_{12}\bQ_{21}=-\bQ_{\emptyset}^+\bQ_{\emptyset}^-$ holds. The solution is given by the following expression:
\beq
\label{eq:tsuboi_appendix_sol_unique}
\bQ_{\emptyset}^-\bP_{\hat a}^{\; a}=-\mathbf{C}^{ab}\bS_{b|k}^-\epsilon^{ki}\bQ_{ij}\bS_{\hat a|l}^-\epsilon^{jl}\,.
\eeq
Inverting the spinors and antispinors in the latter equation, which is made particularly simple thanks to the unit determinant conditions that are the second and third equations in \eqref{def:constr-tsuboi}, we arrive at the useful relation:
\beq
\label{eq:defQtsuboi}
\bQ_{\emptyset}^-\bQ_{ij} =\bS_{a|i}^- \, \bP_{\hat b}^{\, a} \,\, \hat{\mathbf{C}}^{\hat b \hat c}\,\, \bS_{\hat c|j}^-\,.
\eeq
Now take again the first QQ-relation in~\eqref{def:QQ-tsuboi} and multiply by $\bP^A$ and use the normalisation \eqref{eq:constraintP} to get:
\beq
2\bQ_{\emptyset}^+\bQ_{\emptyset}^-\bQ_{ii}=\left(\bS_{a|i}^+\bS_{\hat a|i}^+-\bS_{a|i}^-\bS_{\hat a|i}^-\right)\bP_{\hat b}^{\; a}\hat{\mathbf{C}}^{\hat a\hat b}\,.
\eeq
Now using the fact that $\bP_{b}^{\;\hat a}\mathbf{C}^{ab}=\hat{\mathbf{C}}^{\hat a\hat b}\bP_{\hat b}^{\;a}$ and the relation \eqref{eq:defQtsuboi}, we obtain immediately:
\beq
\bQ_{\emptyset}^+\bS_{a|i}^-\bP_{\hat b}^{\;a}\hat{\mathbf{C}}^{\hat b\hat c}\bS_{\hat c|i}^-=\bQ_{\emptyset}^-\bS_{\hat a|i}^+\bP_{b}^{\;\hat a}\mathbf{C}^{ab}\bS_{a|i}^+\,.
\eeq
It is possible to prove that this expression also holds for $\bQ_{ij}, \,i\neq j$ by using \eqref{eq:tsuboi_appendix_sol_unique}. 
This implies that the fermionic Q-functions admit the alternative expressions:
\begin{equation}
\bQ_{\emptyset}^+\bQ_{ij} =\bS_{\hat a|i}^+\bP_{b}^{\;\hat a}\mathbf{C}^{ab}\bS_{a|j}^+=\bS_{\hat a|i}^+\bP_{\hat b}^{\;a}\hat{\mathbf{C}}^{\hat a\hat b}\bS_{a|j}^+\,.
\end{equation}
Inverting the spinorial Q-functions, substituting $\bQ_{ij}$ using the second QQ-relations in \eqref{def:QQ-tsuboi}, and using the normalisation of the spinors, we finally get to:
\beq
\begin{split}
   & \bP_{\hat b}^{\; c} = \bS^+_{\hat b|k } \epsilon^{ki} \bS_{a|i}^-\mathbf{C}^{ac} \,,\\
   & \bP_{ b}^{\; \hat c} =- \bS^+_{ b|k } \epsilon^{ki} \bS^-_{\hat b|i} \hat{\mathbf{C}}^{\hat b \hat c}\,,
\end{split}
\eeq
which are exactly equations \eqref{eq:F1}. From these equations, it is immediate to retrieve the core relations \eqref{eq:constraintP} by multiplying on the right by respectively $\bS_{c|j}^-$ and $\bS_{\hat c|j}^-$ and using the spinor normalisation conditions.

\subsection{Map with Tsuboi's Q-system}
In this section, we map the Q-functions in our Q-system to the ones of \cite{Tsuboi:2023sfs}.

First, let us establish a connection between the notation used in this paper and that used by Tsuboi. The $\osp(4|2)$ Q-system of the latter contains 4 sets of Q-functions $Q_{A|I}$, where $A$ and $I$ are sets of respectively bosonic and fermionic indices, with cardinality $|A|=0,1,2$ and $|I|=0,1$. The Q-functions with 1 bosonic and 1 fermionic index can be factorised as $Q_{a|i}=S_{ab|i}S_{ab'|i}$; furthermore, it holds that $Q_{ab|i}=S_{ab|i}^+S^-_{ab|i}$. Hence, we will take   the following Q-functions as fundamental objects:
\beq
\mathbb{Q}_{\emptyset},\, Q_{a|\emptyset},\,Q_{\emptyset|i},\, S_{ab|i}.
\eeq
The bosonic indices that we denote with $a,b,\dots $ take values in the set $\{1,2,3,4\}$, while the fermionic indices denoted as $i,j,\dots $ take values in $\{1,2\}$. Furthermore, there exist two `special' fermionic indices which we will denote as $0, 0'$. Writing $a'\equiv 4-a+1$, a bosonic set $A$ containing the index $a$ is not allowed if it also contains $a'$ or another $a$. We also define similarly $i'\equiv 2-i+1$.

The QQ-relations of Tsuboi for $\osp(4|2)$, found in section 4.4.2 of \cite{Tsuboi:2023sfs}, are:
\begin{align}
    &
    \label{qqB1}
    \bQ_{\emptyset|\emptyset}Q_{ab|\emptyset}=Q_{a|\emptyset}^+Q_{b|\emptyset}^--Q_{a|\emptyset}^-Q_{b|\emptyset}^+\,,\\
    &
        \label{qqF1}
Q_{a|\emptyset}Q_{\emptyset|i}=\bQ_{\emptyset}^+S_{ab|i}^-S_{ab'|i}^--\bQ_{\emptyset}^-S_{ab|i}^+S_{ab'|i}^+\,,\\
    &
        \label{qqB2}
    Q_{\emptyset|i}=S_{ab'|i}^+S_{a'b|i}^--S_{ab'|i}^-S_{a'b|i}^+\,,\\
    &
        \label{qqF2}
    Q_{ab|\emptyset}S_{ab'|i}=Q_{a|\emptyset}^+S_{ab|i}^{--}-Q_{a|\emptyset}^-S_{ab|i}^{++}\,,\\
    &
        \label{qqB3}
    Q_{ab|\emptyset}=S_{ab|i}^{--}S_{ab|i'}^{++}-S_{ab|i}^{++}S_{ab|i'}^{--}\,.
\end{align}
It is easy to see that these have the same form as the QQ-relations of section \ref{sec:QQsys} written in components, if we use the metric \eqref{def:sometric}.\footnote{Actually, as explained in section \ref{sec:QQsys}, we have \textit{more} QQ-relations than the ones in \eqref{qqB1}. The extra ones we have also involve Q-functions without a direct analogue in the construction by Tsuboi.}
In particular, within our conventions, including our choice of sigma matrices, the two Q-systems are the same under the following map between Q-functions:
\begin{align}
\label{def:mapTsuboi}
\begin{array}{@{}ccc@{}}
    \{\mathbf{P}_1, \,\mathbf{P}_2,\, \mathbf{P}_3,\,\mathbf{P}_4\} &\leftrightarrow &\{Q_{1|\emptyset},\,Q_{2|\emptyset},Q_{2'|\emptyset},Q_{1'|\emptyset}\}\\
    \{\mathbf{Q}_{11}, \,\mathbf{Q}_{22}\} &\leftrightarrow &\{Q_{\emptyset|1},\,Q_{\emptyset|1'}\}\\
    \{\mathbf{P}_{12}, \,\mathbf{P}_{13},\,\mathbf{P}_{24}\,,  \mathbf{P}_{34}\} &\leftrightarrow &\{Q_{12|\emptyset},\,Q_{12'|\emptyset},\,Q_{1'2|\emptyset}, Q_{1'2'|\emptyset}\}\\
    \{\mathbf{S}_{1|1}, \,\mathbf{S}_{2|1},\, \mathbf{S}_{1|2},\,\mathbf{S}_{2|2}\} &\leftrightarrow &\{S_{12|1},\,S_{1'2'|1},S_{12|2},S_{1'2'|2}\}\\
    \{\mathbf{S}_{\hat 1|1}, \,\mathbf{S}_{\hat 2|1},\, \mathbf{S}_{\hat 1|2},\,\mathbf{S}_{\hat 2|2}\} &\leftrightarrow &\{S_{12'|1},\,S_{1'2|1},S_{12'|2},S_{1'2|2}\}
    \end{array}
\end{align}
Notice that the way the spinors are indexed in Tsuboi's notation precisely embeds the selection rules we gave in tables \ref{tab:goodindices} and \ref{tab:goodindicesstar}. 

We postulate that a combination of our two remaining fermionic Q-functions $\bQ_{12},\bQ_{21}$  in our construction corresponds to the two `special' Q-functions with one index of Tsuboi, i.e. $Q_{\emptyset|0},\,Q_{\emptyset|\bar 0}$.

Finally, we note that the Q-system of Tsuboi contains a twist while the one we use in the context of the QSC is untwisted. Therefore, the two are only equivalent if we take an ``untwisting limit'' of the former. This should be defined with some care, as simply sending the twist parameters to 1 usually results in singular behaviour \cite{Kazakov:2015efa}. Alternatively, we should insert the twist in our QQ-relations by appropriately dressing our Q-functions with twist-dependent prefactors. We leave a precise definition of these two procedures for future work. 

\section{Derivation of the QQ-relations from the core relations}
\label{appendix:QQproofs}
In this section, we write down the derivations of the QQ-relations appearing in section~\ref{sec:proofsQQ} starting from the core relations \eqref{eq:core_relations} and the constraints \eqref{constraint_detP}, \eqref{eq:constraintQai}.

\paragraph{Proof of~\eqref{eq:relQ1} and~\eqref{eq:relQ2}.}
Proving~\eqref{eq:relQ1} is simple: start from~\eqref{eq:defQ} and use the core relations~\eqref{eq:core_relations} to rewrite the combination $\bS_{a|k}^-\bP_{\hat b}^{\;a}$ in terms of the antispinor $\bS_{\hat{b}|k}^+$. For~\eqref{eq:relQ2}, we first employ the trivial identity $
\mathbf{C}^{b c} \bP_{c}^{\; \hat b} = \hat{\mathbf{C}}^{\hat b \hat a} \bP_{\hat{a}}^{\, b}$ on the core relations, and then use them on~\eqref{eq:defQ} to rewrite the combination $\bS_{\hat{a}|k}^-\bP_{ b}^{\;\hat a}$ in terms of the spinor $\bS_{{b}|k}^+$.
\paragraph{Proof of~\eqref{eq:F1}.} 
We start from the core relations in matrix form and multiply them on the left by respectively $(\bS^-)^{-1}$ and $(\hat{\bS}^-)^{-1}$ to obtain:
\beq
\hat{\bP}=\bQ_{\emptyset}^-\,\hat{\bS}^+\cdot (\bS^-)^{-1}\,,\quad {\bP}=\bQ_{\emptyset}^-\,{\bS}^+\cdot (\hat \bS^-)^{-1}\,.
\eeq
Then using the fact that $(\bS^{-})^{-1}=\dfrac{-\epsilon \cdot\bS^t\cdot \mathbf{C}}{\bQ_{\emptyset^-}} $ and $(\hat \bS^{-})^{-1}=\dfrac{\epsilon \cdot\hat \bS^t\cdot \hat{\mathbf{C}}}{\bQ_{\emptyset}^-} $ we arrive immediately at~\eqref{eq:F1}. 
\paragraph{Proof of \eqref{eq:F2QQrel} and \eqref{eq:F2QQrel2}.} These equations can be proven by applying the core relations \eqref{eq:core_relations} to both RHS and LHS and then using the constraints \eqref{eq:constraintP}, \eqref{eq:constraintQai} opportunely shifted.

\paragraph{Proof of \eqref{eq:Z2QQrel}.}We start from the LHS and apply the core relations with appropriate shifts to obtain:
\beq
\frac{2}{\bQ_{\emptyset}^2}\, \bS_{\hat a|i}(\bP^+)_{a}^{\;\hat a}\bS_{\hat b|j}\epsilon^{ij}(\bP^-)_{b}^{\;\hat b}\mathbf{C}^{ac}(\sigma_{[A}\bar \sigma_{B]})_c^{\;b}=\frac{2}{\bQ_{\emptyset}} \,\mathbf{C}_{\hat a \hat b}(\bP^+)_{a}^{\;\hat a}(\bP^-)_{b}^{\;\hat b}\mathbf{C}^{ac}(\sigma_{[A}\bar \sigma_{B]})_c^{\;b}\,,
\eeq
where in the second passage we have used the normalisation condition $\bS_{\hat a|i}\epsilon^{ij}\bS_{\hat b|j}=\bQ_{\emptyset}\,\bar {\mathbf{C}}_{\hat a \hat b}$ which is equivalent to \eqref{eq:constraintQai}. We then rewrite:
\beq
2(\sigma_{[A}\bar \sigma_{B]})_c^{\;b}=(\delta_A^C\delta_B^D-\delta_B^C\delta_A^D-\epsilon_{C'D'AB}\rho^{CC'}\rho^{DD'})(\sigma_C\cdot \bar{\sigma}_D)_c^{\;b}\,.
\eeq
It is then easy to verify that with our definitions:
\beq
 \mathbf{C}_{\hat a \hat b}(\bP^+)_{a}^{\;\hat a}(\bP^-)_{b}^{\;\hat b}\mathbf{C}^{ac}(\sigma_C\cdot \bar{\sigma}_D)_c^{\;b}=\bQ_{\emptyset}\,(\bP_{CD}+\mathbb{M}_{CD})\,,
\eeq
where $\mathbb{M}_{CD}$ is a matrix built from $\bP^+$ and $\bP^-$ that is annihilated by the tensor $(\delta_A^C\delta_B^D-\delta_B^C\delta_A^D-\epsilon_{C'D'AB}\rho^{CC'}\rho^{DD'})$. The proof for \eqref{eq:Z2starQQrel} is almost exactly the same.

\paragraph{Proof of \eqref{eq:QQrelZ}.} These are easily verified by shifting the core relations and using them on the RHS, and then applying the constraint \eqref{eq:constraintP}.


\section{Analytic continuations for the $\bQ-\tau$ and $\bP-\nu$ systems}
\label{appendix:numirror}
In this appendix, we provide the proofs for the various analytic continuations and properties we have presented in section \ref{sec:Pmusystem}. 

\paragraph{Some useful constraints. }
Below we present some constraints that can be easily derived from our main equations in section \ref{sec:QQsys} (setting $\bQ_{\emptyset}=1$) and that will be used in the rest of the appendix.
\beqa
\bS &=& -\mathbf{C}^{-1}\cdot \bS^{-t}\cdot \epsilon^{-1},\\
\hat\bS &=& \hat{\mathbf{C}}^{-1}\cdot \hat\bS^{-t}\cdot \epsilon^{-1},\\
 \bQ &=&(\bS^-)^t\cdot \hat\bP^t\cdot  \hat{\mathbf{C}}\cdot  \hat \bS^-=(\hat\bS^+)^t\cdot  \bP^t \cdot \, {\mathbf{C}}^t \cdot  \bS^+ ,\\
 \bQ &=&(\hat\bS^+)^t\cdot  \bP^t \cdot  (\bS^+ )^{-t} \cdot \epsilon^{-1},\\
 \epsilon \cdot \bQ^t &=&-(\bS^+)^{-1}\cdot  \bP \cdot  (\hat\bS^+ )^{-1}  .\label{eq:appexp1}
\eeqa
\paragraph{Analytic continuation of $\Omega$ matrices.}In this paragraph, we compute the analytic continuation along the path $\gamma$ of the $\Omega$ matrices introduced in \eqref{def:Omegadef}. To do so, we need to continue the following definitions analytically:
\beq
\Omega=(\bS^{-\uparrow})^t\cdot (\bS^{-\downarrow})^{-t}\,,\quad \hat{\Omega}=(\hat{\bS}^{-\uparrow})^t\cdot (\hat{\bS}^{-\downarrow})^{-t}\,.
\eeq
Since LHPA spinors have cuts in the upper half plane starting from $\frac{i}{2}$, it follows that $\bS^{-\uparrow}$ and $\hat\bS^{-\uparrow}$ do not have a branch cut on the real axis. Therefore, they go unchanged under analytic continuation along $\gamma$, and so:
\beq\label{eq:Omegagamma}
(\Omega)^{\gamma}=(\bS^{-\uparrow})^t\cdot \left((\bS^{-\downarrow})^{-t}\right)^{\gamma}\,,\quad (\hat{\Omega})^\gamma=(\hat{\bS}^{-\uparrow})^t\cdot \left((\hat{\bS}^{-\downarrow})^{-t}\right)^{\gamma}\,,
\eeq
Then, inverting the QQ-relations \eqref{eq:relQ1} and \eqref{eq:relQ2} and applying the continuation along $\gamma$ to them (keeping in mind that $\bS^{+\downarrow}$ also do not possess cuts on the real axis), we can determine
\beq
\label{eq:ancontspinors}
(\bS^{\downarrow}\,^-)^{\gamma } = \mathbf{C}^{-1} \cdot (\bS^{\downarrow}\,^+ )^{-t} ({\bQ}^{\downarrow t})^{\gamma} ,\;\;\;\; (\hat{\bS}^{\downarrow}\,^-)^{\gamma } = \hat{\mathbf{C}}^{-1} \cdot (\hat{\bS}^{\downarrow}\,^+ )^{-t} ({\bQ}^{\downarrow })^{ \gamma}.
\eeq
Using these relations in (\ref{eq:Omegagamma}), we can obtain equations where only the $\bQ$ functions are analytically continued:

\beqa
&(\Omega)^{\gamma}=(\bS^{-\uparrow})^t\cdot \mathbf{C}^t\cdot \bS^{+\downarrow}\cdot(\bQ^{\downarrow-1})^{\gamma}=\bQ^{\uparrow}\cdot (\bS^{+\uparrow})^{-1}\cdot \bS^{+\downarrow}\cdot(\bQ^{\downarrow-1})^{\gamma}\,,\\
&(\hat\Omega)^{\gamma}=-(\hat{\bS}^{-\uparrow})^t\cdot \mathbf{C}\cdot \hat{\bS}^{+\downarrow}\cdot(\bQ^{\downarrow-t})^{\gamma}=\bQ^{\uparrow t}\cdot (\hat{\bS}^{+\uparrow})^{-1}\cdot  \hat{\bS}^{+\downarrow}\cdot(\bQ^{\downarrow-t})^{\gamma}\,.
\eeqa
Finally, from \eqref{def:fmatrices} we can see that the combinations of spinors and antispinors appearing in the equations above are simply related to the $\Omega,\hat{\Omega}$ matrices and so:
\beq
(\Omega)^{\gamma}=\bQ^{\uparrow}\cdot (\Omega^{[+2]})^{-t}\cdot(\bQ^{\downarrow-1})^{\gamma}\,,\quad (\hat{\Omega})^\gamma=\bQ^{\uparrow t}\cdot (\hat{\Omega}^{[+2]})^{-t}\cdot(\bQ^{\downarrow-t})^{\gamma}\,.
\eeq
Repeating the calculation on the other wing, we get:
\beq
\label{eq:Omegacont}
(\dot\Omega)^{\gamma}=\slQ^{\uparrow}\cdot (\dot\Omega^{[+2]})^{-t}\cdot(\slQ^{\downarrow-1})^{\gamma}\,,\quad (\hat{\dot{\Omega}})^\gamma=\slQ^{\uparrow t}\cdot (\hat{\dot{\Omega}}^{[+2]})^{-t}\cdot(\slQ^{\downarrow-t})^{\gamma}\,.
\eeq

\paragraph{Analytic continuation of $\tau$ matrices.}In this paragraph we compute the analytic continuation along $\gamma$ of the $\tau$ matrices introduces in \eqref{def:fmatrices}. This is quite simple, given that we have just computed $\Omega^{\gamma}$ and all gluing matrices $\mathcal{G}$ do not have branch cuts. Hence we multiply on the left the first of (\ref{eq:Omegacont}) by $\mathcal{G}$, obtaining an equation linking $\tau^{\gamma}$ with $\tau^{[+2]}=\hat \tau$:
\beq
\label{eq:f_an_cont}
\tau^{\gamma}=(\bQ^{\downarrow})^{\gamma}\cdot \epsilon \cdot \hat \tau \cdot \epsilon^{-1} \cdot (\slQ^{\downarrow -1})^{\gamma}\longrightarrow 
(\dot{\bQ}^{\downarrow t}\cdot \tau^t \cdot \bQ^{\downarrow-t} )^{\gamma} = \eps^{-1} \cdot \hat{\tau}^t \cdot \eps \,,
\eeq
where in the first identity we used both the definition $\mathcal{G} \cdot {\Omega^{++}} = \mathcal{G}\cdot  \hat{\Omega } = \hat{\tau}$ and the constraint $\hat{\tau} \cdot \epsilon \cdot \hat{\tau}^t = \epsilon $. 
To find the analytic continuation of $\hat \tau$ we instead multiply by $\mathcal{G}$ the second equation in \eqref{eq:Omegacont}:
\beq
\hat \tau^{\gamma}=(\bQ^{t\downarrow})^{\gamma}\cdot \epsilon \cdot  \tau \cdot \epsilon^{-1} \cdot (\slQ^{\downarrow -t})^{\gamma}\longrightarrow 
(\dot{\bQ}^{\downarrow}\cdot \hat \tau^t \cdot \bQ^{\downarrow-1} )^{\gamma} = \eps^{-1} \cdot {\tau}^t \cdot \eps \,.
\eeq
In the other wing, we have analogously:
\beq\label{eq:f_an_contdotted}
({\bQ}^{\downarrow t}\cdot \dot\tau^t \cdot \slQ^{\downarrow-t} )^{\gamma} = \eps^{-1} \cdot \hat{\dot\tau}^t \cdot \eps \,,\quad ({\bQ}^{\downarrow}\cdot \hat {\dot\tau}^t \cdot \slQ^{\downarrow-1} )^{\gamma} = \eps^{-1} \cdot {\dot\tau}^t \cdot \eps \,.
\eeq

\paragraph{Mirror periodicity of $\nu$ matrices.}In this section, we prove the main property of the $\nu$ matrices \eqref{def:nu_matrices}, i.e. their mirror periodicities $\nu^{\gamma}=\nu^{++}$. This is easy to do since we are familiar with the properties of the $\tau$ matrices under analytic continuation. We start from the definition \eqref{def:nu_matrices}, which we repeat here for convenience,
\beq
 {\nu} \equiv   {\bS}^-\cdot \dot{ {\tau}}^t \cdot (\dot{ {\bS}}^{-} )^{-1} \,,
 \eeq
 and continue it along the path $\gamma$:
\beq
\nu^{\gamma} = ({\bS}^{-\downarrow})^{\gamma}\cdot (\dot{ {\tau}}^t)^\gamma \cdot (\dot{ {\bS}}^{-\downarrow} )^{-1\,\gamma} \,.
\eeq
Then, using (\ref{eq:ancontspinors}), we can write
\beq
\nu^{\gamma} = \mathbf{C}^{-1}\cdot (\bS^{+\downarrow})^{-t}\cdot(\bQ^{\downarrow t})^\gamma\cdot (\dot{ {\tau}}^t)^\gamma \cdot(\slQ^{\downarrow-t})^\gamma\cdot(\slS^{+\downarrow})^t \cdot \mathbf{C}\,.
\eeq
We recognise then that the three factors in the middle can be rewritten thanks to the first relation in (\ref{eq:f_an_contdotted}), i.e.  $ (\bQ^{\downarrow\,t}\cdot \dot \tau^t \cdot \slQ^{\downarrow\,-t})^{\gamma}=\epsilon^{-1}\cdot \hat{\dot{\tau}}^{t}\cdot \epsilon$, and thus:
\beq
\nu^{\gamma} =\mathbf{C}^{-1}\cdot (\bS^{+\downarrow})^{-t}\cdot \epsilon^{-1}\cdot \hat{\dot{ \tau} }^t\cdot  \epsilon\cdot (\slS^{+\downarrow})^{t}\cdot \mathbf{C}=\bS^{+\downarrow}\cdot \hat{\dot \tau}^t\cdot(\slS^{+\downarrow})^{-1}\,,
\eeq 
where in the second equality we have used the first equation in \eqref{eq:appexp1}.
{Since $\hat{\dot{\tau}}=\dot \tau^{[+2]}$, the RHS is explicitly equal to $\nu^{[+2]}$ . The mirror periodicity of the other matrices $\dot\nu,\,\hat \nu, \hat{\dot \nu}$ follows similarly. 

\paragraph{Analytic continuation of $\bP$ functions}
In this paragraph, we derive equations \eqref{eq:Pgluing1main}. Since we know the analytical continuation of $\bQ,\,\bS,\,\hat\bS$ functions, we use the QQ-relations \eqref{eq:defQ} to express $\bP$ in terms of these quantities (which we assume to be analytic in the upper half plane, omitting the label $\downarrow$):
\beq
\label{eq:PintermsofQ}
\mathbf{C}\cdot \bP=(\bS^+)^{-t}\cdot \bQ^t\cdot (\hat{\bS}^+)^{-1}\,,
\eeq
We analytically continue this equation along $\gamma$; by their analytical properties $\bS^+,\,\hat \bS^+$ do not have cuts on the real axis, and therefore are unchanged under $\gamma$, while $\bQ^{\gamma}$ is given by \eqref{def:gluingQ}. Then it is easy to find that:
\beq
\bP^{\gamma}=\mathbf{C}^{-1}\cdot\underbrace{(\bS^+)^{-t}\cdot \tau^{[+2]} \cdot (\slS^+)^t}_{(\dot \nu^{[+2]})^t}\cdot (\slS^+)^{-t}\cdot \slQ^t \cdot (\hat{\slS}^+)^{-1} \cdot \underbrace{\hat{\slS}^+ \cdot (\hat\tau^{[+2]})^t \cdot (\hat\bS^+)^{-1}}_{\hat{\dot\nu}^{[+2]}}\,,
\eeq
where we have used the definitions \eqref{def:nu_matrices} and the properties of the $\tau$ matrices \eqref{eq:perf}. Then employing again \eqref{eq:PintermsofQ} we arrive at:
\beq
\bP^{\gamma}=\mathbf{C}^{-1}\cdot(\dot \nu^{[+2]})^t\cdot \mathbf{C}\cdot \bP \cdot \hat{\dot\nu}^{[+2]}\,,
\eeq
and using the constraint \eqref{eq:Pnusecond} we finally get
\beq
\label{eq:Pgluing1}
(\bP)^{\gamma}=(\dot{\nu}^{[+2]})^{-1}\cdot \slP\cdot \hat{\dot{\nu}}^{[+2]}\,,\qquad (\slP)^{\gamma}=({\nu}^{[+2]})^{-1}\cdot \bP\cdot \hat{{\nu}}^{[+2]}\,,
\eeq
The same calculation is easily repeated in the other wing.
\section{On the derivation of the linear integral equation}\label{appendix:detailsLIE}
We start from equation \eqref{eq:VVgamma}, which we repeat here for convenience:
\beq
\label{eq:VVgamma2}
{V}(u) {V}^{\gamma^{-1}}(u) = \mathcal{A}(u) \times \left(\frac{ {W}^+(u)}{ W^-(u) }\right)^2 \; \mathcal{E} ,
\eeq
and we will study this equation taking $u\in (-2 h, 2h) + i 0^+$.  

For such values of $u$, the left-hand side of \eqref{eq:VVgamma2} represents a symmetrised product of $V$ above and below the cut on the real axis. Since $\log V$ is expressed in terms of an integral kernel with a pole exactly at $z = u$, such symmetrisation is simply given by the principal value integral
\beq
 \label{eq:VPV}
\log\left( {V}(u) {V}^{\gamma^{-1}}(u) \right) = 2 P.V.\left[\int_{-2 h}^{2 h} \mathcal{D}_V(z) \frac{d z }{2 i \sinh(\pi (z - u) )} \right] .
 \eeq
Next, to evaluate the combination  $W^+/W^-$, appearing on the rhs of \eqref{eq:VVgamma2}, we use the integral representation \eqref{eq:logW}, and the identity for the shifted kernels
\beq
\mathbb{K}(u+i,z) - \mathbb{K}(u,z) = 2 \mathbb{K}(u+i,z) + \frac{1}{z-u}.
\eeq
The resulting kernel on the rhs has a pole at $z=u$, which will lie just above the integration contour for the values of $u$ we are considering. We can rewrite such an integral by isolating a principal value contribution plus half a residue, to write
\beq\label{eq:LIEapp0}
\log\left(\frac{W^+(u)}{W^-(u)}\right)^2 = 
 4 \int_{-2 h}^{2 h} \mathcal{D}_V(z)  \frac{dz}{2i\pi} \mathbb{K}(u + i , z) + 2 P. V.  \int_{-2 h}^{2 h} \mathcal{D}_V(z)  \frac{dz}{2i\pi (z-u)} + \mathcal{D}_V(u) .
\eeq
Thanks to (\ref{eq:VPV}) and (\ref{eq:LIEapp0}), the constraint \eqref{eq:VVgamma2} can now be written as:
\beq\label{eq:LIEapp2}
P.V.\int_{-2 h}^{2 h} \mathcal{D}_V(z) \frac{d z }{2 \pi i} \left[ \underbrace{  \frac{2 \pi }{\sinh(\pi (z - u) )}-\frac{2}{(z - u) } - 4 \mathbb{K}(u+i, z) }_{\equiv 2 \pi i \mathbb{G}(u-z)}\right]- \mathcal{D}_V(u) =
 \log\mathcal{A}(u) + \log\mathcal{E} ,
\eeq
but now we can notice that the kernel appearing in the square brackets $\mathbb{G}(u-z)$ has no pole at $z = u$, thanks to a cancellation between different terms!
The equation can thus be written without a principal value. The kernel, symmetric and of difference type, $\mathbb{G}(u-z)=\mathbb{G}(z-u)$, takes the form
\beqa
\mathbb{G}(u,v)\equiv \mathbb{G}(u-v)&\equiv&\frac{1}{2i\pi}\left(\frac{2 \pi }{\sinh(\pi (u - v) )}-\frac{2}{(u - v) } - 4 \mathbb{K}(v+i, u) \right)\\&=&\frac{1}{i\pi}\sum_{n=1}^{\infty} (-1)^n\left(\frac{1}{u-v+i n}-\frac{1}{u-v-i n}\right),\eeqa
where the expression as an infinite sum makes it obvious that it has no singularities on the real axis and is symmetric, i.e. $\mathbb{G}(u,v)=\mathbb{G}(v,u)$. 

In conclusion, the linear integral equation can be written  in the form we reported in the main text, i.e.
\beq\label{eq:LIE33}
\int_{-2h}^{2h} \frac{dz}{2i\pi}\mathcal{D}_V(z)\left[2\sum_{n=1}^{\infty} (-1)^n\left(\frac{1}{z-u+i n}-\frac{1}{z-u-i n}\right)\right] -\mathcal{D}_V(u)=\log{\mathcal{A}(u)} + \log\mathcal{E}\,.
\eeq

\section{Some useful quantities in the ABA limit}\label{appendix:dressing}
In this appendix, we collect the notation for the functions emerging naturally in the Asymptotic Bethe Ansatz limit. We leave aside the discussion of the new piece of the dressing phases emerging in the non-symmetric sector, which is discussed separately in section \ref{sec:dressingnew}.

\subsection{Basic notations }
\paragraph{Zhukovski variables.}
The Zhukovski variables are functions of $u$ defined as solutions to $x(u)+x(u)^{-1}=u/h$. We use the solution with a single branch cut at $u\in[-2h,2h]$, which is the following
\beq
x(u)\equiv \frac{u+\sqrt{u+2h}\,\sqrt{u-2h}}{2h}.
\eeq
Note that under analytic continuation across their branch cut, the Zhukovski variables change as $x(u^{\gamma})= x(u^{\gamma^{-1}})=x(u)^{-1}.$
\paragraph{Zhukovski polynomials.}From the Zhukovski variables, we can define the following functions of two spectral parameters:
\beq
\label{def:Appendix_RB_unshifted}
B(u,v) \equiv {\frac{1}{x(u)} - x(v) },\;\;\;R(u,v) \equiv{x(u)} - x(v) ,
\eeq
and we denote two versions with shifts on the second variable as 
\beq
\label{def:Appendix_RB}
B_{(\pm)}(u,v) \equiv \frac{\frac{1}{x(u)} - x^{\mp}(v) }{\sqrt{x^{\mp}(v) }},\;\;\;R_{(\pm)}(u,v) \equiv \frac{{x(u)} - x^{\mp}(v) }{\sqrt{x^{\mp}(v) }},
\eeq
which satisfy
$$
B_{(\pm)}(u,v) R_{(\pm)}(u,v) = v - u \mp \frac{i}{2}.
$$
We can evaluate \eqref{def:Appendix_RB_unshifted} and \eqref{def:Appendix_RB} on a given set of Bethe roots $\{v_{a,k}\}_{k=1}^{K_{a}}$ associated to some Q-functions to obtain the Zhukovski polynomials in $u$: 
\beq\label{eq:def_B_R}
B_{a}(u) \equiv \prod_{k=1}^{K_a}B(u,v_{a,k})=\prod_{k=1}^{K_a}\left(\frac{1}{x(u)} - x_k \right) , \;\;\; R_a(u) \equiv \prod_{k=1}^{K_a}R(u,v_{a,k})=\prod_{k=1}^{K_a} ( {x(u)} - x_k ).
\eeq
Moreover we define
\beq \label{def:Zhukovsky_B}
B_{a,(\pm)} \equiv \prod_{i=1}^{K_{a}}\frac{\frac{1}{x(u)} - x_{a, i}^{\mp} }{\sqrt{x_{a,i}^{\mp}}},\;\;\;R_{a,(\pm)} \equiv \prod_{i=1}^{K_{a}}\frac{{x(u)} - x_{a, i}^{\mp} }{\sqrt{x_{a,i}^{\mp}}}.
\eeq
The Zhukovski polynomials $B_a$ are free of zeros on the first sheet, while $R_a$ are free of zeros on the second sheet. It is obvious from the properties of $x(u)$ that $R_{a}(u^{\gamma})=B_a(u)$ and $B_{a}(u^{\gamma})=R_a(u)$, with the same relation connecting the functions in \eqref{def:Zhukovsky_B}. Furthermore, in terms of the polynomials $\mathbb{Q}_a(u)\equiv\prod_{i=1}^{K_a}(u-u_i)$ introduced in the main text, it holds that \beq\mathbb{Q}^{\pm}_a(u)=
(-1)^{K_a} B_{a,(\pm)}(u)R_{a,(\pm)}(u).\eeq

\paragraph{Functions with infinitely many cuts.}
We define
\beq
f(u,v) \equiv \text{exp}\left[ \int_{-2 h}^{+2 h} \frac{dz}{2 \pi i} \log\frac{B_{(-)}(z+i 0^+,v) R_{(+)}(z+i 0^+,v) }{ B_{(+)}(z+i 0^+,v) R_{(-)}(z+i 0^+,v)  } \partial_z \log{\Gamma(iz - i u) }\right],
\eeq
where the integrand can be written more explicitly as
\beq
f(u,v) \equiv \text{exp}\left[ \int_{-2 h}^{+2 h} \frac{dz}{2 \pi i} \log\frac{\left(\frac{1}{x(z+i 0^+)} - x^+(v) \right) ({x(z+i 0^+)} - x^-(v) ) }{ \left(\frac{1}{x(z+i 0^+)} - x^-(v) \right) ({x(z+i 0^+)} - x^+(v) ) } \partial_z \log{\Gamma(iz - i u) }\right].
\eeq
This function has no zeros, poles or cuts in the upper half plane, and it solves the difference equation:
\beq\frac{f(u+i, v)}{f(u,v) } = \frac{ {B}_{(-)}(u,v) }{ {B}_{(+)}(u,v)}.
\eeq
As we can see from this equation or from the integral representation above, it has an infinite ladder of cuts in the lower half plane. As we explain in the following paragraph, it is connected to the BES dressing factor.

Similarly, we define
\beq
\bar{f}(u,v) \equiv \text{exp}\left[ -\int_{-2 h}^{+2 h} \frac{dz}{2 \pi i} \log\frac{B_{(-)}(z+i 0^+,v) R_{(+)}(z+i 0^+,v) }{ B_{(+)}(z+i 0^+,v) R_{(-)}(z+i 0^+,v)  } \partial_z \log{\Gamma(i u -iz ) }\right] ,
\eeq
which satisfies
\beq\frac{\bar{f}(u, v)}{\bar{f}(u-i,v) } = \frac{ {B}_{(-)}(u,v) }{ {B}_{(+)}(u,v)},
\eeq
and which is now analytic in the lower half plane. This function is the complex conjugate of $f(u,v)$ when both arguments are real. 

In the main text, we introduced the function $\mathbf{f}(u)$, which is simply a certain product of the functions $f(u, v)$ over $v$ taking Bethe roots values. In particular, dividing the roots into the four families labelled by $s$, $\hat{s}$, $\dot{s}$ and $\dot{\hat s}$, we split
\beq
\mathbf{f} \equiv \mathbf{f}_s \mathbf{f}_{\hat{s}} \mathbf{f}_{\dot s} \mathbf{f}_{\hat{\dot s}} ,
\eeq
with $\mathbf{f}_{\bullet} $ solving the difference equation:
\beq
\frac{\mathbf{f}_{\bullet}^{[+2]}}{\mathbf{f}_{\bullet}}={B_{\bullet, (-)}\over B_{\bullet, (+)}}\,,
\eeq
where now $B_{\bullet,(\pm)}$ are the Zhukovski polynomials \eqref{def:Zhukovsky_B}.
Therefore, we simply have
$$
\mathbf{f}_{\bullet}(u)\equiv \prod_{i=1}^{K_{\bullet}} f(u, v_{\bullet, i}).
$$
An analogous definition holds for $\bar{\mathbf{f}}_{\bullet}$. 
\subsection{Building blocks for the dressing factors. }
In this section, we provide details about the building blocks for the dressing factors introduced in section \ref{sec:dressings}. Our notation follows closely that of \cite{Cavaglia:2021eqr, Ekhammar:2024kzp}.

\paragraph{The BES dressing factor and related functions. }
Borrowing the notation from \cite{Gromov:2014caa}, we now discuss how to define the function $\sigma^{1,\text{BES}}(u,v),
$
 which can be seen as a  building block of the Beisert-Eden-Staudacher $\Sigma_{\text{BES}}$ dressing factor, i.e. 
\beq
\Sigma_{\text{BES}}(u,v)=\frac{\sigma^{1,\text{BES}}(u+i/2,v)}{\sigma^{1,\text{BES}}(u-i/2,v)}.
\eeq
We can define this function through the following properties.  First, it has a single cut on the real axis in $u$, and it has neither other singularities nor zeros on the first sheet. Moreover, it satisfies the Riemann-Hilbert equation
\beq
\sigma^{1,\text{BES}}(u^{\gamma},v) \, \sigma^{1,\text{BES}}(u,v)  = f(u+i, v) \bar{f}(u-i , v) ,
\eeq
where $f$, $\bar{f}$ are the functions defined in the previous section. 

To construct the solution with the above properties it is useful to break it as follows:
$$
\sigma^{1,\text{BES}}(u,v) \equiv \text{exp}\left(i \chi_{\text{BES}}(u, v^+) - i \chi_{\text{BES}}(u, v^-)\right),
$$
where $\chi_{\text{BES}}(u,v)$ should have the same analytic properties  in $u$ discussed above for $\sigma^{1,\text{BES}}$, and should satisfy 
\beq\label{eq:BESRH}
\chi_{\text{BES}}(u^{\gamma}, v) + \chi_{\text{BES}}(u, v) = -i\log\left( g(u+i, v) \bar{g}(u - i , v) \right) ,
\eeq
where 
\beqa
g(u,v) &\equiv& \text{exp}\left[ \int_{-2 h}^{+2 h} \frac{dz}{2 \pi i} \log\frac{ B(z+i 0^+,v) }{ R(z+i 0^+,v) } \partial_z \log{\Gamma(iz - i u) }\right],\\
\bar{g}(u,v) &\equiv& \text{exp}\left[ -\int_{-2 h}^{+2 h} \frac{dz}{2 \pi i} \log\frac{ B(z+i 0^+,v) }{ R(z+i 0^+,v)   } \partial_z \log{\Gamma(i u -iz ) }\right] .
\eeqa
The Riemann-Hilbert  problem (\ref{eq:BESRH}) then has the solution
\beqa
&&\chi_{\text{BES}}(u,v)= \\&&- \int_{-2 h}^{2 h} \frac{dw \, \sqrt{(u - 2 h)} \sqrt{(u+2 h)}}{2 \pi i \sqrt{ 4 h^2 - w^2} (w - u)} \int_{- 2h}^{+2 h} \frac{dz}{2 \pi i} 
 \partial_z \log\frac{\Gamma( 1 + i (z - w) )}{\Gamma( 1-i (z-w) )} \log\left(\frac{\frac{1}{x(z+i0^+)} - x(v)}{x(z+i 0^+) - x(v)}\right).\nonumber
\eeqa
This expression is not exactly the same as the form of $\chi_{\text{DHM}}$ given by the DHM formula~\cite{Dorey:2007xn,Vieira:2010kb}: 
$$
\chi_{\text{DHM}}(u,v) = i \oint_{|x_w|=1} \frac{dx_w}{2 \pi i} \oint_{|x_z|=1} \frac{dx_z }{2 \pi i} \frac{1}{(x_w - x(u)) (x_z - x(v) )} \frac{\Gamma(1 + i (w - z) )}{\Gamma(1 - i (w - z) )},
$$ 
however it is simple to verify explicitly that, after an integration by parts, the two differ by simple terms that cancel in the combinations with shifts appearing in the dressing phase. In fact, with the notations above,
\beqa
&&\chi_{\text{DHM}}(u,v) - \chi_{\text{BES}}(u,v) \\
&=& -\int_{-2 h}^{2 h} \frac{dw \, }{2 \pi i \sqrt{ 4 h^2 - w^2} } \int_{- 2h}^{+2 h} \frac{dz}{2 \pi i} 
 \partial_z \log\frac{\Gamma( 1 + i (z - w) )}{\Gamma( 1-i (z-w) )} \log\left(\frac{\frac{1}{x(z+i0^+)} - x(v)}{x(z+i 0^+) - x(v)}\right),\nonumber
\eeqa
which does not depend on $u$, and therefore
\beqa
&&\chi_{\text{DHM}}(u^{+}, v^+)+\chi_{\text{DHM}}(u^{-}, v^-)-\chi_{\text{DHM}}(u^{+}, v^-)-\chi_{\text{DHM}}(u^{-}, v^+) \\
&=&\chi_{\text{BES}}(u^{+}, v^+)+\chi_{\text{BES}}(u^{-}, v^-)-\chi_{\text{BES}}(u^{+}, v^-)-\chi_{\text{BES}}(u^{-}, v^+) .
\eeqa
In summary we showed that the BES dressing phase can be written as
$$
\Sigma_{\text{BES}}(u,v)=\frac{\sigma^{1,\text{BES}}(u+i/2,v)}{\sigma^{1,\text{BES}}(u-i/2,v)} = e^{i\left(\chi_{\text{BES}}(u^{+}, v^+)+\chi_{\text{BES}}(u^{-}, v^-)-\chi_{\text{BES}}(u^{+}, v^-)-\chi_{\text{BES}}(u^{-}, v^+)\right)},
$$
in terms of the functions defined explicitly above, and this is equivalent to the DHM representation.

\paragraph{Dressing factors appearing in the AdS$_3\times $S$_3\times$T$^4$ model. }
We then introduce the functions $\sigma^{1,\text{extra}},\,\tilde{\sigma}^{1,\text{extra}}$,
which are now related to the dressing factors 
$\Sigma^{\text{extra}}$ and $\tilde{\Sigma}^{\text{extra}}$ which already appeared in the $AdS_3\times S^3 \times T^4$ model in the standard way:
\begin{equation}
\Sigma^{\text{extra}}(u, v) = \frac{\sigma^{1,\text{extra}}(u+i/2, v)}{ \sigma^{1,\text{extra}}(u-i/2, v)} ,\;\;\; \tilde{\Sigma}^{\text{extra}}(u, v) = \frac{\tilde{\sigma}^{1,\text{extra}}(u+i/2, v)}{ \tilde{\sigma}^{1,\text{extra}}(u-i/2, v)} .
\end{equation}
Let us review the definition of the building blocks $\sigma^{1,\text{extra}},\,\tilde{\sigma}^{1,\text{extra}}$. For a fixed $v$, we require them to be real and analytic without zeros on the first sheet, with only a single cut on the real axis. Moreover, they must satisfy the following functional relations:
\beqa\label{eq:extratermBudapest}
\sigma^{1,\text{extra}}(u^{\gamma},v)\tilde{\sigma}^{1,\text{extra}}(u,v)&=&\sqrt{\frac{R_{(+)}(u,v)}{R_{(-)}(u,v)}}\, \xi(v),\\\tilde{\sigma}^{1,\text{extra}}(u^{\gamma},v) {\sigma}^{1,\text{extra}}(u,v)&=&\sqrt{\frac{B_{({-})}(u,v)}{B_{({+})}(u,v)}}\,\xi^{-1}(v),
\eeqa
where
\beq
\log\xi(v) = \frac{1}{8}  \log\frac{(v^+ - 2 h) (v^++ 2 h)}{(v^- - 2 h) (v^-+ 2 h)}.
\eeq
These relations then imply the correct crossing relations for the full dressing factors $\Sigma^{\text{extra}},\,\tilde{\Sigma}^{\text{extra}}$. 
The presence of the $\xi(v)$ factors does not affect the form of the crossing equations, but is crucial for the braiding unitarity of the resulting S-matrix elements. 

The solution to the Riemann-Hilbert equations (\ref{eq:extratermBudapest}) with the required analytic properties can be written, as pointed out in (4.84)-(4.85) of~\cite{Ekhammar:2024kzp} (here $x\equiv x(u), \,y\equiv x(v)$), as follows: 
{\footnotesize
\beqa
\sigma^{1,\text{extra}} &\propto&\exp\Bigg(
-\frac{i \text{Li}_2\frac{(1-x) \left(y^-+1\right)}{(x+1) \left(y^--1\right)}}{2 \pi }+\frac{i \text{Li}_2\frac{(1-x) \left(y^++1\right)}{(x+1) \left(y^+-1\right)}}{2 \pi }
-\frac{i \log \frac{x+1}{x-\frac{1}{y^-}} \log \frac{y^--1}{y^-+1}}{2 \pi }+\frac{i \log \frac{x+1}{x-\frac{1}{y^+}} \log \frac{y^+-1}{y^++1}}{2 \pi }\nonumber\\ &&
+\frac{i \log \frac{x-1}{x+1} \left(\log \left(1-\frac{1}{(y^-)^2}\right)-\log \left(1-\frac{1}{(y^+)^2}\right)
-2 \log ({1-\frac{1}{x y^-}})
+2 \log ({1-\frac{1}{x y^+}})
\right)}{4 \pi }\Bigg)\;,\nonumber 
\eeqa
}
{\footnotesize
\beqa
\tilde{\sigma}^{1,\text{extra}} &\propto&\exp\Bigg(
\frac{i \text{Li}_2\left(\frac{2 \left(y^+-x\right)}{(x+1) \left(y^+-1\right)}\right)}{2 \pi }-\frac{i \text{Li}_2\left(\frac{2 \left(y^--x\right)}{(x+1) \left(y^--1\right)}\right)}{2 \pi }
-\frac{i \log \frac{x-1}{x+1} \left(\log 
\frac{y^-+1}{y^--1}-
\log 
\frac{y^++1}{y^+-1}
\right)}{4 \pi }\Bigg)\; ,\nonumber
\eeqa} 
where we omitted proportionality factors that are not crucial since they cancel in the full dressing factors. 
The same quantities can also be written in a perhaps more transparent way through integral representations. In particular, the ratio of these two dressing factors is the piece that is affected by the presence of the factors $\xi(v)$ above. The Riemann-Hilbert equation satisfied by the ratio has a very simple solution, given by 
\beqa
\log\frac{\sigma^{1,\text{extra}}(u,v)}{\tilde{\sigma}^{1,\text{extra}}(u,v)} &=& \frac{1}{4\pi i} \int_{-2 h}^{+ 2 h} dz \log\frac{v^{ {+}}-z}{v^{ {-}}-z} \partial_z \log{(u-z)} \\
&&+\frac{1}{8 \pi i} \log\frac{(v^+ - 2 h) (v^++ 2 h)}{(v^- - 2 h) (v^-+ 2 h)} \,{\log{\frac{(u - 2 h) }{(u+ 2 h)}}} ,
\eeqa
translating into the following explicit result for the ratio of dressing phases
\beqa
\log\frac{\Sigma^{\text{extra}}(u,v)}{\tilde{\Sigma}^{\text{extra}}(u,v)} &=& \frac{1}{4\pi i} \int_{-2 h}^{+ 2 h} dz \log\frac{v^{{+}}-z}{v^{ {-}}-z}   \partial_z \log\frac{u^+-z}{u^--z} \\
&&+\frac{1}{8 \pi i} \log\frac{(v^+ - 2 h) (v^++ 2 h)}{(v^- - 2 h) (v^-+ 2 h)} \,\log\frac{(u^+ - 2 h) (u^{{-}}+ 2 h)}{(u^- - 2 h) (u^{{+}}+ 2 h)} ,
\eeqa
which satisfies braiding unitarity and was first written in a rather different but equivalent parametrisation in \cite{Frolov:2021fmj}. 
The product, instead, is given explicitly by
\beq\label{eq:productSigmasapp}
\log\left[{\sigma^{1,\text{extra}}(u,v)}{\tilde{\sigma}^{1,\text{extra}}(u,v)} \right]= \frac{\sqrt{(u-2 h)} \sqrt{(u+2 h)}}{4\pi } \int_{-2 h}^{+ 2 h} dz \frac{
\log\frac{R_{(+)}(z+i 0^+, v) B_{(-)}(z+i 0^+, v)}{R_{(-)}(z+i 0^+, v) B_{(+)}(z+i 0^+, v)} 
}{(z - u ) \sqrt{4 h^2 - z^2 }}.
\eeq
 We can build the product of the full dressing factors ${\Sigma^{\text{extra}}(u,v)}{\tilde{\Sigma}^{\text{extra}}(u,v)}$ by taking the appropriate combinations with shifts. 

It is also notable that, as already noticed in \cite{Borsato:2013hoa},  this product can be identified with the well-known Hernandez-Lopez phase~\cite{Hernandez:2006tk}
\beq\label{eq:HLrel}
\Sigma^{\text{extra}} \times \tilde{\Sigma}^{\text{extra}} = \frac{1}{\Sigma_{\text{HL}}} ,
\eeq
where $\Sigma_{\text{HL}}$ is defined as
$$-i \log{\Sigma_{\text{HL}}}(u,v)=\chi_{\text{HL}}(u^{+}, v^+)+\chi_{\text{HL}}(u^{-}, v^-)-\chi_{\text{HL}}(u^{+}, v^-)-\chi_{\text{HL}}(u^{-}, v^+), $$
with~\cite{Beisert:2006ib}
\beqa
\chi_{\text{HL}}(u,v) &\equiv& \frac{\pi}{2}\oint\frac{d\omega_1}{2 i \pi}\oint\frac{d\omega_2}{2 i \pi}\frac{\text{sign}(\omega_2+\frac{1}{\omega_2}-\omega_1-\frac{1}{\omega_1})}{(\omega_1-x_u)(\omega_2-x_v)}\\
&=& \left( \int_{C^+} - \int_{C^-} \right) \frac{dx_w}{4 \pi (x_w - x_u)} \log\left(\frac{\frac{1}{x_w} - x_v}{x_w - x_v}\right),
\eeqa
with $C^{\pm}$ representing semi-circle contours with $x_w$ = 1 running anti-clockwise in the upper and lower half planes, respectively. The second, single integral representation can be used to prove (\ref{eq:HLrel}) using (\ref{eq:productSigmasapp}). 

The HL phase can also be computed explicitly in terms of polylogarithms, see for example \cite{Beisert:2006ib}.

In section \ref{sec:ABA}, we find the same building blocks, but with the variable $v$ evaluated at some set of Bethe roots $\{v_{a,i}\}_{i=1}^{K_a}$. Thus, we define the following functions of one variable:
\beq
\label{appendix:eq_bes}
\sigma^{1,\text{BES}}_{a}(u)\equiv \prod_{i=1}^{K_{a}}\sigma^{1,\text{BES}}(u,v_{a,i}),
\eeq
\beq\sigma^{1,\text{extra}}_{a}(u)\equiv \prod_{i=1}^{K_{a}}\sigma^{1,\text{extra}}(u,v_{a,i}),\;\;\;\tilde{\sigma}^{1,\text{extra}}_{a}(u)\equiv \prod_{i=1}^{K_{a}}\tilde{\sigma}^{1,\text{extra}}(u,v_{a,i}).
\eeq
The functions \eqref{appendix:eq_bes} satisfy the following useful functional equations, where $\bullet=s,\hat s,\dot s,\hat{\dot s}$:
\beq
(\sigma^{1,\text{BES}}_{\bullet})^{\gamma}\sigma^{1,\text{BES}}_{\bullet}=(\sigma^{1,\text{BES}}_{\bullet})^{\gamma^{-1}}\sigma^{1,\text{BES}}_{\bullet}=\label{appendix:bes_functional}
\mathbf{f}^{++}_{\bullet}\bar{\mathbf{f}}^{--}_{\bullet}.
\eeq
\paragraph{Crossing relations. }
Finally, we write down the crossing equations along the crossing path $\gamma_{\text{cross}}$ for the various pieces of the dressing factors. These are given by 
\beqa
\label{appendix:crossing_blocks}
\Sigma_{\text{BES}}(u^{\gamma_{\text{cross}}},v)\Sigma_{\text{BES}}(u,v)&=&\dfrac{x_2^-}{x_2^+}\dfrac{R^-_{(-)}}{R^-_{(+)}}\dfrac{B^+_{(-)}}{B^+_{(+)}},\\
\Sigma^{\text{extra}}(u^{\gamma_{\text{cross}}},v)^2\tilde{\Sigma}^{\text{extra}}(u,v)^2&=&\dfrac{R^+_{(-)}R^-_{(+)}}{R^-_{(-)}R^+_{(+)}}\,,\\
\tilde{\Sigma}^{\text{extra}}(u^{\gamma_{\text{cross}}} , v)^2 {\Sigma}^{\text{extra}}(u , v)^2  &=& \dfrac{B^+_{(-)}B^-_{(+)}}{B^-_{(-)}B^+_{(+)}}\,,
\eeqa
where we used the shorthands $x_1\equiv x(u)$ and $x_2\equiv x(v)$.
\section{Comparison with the worldsheet S-matrix bootstrap}
\label{sec:comparison}
\subsection{Asymptotic Bethe Ansatz}
In this section, we write down the identification between our dressing phases $\Sigma_{\bullet\bullet}$, written in \eqref{def:ABA_Sigma_1} and \eqref{def:ABA_Sigma_2}, and those of \cite{Borsato:2012ss}, which we denote as $S_{\bullet\bullet}$. This map is needed to match our Asymptotic Bethe equations, the ones in equations (4.5 - 4.10) of \cite{Borsato:2012ss}, and it is very simple, as it only contains some rational factors in the Zhukovski variables.

For the dressing phases with first index in the L wing, i.e. $\Sigma_{s\bullet}$ and $\Sigma_{\hat s \bullet}$, we have:
\beqa
S_{11}(u,v) &\equiv& \Sigma_{ss}(u,v) \times \frac{1 - \frac{1}{x^+(u) x^-(v)}}{1 - \frac{1}{x^-(u) x^+(v)}} ,\\
S_{13}(u,v) &\equiv& \Sigma_{s \hat{s}}(u,v)   ,\\
S_{1\bar{1}}(u,v) &\equiv& \Sigma_{s\dot{s}}(u,v) \times \left(\frac{1 - \frac{1}{x^+(u) x^-(v)}}{1 - \frac{1}{x^-(u) x^+(v)}} \right)^{\frac{1}{2}},\\
S_{1\bar{3}}(u,v) &\equiv& \Sigma_{s\hat{\dot{s}}}(u,v) \times \left(\frac{1 - \frac{1}{x^+(u) x^-(v)}}{1 - \frac{1}{x^-(u) x^+(v)}} \right)^{\frac{1}{2}},\\
S_{33}(u,v) &\equiv& \Sigma_{\hat s \hat s}(u,v) \times \frac{1 - \frac{1}{x^+(u) x^-(v)}}{1 - \frac{1}{x^-(u) x^+(v)}} ,\\
S_{31}(u,v) &\equiv& \Sigma_{ \hat{s} s}(u,v)   ,\\
S_{3\bar{3}}(u,v) &\equiv& \Sigma_{\hat s\hat{\dot s}}(u,v) \times \left(\frac{1 - \frac{1}{x^+(u) x^-(v)}}{1 - \frac{1}{x^-(u) x^+(v)}} \right)^{\frac{1}{2}},\\
S_{3\bar{1}}(u,v) &\equiv& \Sigma_{\hat s\dot{{s}}}(u,v) \times \left(\frac{1 - \frac{1}{x^+(u) x^-(v)}}{1 - \frac{1}{x^-(u) x^+(v)}} \right)^{\frac{1}{2}}.
\eeqa
For those with first index in the R wing, i.e. $\Sigma_{\dot s\bullet}$ and $\Sigma_{\hat{\dot s} \bullet}$, we have instead:
\beqa
S_{\bar 1 \bar 1}(u,v) &\equiv& \Sigma_{\dot s \dot s}(u,v) \times \frac{1 - \frac{1}{x^+(u) x^-(v)}}{1 - \frac{1}{x^-(u) x^+(v)}} ,\\
S_{\bar 1 \bar 3}(u,v) &\equiv& \Sigma_{\dot s \hat{\dot{s}}}(u,v)   ,\\
S_{\bar 1 {1}}(u,v) &\equiv& \Sigma_{\dot s {s}}(u,v) \times \left(\frac{1 - \frac{1}{x^+(u) x^-(v)}}{1 - \frac{1}{x^-(u) x^+(v)}} \right)^{\frac{1}{2}},\\
S_{\bar 1 {3}}(u,v) &\equiv& \Sigma_{\dot s {\hat{s}}}(u,v) \times \left(\frac{1 - \frac{1}{x^+(u) x^-(v)}}{1 - \frac{1}{x^-(u) x^+(v)}} \right)^{\frac{1}{2}},\\
S_{\bar 3 \bar 3}(u,v) &\equiv& \Sigma_{\hat{\dot  s} \dot{ \hat s} }(u,v) \times \frac{1 - \frac{1}{x^+(u) x^-(v)}}{1 - \frac{1}{x^-(u) x^+(v)}} ,\\
S_{\bar 3 \bar 1}(u,v) &\equiv& \Sigma_{\hat{\dot{s}} \dot s }(u,v)   ,\\
S_{\bar 3 {3}}(u,v) &\equiv& \Sigma_{\hat{\dot  s} {\hat s}}(u,v) \times \left(\frac{1 - \frac{1}{x^+(u) x^-(v)}}{1 - \frac{1}{x^-(u) x^+(v)}} \right)^{\frac{1}{2}},\\
S_{\bar 3 {1}}(u,v) &\equiv& \Sigma_{\hat{\dot  s} {{s}}}(u,v) \times \left(\frac{1 - \frac{1}{x^+(u) x^-(v)}}{1 - \frac{1}{x^-(u) x^+(v)}} \right)^{\frac{1}{2}}.
\eeqa
It is then easy to see that, with these definitions, our ABA equations are identical to (4.5 - 4.10) of \cite{Borsato:2012ss} when taken at the point $\alpha=1/2$. In appendix \ref{app:square_roots}, we will see that the square roots in the above S-matrix elements are exactly compensated by analogous factors in the dressing phases.
\subsection{Crossing equations}
In this section, we compare our crossing equations \eqref{eq:crossings} with the crossing equations (5.28) of \cite{Borsato:2015mma}, where the dressing phases are described in detail. We shall denote the dressing phases of \cite{Borsato:2015mma} as $\mathbb{S}_{ab}$. These objects differ from our dressing phases, denoted as $\Sigma_{ab}$, by rational functions of $x(u)$ and $x(v)$ which we need to reconstruct\footnote{Unfortunately, the notation used in this paper is also different from the one of \cite{Borsato:2012ss}.}.
To do so, we build from \cite{Borsato:2015mma} the S-matrix elements that would appear in the Asymptotic Bethe Ansatz, and confront them with our results in section \ref{sec:ABA}. The S-matrix elements we need are:
\beq
\mathcal{S}_{11}, \,\, \mathcal{S}_{13},\,\,\mathcal{S}_{\bar 1 1}, \,\, \mathcal{S}_{\bar 1 3}.
\eeq
In \cite{Borsato:2015mma}, these are formed by the blocks found in the Faddeev-Zamolodchikov algebra and the dressing phases:
\beq
\mathcal{S}_{ab}= FZ^{-1}_{ab}  \times \mathbb{S}^{-1}_{ab}\,,
\eeq
where the dressing factors $\mathbb{S}$ satisfy the crossing equations (5.28) of \cite{Borsato:2015mma}:
 \beqa
 \label{def:crossing_BSS}
 \mathbb{S}_{ab}(u^{\gamma_\text{cross}}, v) \mathbb{S}_{\bar ab}(u,v) &=&\left(\frac{x_u^+}{x_u^-}\right)^{-1/4}\left(\frac{x_v^+}{x_v^-}\right)^{1/4}\frac{1-\frac{1}{x_u^-x_v^+}}{1-\frac{1}{x_u^-x_v^-}}\sqrt{\frac{1-\frac{1}{x_u^-x_v^-}}{1-\frac{1}{x_u^+x_v^+}}}\,,\\
 \mathbb{S}_{\bar ab}(u^{\gamma_\text{cross}}, v) \mathbb{S}_{ ab}(u,v) &=&\left(\frac{x_u^+}{x_u^-}\right)^{1/4}\left(\frac{x_v^+}{x_v^-}\right)^{-1/4}\frac{x_u^--x_v^-}{x_u^+-x_v^-}\sqrt{\frac{x_u^+-x_v^+}{x_u^--x_v^-}}\,,\\
 \eeqa
and the elements of the Faddeev-Zamolodchikov algebra can be found in Appendix G of \cite{Borsato:2015mma}.  In particular, those that we need are the following: 
\beqa
 (FZ)_{11}&=&(FZ)_{13}=A^{LL}=1\,,\\
 (FZ)_{\bar11}&=&(FZ)_{\bar13}=D^{RL}=\left(\frac{x_u^+x_v^+}{x_u^-x_v^-}\right)^{1/4}\sqrt{\frac{1-\frac{1}{x_u^+x_v^+}}{1-\frac{1}{x_u^-x_v^-}}}\,.
 \eeqa
We also notice that our equations and the ones of \cite{Borsato:2015mma} differ in the choice of \emph{framing}\footnote{We thank R. Borsato and A. Sfondrini for suggestions on this point.}: in particular, this amounts to a different definition of the length $L$ appearing in the Bethe equations. The correct map is $L_{\text{here}}=L_{\text{BOSS}}-1$. When accounted for, this discrepancy is equivalent to distributing an extra factor of $\left(\frac{x_u^{+}}{x_u^-}\right)^{1/4}$ among all the dressing phases of the S-matrix elements $\Sigma_{\bullet\bullet}$ appearing in the momentum-carrying Bethe equations. To keep unitarity of the S-matrix, we also need to add to the same dressings factors of $\left(\frac{x_v^-}{x_v^+}\right)^{1/4}$. It is always possible to do so in the Bethe equations thanks to the zero-momentum condition \eqref{eq:zero_momentum_condition}. 

Confronting with the S-matrix elements appearing in our ABA equations \eqref{eq:ABA1}, it is easy to see that the map between $\Sigma_{ab}$ and $\mathbb{S}_{ab}$ is the following:
\beqa
\label{def:map_Sf}
\mathbb{S}^{-1}_{11}(u,v)&=&\left(\frac{x_u^-}{x_u^+}\frac{x_v^+}{x_v^-}\right)^{1/4}{\frac{x_u^+-x_v^-}{x_u^--x_v^+}}{\frac{1-\frac{1}{x_u^+x_v^-}}{1-\frac{1}{x_u^-x_v^+}}}\,\Sigma_{ss}(u,v)\,,\\
\mathbb{S}^{-1}_{13}(u,v)&=&\left(\frac{x_u^-}{x_u^+}\frac{x_v^+}{x_v^-}\right)^{1/4}\Sigma_{s\hat s}(u,v)\,,\\
\mathbb{S}^{-1}_{\bar 11}(u,v)&=&\left(\frac{x_v^+}{x_v^-}\right)^{1/2}\sqrt{\frac{1-\frac{1}{x_u^+x_v^-}}{1-\frac{1}{x_u^-x_v^+}}}
\,\Sigma_{\dot s s}(u,v)\,,\\
\mathbb{S}^{-1}_{\bar 13}(u,v)&=&\left(\frac{x_v^+}{x_v^-}\right)^{1/2}\sqrt{\frac{1-\frac{1}{x_u^+x_v^-}}{1-\frac{1}{x_u^-x_v^+}}}
\,\Sigma_{\dot s \hat s}(u,v)\,.
\eeqa

So in terms of our dressing phases $\Sigma$, equations \eqref{def:crossing_BSS} predicts:
\begin{small}
\beqa
\Sigma_{ss}(u^{\gamma_{\text{cross}}},v)\Sigma_{\dot ss}(u,v)&=&{\frac{x_2^-}{x_2^+}}\frac{x_1^--x_2^+}{x_1^+-x_2^-}\frac{\sqrt{\left(1-\frac{1}{x_1^-x_2^+}\right)\left(1-\frac{1}{x_1^+x_2^+}\right)\left(1-\frac{1}{x_1^-x_2^-}\right)}}{\left(1-\frac{1}{x_1^+x_2^-}\right)^{3/2}}\,,\\
\Sigma_{s\hat s}(u^{\gamma_{\text{cross}}},v)\Sigma_{\dot s\hat s}(u,v)&=&{\frac{x_2^-}{x_2^+}}\sqrt{\frac{\left(1-\frac{1}{x_1^+x_2^+}\right)\left(1-\frac{1}{x_1^-x_2^-}\right)}{\left(1-\frac{1}{x_1^-x_2^+}\right)\left(1-\frac{1}{x_1^+x_2^-}\right)}}\,,\\
\Sigma_{\dot ss}(u^{\gamma_{\text{cross}}},v)\Sigma_{ ss}(u,v)&=&{\frac{x_2^-}{x_2^+}}\frac{\left(x_1^--x_2^+\right)^{3/2}}{\sqrt{\left(x_1^+-x_2^-\right)\left(x_1^+-x_2^+\right)\left(x_1^--x_2^-\right)}}\frac{1-\frac{1}{x_1^-x_2^+}}{1-\frac{1}{x_1^+x_2^-}}\,,\\
\Sigma_{\dot s\hat s}(u^{\gamma_{\text{cross}}},v)\Sigma_{ s\hat s}(u,v)&=&{\frac{x_2^-}{x_2^+}}\sqrt{\frac{\left(x_1^--x_2^+\right)\left(x_1^+-x_2^-\right)}{\left(x_1^+-x_2^+\right)\left(x_1^--x_2^-\right)}}\,.
\eeqa
\end{small}
Compared with our crossing equations \eqref{eq:crossings}, we notice that they differ by the function $\mathbf{e}(v)$ appearing in \eqref{eq:crossings}. Therefore, we only have a match if $\mathbf{e}(v)=0$. As explained in section \ref{sec:braiding}, this choice seems incompatible with the braiding unitarity of the dressing phase $\Sigma_{\text{new}}$, which instead demands fixing $\mathbf{e}(v)$ in a different way.
\section{Square root singularities in the dressing phases }
\label{app:square_roots}
In this appendix, we deal with the fact that the S-matrix elements appearing in our ABA in section \ref{sec:ABA} and \ref{sec:comparison} possess apparent square-root singularities on the second sheet, which would create unwanted branch cuts.\footnote{We thank S. Frolov, D. Polvara and A. Sfondrini for pointing this out and for related discussions.} We show here that in our construction these singularities are exactly cancelled by similar factors which are hidden in the dressing phases ${\Sigma}_{\text{extra}}$ and $\Sigma_{\text{new}}$.
\paragraph{${\Sigma}_{\text{extra}}$.}
By definition, the building blocks $\sigma^{1,\text{extra}}$ and $\tilde{\sigma}^{1,\text{extra}}$ do not have branch points on the first sheet outside $u\in (-2h, 2h)$. However, equations (\ref{eq:extratermBudapest}) imply the following feature: of these two functions, $\sigma^{1,\text{extra}}$ has a branch point on the second sheet, at location depending on $v$. In fact, the equation implies
$$
\sigma^{1,\text{extra}}(u^{\gamma},v) = \sqrt{\frac{u-v + \frac{i}{2}}{u-v-\frac{i}{2}} }\times \texttt{regular}(u,v),
$$
where the regular part has no singularity at $u = v \pm i/2$. Instead, $\tilde{\sigma}^{1, \text{extra}}$ has no such rapidity-dependent singularities on the second sheet.

It follows that $$\Sigma^{\text{extra}}(u,v)\equiv \dfrac{\sigma^{1,\text{extra}}(u+i/2,v)}{\sigma^{1,\text{extra}}(u-i/2,v)}$$ possesses square root branch points on the second sheet, at $u=v\pm i$ or $u = v$.

\paragraph{$\Sigma_{\text{new}}$.}
In this paragraph, we will show that our new dressing phase \eqref{def:Sigma_new} also possesses $v$-dependednt branch points in the second sheet.
Without loss of generality, we consider here $v$ on the real axis. In terms of our density, the building block of the new dressing phase $\Sigma_{\text{new}}\equiv \frac{\sigma_{\text{new}}^+}{\sigma_{\text{new}}^-}$ is defined as
$$
\log\sigma_{\text{new}}(u,v)=\frac{1}{2i \pi}\int_{-2h}^{2h}\frac{dz}{u-z}\,\mathbf{d}(z|v).
$$
Under analytic continuation across the branch cut on the real axis, we get that:
\beqa
\log\sigma_{\text{new}}(u^{\gamma},v)&=&\log\sigma_{\text{new}}(u,v)+\mathbf{d}(u|v)=\\&&\log\sigma_{\text{new}}(u,v)+\log{\sqrt{\frac{u-v-i/2}{u-v+i/2}}}-\mathbf{e}(v)+\int_{-2h}^{2h} \dots ,
\eeqa
where we have plugged in the linear integral equation, in particular the first square root term comes directly from the source term in this equation. The term omitted (under $\dots$) is a regular integral, which cannot generate any singularity for $|\text{Im}(u)|<2$, in particular it is regular at $u = v \pm i/2$. 

Therefore, if we focus on the singularities on the second sheet at $u=v\pm i/2$, we get
$$
\sigma_{\text{new}}(u^{\gamma},v)=\sqrt{\frac{u-v - \frac{i}{2}}{u-v+\frac{i}{2}} }\times \texttt{``regular''}(u,v),
$$
where now \texttt{``regular''} has no branch cuts at $u=v\pm i/2$ (although it has other branch cuts at different  positions, which do not depend on the rapidity $v$).

Then, given that $$
\Sigma_{\text{new}}(u,v)
\equiv \dfrac{\sigma_{\text{new}}(u+i/2,v)}{\sigma_{\text{new}}(u-i/2,v)} ,$$
it follows that $\Sigma_{\text{new}}$ possesses square root branch points on the second sheet, at $u=v\pm i$ or $u = v$.

It is then easy to see that all the $v$-dependent singularities in the S-matrix elements of section \ref{sec:comparison} are compensated exactly by the ones in the dressing phases; furthermore, no new $v$-dependent singularity is produced by the dressing phases.


\bibliography{bibliography}

\end{document}